\documentclass[times,12pt,letterpaper,oneside,notitlepage,openright]{report}
\usepackage{graphicx, wrapfig, epsfig, vmargin, setspace, fancyhdr, amsfonts, amssymb, hyperref}

 \setpapersize{USletter}
\setmarginsrb{1.5in}{0.65in}{1in}{1.0in}{0pt}{0.05in}{0pt}{0mm}
\fancyhfoffset[]{0.25In}

\begin{document}

\pagenumbering{roman}
\pagestyle{empty}
\begin{center}
\ \\
\ \\
\ \\
TOPICS IN INFLATIONARY COSMOLOGY AND ASTROPHYSICS\\
\ \\
by\\
\ \\
Matthew M. Glenz\\
\ \\
A Dissertation Submitted in\\
Partial Fulfillment of the\\
Requirements for the Degree of\\
\ \\
Doctor of Philosophy\\
in Physics\\
\ \\
at\\
The University of Wisconsin-Milwaukee\\
December 2008\\
\end{center}

\newpage
\pagestyle{plain}
\begin{center}
\ \\
\ \\
\ \\
TOPICS IN INFLATIONARY COSMOLOGY AND ASTROPHYSICS\\
\ \\
by\\
Matthew M. Glenz\\
\ \\
A Dissertation Submitted in\\
Partial Fulfillment of the\\
Requirements for the Degree of\\
\ \\
Doctor of Philosophy\\
in Physics\\
\ \\
at\\
The University of Wisconsin-Milwaukee\\
December 2008\\
\ \\
\ \\
\hrule \vskip 0.2em Major Professor \hfill Date\ \ \ \ \ \ \ \ \ \\
\ \\
\ \\
\hrule \vskip 0.2em Graduate School Approval \hfill Date\ \ \ \ \ \
\ \ \
\end{center}

\newpage
\begin{center}
\ \\
\ \\
\ \\
ABSTRACT\\
TOPICS IN INFLATIONARY COSMOLOGY AND ASTROPHYSICS\\
by\\
Matthew M. Glenz\\
\ \\
{\onehalfspacing The University of Wisconsin-Milwaukee, 2008\\
\onehalfspacing Under the Supervision of Distinguished Professor
Leonard Parker\\}
\end{center}

\ \\

We introduce a general way of modeling inflation in a framework that
is independent of the exact nature of the inflationary potential.
Because of the choice of our initial conditions and the continuity
of the scale factor in its first two derivatives, we obtain
non-divergent results without the need of any renormalization beyond
what is required in Minkowski space.  In particular, we assume
asymptotically flat initial and final values of our scale factor
that lead to an unambiguous measure of the number of particles
created versus frequency.  We find exact solutions to the evolution
equation for inflaton perturbations when their effective mass is
zero and approximate solutions when their effective mass is
non-zero.  We obtain results for the scale invariance of the
inflaton spectrum and the size of density perturbations.  Finally,
we show that a substantial contribution to reheating occurs due to
gravitational particle production during the exit from the
inflationary stage of the universe.

The second part of this dissertation deals with a post-Minkowski
approximation to a binary point mass system with helical symmetry.
Numerical solutions for particles of unequal masses are examined in
detail for two types of Fokker actions, and these solutions are
compared with predictions from the full theory of General Relativity
and with post-Newtonian approximations.  Analytic solutions are
derived for the Extreme Mass Ratio case.

The third part of this dissertation discusses the detection
sensitivity of the IceCube Neutrino Telescope for observing
interactions involving TeV-scale black holes produced by an incoming
high-energy cosmic neutrino colliding with a parton in the Antarctic
ice of the south pole.  Parton Distribution Functions and the black
hole interaction cross section are computed numerically. Our
computation shows that IceCube could detect such black hole events
at the 5-sigma level for a ten-dimensional Planck mass of 1.3 TeV.
\ \\
\ \\
\ \\
\ \\
\ \\
\ \\
\begin{center}
\hrule \vskip 0.2em Major Professor \hfill Date\ \ \ \ \ \ \ \ \ \\
\end{center}

\newpage
\renewcommand\contentsname{\ \ \ \ \ \ \ \ \ \ Table of Contents}
\tableofcontents

\newpage
\listoffigures
\addcontentsline{toc}{chapter}{List of Figures}

\newpage
\listoftables
\addcontentsline{toc}{chapter}{List of Tables}

\newpage
\addcontentsline{toc}{chapter}{Acknowledgments}
\begin{center}
ACKNOWLEDGMENTS
\end{center}
I wish to thank my advisor, Distinguished Professor Leonard Parker,
for suggesting Part~I of this dissertation.  I appreciate his
patience, his trust, and his guidance.  Without his pioneering work
on gravitational particle production, this dissertation would not
have been possible.

I am also thankful for my other collaborators on Parts~II and~III,
K\={o}ji Ury\={u} and Luis Anchordoqui.  K\={o}ji graciously let me
contribute to his research, even though he could have calculated my
results faster by himself. I appreciate Luis's generosity and his
sincere desire to see me succeed in physics and life.

I am grateful for the support of the Lynde and Harry Bradley
Foundation, and for the support of the National Space Grant College
and Fellowship Program and the Wisconsin Space Grant Consortium.

My wife, Alyson, sacrificed her own scholarships so that I might
attend the University of Wisconsin|Milwaukee.  Thank you.

\newpage
\chapter{Introduction}
\pagenumbering{arabic}
\thispagestyle{fancy}
\pagestyle{fancy}

This dissertation is an exploration of space on scales that are
small (quantum fluctuations, TeV-scale black holes, vacuum particle
creation); scales that are big (anisotropies in the Cosmic Microwave
Background, seeding of large-scale structure, the Hubble radius);
and scales that are in between (Extreme Mass Ratio binary black
holes, temperatures associated with horizons, Innermost Stable
Circular Orbits). The first part of this dissertation is an
outgrowth of methods developed by my thesis advisor in the following
works \cite{Parker1,Parker2,Parker3,Parker4}.  These methods are
applicable to the creation of quantized perturbations of the
inflaton field, which is the topic we explore in Part~I.  The new
results that appear in this dissertation are based primarily on the
work of three papers. The first of these papers, ``Study of the
Spectrum of Inflaton Perturbations," examines an exact calculation
of the evolution of quantum fluctuations and the subsequent particle
creation in a model of the early expansion of the universe that is
relevant to a wide range of inflationary potentials consistent with
observations and that does not depend on renormalization in curved
spacetime~\cite{Glenz1}. The second of these papers, ``Circular
solution of two unequal mass particles in Post-Minkowski
approximation," computes numerically a set of solutions to a
helically symmetric binary system of point masses in a particular
approximation to General Relativity and presents analytical formulas
for the limit that the mass of the lighter particle is negligible
with respect to that of the more massive particle~\cite{Glenz2}. The
third of these papers, ``Black Holes at the IceCube neutrino
telescope," calculates the experimental sensitivity for observing
TeV-scale black holes produced by a gravitational interaction
between a cosmic neutrino and an elementary particle within the
atomic nuclei of ice molecules~\cite{Anchordoqui1}.  This
dissertation is divided into three main parts corresponding to these
three papers.

In Part I, ``New Aspects of Inflaton Fluctuations," we begin with a
brief summary of early universe cosmology.  Two of the most
important cosmological theories of the twentieth century are the Big
Bang theory and the theory of Inflation.  The Big Bang theory
supposes that our universe was once much smaller and much hotter
that it is today.  It explains the expansion of the universe, the
presence of the Cosmic Microwave Background Radiation, and the
primordial abundances of light elements.  Cosmological Inflation
supposes that the early universe underwent an extremely large
increase in size in a very small amount of time.  This explains why
the density of our universe today is so close to the critical
density that separates a universe that expands forever from one that
eventually recollapses, it explains the near homogeneity and
isotropy of the universe, and it explains why we don't observe
magnetic monopoles.  Most importantly of all, however, inflation
explains the origins of those anisotrophies that do exist in our
universe.  Although a key ingredient of the Big Bang theory is a
high energy density in the early universe and a correspondingly high
temperature, the classical theory of inflation predicts an extreme
cooling of the universe as it expands| much like the air in a piston
cools as it expands to do work on its surroundings.  We consider
Reheating, and specifically the energy density of particles created
by an expanding universe, as a means of preserving both theories
without sacrificing any of their successes. We give a general
overview of the amplification of quantum fluctuations into
large-scale density perturbations during inflation, and we describe
some of the ways of relating theoretical predictions to
observations.  We then list some of the observational findings of
experiments.

We continue with the details of the method we use to model
inflation.  Instead of specifying an inflationary potential, as is
usually done, we specify directly the change in the scale factor,
which is a measure of the size of the universe, versus time. We
consider a scale factor that accommodates several parameters, but
its most important features are that it asymptotically approaches a
constant values at early times, that it approaches a different
constant value at late times, and that its first two derivatives
with respect to time are continuous. The asymptotically flat regions
of our scale factor allow us to associate our model with Minkowski
spacetime at early and late times.  Identification with a Minkowski
vacuum at early times leads us to initial conditions that contain no
infrared divergences, and comparison with a Minkowski spacetime at
late times leads us to an unambiguous measure of the
frequency-dependent density of particles created by the expansion of
the universe. That our scale factor is continuous up to its second
derivative with respect to time ensures we have no ultraviolet
divergences, in addition to the prevention of infrared divergencies
mentioned before.  We choose for our scale factor a composite of
three segments.  The initial and final segments are each associated
with a particular form of asymptotically flat scale factor with
different choices of parameters.  The middle segment of the scale
factor, where most of the expansion takes place, is a region that
grows exponentially with respect to proper time.  Such an
exponential growth is indicated by experimental observations.  We
solve for the matching conditions necessary to maintain the desired
continuity of our composite scale factor. For each of our scale
factor segments we have exact solution to the evolution equation for
fluctuations of a massless, minimally-coupled scalar field. We also
describe two different approximations to the case of a constant
mass.  We match up our solutions to the evolution equation at the
interfaces between the segments of our composite scale factor, and
at late times we are able to determine the particle production due
to the expansion of the universe.  From here we discuss the
dispersion spectrum.  We note the scale-invariance of the scalar
index, provided the requirement is met that each mode be converted
into a curvature perturbation at a time related to when it crosses
the Hubble radius, and that all modes not be converted at once after
the end of inflation.  Using a hybrid combination of our method with
the slow roll approximation, we describe a way of calculating the
density perturbations produced by inflation. Finally, we show how
Reheating, or a return to the hot Big Bang conditions after the end
of inflation, can accompany inflation.  We discuss possible
consequences of Reheating and its relationship to constraints on
predictions for exotic particles and high energy physics.

In Part II, ``Binary System of Compact Masses," we examine a
post-Minkowski approximation to a helically symmetric binary system
of point masses.  The helical symmetry is maintained through the
presence of half-advanced and half-retarded fields.  The equations
of motion are given for one of two Fokker actions|
parametrization-invariant and affine| by Friedman and Ury\=u in
\cite{Friedman}, and from their results we calculate numerically the
solutions in the case of unequal masses. We also derive analytical
formulas for the Extreme Mass Ratio limit where the ratio of the
smaller mass divided by the larger mass goes to zero.  This limit
would be applicable to the inspiral of a solar-mass black hole into
a billion-solar-mass black hole, such as is predicted to exist at
the centers of many galaxies.  For both the numerical computations
and the analytic equations, we plot three graphs: the angular
momentum versus the velocity of the lighter particle, the unit
energy of the lighter particle versus the angular momentum, and the
unit angular momentum of the lighter particle versus the angular
momentum.  These plots are given for four mass ratios and for both
types of Fokker action.  For the parametrization-invariant case we
include one of two different correction terms that generates
solutions that agree with the first post-Newtonian approximation,
and we demonstrate this in the Extreme Mass Ratio limit.  We discuss
the locations of Innermost Stable Circular Orbits, and we compare
the predictions of this post-Minkowski approximation with both those
of the post-Newtonian approximation and those of the full theory of
General Relativity.

In Part III, ``Production and Decay of Small Black Holes at the
TeV-Scale," we investigate the possibility of using the IceCube
Neutrino Telescope to detect TeV-scale black holes.  In the physics
of the Standard Model, it is not impossible that a cosmic neutrino
could come close enough to an elementary particle in the cubic
kilometer of ice in the IceCube experiment to form a black hole.
Such interactions involving gravity, however, are so much less
likely than interactions involving the weak force, that IceCube
would never differentiate their signal from the background noise of
weak-interaction event rates.  Many theories of physics beyond the
Standard Model, such as string theory, require additional dimensions
of spacetime beyond the 3+1 dimensions of our common experience.
These additional dimensions might not have been noticed before if
they were compactified, or curled up, with a simple example being
the topology of a higher-dimensional torus.  At the compactification
scales, then, gravity would be much stronger than in a
3+1-dimensional theory, whereas at macroscopic scales gravity would
appear to be much weaker than the strong and electroweak forces. In
addition, if only gravitons propagated into the compactified
dimensions, then the scale of compactification could be anything
small enough not to conflict with observations.  On distances
smaller than this scale, gravity would grow stronger with decreasing
separation faster than an inverse-square law would predict.  If the
strength of gravity were equal to the strength of the
electromagnetic force around energies of roughly one TeV, or
$10^{-19}$ meters, the scale at which the electromagnetic and weak
forces unify into the electroweak force, then gravity could be
sufficiently strong that the IceCube detector could observe the
production of TeV-scale black holes in the interactions between
cosmic neutrinos and partons, which are the fundamental particles|
both quarks and gluons| that are found within nucleons in atoms. For
the high energies of interest for this experiment, the nucleons
cannot be treated as single particles, which is why we treat them as
collections of partons.  At any moment, a parton can have an energy
ranging from nothing to the entire rest mass energy of the nucleon,
and parton distribution functions describe the probabilities of
finding each parton with a given energy.  We develop simple fits to
a specific model of the parton distribution function, and with this
information we are able to numerically integrate an expression
giving us the cross section for the gravitational interaction.  The
black holes formed by these interactions would decay almost
immediately via Hawking radiation, or particles produced by the
strong curvature of spacetime outside of black holes.  The Cherenkov
light of these events could be measured by the photomultiplier tubes
of IceCube, and signals could be picked out from the background
event rate by searching for muon-daughter particles with less than
20\% of the total energy, which is sufficiently unlikely in Standard
Model physics that we would be able to discern TeV-scale black hole
events from interactions through the weak force.  We find that the
IceCube detector could measure TeV-scale black holes at a
statistically significant 5$\sigma$ excess for a 10-dimensional
Planck scale of 1.3 TeV.

The relationship between space at the smallest and largest scales
is, perhaps, nowhere so evident as the inflation of quantum
fluctuations from below the Planck length to sizes beyond our
observable universe in what follows: Part I - New Aspects of
Inflaton Fluctuations.

\

\newpage

\

\

\

\

\

\

\

\

\

\

\

\noindent\textbf{\Huge Part I:}

\

\noindent\textbf{\huge New Aspects of}
\addcontentsline{toc}{chapter}{Part I - New Aspects of Inflaton
Fluctuations}

\

\noindent\textbf{\huge Inflaton Fluctuations}

\newpage
\thispagestyle{fancy}
\chapter{Inflationary Cosmology}
\thispagestyle{fancy}
\pagestyle{fancy}

At the beginning of the twentieth century, most scientists believed
that the universe was infinite and eternal.  Such a situation is not
compatible with cosmology governed by the theory of General
Relativity, which predicts that a static universe would be unstable
to perturbations. From this it follows that our expanding universe
started from a singularity of infinite density and temperature. This
Big Bang theory of the universe successfully explains several
observational phenomena. One of these is the expansion of the
universe and Olber's paradox, which asks| if the universe is
infinite, then why do we not observe stars in every direction; why
do we see dark space between stars? With help from Hubble, Einstein
and others came to realize that the universe is not only expanding,
but it must also have a finite age. Thus, not all of the light from
stars in the universe has had time to reach us, and for distant
stars this light is redshifted by the expansion of the universe.
Another question resolved by the Big Bang theory is that of the
primordial abundances of the light elements: hydrogen, deuterium,
tritium, helium-3, helium-4, and lithium. Stars convert hydrogen to
heavier elements through nuclear fusion, but the light elements are
found in definite ratios in galactic dust thought never to have been
part of any star.  This is explained by looking back to the high
temperatures and pressures of the universe when it was much more
dense, shortly after the Big Bang.  The universe was hotter than any
star, and a series of calculations involving the thermal-equilibrium
ratio of protons to neutrons, the ratio of baryons to photons, the
half-life for a free neutron, and the cross section for neutrons to
become bound in nuclei~\cite{Kolb,Dodelson}; predicts ratios of
primordial abundances of the light elements that agree very well
with observations.  A final success of the Big Bang theory is the
explanation of the observed Cosmic Microwave Background Radiation
(CMBR) at a temperature of approximately 2.7 Kelvin.  This was first
discovered by Penzias and Wilson in 1965 while they were working at
Bell Labs, and for this discovery they were awarded a Nobel Prize in
1978. This background noise is the red-shifted relic of the early
universe's radiation dominance. Although the Big Bang theory
explained some questions about our universe, Cosmological Inflation
was necessary to explain other observed properties of our universe.

Inflation was originally conceived to explain three primary
phenomena.  The first of these was the flatness problem. The density
of our universe is surprisingly close to the critical density needed
to close the universe, above which a closed universe would
eventually re-collapse into a Big Crunch and below which an open
universe would expand forever| neglecting acceleration caused by the
presence of dark energy. Surprisingly close, because unless our
universe's density is precisely equal to the critical density| and
there is no reason to assume it must be| the ratio between the two
drifts rapidly away from 1 in a Big-Bang-only universe. Inflation
solves this problem by very rapidly driving this ratio exceedingly
close to 1 during a short period of enormous growth of the universe.
The second argument for inflation is that all the CMBR is, to
excellent approximation of within about one part in ten thousand, in
thermal equilibrium.  Just as the resolution to Olber's paradox
involves light taking a finite time to reach the Earth, so does this
present a problem for early-universe light, emanating from different
directions, that is just now reaching us. In a Big-Bang-only model,
widely separated regions of the currently observable universe
weren't previously in causal contact, and that they should be in
thermal equilibrium now is a mystery. This problem is resolved by
explaining how the space in minute regions of our universe that were
once in thermal contact expanded sufficiently rapidly during
inflation to remove the different parts of the equilibrated sections
to causally disconnected parts of the universe: the space between
points within equilibrated regions of the universe grew much faster
than signals could travel across the distance between those points.
Thus, the CMBR reaching the Earth today, even from different
directions, has come from regions of the universe that were
previously in thermal equilibrium.  The third issue that motivated
inflation is the observed absence of magnetic monopoles, which may
have been created in the very early universe. Inflation resolves
this by showing how monopoles could be inflated away with the
expansion of space such that| unless monopoles were produced after
inflation| on average there shouldn't be any monopole close enough
to us to detect after inflation.

Inflation has come up with an unforseen prediction that has since
turned out to be more important than any of the historical
justifications for its existence: the creation of fluctuations
during inflation that lead to the anisotropies of our present-day
universe.  For NASA-COBE's (Cosmic Background Explorer) 1989
detection of these anisotropies in the CMBR, Mather and Smoot were
awarded a Nobel Prize in 2006. In the most widely used models of
inflation, this expansion is driven by the inflaton field, which is
a scalar quantum field, and the perturbations of the inflaton field
seed galaxy formation and are responsible for large-scale structure
of our universe today.

\section{Cosmology in General Relativity}

In units of $c=\hbar=1$ Einstein's equation is \cite{Einstein,Wald1}
\begin{equation}
G_{ab}\equiv R_{ab}-\frac{1}{2}Rg_{ab}=8\pi G\,T_{ab}.
\label{eq:einstein}
\end{equation}
On large enough scales, our universe appears to be of a fairly
uniform density in all directions.  If the Earth is not in a
privileged position in the universe, this implies that the universe
is homogeneous and isotropic.  Following the example of
\cite{Wald1,Ohanian}, if we assume no distinction between the
spatial directions, we can write the
Friedmann-Robertson-Walker-Lema\^{i}tre (FRWL) metric as
\begin{equation}
ds^{2}=-dt^{2}+a(t)^{2}\left[\frac{dr^{2}}{1-kr^{2}}+r^{2}(d\theta^{2}+\sin^{2}\theta\
d\phi^{2})\right],
\label{eq:FRWL}
\end{equation}
where $a(t)$ is the scale factor that relates the chosen coordinate
scale to the proper time $t$, and the variable $k$ describes the
topology of the universe: $k>0$ corresponds to positive curvature
(closed universe), $k=0$ corresponds to zero intrinsic curvature
(flat universe), and $k<0$ corresponds to negative curvature
(hyperbolic, open universe). We then have
\begin{equation}
g_{ab}=\left[ \begin{array}{cccc} -1 & 0 & 0 & 0
\\ 0 & \frac{a(t)^{2}}{1-kr^{2}} & 0 & 0
\\ 0 & 0 & a(t)^{2}r^{2} & 0
\\ 0 & 0 & 0 & a(t)^{2}r^{2}\sin^{2}\theta \end{array} \right],
\end{equation}
\begin{equation}
g^{ab}=\left[ \begin{array}{cccc} -1 & 0 & 0 & 0
\\ 0 & \frac{1-kr^{2}}{a(t)^{2}} & 0 & 0
\\ 0 & 0 & a(t)^{-2}r^{-2} & 0
\\ 0 & 0 & 0 & a(t)^{-2}r^{-2}\sin^{-2}\theta \end{array} \right].
\end{equation}
In this section, only, we will not use the Einstein summation
convention. In the basis of $\{t,r,\theta,\phi\}$, the Christoffel
symbols are given by
\begin{equation}
\Gamma^{c}{}_{ab}=\sum_{d}\left[\frac{1}{2}g^{cd}\left(\partial_{a}g_{bd}+\partial_{b}g_{ad}-\partial_{d}g_{ab}\right)\right],
\end{equation}
where $\nabla_{a}V^{c}=\partial_{a}V^{c}+\Gamma^{c}{}_{ab}V^{b}$ and
$\nabla_{a}W_{c}=\partial_{a}W_{c}-\Gamma^{b}{}_{ac}W_{b}$, with
$\partial_{a}$ the covariant derivative operator of the flat metric
\cite{Hughston}. For the metric given by Eq.~(\ref{eq:FRWL}), we see
that $g^{cd}=\delta_{c}^{d}g^{cc}$ and
$g_{cd}=\delta_{c}^{d}g_{cc}$, where $\delta^{c}_{d}$ is the
Kronecker delta, so we have
\begin{equation}
\Gamma^{c}{}_{ab}=\frac{1}{2}g^{cc}\left(\delta^{b}_{c}\partial_{a}g_{cc}+\delta^{a}_{c}\partial_{b}g_{cc}-\delta^{a}_{b}\partial_{c}g_{aa}\right).
\end{equation}
In the set of coordinates defined by $\{t,r,\theta,\phi\}$, we
consider the four cases of $a=b=c$, $a=b\ne c$, $a\ne b=c$, and
$a\ne b\ne c$ (each of the indices is different in this last case)
to get
\begin{eqnarray}
a=b=c&:&\Gamma^{c}{}_{cc}=\frac{1}{2}g^{cc}\partial_{c}g_{cc},\\
a=b\ne c&:&\Gamma^{c}{}_{aa}=-\frac{1}{2}g^{cc}\partial_{c}g_{aa},\\
a\ne b=c&:&\Gamma^{c}{}_{ca}=\frac{1}{2}g^{cc}\partial_{a}g_{cc},\\
a\ne b\ne c&:&\Gamma^{c}{}_{ab}=0.
\end{eqnarray}
The non-zero derivatives are $\partial_{t}g_{rr}$,
$\partial_{t}g_{\theta\theta}$, $\partial_{t}g_{\phi\phi}$,
$\partial_{r}g_{rr}$, $\partial_{r}g_{\theta\theta}$,
$\partial_{r}g_{\phi\phi}$, and $\partial_{\theta}g_{\phi\phi}$.
Thus, the non-vanishing Christoffel symbols are $\Gamma^{t}{}_{rr}$,
$\Gamma^{t}{}_{\theta\theta}$, $\Gamma^{t}{}_{\phi\phi}$,
$\Gamma^{r}{}_{rr}$, $\Gamma^{r}{}_{\theta\theta}$,
$\Gamma^{r}{}_{\phi\phi}$, $\Gamma^{r}{}_{rt}=\Gamma^{r}{}_{tr}$,
$\Gamma^{\theta}{}_{\phi\phi}$, $\Gamma^{\theta}{}_{\theta
t}=\Gamma^{\theta}{}_{t\theta}$, $\Gamma^{\theta}{}_{\theta
r}=\Gamma^{\theta}{}_{r\theta}$, $\Gamma^{\phi}{}_{\phi
t}=\Gamma^{\phi}{}_{t\phi}$, $\Gamma^{\phi}{}_{\phi
r}=\Gamma^{\phi}{}_{r\phi}$, and $\Gamma^{\phi}{}_{\phi
\theta}=\Gamma^{\phi}{}_{\theta\phi}$.

When we write the Ricci tensor as \cite{Wald1}
\begin{equation}
R_{ab}=\sum_{c}\left(\partial_{c}\Gamma^{c}{}_{ab}-\partial_{a}\Gamma^{c}{}_{cb}\right)+\sum_{c,d}\left(\Gamma^{d}{}_{ab}\Gamma^{c}{}_{cd}-\Gamma^{d}{}_{cb}\Gamma^{c}{}_{da}\right),
\end{equation}
we find, using an underline to indicate terms that cancel, using an
overline to indicate terms to be consolidated, and using $a=a(t)$,
$\dot{a}=da/dt$, and $\ddot{a}=d\dot{a}/dt$, that
\begin{eqnarray}
R_{tt}&=&-\partial_{t}\left(\Gamma^{r}{}_{rt}+\Gamma^{\theta}{}_{\theta
t}+\Gamma^{\phi}{}_{\phi
t}\right)-\left(\Gamma^{r}{}_{rt}\Gamma^{r}{}_{rt}+\Gamma^{\theta}{}_{\theta
t}\Gamma^{\theta}{}_{\theta t}+\Gamma^{\phi}{}_{\phi
t}\Gamma^{\phi}{}_{\phi t}\right)\nonumber\\
&=&-\left[\left(\frac{\ddot{a}}{a}-\frac{\dot{a}^{2}}{a^{2}}\right)+\left(\frac{\ddot{a}}{a}-\frac{\dot{a}^{2}}{a^{2}}\right)+\left(\frac{\ddot{a}}{a}-\frac{\dot{a}^{2}}{a^{2}}\right)\right]-\left[\left(\frac{\dot{a}}{a}\right)^{2}+\left(\frac{\dot{a}}{a}\right)^{2}+\left(\frac{\dot{a}}{a}\right)^{2}\right]\nonumber\\
&=&-3\frac{\ddot{a}}{a},
\end{eqnarray}
\begin{eqnarray}
R_{rr}&=&\left(\partial_{t}\Gamma^{t}{}_{rr}+\underline{\partial_{r}\Gamma^{r}{}_{rr}}\right)-\left(\underline{\partial_{r}\Gamma^{r}{}_{rr}}+\partial_{r}\Gamma^{\theta}{}_{\theta
r}+\partial_{r}\Gamma^{\phi}{}_{\phi
r}\right)\nonumber\\
&&+\bigg[\Gamma^{t}{}_{rr}\left(\overline{\Gamma^{r}{}_{rt}}+\Gamma^{\theta}{}_{\theta
t}+\Gamma^{\phi}{}_{\phi
t}\right)+\Gamma^{r}{}_{rr}\left(\underline{\Gamma^{r}{}_{rr}}+\Gamma^{\theta}{}_{\theta
r}+\Gamma^{\phi}{}_{\phi r}\right)\nonumber\\
&&-\left(\underline{\Gamma^{r}{}_{rr}\Gamma^{r}{}_{rr}}+\Gamma^{\theta}{}_{\theta
r}\Gamma^{\theta}{}_{\theta r}+\Gamma^{\phi}{}_{\phi
r}\Gamma^{\phi}{}_{\phi r}+2\overline{\Gamma^{r}{}_{tr}\Gamma^{t}{}_{rr}}\right)\bigg]\nonumber\\
&=&\left(\frac{a\ddot{a}+\dot{a}^{2}}{1-kr^{2}}\right)+\left(\frac{1}{r^{2}}+\frac{1}{r^{2}}\right)+\bigg[\frac{a\dot{a}}{1-kr^{2}}\left(-\frac{\dot{a}}{a}+\frac{\dot{a}}{a}+\frac{\dot{a}}{a}\right)\nonumber\\
&&+\frac{kr}{1-kr^{2}}\left(\frac{1}{r}+\frac{1}{r}\right)-\left(\frac{1}{r^{2}}+\frac{1}{r^{2}}\right)\bigg]\nonumber\\
&=&\frac{a^{2}}{1-kr^{2}}\left(\frac{\ddot{a}}{a}+2\frac{\dot{a}^{2}}{a^{2}}+2\frac{k}{a^{2}}\right),
\end{eqnarray}
\begin{eqnarray}
R_{\theta\theta}&=&\left(\partial_{t}\Gamma^{t}{}_{\theta\theta}+\partial_{r}\Gamma^{r}{}_{\theta\theta}
-\partial_{\theta}\Gamma^{\phi}{}_{\phi\theta}\right)+\bigg[\Gamma^{t}{}_{\theta\theta}\left(\Gamma^{r}{}_{rt}+\overline{\Gamma^{\theta}{}_{\theta
t}}+\Gamma^{\phi}{}_{\phi t}\right)\nonumber\\
&&+\Gamma^{r}{}_{\theta\theta}\left(\Gamma^{r}{}_{rr}+\overline{\Gamma^{\theta}{}_{\theta
r}}+\Gamma^{\phi}{}_{\phi
r}\right)-\left(\Gamma^{\phi}{}_{\phi\theta}\Gamma^{\phi}{}_{\phi\theta}+2\overline{\Gamma^{t}{}_{\theta\theta}\Gamma^{\theta}{}_{t\theta}}+2\overline{\Gamma^{r}{}_{\theta\theta}\Gamma^{\theta}{}_{r\theta}}\right)\bigg]\nonumber\\
&=&\left(r^{2}\left\{a\ddot{a}+\dot{a}^{2}\right\}-\left\{1-3kr^{2}\right\}+\left\{1+\frac{\cos^{2}\theta}{\sin^{2}\theta}\right\}\right)\nonumber\\
&&+\bigg[r^{2}a\dot{a}\left(\frac{\dot{a}}{a}-\frac{\dot{a}}{a}+\frac{\dot{a}}{a}\right)-r\left\{1-kr^{2}\right\}\left(\frac{kr}{1-kr^{2}}-\frac{1}{r}+\frac{1}{r}\right)-\frac{\cos^{2}\theta}{\sin^{2}\theta}\bigg]\nonumber\\
&=&a^{2}r^{2}\left(\frac{\ddot{a}}{a}+2\frac{\dot{a}^{2}}{a^{2}}+2\frac{k}{a^{2}}\right).
\end{eqnarray}
\begin{eqnarray}
R_{\phi\phi}&=&\left(\partial_{t}\Gamma^{t}{}_{\phi\phi}+\partial_{r}\Gamma^{r}{}_{\phi\phi}+\partial_{\theta}\Gamma^{\theta}{}_{\phi\phi}\right)+\bigg[\Gamma^{t}{}_{\phi\phi}\left(\Gamma^{r}{}_{rt}+\Gamma^{\theta}{}_{\theta
t}+\overline{\Gamma^{\phi}{}_{\phi
t}}\right)+\left(\overline{\Gamma^{\theta}{}_{\phi\phi}\Gamma^{\phi}{}_{\phi\theta}}\right)\nonumber\\
&&+\Gamma^{r}{}_{\phi\phi}\left(\Gamma^{r}{}_{rr}+\Gamma^{\theta}{}_{\theta
r}+\overline{\Gamma^{\phi}{}_{\phi r}}
\right)-\left(2\overline{\Gamma^{t}{}_{\phi\phi}\Gamma^{\phi}{}_{t\phi}}+2\overline{\Gamma^{r}{}_{\phi\phi}\Gamma^{\phi}{}_{r\phi}}+2\overline{\Gamma^{\theta}{}_{\phi\phi}\Gamma^{\phi}{}_{\theta\phi}}\right)\bigg]\nonumber\\
&=&\left(\left\{a\ddot{a}+\dot{a}^{2}\right\}r^{2}\sin^{2}\theta-\left\{1-3kr^{2}\right\}\sin^{2}\theta+\left\{\sin^{2}\theta-\cos^{2}\theta\right\}\right)\nonumber\\
&&+\bigg[a\dot{a}r^{2}\sin^{2}\theta\left(\frac{\dot{a}}{a}+\frac{\dot{a}}{a}-\frac{\dot{a}}{a}\right)-\left(-\cos^{2}\theta\right)\nonumber\\
&&-\left\{r-kr^{3}\right\}\sin^{2}\theta\left(\frac{kr}{1-kr^{2}}+\frac{1}{r}-\frac{1}{r}\right)\nonumber\\
&=&a^{2}r^{2}\sin^{2}\theta\left(\frac{\ddot{a}}{a}+2\frac{\dot{a}^{2}}{a^{2}}+2\frac{k}{a^{2}}\right).
\end{eqnarray}

The Ricci Scalar Curvature is
\begin{eqnarray}
R&\equiv&\sum_{ab}g^{ab}R_{ab}\nonumber\\
&=&g^{tt}R_{tt}+g^{rr}R_{rr}+g^{\theta\theta}R_{\theta\theta}+g^{\phi\phi}R_{\phi\phi}\nonumber\\
&=&6\left(\frac{\ddot{a}}{a}+\frac{\dot{a}^{2}}{a^{2}}+\frac{k}{a^{2}}\right).
\label{eq:ricciscalar}
\end{eqnarray}
The most general stress tensor associated with homogeneity and
isotropy is that of a perfect fluid \cite{Wald1}, given by
\begin{equation}
T_{ab}=\rho U_{a}U_{b}+P\left(g_{ab}+U_{a}U_{b}\right),
\end{equation}
where $\rho$ is the energy-density, $P$ is the pressure, and in
these coordinates $U^{a}=(-1,0,0,0)$ is the four-velocity of a
comoving observer, and
\begin{equation}
U_{b}=\sum_{a} g_{ab}U^{a}.
\end{equation} The time-time components
of the Einstein Equation, Eq.~(\ref{eq:einstein}), give us the
Friedmann equation:
\begin{equation}
G_{tt}=-3\frac{\ddot{a}}{a}-\frac{1}{2}\left[6\left(\frac{\ddot{a}}{a}+\frac{\dot{a}^{2}}{a^{2}}+\frac{k}{a^{2}}\right)\right]\left(-1\right)=3\frac{\dot{a}}{a}+3\frac{k}{a^{2}}=8\pi
G\,\rho,
\end{equation}
or,
\begin{equation}
H(t)^{2}=\frac{8\pi G}{3}\rho-\frac{k}{a^{2}},
\label{eq:friedmann}
\end{equation}
where the Hubble constant is defined by
\begin{equation}
H(t)\equiv\frac{d\,a(t)/d\, t}{a(t)}.
\end{equation}
Any same space-space components of the Einstein equation, for which
we will use $r$-$r$, give us the Raychaudhuri equation:
\begin{equation}
G_{rr}=g_{rr}\left(\frac{\ddot{a}}{a}+2\frac{\dot{a}^{2}}{a^{2}}+2\frac{k}{a^{2}}\right)-\frac{1}{2}\left[6\left(\frac{\ddot{a}}{a}+\frac{\dot{a}^{2}}{a^{2}}+\frac{k}{a^{2}}\right)\right]g_{rr}=8\pi
G\,P\,g_{rr},
\end{equation}
or,
\begin{equation}
2\frac{\ddot{a}}{a}+\frac{\dot{a}^{2}}{a^{2}}+\frac{k}{a^{2}}=-8\pi
G\, P,
\end{equation}
which, when we use $H=H(t)$ and
$\dot{H}=d\,H/d\,t=a^{-1}\ddot{a}-a^{-2}\dot{a}^{2}$, can be written
\begin{equation}
2\dot{H}+3H^{2}+\frac{k}{a^{2}}=-8\pi G\,P,
\end{equation}
which we rewrite, using Eq.~(\ref{eq:friedmann}), as either
\begin{equation}
\dot{H}=-4\pi G(\rho+P)+\frac{k}{a^{2}},
\label{eq:preraychaudhuri}
\end{equation}
or as the Raychaudhuri equation, which is
\begin{equation}
\dot{H}+H^{2}=-\frac{4\pi G}{3}(\rho+3P).
\end{equation}
We get the continuity equation by taking the time derivative of
Eq.~(\ref{eq:friedmann}) and then inserting
Eq.~(\ref{eq:preraychaudhuri}) to find
\begin{equation}
\frac{8\pi G}{3}\dot{\rho}=2H\dot{H}=2H\left[-4\pi
G(\rho+P)+\frac{k}{a^{2}}\right],
\end{equation}
which becomes
\begin{equation}
\dot{\rho}=-3H(\rho+P)+\frac{3H}{8\pi G}\frac{k}{a^{2}}.
\end{equation}
In a flat universe, where $k/a^{2}$ can be neglected and the metric
can be written as $ds^{2}=-dt^{2}+a(t)^{2}(dx^{2}+dy^{2}+dz^{2})$,
the continuity equation becomes
\begin{equation}
\dot{\rho}=-3H(\rho+P).
\end{equation}
A simpler way of deriving this equation would be to use conservation
of energy in a comoving reference frame to show, in units where
$E=mc^{2}=m$, that
\begin{equation}
d\left(\frac{E}{V}\right)=-\frac{M}{V}dV-\frac{P}{V}dV,
\end{equation}
where $M=\rho V$ and $V\propto a^{3}$.  If there were no pressure,
as is the case for what is referred to as dust, then in the
coordinates $\{t,x,y,z\}$ this would reduce to conservation of a
density current:
\begin{eqnarray}
0&=&\sum_{a}\left[\nabla_{a}\left(\rho U^{a}\right)\right]\nonumber\\
&=&\sum_{a}\left[U^{a}\partial_{a}\rho+\rho\nabla_{a}U^{a}\right]\nonumber\\
&=&U^{t}\partial_{t}\rho+\rho\left(U^{t}\Gamma^{x}{}_{xt}+U^{t}\Gamma^{z}{}_{zt}+U^{t}\Gamma^{z}{}_{zt}\right)\nonumber\\
&=&-\partial_{t}\rho-3H\rho.
\end{eqnarray}
For dust, which is the term for matter that satisfies $P=0$, such as
cold dark matter and| to good approximation| galaxies, we can solve
the differential equation
\begin{equation}
\frac{\dot{\rho}}{\rho}=-3\frac{\dot{a}}{a},
\end{equation}
by integrating both sides with respect to time to get
\begin{equation}
\ln\rho\propto-3\ln a,
\end{equation}
or
\begin{equation}
\rho\propto a^{-3}.
\end{equation}
We combine this with Eq.~(\ref{eq:friedmann}) to get
\begin{equation}
\frac{\dot{a}^{2}}{a^{2}}\propto a^{-3},
\end{equation}
which leads to
\begin{equation}
\dot{a}\propto a^{-1/2},
\end{equation}
and (with $k=0$)
\begin{equation}
a_{\rm dust}(t)\propto t^{2/3}.
\end{equation}
We refer to this as a matter-dominated universe.  For the case of a
radiation-dominated universe, where radiation obeys the equation of
state
\begin{equation}
P=\frac{1}{3}\rho,
\end{equation}
we would have
\begin{equation}
\rho\propto a^{-4},
\end{equation}
\begin{equation}
\dot{a}\propto a^{-1},
\end{equation}
and (with $k=0$)
\begin{equation}
a_{\rm radiation}(t)\propto t^{1/2}.
\end{equation}
In the next section we will show that a slowly-changing scalar field
displaced from its minimum potential energy obeys the equation of
state
\begin{equation}
P\simeq-\rho,
\end{equation}
for which we have from Eq.~(\ref{eq:preraychaudhuri})
\begin{equation}
\dot{H}_{\rm inflation}\simeq0.
\end{equation}
We discuss inflation in more detail in the next section, but first
we mention that with a time-invariant Hubble constant, we would have
(in a flat universe) a de Sitter metric given by
\begin{equation}
ds^{2}=-dt^{2}+e^{2Ht}(dx^{2}+dy^{2}+dz^{2}).
\label{eq:exponential}
\end{equation}

Whether $k=0$ in Eq.~(\ref{eq:friedmann}), or not, we may define a
critical density that would produce an equivalent Hubble constant if
$k$ were 0. This we define as
\begin{equation}
\rho_{c}=\frac{3H^{2}}{8\pi G}.
\end{equation}
We define the density parameter as
\begin{equation}
\Omega\equiv\frac{\rho}{\rho_{c}}=\frac{\frac{3}{8\pi
G}\left(H^{2}+\frac{k}{a^{2}}\right)}{\frac{3H^{2}}{8\pi
G}}=1+\frac{k}{a(t)^{2}H(t)^{2}},
\end{equation}
where a in a flat universe ($k=0$), we would have $\Omega=1$.  One
of the primary motivations for inflation was reconciling
observations that in our universe $\Omega\simeq1$, when there was no
reason to expect that it necessarily would be.  In fact, in either a
radiation- or matter-dominated universe (for both $H\propto t^{-1}$
when $k\simeq0$), we should expect
\begin{eqnarray}
\Omega_{\rm rad}=1+\frac{k}{a(t)^{2}H(t)^{2}}=1+\tilde{k}\,t,\\
\Omega_{\rm mat}=1+\frac{k}{a(t)^{2}H(t)^{2}}=1+\tilde{k}\,t^{2/3},
\end{eqnarray}
where $\tilde{k}\propto k$. The Big Bang theory predicts| based on
the presence of the approximately $2.7$K CMBR and the relationship
between the current matter density and Hubble constant| that our
universe was radiation-dominated until it was about 300,000 years
old and has been roughly matter-dominated (neglecting any recent
acceleration of the universe due to dark energy) since then.  Thus,
$\Omega$ in our universe should diverge rapidly from 1, unless the
value of $k$ was very nearly zero at early times in our universe.
One mechanism for driving $\Omega$ close to 1 is inflation.  When
$a(t)=e^{Ht}$ and $H={\rm constant}$, we have
\begin{equation}
\Omega_{\rm infl}=1+\frac{k}{a(t)^{2}H(t)^{2}}=1+kH^{-2}e^{-2Ht}.
\end{equation}
Inflation very rapidly drives the value of $\Omega$ towards 1.  With
enough inflation, an initial value of $\Omega$ that may have
differed from 1 by orders of magnitude, could have been driven close
enough to 1 that it would still be approximately equal to 1 in our
universe today. For the rest of this dissertation we will assume
that the universe is flat, in the sense that we will take the
curvature constant $k$ to be zero.  From now on we will not make use
of this variable and will reserve $k$ for other quantities, namely
the Fourier mode-number.

\section{Inflation}

For the rest of this dissertation, we will adopt the Einstein
summation convention.  The Lagrangian density of a scalar field with
metric signature of +2 is \cite{Birrell1,Parker5,Misner}
\begin{equation}
\mathcal{L}=\frac{1}{2}\left|g\right|^{1/2}(-g^{ab}\partial_{a}\phi\partial_{b}\phi-m^{2}\phi^{2}-\xi
R\phi^{2}), \label{eq:lagrangian1}
\end{equation}
where $g\equiv{\rm det}(g_{ab})$.  A massless ($m=0$), uncoupled
($\xi=0$) field with a $\phi$-dependent potential, where the
potential may incorporate a non-zero scalar field mass, becomes
\begin{equation}
\mathcal{L}=-\frac{1}{2}\left|g\right|^{1/2}g^{ab}\partial_{a}\phi\partial_{b}\phi-\left|g\right|^{1/2}V(\phi).
\label{eq:lagrangian2}
\end{equation}
The origin of this potential depends on the various models being
considered, but the main prerequisites are that $\phi$ initially be
displaced from the true minimum of the potential, and that some
portion of the slope of the potential must be relatively flat with
respect to changes in $\phi$ during the slow roll approximation, for
which see Sec.~\ref{sec:relationtoobservations}. If we were to
retain the Ricci curvature scalar in Eq.~(\ref{eq:lagrangian1}),
then the variation of the action would lead to the Einstein
Eq.~(\ref{eq:einstein}) in the calculation below
\cite{Parker5}\cite[pp. 491-505]{Misner}. The action is
\cite{Birrell1}
\begin{equation}
\mathcal{S}=\int d^{4}x'\ \mathcal{L}=\int d^{4}x'\
\left[\frac{1}{2}\left|g\right|^{1/2}(-g^{a'b'}\partial_{a'}\phi\partial_{b'}\phi-2V)\right],
\end{equation}
and the stress-energy tensor is \cite{Birrell1}
\begin{equation}
T_{ab}=\frac{2}{\left|g\right|^{1/2}}\frac{\delta
\mathcal{S}}{\delta g^{ab}}.
\end{equation}
Using the identities \cite{Parker5}
\begin{equation}
\delta g^{ab} = -g^{a c}g^{b d}\delta g_{cd},
\end{equation}
\begin{equation}
\delta\left|g\right|^{1/2}=\frac{1}{2}\left|g\right|^{1/2}g^{ab}\delta
g_{ab},
\end{equation}
leads to
\begin{eqnarray}
T_{ab}&=&\frac{2}{\left|g(x)\right|^{1/2}}\frac{\delta \int d^{4}x'\
[\frac{1}{2}\left|g(x')\right|^{1/2}(-g^{a'b'}(x')\partial_{a'}\phi\partial_{b'}\phi-2V)]}{\delta
g^{ab}(x)}\nonumber\\[2mm]
&=&\frac{\delta \int
d^{4}x'[\left|g(x')\right|^{1/2}(-g^{a'b'}(x')\partial_{a'}\phi\partial_{b'}\phi-2V)]}{\left|g(x)\right|^{1/2}\delta
g^{ab}(x)}\nonumber\\[2mm]
&=&\int
d^{4}x'\frac{\left|g(x')\right|^{1/2}}{\left|g(x)\right|^{1/2}}\frac{\delta
g^{a'b'}(x')}{\delta
g^{ab}(x)}\left[g_{a'b'}(x')(-\frac{1}{2}\partial^{c}\phi\partial_{c}\phi-V)+\partial_{a'}\phi\partial_{b'}\phi\right].\nonumber\\
\
\end{eqnarray}
Finally, using the delta function identity \cite{Parker5}
\begin{equation}
\frac{\delta g^{a'b'}(x')}{\delta g^{ab}(x)}=g_{a}^{\ a'}g_{b}^{\
b'}\delta^{4}(x',x),
\end{equation}
the stress tensor is
\begin{equation}
T_{ab}=g_{ab}\left(-\frac{1}{2}\partial^{c}\phi\partial_{c}\phi-V\right)+\partial_{a}\phi\partial_{b}\phi,
\end{equation}
and
\begin{equation}
T^{a}_{\ \ b}=g^{a}_{\ \
b}\left(-\frac{1}{2}\partial^{c}\phi\partial_{c}\phi-V\right)+\partial^{a}\phi\partial_{b}\phi.
\end{equation}
The spatial slicing and coordinate threading of time is chosen such
that \begin{math}\phi=\phi(t)\end{math}.  In absence of
perturbations, space-time is homogeneous and isotropic:
\begin{equation}
T^{a}_{\ \ b}=g^{a}_{\ \
b}\left(\frac{1}{2}\dot{\phi}^{2}-V\right)-\delta^{a}_{0}\
\delta^{0}_{b}\ \dot{\phi}^{2},
\end{equation}
where a dot represents derivatives with respect to time. Because of
homogeneity and isotropy, the stress tensor is described by a
perfect fluid,
\begin{equation}
T^{a}{}_{b}=\left[ \begin{array}{cccc} -\rho & 0 & 0 & 0
\\ 0 & P & 0 & 0
\\ 0 & 0 & P & 0
\\ 0 & 0 & 0 & P \end{array} \right],
\end{equation}
where $\rho$ is the energy density and $P$ is the pressure. It is
now possible to solve for the energy density and pressure: the
energy density is equal to minus the time-time component of the
stress tensor; and the pressure is equal to any of the three
diagonal space-space components of the stress tensor
\cite{Dodelson}.
\begin{equation}
\rho=-T^{0}_{\ \
0}=-\left[\left(\frac{1}{2}\dot{\phi}^{2}-V\right)-\dot{\phi}^{2}\right]=\frac{1}{2}\dot{\phi}^{2}+V(\phi),
\end{equation}
\begin{equation}
P=T^{1}_{\ \ 1}=T^{2}_{\ \ 2}=T^{3}_{\ \
3}=\frac{1}{2}\dot{\phi}^{2}-V(\phi).
\end{equation}
The Friedmann equation,
\begin{equation}
H^{2}=\frac{8\pi G}{3}\rho,
\end{equation}
and the continuity equation,
\begin{equation}
\dot{\rho}=-3H(\rho+P),
\end{equation}
become
\begin{equation}
H^2=\frac{8\pi G}{3}\left(\frac{1}{2}\dot{\phi}^{2}+V(\phi)\right),
\label{eq:inflationeq1}
\end{equation}
and
\begin{equation}
\dot{\phi}\ddot{\phi}+\dot{V}(\phi)=-3H\dot{\phi}^{2},
\end{equation}
\begin{equation}
\ddot{\phi}+\frac{dV/dt}{d\phi/dt}=-3H\dot{\phi},
\end{equation}
\begin{equation}
\ddot{\phi}+V'=-3H\dot{\phi},
\label{eq:inflationeq2}
\end{equation}
where a dot represents a derivative with respect to time and a prime
represents a derivative with respect to $\phi$.  The curvature term
in the Friedmann equation is here set to zero. Whether or not this
is precisely the case, soon after inflation begins the curvature of
the universe will become negligible.

\section{Quantum Fluctuations of a Scalar Field}

Well after inflation has begun, the scalar field can be treated as a
homogeneous, isotropic classical field with the fluctuations
consisting of quantum perturbations. Inflation smooths out all other
perturbations to the point that quantum fluctuations are all that
remain. For models of inflation driven by a single scalar field,
perturbations can be expressed as time-dependent, location-dependent
fluctuations on a homogeneous, time-dependent background:
\begin{equation}
\phi(\vec{x},t)=\phi(t)+\delta\phi(\vec{x},t).
\label{eq:inflperturb}
\end{equation}
The Euler-Lagrange equation,
\begin{equation}
\partial_{\phi}\mathcal{L}-\partial_{a}\left[\frac{\partial\mathcal{L}}{\partial(\partial_{a}\phi)}\right]=0,
\end{equation}
with Eqs.~(\ref{eq:lagrangian1}) and~(\ref{eq:lagrangian2}), becomes
\begin{equation}
-\sqrt{-g}\,V'(\phi)+\frac{1}{2}\partial_{a}\left(\sqrt{-g}\,g^{ab}\partial_{b}\phi\right)+\frac{1}{2}\partial_{b}\left(\sqrt{-g}\,g^{ab}\partial_{a}\phi\right)=0,
\end{equation}
or
\begin{equation}
\frac{1}{\sqrt{-g}}\partial_{a}\left(\sqrt{-g}\,g^{ab}\partial_{b}\phi\right)-V'(\phi)=0,
\end{equation}
which is equivalent to \cite[p. 38]{Birrell1}\cite[p. 542]{Misner}
\begin{equation}
\Box\phi-V'(\phi)=0.
\label{eq:inflzeroperturb}
\end{equation}
If we perturb this with Eq.~(\ref{eq:inflperturb}), then we get
\begin{equation}
\Box(\phi+\delta\phi)-V'(\phi+\delta\phi)=0.
\end{equation}
To first order in $\delta\phi$, we write this as
\begin{equation}
\Box\phi+\Box\delta\phi-\left[V'(\phi)+\delta\phi
V''(\phi)\right]=0,
\end{equation}
and we then use Eq.~(\ref{eq:inflzeroperturb}) to show
\begin{equation}
\Box\delta\phi-\delta\phi V''(\phi)=0.
\end{equation}
We can see from Eq.~{\ref{eq:lagrangian1}) that for a free field we
may make the association
\begin{equation}
V''(\phi)=m^{2}+\xi R,
\label{eq:vprimeprime}
\end{equation}
where $m$ is the scalar mass, $\xi$ is the coupling constant, and
$R$ is the Ricci scalar curvature.

The perturbation of Eq.~(\ref{eq:inflperturb}) expanded in terms of
creation and annihilation operators is \cite{Parker5}
\begin{equation}
\delta\phi=({\rm
volume})^{-1/2}\sum_{\vec{k}}[a_{\vec{k}}g_{k}(t)e^{i\vec{k}\cdot\vec{x}}+H.C.].
\label{eq:Fourier}
\end{equation}
where
\begin{equation}
{\rm volume}=[L\,a(t)]^{3},
\end{equation}
which is the physical length found from multiplying the coordinate
length times the scale factor.  The time dependent part of the
fluctuations is $\psi_{k}$, where
\begin{equation}
\psi_{k}\equiv a(t)^{-\frac{3}{2}}g_{k},
\end{equation}
and
\begin{equation}
\left|\delta\phi_{k}\right|^{2}=L^{-3}\left|\delta\psi_{k}\right|^{2}.
\end{equation}
The solution thus far has periodic boundary conditions, but in the
limit that L
\begin{math}\rightarrow\infty\end{math}, a volume even as large as
the observable universe will not be affected by this choice of
boundary conditions.  Combining the metric
\begin{equation}
ds^{2}=-dt^{2}+a(t)^{2}(dx^{2}+dy^{2}+dz^{2});
\end{equation}
where
\begin{equation}
g_{ab}=\left[ \begin{array}{cccc} -1 & 0 & 0 & 0
\\ 0 & a(t)^{2} & 0 & 0
\\ 0 & 0 & a(t)^{2} & 0
\\ 0 & 0 & 0 & a(t)^{2} \end{array} \right],
\end{equation}
\begin{equation}
g^{ab}=\left[ \begin{array}{cccc} -1 & 0 & 0 & 0
\\ 0 & a(t)^{-2} & 0 & 0
\\ 0 & 0 & a(t)^{-2} & 0
\\ 0 & 0 & 0 & a(t)^{-2} \end{array} \right],
\end{equation}
and
\begin{equation}
\sqrt{\left|g\right|}=\sqrt{\left|[-1][a(t)^{2}][a(t)^{2}][a(t)^{2}]\right|}=a(t)^{3},
\end{equation}
with the massless, uncoupled scalar field equation \cite{Birrell1}
\begin{equation}
\Box\delta\phi-\delta\phi
V''(\phi)=\frac{1}{|g|^{1/2}}\partial_{a}(|g|^{1/2}g^{ab}\partial_{b}\delta\phi)-\delta\phi
V''(\phi)=0, \label{eq:kleingordon}
\end{equation}
yields
\begin{eqnarray}
0&=&a(t)^{-3}\partial_{t}\left[a(t)^{3}(-1)\partial_{t}\delta\phi\right[+a(t)^{-3}\partial_{i}\left[a(t)^{3}\left(a(t)^{-2}\right)\partial^{i}\delta\phi\right]-\delta\phi
V''(\phi)\nonumber\\
&=&\partial^{2}_{t}\delta\phi+3H(t)\partial_{t}\delta\phi-a(t)^{-2}\partial_{i}\partial^{i}\delta\phi+\delta\phi
V''(\phi),
\end{eqnarray}
where
\begin{equation}
H\equiv\frac{da/dt}{a}.
\end{equation}
With the spatial dependence given by Eq.~(\ref{eq:Fourier}), the
evolution equation for mode-$k$ becomes
\begin{equation}
\partial^{2}_{t}\delta\phi+3H(t)\partial_{t}\delta\phi+\frac{k^{2}}{a(t)^{2}}\delta\phi+\delta\phi
V''(\phi)=0.
\label{eq:evolutionequation}
\end{equation}

Using the scale factor associated with the de Sitter universe given
by Eq.~(\ref{eq:exponential}),
\begin{equation}
a=e^{Ht},
\end{equation}
and assuming a constant value of $V''(\phi)$ to simplify the
calculation, leads to an evolution equation for mode-$k$ of
\begin{equation}
\partial^{2}_{t}\delta\phi+3H\partial_{t}\delta\phi+\frac{k^{2}}{e^{2Ht}}\delta\phi+\delta\phi
V''=0.
\end{equation}
Combining this with Eq.~(\ref{eq:Fourier}) leads to
\begin{eqnarray}
0&=&\bigg(\left[\frac{9}{4}H^{2}e^{-\frac{3}{2}Ht}g_{k}-3He^{-\frac{3}{2}Ht}\partial_{t}g_{k}+e^{-\frac{3}{2}Ht}\partial^{2}_{t}g_{k}\right]\nonumber\\
&&+3H\left[-\frac{3}{2}He^{-\frac{3}{2}Ht}g_{k}+e^{-\frac{3}{2}Ht}\partial_{t}g_{k}\right]+\left[k^{2}e^{-\frac{7}{2}Ht}g_{k}\right]+V''\left[e^{-\frac{3}{2}Ht}g_{k}\right]\bigg)\nonumber\\
&=&e^{-\frac{3}{2}Ht}\partial^{2}_{t}g_{k}+k^{2}e^{-\frac{7}{2}Ht}g_{k}-\frac{9}{4}H^{2}e^{-\frac{3}{2}Ht}g_{k}+V''e^{-\frac{3}{2}Ht}g_{k}.
\label{eq:evopsi}
\end{eqnarray}
Using the change of variables,
\begin{equation}
u\equiv-\frac{k}{H}e^{-Ht},
\end{equation}
which is $k$ times the conformal time, we then have
\begin{equation}
\partial_t=\frac{du}{dt}\frac{d}{du}=ke^{-Ht}\partial_{u},
\end{equation}
and
\begin{eqnarray}
\partial_t^{2}&=&ke^{-Ht}\partial_{u}ke^{-Ht}\partial_{u}=ke^{-Ht}\partial_{u}\left[-Hu\partial_{u}\right]\nonumber\\
&=&-kHe^{-Ht}\partial_{u}-ukHe^{-Ht}\partial_{u}^{2},
\end{eqnarray}
so the evolution equation Eq.~(\ref{eq:evopsi}) for mode-$k$ in
terms of $u$ is
\begin{eqnarray}
0&=&e^{-\frac{3}{2}Ht}\left[-kHe^{-Ht}\partial_{u}-ukHe^{-Ht}\partial_{u}^{2}\right]g_{k}+k^{2}e^{-\frac{7}{2}Ht}g_{k}-\frac{9}{4}H^{2}e^{-\frac{3}{2}Ht}g_{k}+V''e^{-\frac{3}{2}Ht}g_{k}\nonumber\\
&=&H^{2}e^{-\frac{3}{2}Ht}\left\{-\frac{k}{H}e^{-Ht}\partial_{u}g_{k}-u\frac{k}{H}e^{-Ht}\partial_{u}^{2}g_{k}+\frac{k^{2}}{H^{2}}e^{-2Ht}g_{k}-\frac{9}{4}g_{k}+\frac{V''}{H^{2}}g_{k}\right\}\nonumber\\[2mm]
&=&u^{2}\partial_{u}^{2}g_{k}+u\partial_{u}g_{k}+\left[u^{2}-\left(\frac{9}{4}-\frac{V''}{H^{2}}\right)\right]g_{k}.
\label{eq:Bessel}
\end{eqnarray}
Eq.~(\ref{eq:Bessel}) is Bessel's equation.  The most general
solution for a given $k$-component, $g_{k}$, is \cite{Vilenkin}
\begin{equation}
g_{k}(t)=\frac{1}{2}\sqrt{\pi/H}\left\{c_{1}H^{(1)}_{\sqrt{\frac{9}{4}-\frac{V''}{H^{2}}}}(u)+c_{2}H^{(2)}_{\sqrt{\frac{9}{4}-\frac{V''}{H^{2}}}}(u)\right\}.
\label{eq:BesselSol}
\end{equation}
We then have
\begin{equation}
\psi_{k}(t)=a(t)^{-\frac{3}{2}}\,\frac{1}{2}\sqrt{\pi/H}\left\{c_{1}H^{(1)}_{\sqrt{\frac{9}{4}-\frac{V''}{H^{2}}}}(u)+c_{2}H^{(2)}_{\sqrt{\frac{9}{4}-\frac{V''}{H^{2}}}}(u)\right\},
\label{eq:bunchdavies}
\end{equation}
but for the $k=0$ mode of the massless, minimally coupled case in a
purely de Sitter universe, a universe that has an infinite history
and future that is at all times described by the metric of
Eq.~(\ref{eq:exponential}), see also Refs.~\cite{Allen1,Allen2}.

For sufficiently large $k$-modes the solution should be
asymptotically insensitive to the de Sitter curvature, as this
corresponds to very small wavelengths.  On a very small scale that
locally appears nearly flat, the curvature becomes negligible.  For
these large $k$-modes, the solution we expect| due to the rapid
attenuation of matter and radiation in a de Sitter universe| is that
of the positive frequency WKB vacuum solution \cite{Parker5}
\begin{equation}
\psi_{k}(t)\sim\frac{1}{\sqrt{2\omega_{k}(t)a(t)^{3}}}e^{-i\int\omega_{k}(t')dt'}=\frac{1}{\sqrt{-2a(t)^{3}Hu}}e^{-iu},
\label{eq:evoasymp}
\end{equation}
where the frequency is
\begin{equation}
\omega_{k}(t)\equiv\sqrt{\frac{k^{2}}{a(t)^{2}}+m^{2}}.
\end{equation}
See also Sec.~\ref{sec:evoasymp}.  To match our constants, $c_{1}$
and $c_{2}$, when $k\rightarrow\infty$, we use the large argument
expansion of the Hankel functions \cite{Abramowitz}
\begin{eqnarray}
H_{\nu}^{(1)}(z)&\sim&\sqrt{2/(\pi
z)}e^{i(z-\frac{1}{2}\nu\pi-\frac{1}{4}\pi)}\nonumber\\
H_{\nu}^{(2)}(z)&\sim&\sqrt{2/(\pi
z)}e^{-i(z-\frac{1}{2}\nu\pi-\frac{1}{4}\pi)},
\label{eq:hanklargarg}
\end{eqnarray}
which means that, to within a phase,
\begin{equation}
\psi_{k}(t)=\frac{1}{\sqrt{-2a(t)^{3}Hu}}\left\{c_{1}e^{iu}+c_{2}e^{-iu}\right\}.
\end{equation}
To match to the positive-frequency, vacuum solution given by
Eq.~(\ref{eq:evoasymp}) we choose \cite{Parker5,Vilenkin}
\begin{eqnarray}
\lim_{k\rightarrow\infty}c_{1}(k)&\sim&0,\nonumber\\
\lim_{k\rightarrow\infty}c_{2}(k)&\sim&1.
\end{eqnarray}
The de Sitter metric and the physical volume are symmetric under the
transformation~\cite{Parker5}
\begin{equation}
t\rightarrow t+t_{0}\ {\rm and}\ \vec{x}\rightarrow
e^{-Ht_{0}}\vec{x}.
\end{equation}
The Killing vector generating this isometry, \cite{Deruelle}
\begin{equation}
\xi^{0}=1,\ \ \ \ \ \xi^{i}=-Hx^{i}.
\end{equation}
corresponds to conservation of energy.  Since the vacuum
fluctuations can be expected to share this symmetry of space-time,
provided| as will be explained in Sec.~\ref{sec:KBSENS}| there is an
infinite expansion and the universe is de Sitter in the infinite
past and infinite future, the variable $u$ is thus invariant under
\begin{equation}
t\rightarrow t+t_{0}\ {\rm and}\
\vec{k}\rightarrow\vec{k}e^{Ht_{0}}.
\end{equation}
Then, with $k'\equiv ke^{Ht_{0}}$,
\begin{equation}
\psi_{k'}(t+t_{0})=\psi_{k}(t)
\end{equation}
requires
\begin{equation}
c_{1}(k')=c_{1}(k)\ {\rm and}\ c_{2}(k')=c_{2}(k).
\end{equation}
Thus, because \begin{math} t_{0} \end{math} is arbitrary, we have
\cite{Bunch},
\begin{equation}
\psi_{k}(t)=\frac{1}{2}a(t)^{-3/2}\sqrt{\pi/H}\
H^{(2)}_{\sqrt{\frac{9}{4}-\frac{V''}{H^{2}}}}(u).
\label{eq:desitterpsi}
\end{equation}
We note for future reference that changing the sign of the argument
in Eq.~(\ref{eq:BesselSol}) also yields a linearly independent
solution to Eq.~(\ref{eq:Bessel}) under the transformation
$u\rightarrow\tilde{u}=-u$, because the Hankel functions of the
first and second kind form an orthogonal and complete set. The
coefficients $c_{1}(k)$ and $c_{2}(k)$ will, in general, change
under the transformation $u\rightarrow\tilde{u}$, but the procedure
outlined above for finding these coefficients in the
$k\rightarrow\infty$ limit, leads to $c_{1}(k)=-i$ and $c_{2}(k)=0$.
A simpler way of seeing this, once we have
Eq.~(\ref{eq:desitterpsi}), is to change the sign of $H$.  Although
we will later take $H$ to be real and positive, we have not yet made
this assumption, so changing the sign of $H$ should leave
Eq.~(\ref{eq:desitterpsi}) intact in the flat-space limit of
$k\rightarrow\infty$, where again a mode should not see the
curvature of space.  Using Eq.~(\ref{eq:hanklargarg}), we see that
this large argument limit of the Hankel functions takes| to within a
phase| $H^{(2)}_{v}(z)\rightarrow-iH^{(1)}_{v}(-z)$.

\subsection{Relation to Observations}
\label{sec:relationtoobservations}

In this section we will focus on defining the slow roll
approximation, the slow roll parameters, the number of e-folds, the
curvature perturbation, the spectrum of curvature perturbations, and
the spectral index.

In the slow roll
approximation~\cite{Guth1,Sato,Starobinsky1,Starobinsky2,Guth2,Bardeen1,Boerner,Brout,Mukhanov}
\begin{equation}
\dot{\phi}^{2}\ll V(\phi)
\end{equation}
and
\begin{equation}
|\ddot{\phi}|\ll |V'|.
\end{equation}
This means Eqs.~(\ref{eq:inflationeq1}) and~(\ref{eq:inflationeq2})
become
\begin{equation}
H^{2}\simeq\frac{8\pi G}{3}V(\phi) \label{eq:slowroll1}
\end{equation} and
\begin{equation}
\dot{\phi}\simeq-\frac{V'}{3H}. \label{eq:slowroll2}
\end{equation}
These conditions ensure that
\begin{math}P\simeq-\rho\end{math}, which is the property of a
space-time dominated by a cosmological constant, or de Sitter space;
and that the kinetic term does not grow appreciably since the
potential is assumed to be flat and H is large.  During inflation,
the slow roll parameters must satisfy \cite{Peiris}
\begin{equation}
\epsilon\ll1\ {\rm and}\ |\eta |\ll1,
\end{equation}
where the slow roll parameters are defined by \cite{Peiris}
\begin{equation}
\epsilon\equiv\frac{1}{16\pi
G}\left(\frac{V'}{V}\right)^{2}\simeq-\frac{\dot{H}}{H^{2}},
\label{eq:slowrolleps}
\end{equation}
\begin{equation}
\eta\equiv\frac{1}{8\pi G}\left(\frac{V''}{V}\right),
\label{eq:slowrolleta}
\end{equation}

Using the slow-roll equations~(\ref{eq:slowroll1})
and~(\ref{eq:slowroll2}), we can express the number of e-folds of
inflation as \cite{Liddle1}
\begin{equation}
N_{e}\equiv\ln\left[\frac{a(t_{\rm final})}{a(t_{\rm
initial})}\right]=\int_{t_{\rm initial}}^{t_{\rm
final}}H\,dt\simeq8\pi G\int_{\phi_{\rm final}}^{\phi_{\rm
initial}}\,\frac{V(\phi)}{V'(\phi)}\,d\phi. \label{eq:nefolds}
\end{equation}

We define a mode to be crossing the Hubble radius when the mode's
wavelength, $a(t)/k$, is the same size as the Hubble radius,
$H^{-1}$, which would be the horizon size in a purely de Sitter
universe.  During inflation, when the scale factor is growing
exponentially and $k$ and $H$ are both constant, a mode exits the
Hubble radius when $k/[a(t)\,H]=1$. After inflation, when the scale
factor is given by either a radiation-dominated $a(t)\propto
t^{1/2}$ growth or by a matter-dominated $a(t)\propto t^{2/3}$
growth, where for both cases $H\propto t^{-1}$, then
$k/[a(t)\,H(t)]=1$ defines the time when a mode re-enters the Hubble
radius.

We can apply the small argument limit of the Hankel functions
\cite[Eq.~9.1.9]{Abramowitz},
\begin{equation}
\left|H_{v}^{(1)}(z)\right|^{2}\simeq\left|H_{v}^{(2)}(z)\right|^{2}\simeq\left(\frac{\Gamma(v)}{\pi}\right)^{2}\left(\frac{1}{2}\left|z\right|\right)^{-2v},
\end{equation}
when the real part of the parameter $v$ is positive and non-zero, to
Eq.~(\ref{eq:desitterpsi}), to get
\begin{equation}
\left|\psi_{k}\right|^{2}\simeq\frac{\pi}{4H}a(t)^{-3}\left(\frac{\Gamma(v)}{\pi}\right)^{2}\left(\frac{1}{2}\frac{k}{a(t)\,H}\right)^{-2v}.
\end{equation}
In the massless, minimally-coupled case, $v=3/2$, and we find
\begin{equation}
\left|\psi_{k}\right|^{2}\simeq\frac{\pi}{4H}a(t)^{-3}\left(\frac{\sqrt{\pi}/2}{\pi}\right)^{2}\left(\frac{k}{2a(t)\,H}\right)^{-3}.
\end{equation}
Late enough into inflation for a given mode to be well outside the
Hubble radius, we then have
\begin{equation}
\left|\psi_{k}\right|^{2}\simeq\frac{H^{2}}{2k^{3}},
\label{eq:superhorizon}
\end{equation}
which is approximately half the value of $\left|\psi_{k}\right|^{2}$
at the time it exits the Hubble radius| see Sec.~\ref{sec:spec}.
Although this perturbation of the inflaton field is not a
gauge-invariant quantity, there is a gauge-invariant quantity, a
curvature perturbation that we call $\mathcal{R}_{k}$, that is
approximately conserved outside of the Hubble radius, and we can use
it to relate the inflaton fluctuations to density perturbations at
the time of re-entry as follows:
\cite{Kolb,Dodelson,Guth1,Starobinsky2,Bardeen1,Liddle1,Bardeen2,Sasaki,Hawking1}
\begin{equation}
\frac{\delta\phi_{k}}{\dot{\phi}}H\simeq\mathcal{R}_{k,{\rm
exit}}\simeq\mathcal{R}_{k,{\rm
re-entry}}\propto\delta_{k}\equiv\frac{\delta\rho_{k}}{\rho},
\label{eq:curvpert}
\end{equation}
where for re-entry into a matter-dominated universe
$\delta_{k}\simeq\frac{2}{5}\mathcal{R}_{k}$, and for re-entry into
a radiation-dominated universe
$\delta_{k}\simeq\frac{4}{9}\mathcal{R}_{k}$.  The value of
$\delta\phi_{k}$ is usually taken (neglecting the coordinate length
$L$) to be the unrenormalized value $H^{2}/k^{3}$ obtained at the
time of exiting the Hubble radius. The justification for using an
unrenormalized value of $\delta\phi_{k}$, when it is well known that
the Bunch-Davies state given by Eq.~(\ref{eq:desitterpsi}) leads to
a divergent $\delta\phi$ when summed over all modes, is usually
given as implicit large and small cutoff frequencies.  It is often
assumed that the infrared and ultraviolet divergences come from
infrared and ultraviolet frequencies that do not affect the
treatment of modes exiting the Hubble radius during inflation.
Parker \cite{Parker6}, however, has shown that the divergences
affect every mode, and that neglecting a proper renormalization
drastically alters the results that are obtained.

We use the definition of a spectrum given by Liddle and Lyth
\cite{Liddle1}:
\begin{equation}
\mathcal{P}_{f}(k)\equiv\left(\frac{L}{2\pi}\right)^{3}4\pi
k^{3}\langle\left|f_{k}\right|^{2}\rangle.
\label{eq:specdef}
\end{equation}
Thus, under the standard assumption that it is not necessary to
renormalize the inflaton fluctuations as they are exiting the Hubble
radius, we could show
\begin{equation}
\mathcal{P}_{\delta}\propto\mathcal{P}_{\mathcal{R}}=\left(\frac{H}{\dot{\phi}}\right)^{2}\mathcal{P}_{\delta\phi}=\left(\frac{H}{\dot{\phi}}\right)^{2}\left(\frac{H}{2\pi}\right)^{2},
\label{eq:inflationspectra}
\end{equation}
from the super-Hubble radius behavior given by
Eq.~(\ref{eq:superhorizon}).  The renormalization of \cite{Parker6},
however, changes this: the renormalized spectrum of inflaton
perturbations, at the time of exiting the Hubble radius when
$k/[a(t)\,H]=1$, is
\begin{equation}
\mathcal{P}_{\delta\phi}=\left(\frac{H}{2\pi}\right)^{2}\left(\frac{\pi}{2}\left|H_{n}^{(1)}(1)\right|^{2}-\frac{m_{H}{}^{6}+\frac{33}{8}m_{H}{}^{4}+\frac{23}{4}m_{H}{}^{2}+2}{(m_{H}^{2}+1)^{7/2}}\right),
\label{eq:parkerrenorm}
\end{equation}
where $m_{H}\equiv m/H$ and $n\equiv\sqrt{9/4-m_{H}{}^{2}}$.  The
renormalized inflaton fluctuation depends critically on the mass and
when the magnitude of the fluctuation is evaluated.  In the massless
case, $\left|H_{n}^{(1)}(1)\right|^{2}=4/\pi$, and the renormalized
fluctuation is precisely zero.  Well outside the horizon, the
renormalized $\mathcal{P}_{\delta\phi}$ also goes to zero, but this
is perhaps not a problem, as $\mathcal{R}_{k}$ is the conserved
quantity, not $\delta\phi_{k}$, and the value of$\mathcal{R}_{k}$
given by Eq~(\ref{eq:curvpert}) is typically evaluated at the time a
mode crosses the Hubble radius. Thus, renormalization has the
potential to greatly alter the character of the spectrum of
perturbations.

The scalar spectral index, $n_{s}$, is a measure of how the
magnitude of density perturbations changes with scale.  A value of
$n_{s}=1$ indicates scale-invariance.  A value less than one is
called a red-tilted spectrum, and a value greater than one is called
a blue-tilted spectrum. It is defined as
\begin{equation}
n_{s}(k)-1\equiv\frac{d\ln\mathcal{P}_{\mathcal{R}}}{d\ln k},
\label{eq:scalarspectralindex}
\end{equation}
where the value of $n_{s}$ is given for a specific value of $k$,
called the pivot value, which is normally either of $k=0.05\,{\rm
Mpc}^{-1}$ \cite{Spergel1} or $k=0.002\,{\rm Mpc}^{-1}$
\cite{Komatsu}, relative to the value of the scale factor fixed to
be such that $a(t_{\rm now})=1$.  There is little running, or change
in $n_{s}(k)$ with changing scales, so the choice of $k_{\rm pivot}$
is somewhat arbitrary. We can relate the scalar spectral index to
the slow roll parameters given in Eqs.~(\ref{eq:slowrolleps})
and~(\ref{eq:slowrolleta}) \cite{Liddle1,Liddle2}.  Because the
curvature perturbations are evaluated at the time of Hubble radius
crossing, when $k=a(t)H\simeq He^{Ht}$, we see that with a nearly
constant value of $H$ during inflation $d\ln k=d[\ln(H)+Ht]\simeq
H\,dt$. This leads to, with Eq.~(\ref{eq:slowroll2}) rewritten as
$dt=-3H/V'\,d\phi$,
\begin{equation}
\frac{d}{d\ln
k}\simeq-\frac{V'}{3H^{2}}\frac{d}{d\phi}\simeq-\frac{1}{8\pi
G}\frac{V'}{V}\frac{d}{d\phi}. \label{eq:slowrollchainrule}
\end{equation}
Again using Eq.~(\ref{eq:slowroll2}), the spectrum of curvature
perturbations given by Eq.~(\ref{eq:inflationspectra}) becomes
\begin{equation}
\mathcal{P}_{\mathcal{R}}=\left(\frac{3H^2}{V'}\right)^{2}\left(\frac{H}{2\pi}\right)^{2}.
\end{equation}
With Eq.~(\ref{eq:slowroll1}), this becomes
\begin{equation}
\mathcal{P}_{\mathcal{R}}=\left(\frac{8\pi G\,
V}{V'}\right)^{2}\frac{8\pi G\, V}{12\pi^{2}}=\frac{(8\pi
G)^{3}}{12\pi^{2}}\frac{V^{3}}{V'^{2}},
\end{equation}
where the observed value of $\mathcal{P}_{\mathcal{R}}$ is typically
listed for the specific value of $k=0.002\,{\rm Mpc}^{-1}$, which is
different from the value of $k$ used with the scalar spectral index
\cite{Spergel1}.  In \cite{Komatsu}, the pivot scale for the
spectrum of curvature perturbations is chosen to be $k=0.02\,{\rm
Mpc}^{-1}$, as this is a scale that puts tighter constraints on the
magnitude of the curvature perturbation spectrum for a wider array
of model assumptions.  Within the assumptions of various models,
there is still a relatively scale-invariant spectrum of curvature
perturbations.

With Eq.~(\ref{eq:slowrollchainrule}), Liddle and Lyth find
\begin{eqnarray}
\frac{d\ln\mathcal{P}_{\mathcal{R}}}{d\ln k}&\simeq&-\frac{1}{8\pi
G}\frac{V'}{V}\frac{d}{d\phi}\ln\left(\frac{(8\pi
G)^{3}}{12\pi^{2}}\frac{V^{3}}{V'^{2}}\right)\nonumber\\
&\simeq&-\frac{1}{8\pi
G}\frac{V'}{V}\frac{d}{d\phi}\left(3\ln V-2\ln V'\right)\nonumber\\
&\simeq&-\frac{1}{8\pi
G}\frac{V'}{V}\left(3\frac{V'}{V}-2\frac{V''}{V'}\right)\nonumber\\
&\simeq&-6\frac{1}{16\pi
G}\left(\frac{V'}{V}\right)^{2}+2\frac{1}{8\pi G}\left(\frac{V''}{V}\right)\nonumber\\
&\simeq&-6\epsilon+2\eta,
\end{eqnarray}
where the slow roll parameters are given by
Eqs.~(\ref{eq:slowrolleps}) and~(\ref{eq:slowrolleta}).  Thus,
\begin{equation}
n_{s}-1=-6\epsilon+2\eta.
\label{eq:slowrollspectralindex}
\end{equation}
See also the end of Sec.~\ref{sec:spec} for a slightly different
derivation.

Finally, we note that when we define the mass by $m^{2}\equiv
d^{2}V/d\phi^{2}$, we find that
\begin{equation}
m_{H}\equiv m/H=\sqrt{m^{2}/H^{2}}=\sqrt{3 V''/(8 \pi G\,
V)}=\sqrt{3\eta},
\label{eq:slowrollmh}
\end{equation}
and thus the effective inflaton mass is related to the Hubble
constant during inflation through the slow roll parameter $\eta$.

\subsection{Findings of WMAP and SDSS Experiments}
\label{sec:WMAPSDSS}

The Five-Year Wilkinson Microwave Anisotropy Probe (WMAP) data
measures a scalar spectral index of $n_{s}(0.002/{\rm
Mpc})\simeq0.96$ \cite{Dunkley}. The Sloan Digital Sky Survey (SDSS)
measures a scalar spectral index of $n_{s}(0.05/{\rm
Mpc})\simeq0.95$ \cite{Tegmark}. Because the WMAP experiment
measures fluctuations in the CMBR, while the SDSS observes the
locations of galaxies and large-scale structure in our universe,
there is good, independent accord for the red-tilted spectral index
measured by these different approaches. The Five-Year WMAP data
finds a curvature perturbation spectrum of
\begin{equation}
\mathcal{P}_{\mathcal{R}}(0.002/Mpc)\simeq2.4\times10^{-9}.
\label{eq:5wmapspec}
\end{equation}
What follows in this section, where we apply these observations to
two particular models, is based upon work done by \cite{Glenz3}.
The first model we consider, the quadratic chaotic inflationary
potential \cite{Linde,Habib}, is in good agreement with the
Three-Year WMAP data \cite{Spergel2}.  The second model, a type of
Coleman-Weinberg model \cite{Bardeen1,Coleman}, is in good agreement
with the Five-Year WMAP data \cite{Komatsu,Shafi}.

The quadratic chaotic inflationary potential is given by
\cite{Linde,Habib}
\begin{equation}
V(\phi)=\frac{1}{2}m^{2}\phi^{2}.
\label{eq:linde}
\end{equation}
In Fig.~\ref{fig:linde},
\begin{figure}[hbtp]
\includegraphics[scale=2.5]{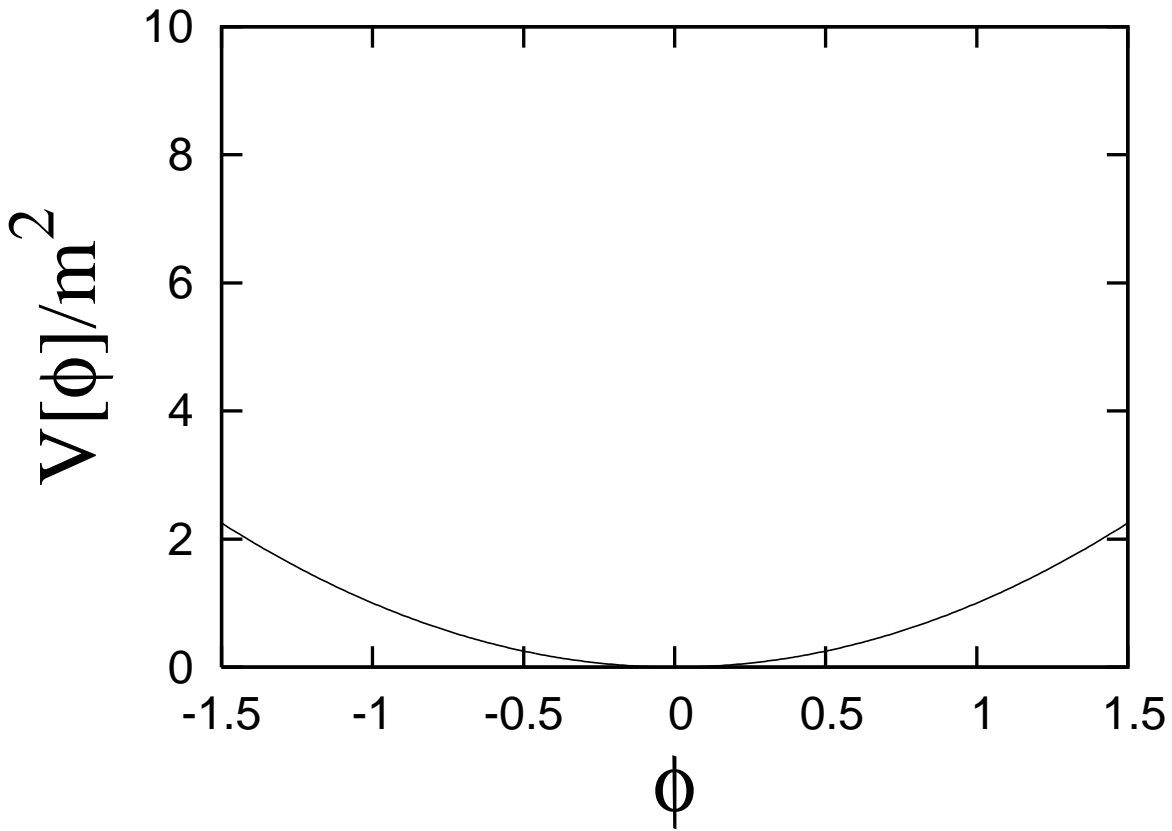}
\caption{\label{fig:linde} Quadratic Chaotic Potential.}
\end{figure}
we plot the potential given by Eq.~(\ref{eq:linde}) versus $\phi$.
In chaotic inflation, the value of $\phi$ is initially perturbed
away from the minimum and rolls slowly| provided the slope of the
potential is sufficiently gradual| down the potential to the minimum
at $\phi=0$. To be contrasted with chaotic inflation is new
inflation, in which $\phi$ begins near the maximum value of the
potential located at $\phi=0$ and rolls slowly to a minimum of the
potential \cite{Lyth}. The Coleman-Weinberg potential, which was
actually one of the earlier models considered for an inflationary
potential that did not involve tunneling through a potential barrier
and its associated problems with bubbles of inflation not
coalescing, is an example of new inflation.

The one-loop Coleman-Weinberg potential is given in the
zero-temperature limit by \cite{Bardeen1,Coleman}
\begin{equation}
V(\phi,T)=\frac{1}{2}B\sigma^{4}+B\phi^{4}\left[\ln(\phi^{2}/\sigma^{2})-\frac{1}{2}\right].
\label{eq:cw}
\end{equation}
In Fig.~\ref{fig:cw},
\begin{figure}[hbtp]
\includegraphics[scale=2.5]{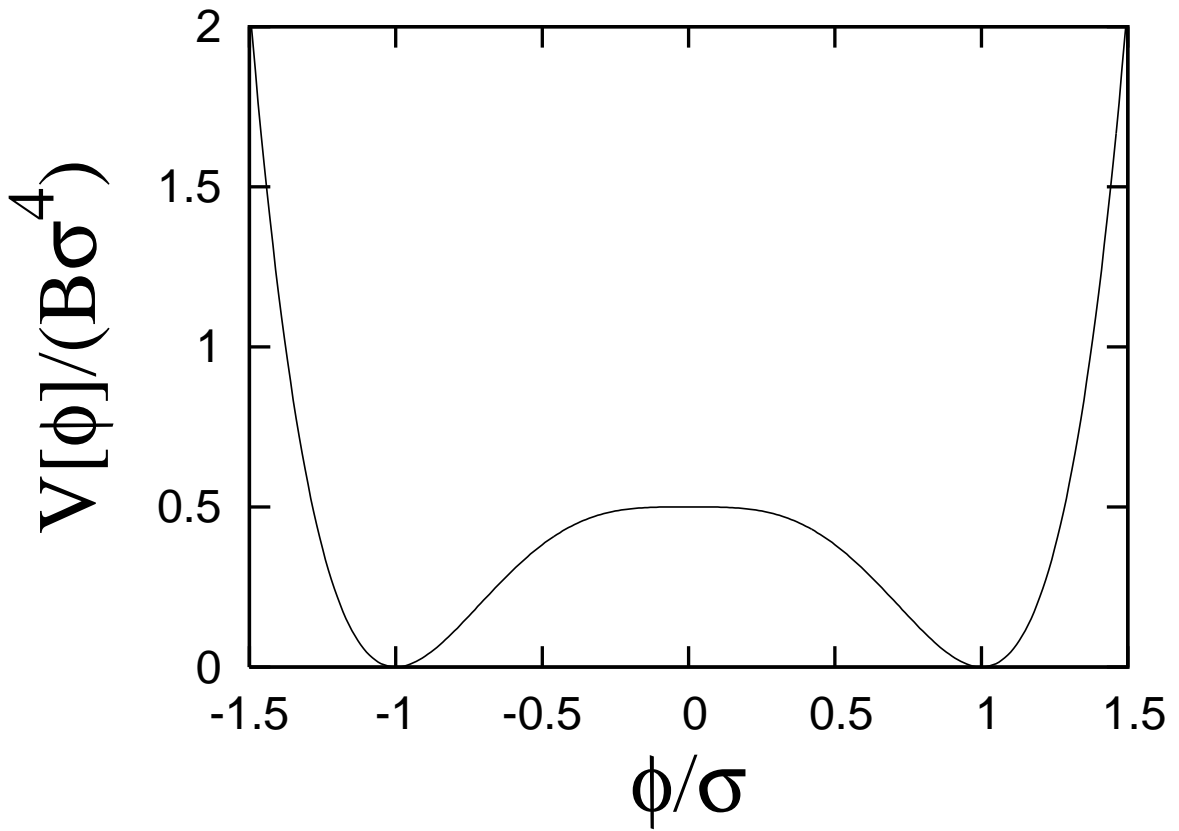}
\caption{\label{fig:cw} Coleman-Weinberg Potential.}
\end{figure}
we plot a dimensionless potential $V(\phi)/(B\sigma^{4})$ versus the
dimensionless parameter $\phi/\sigma$.  In the low temperature
limit, the stable minima of the potential are located at
$\phi=\pm\sigma$.  At the beginning of inflation $\phi\simeq0$,
where the slow roll conditions are satisfied, and $\phi$ rolls to
either of two (in the low temperature limit) stable minima.
Classically, inflation is a period of super-cooling, so the
low-temperature limit should be justified, but see also
Sec.~\ref{sec:reheat}.

For the quadratic chaotic inflationary potential, the slow roll
parameters of Eqs.~(\ref{eq:slowrolleps}) and~(\ref{eq:slowrolleta})
are equal to each other, and we have
\begin{equation}
\epsilon=\eta=\frac{1}{4\pi G\phi^{2}}.
\label{eq:epseta1}
\end{equation}
From Eq.~(\ref{eq:slowrollspectralindex}) and the Five-Year WMAP
spectral index of $n_{s}-1\simeq-0.04$, we find
\begin{equation}
n_{s}-1=-6\epsilon+2\eta=-4\eta\simeq-0.04,
\end{equation}
or
\begin{equation}
\epsilon=\eta\simeq0.01.
\label{eq:epseta2}
\end{equation}
From Eq.~(\ref{eq:slowrollmh}), we have
\begin{equation}
\frac{m}{H}\simeq\sqrt{0.03}\simeq0.2.
\label{eq:quadmh}
\end{equation}
From Eqs.~(\ref{eq:epseta1}) and~(\ref{eq:epseta2}),
\begin{equation}
\frac{1}{4\pi G\phi^{2}}\simeq0.01,
\end{equation}
or
\begin{equation}
\phi_{\rm cmb}\simeq\frac{G^{-1/2}}{\sqrt{0.04\pi}},
\end{equation}
where $\phi_{\rm cmb}$ corresponds roughly to that range of $\phi$
at which the modes observed by WMAP were exiting the Hubble radius
during inflation.  Using Eq.~(\ref{eq:nefolds}), we find the number
of e-folds before the end of inflation at which these modes were
exiting the Hubble radius:
\begin{equation}
N_{e}\simeq8\pi G\int_{0}^{\phi_{\rm
cmb}}\frac{V}{V'}d\phi\simeq8\pi G\int_{0}^{\phi_{\rm
cmb}}\frac{1}{2}\phi\, d\phi\simeq8\pi G\frac{\phi_{\rm
cmb}^{2}}{4}\simeq\frac{2}{0.04}\simeq50.
\label{eq:quadpote}
\end{equation}
For the value of $m_{H}\simeq0.2$ given by Eq.~(\ref{eq:quadmh}),
the renormalized spectrum of inflaton fluctuations given by
Eq.~(\ref{eq:parkerrenorm}) is
\begin{equation}
\mathcal{P}_{\delta\phi}\simeq\left(\frac{H}{2\pi}\right)^{2}\left(1.968-\frac{2.237}{1.147}\right)\simeq\left(\frac{H}{2\pi}\right)^{2}(0.019)\simeq0.00047H^{2}.
\end{equation}
Using the relation given in Eq.~(\ref{eq:inflationspectra}) and the
slow roll approximation given in Eq.~(\ref{eq:slowroll2}), we have
\begin{equation}
\mathcal{P}_{\mathcal{R}}=\left(\frac{H}{\dot{\phi}}\right)^{2}\mathcal{P}_{\delta\phi}\simeq\left(\frac{3H^{2}}{m^{2}\phi}\right)^{2}0.00047H^{2}\simeq\frac{m_{H}{}^{-4}}{\phi^{2}}0.0042H^{2}\simeq\frac{4.7\,H^{2}}{\phi^{2}},
\end{equation}
then, as a rough estimate of the general order of magnitude, we use
$\phi_{\rm cmb}$ to get
\begin{equation}
\mathcal{P}_{\mathcal{R}}\simeq\frac{4.7\,H^{2}}{0.04\pi(G^{-1/2})^{2}}\simeq37\left(\frac{H}{G^{-1/2}}\right)^{2}.
\end{equation}
We can equate this with the amplitude of the spectrum found in the
Five-Year WMAP data to write
\begin{equation}
37\left(\frac{H}{G^{-1/2}}\right)^{2}\simeq2.4\times10^{-9},
\end{equation}
and
\begin{equation}
\frac{H}{G^{-1/2}}\simeq8\times10^{-6}.
\end{equation}
Using the Planck scale, $G^{-1/2}\simeq1.22\times10^{19}{\rm\ GeV}$,
finally we have
\begin{equation}
H\simeq7\times10^{13}{\rm\ GeV}, \label{eq:quadpoth}
\end{equation}
which can be seen as an upper limit on $H$ near the beginning of
inflation, around the time the modes observed by WMAP were exiting
the Hubble radius; as $\phi$ rolls down the potential towards zero,
the size of $H$ decreases.

For the Coleman-Weinberg potential given by Eq.~(\ref{eq:cw}), we
have
\begin{eqnarray}
V'&=&4B\phi^{3}\ln\frac{\phi^{2}}{\sigma^{2}},\\
V''&=&12B\phi^{2}\left(\frac{2}{3}+\ln\frac{\phi^{2}}{\sigma^{2}}\right).
\end{eqnarray}
The slow roll parameters are
\begin{eqnarray}
\epsilon&=&\frac{1}{16\pi
G}\left(\frac{4B\phi^{3}\ln\frac{\phi^{2}}{\sigma^{2}}}{\frac{1}{2}B\sigma^{4}+B\phi^{4}\left[\ln(\phi^{2}/\sigma^{2})-\frac{1}{2}\right]}\right)^{2}\nonumber\\
&=&\frac{(G^{-1/2})^{2}}{16\pi \sigma^{2}}\left(\frac{4r^{3}\ln r^{2}}{\frac{1}{2}+r^{4}\left[\ln(r^{2})-\frac{1}{2}\right]}\right)^{2},\\
\eta&=&\frac{1}{8\pi
G}\left(\frac{12B\phi^{2}\left(\frac{2}{3}+\ln\frac{\phi^{2}}{\sigma^{2}}\right)}{\frac{1}{2}B\sigma^{4}+B\phi^{4}\left[\ln(\phi^{2}/\sigma^{2})-\frac{1}{2}\right]}\right)\nonumber\\
&=&\frac{(G^{-1/2})^{2}}{8\pi
\sigma^{2}}\left(\frac{12r^{2}\left(\frac{2}{3}+\ln
r^{2}\right)}{\frac{1}{2}+r^{4}\left[\ln(r^{2})-\frac{1}{2}\right]}\right),
\end{eqnarray}
where $r\equiv\phi/\sigma$.  We assume the values given by \cite[p.
292]{Kolb} of
\begin{eqnarray}
\sigma&\simeq&2\times10^{15}{\rm\ GeV},\nonumber\\
B&\simeq&10^{-3}.
\label{eq:cwkt}
\end{eqnarray}
With those values and
$G^{-1/2}\simeq1.22\times10^{19}{\rm\ GeV}$, taking $r\ll1$ we find
\begin{eqnarray}
\epsilon&\simeq&\left(7.4\times10^{5}\right)64r^{6}\left(\ln r^{2}\right)^{2},\\
\eta&\simeq&\left(1.5\times10^{6}\right)24r^{2}\ln r^{2},
\label{eq:etar}
\end{eqnarray}
and we find in the limit $\phi\ll\sigma$, that $\epsilon\ll\eta$.
Using the WMAP value of $0.96$ for the spectral index, this leads to
$-6\epsilon+2\eta\simeq2\eta\simeq-0.04$, or
\begin{equation}
\eta\simeq-0.02.
\label{eq:etacw}
\end{equation}
Then we have $m_{H}{}^{2}\simeq-0.06$, or
\begin{equation}
m_{H}\simeq 0.245i\simeq i/4.
\label{eq:tachmass}
\end{equation}
An imaginary physical mass could lead to tachyonic behavior
\cite{Anderson1}, however in this case, recall we are dealing with
an effective mass. To find $r$, which we assume to be much less than
one, we combine Eqs.~(\ref{eq:etar}) and~(\ref{eq:etacw}) to get
\begin{equation}
r\simeq\pm5.3\times10^{-6}.
\label{eq:cwr}
\end{equation}
With Eq.~(\ref{eq:nefolds}), we have
\begin{equation}
N_{e}\simeq8\pi\frac{\sigma^{2}}{(G^{-1/2})^{2}}\int_{1}^{5.3\times10^{-6
}}\left(\frac{\frac{1}{2}+r^{4}\left[\ln(r^{2})-\frac{1}{2}\right]}{4r^{3}\ln
r^{2}}\right)dr\simeq64.
\end{equation}
Finally, Eqs.~(\ref{eq:slowroll2}), (\ref{eq:inflationspectra}),
(\ref{eq:parkerrenorm}), (\ref{eq:5wmapspec}),
and~(\ref{eq:tachmass}) lead us to
\begin{equation}
2.4\times10^{-9}\simeq\left(\frac{H}{\dot{\phi}}\right)^{2}\left(\frac{H}{2\pi}\right)^{2}\left(0.012\right)\simeq\left(\frac{9H^{6}}{4\pi^{2}(4B\phi^{3}\ln\frac{\phi^{2}}{\sigma^{2}})^{2}}\right)\left(0.012\right).
\end{equation}
Then, using the values given in Eqs.~(\ref{eq:cwkt})
and~(\ref{eq:cwr}), we have
\begin{equation}
H=4.7\times10^{8}{\rm\ GeV}.
\end{equation}
This value of $H$ listed here for the Coleman-Weinberg potential can
be compared with that found in Eq.~(\ref{eq:quadpoth}) to see how
discrepancies can arise when choosing between different models
consistent with observations.

The usual method of describing inflation by first specifying a
potential and then calculating observable quantities is thus in some
ways not very constraining in its predictions for the early
universe. In the next chapter we will discuss a means of modeling
inflation in a potential-independent way by specifying the evolution
of a scale factor consistent with inflation instead of attempting to
discern between individual models of potentials consistent with
inflation.

\newpage
\thispagestyle{fancy}
\chapter{Spectrum of Inflaton Fluctuations}
\thispagestyle{fancy}
\pagestyle{fancy}

In \cite{Parker6}, Parker showed how to renormalize fluctuations in
the inflaton field in curved spacetime using adiabatic
regularization, for which see also
\cite{Parker7,Fulling1,Anderson2}. Other papers
\cite{Finelli,Agullo} have since found similar disagreement with the
standard treatment of the dispersion. The technique used in
\cite{Parker6} has been shown to give the same results in
homogeneous and isotropic universes as other methods of
renormalization, such as point-splitting, and to be related to the
Hadamard condition in curved space time
\cite{Glenz1,Birrell2,Luders,Pirk,Junker}, which states that the
two-point function $\langle0\left|\phi(x),\phi(x')\right|0\rangle$,
in the limit $x'\rightarrow x$ takes the form of a Hadamard Solution
\cite{Birrell1,Pirk}
\begin{equation}
S(x,x')=\frac{\Delta^{1/2}}{8\pi^{2}}\left(\frac{2}{\sigma}+v\ln\sigma+w\right),
\end{equation}
where $\sigma$ is the proper distance of interval of spacetime
between $x$ and $x'$, $\Delta\equiv-{\rm
det}[\partial_{a}\partial_{b}\sigma][g(x)g(x')]^{-1/2}$ and reduces
to $[-g(x)]^{-1/2}$ as $x'\rightarrow x$, and
\begin{eqnarray}
v&\equiv&\sum_{l=0}^{\infty}v_{l}\sigma^{k},\\
w&\equiv&\sum_{l=0}^{\infty}w_{l}\sigma^{k}.
\end{eqnarray}
As an additional check on adiabatic regularization, we examine the
spectrum of inflaton perturbations in spacetimes that asymptotically
approach Minkowski space at early and late times. This is a method
introduced and used in Parker's analysis of particle creation by an
expanding universe \cite{Parker1,Parker2,Parker3}, and it requires
no renormalization beyond that already known in Minkowski space. To
make use of Minkowski space in the analysis of the spectrum of
inflaton perturbations coming from inflation, we investigate a scale
factor, which is a measure of the size of the universe, that is
composed of different scale factor segments joined together, similar
to the treatments of \cite{Ford1,Allen3}. We first tried evolving
forward the inflaton perturbations using a fourth order Runge-Kutta
numerical integration routine in C++ code, but we realized that we
would need to use greater precision for our computation. We decided
instead to use an analytical calculation by matching known solutions
to the evolution equation at the boundary conditions joining the
different segments of the scale factor.  Our calculations were
performed using 500 digit precision in Mathematica.

\section{Composite Scale Factor}

We consider the metric
\begin{equation}
ds^{2}=dt^{2}-a^{2}(t)\left[(dx)^{2}+(dy)^{2}+(dz)^{2}\right].
\label{eq:metric}
\end{equation}
The time $t$ will run continuously from $-\infty$ to $\infty$. The
scale factor $a(t)$ will be composed of three segments.  Our scale
factor will generally be $C^2$, i.e., a continuous function with
continuous first and second derivatives everywhere, including at the
joining points between segments. Briefly, we will consider scale
factors that are only $C^1$ or $C^0$ at the joining points. The
initial and final segments are asymptotically Minkowskian in the
distant past and future, respectively. The middle segment is an
exponential expansion with respect to the time $t$. We choose
specific forms for $a(t)$ in these segments that have exact
solutions of the evolution equations for inflaton quantum
fluctuations of zero effective mass.

Fig.~\ref{fig:diagram} shows an example our composite
\begin{figure}[hbtp]
\includegraphics[scale=2.5]{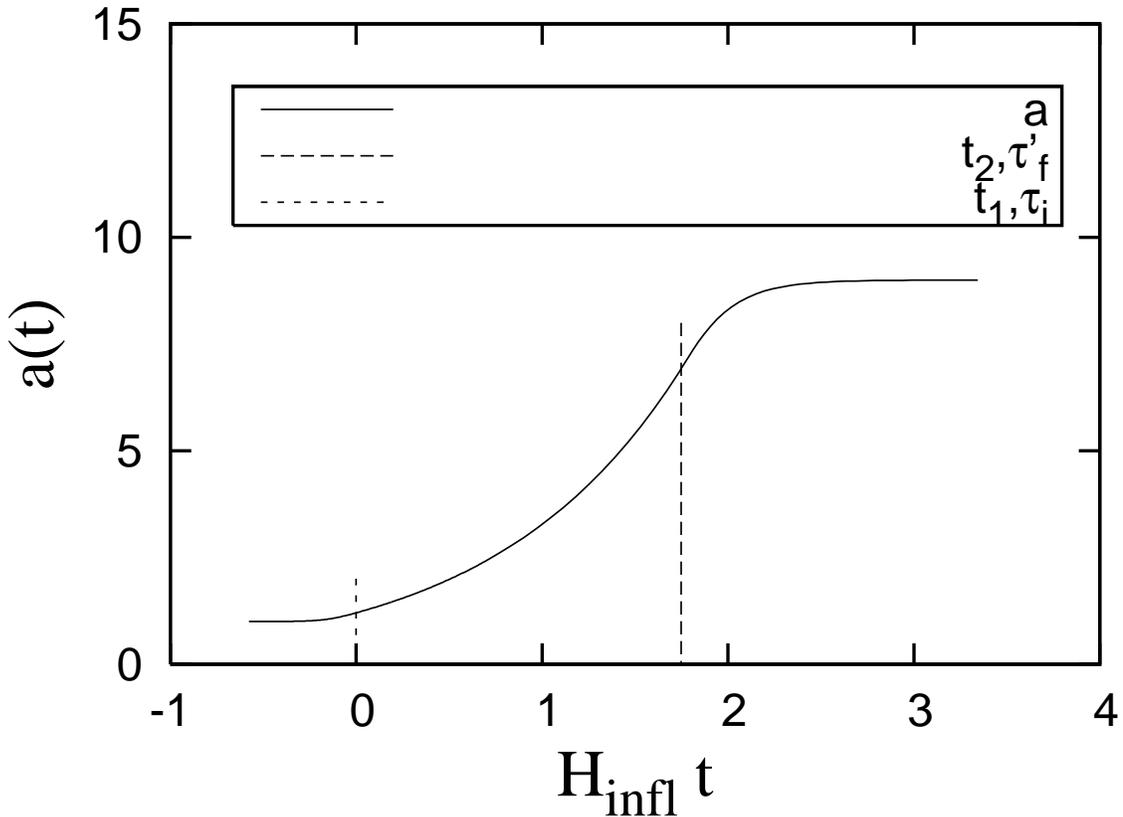}
\caption{\label{fig:diagram} Composite Scale Factor.}
\end{figure}
scale factor plotted versus dimensionless time.  This illustrative
example summarizes our notation using a moderate expansion of
$\sim2$ e-folds. The scale factor, $a(t)$, is continuous, as are
$\dot{a}(t)$ and $\ddot{a}(t)$.  In this case, the parameters for
the initial asymptotically flat segment are $a_{1i}=1$, $a_{2i}=2$,
and $s_{i}=1$.  The free parameters of the final asymptotically flat
segment are $a_{2f}=9$ and $a_{1f}=6$. Both asymptotically flat
scale factors are given by different parameter choices of
Eq.~(\ref{eq:aHG}) with the parameter $b$ in both cases equal to
zero. The asymptotically flat scale factor of the initial region
joins the exponentially expanding scale factor of the middle region
at a time $t_{1}$ in $t$-time and $\tau_{i}$ in $\tau$-time. The
exponentially expanding scale factor of the middle region joins the
asymptotically flat scale factor of the final region at a time
$t_{2}$ in $t$-time and $\tau'_{f}$ in $\tau'$-time of the final
segment, where a prime is used to distinguish between the
$\tau$-times of the initial and final segments.

The equation for the middle (inflationary) segment of our composite
scale factor is given in terms of proper time by
\begin{equation}
a(t)=a(t_1) e^{H_{\rm infl}(t-t_1)},
\label{eq:infl}
\end{equation}
where $H_{\rm infl}$ is the constant value of $H(t)\equiv a^{-1}
da/dt$ during the exponential expansion of the middle segment.

We define the quantity of Eq.~(\ref{eq:nefolds}), $N_{e}\equiv
\ln\left(a_{2f}/a_{1i}\right)$, in terms of $a(t_{\rm
initial})=a_{1i}$ and $a(t_{\rm final})=a_{2f}$. When there is a
long period of exponential growth, $N_{e}$ is essentially the number
of e-foldings of inflation. Typically, $N_{e}$ will be about $60$.
Within the final asymptotically flat scale factor, the ratio of
$a_{2f}$ to $a_{1f}$ determines how gradually the exponential
expansion transitions to the asymptotically flat late-time region.
(For example, this ratio might be 1 e-fold, which we would consider
to be relatively gradual, or it might be 1.0001, which we would
consider to be relatively abrupt.)

\subsection{Asymptotically Minkowski}

The initial and final asymptotically flat regions permit us to
unambiguously interpret our results for free fields without having
to perform any renormalization in curved spacetime. The final
asymptotically flat region will not significantly affect the result
obtained for the spectrum of inflaton perturbations created by the
inflationary segment of the expansion. The initial asymptotically
flat region should have a negligible effect on the spectrum
resulting from a long period of inflation, although we do find
remnants of the early initial conditions in the late-time inflaton
dispersion spectrum, which we will discuss in Sec.~\ref{sec:KBSENS}.

We base each asymptotic segment on a scale factor of the form,
\begin{equation}
a(t(\tau))=\left\{a_{1}^{\ 4}+e^{\tau/s}[(a_{2}^{\ 4}-a_{1}^{\
4})(e^{\tau/s}+1)+b](e^{\tau/s}+1)^{-2}\right\}^{\frac{1}{4}} ,
\label{eq:aHG}
\end{equation}
where $\tau$ is related to the proper time $t$ by
\begin{equation}
d\tau\equiv a(t)^{-3}dt. \label{eq:tvtau}
\end{equation}
\begin{figure}[hbtp]
\includegraphics[scale=2.5]{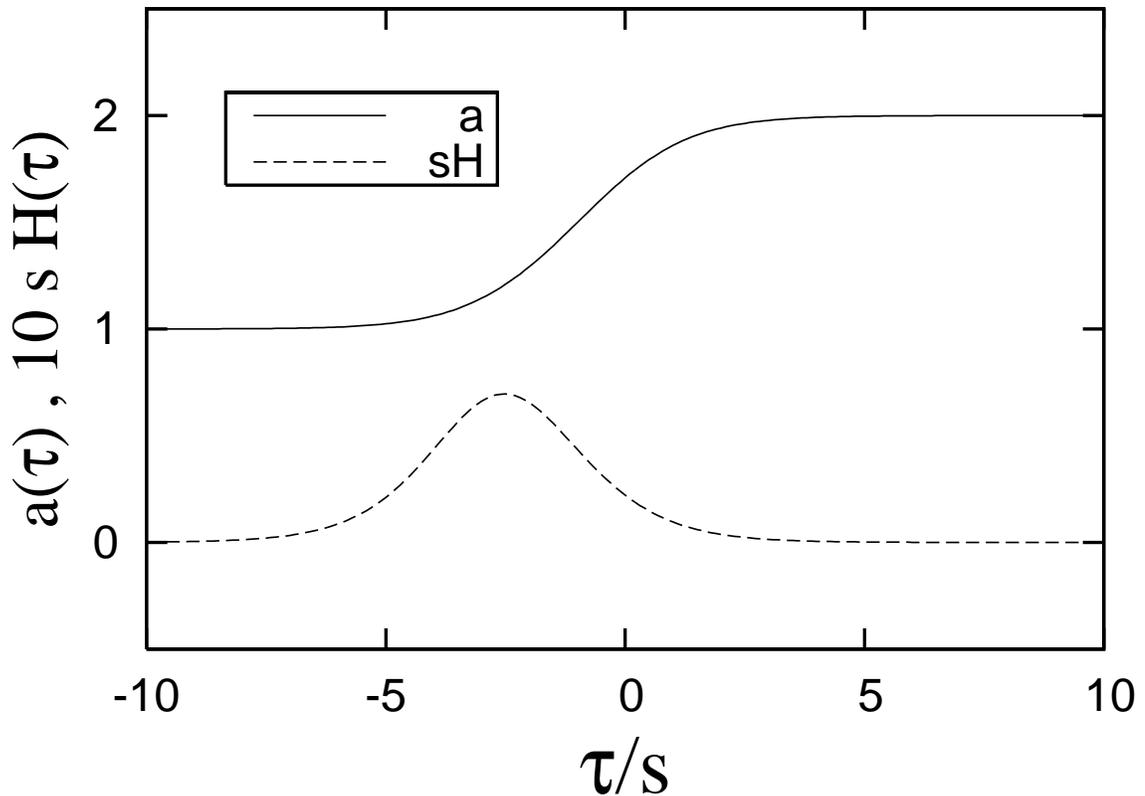}
\caption{\label{fig:aAndHvTau} Asymptotically Flat Scale Factor.}
\end{figure}
See Fig.~\ref{fig:aAndHvTau}.  This figure shows the asymptotically
flat scale factor, $a\left(t(\tau)\right)$, and the associated
dimensionless Hubble parameter, $s H(t(\tau))= s
a^{-1}{da/dt}=sa^{-4}{da/d\tau}$, of Eq.~(\ref{eq:aHG}) with
$a_{1}=1$, $a_{2}=2$, $b=0$, and $s=1$. Note in the graph that the
maximum of $H$ occurs at a value of $a(t(\tau))$ closer to $a_{1}$
than to $a_{2}$.  In both the case where $a_{2}\gg a_{1}$ and the
case where $a_{2}\simeq a_{1}$, $H_{\rm max}$ occurs at a value of
the scale factor where $a(t(\tau))\simeq a_{1}$.

The form of the scale factor in Eq.~(\ref{eq:aHG}) is based on the
form of the index of refraction used by Epstein to model the
scattering of radio waves in the upper atmosphere and by Eckart to
model the potential energy in one-dimensional scattering in quantum
mechanics \cite{Epstein, Eckart}. It was first used in the
cosmological context by Parker \cite{Parker4,Parker8,Parker9} to
model $a(t)$. As can be seen from Fig.~\ref{fig:aAndHvTau}, this
scale factor approaches the constant $a_{1}$ at early times and the
constant $a_{2}$ at late times, and the constant $s$ determines
roughly the interval of $\tau$-time for $a(t)$ to go from $a_1$ to
$a_2$. A sufficiently large magnitude of $b$ would produce a bump or
valley in $a(t)$, but unless otherwise noted, we will take the value
of $b$ to be zero. The parameters $a_{1}$, $a_{2}$, $b$, and $s$ are
different in the initial and final asymptotically flat segments.
Where confusion would arise we will include subscripts $i$ in the
initial set of parameters and $f$ in the final set of parameters.

\subsection{Continuity of Joining Conditions}
\label{sec:joinc2}

With our choices of $a(t)$ in the three segments, we are able to
join them so that $a(t)$ and its first and second derivatives with
respect to time are everywhere continuous.  This requires that we
join the exponentially expanding segment, in which $H(t)$ has the
constant value $H_{\rm infl}$, to the initial and final segments at
the times when $H(t)$ is an extremum. This is a maximum value, when
$b=0$, and we equate this maximum value of $H(t)$ with $H_{\rm
infl}$. A simple power law form of the scale factor, such as that of
a radiation-dominated universe, could not be used to simultaneously
maintain the continuity of the scale factor and its first and second
derivatives when matched directly to the inflationary segment of
exponential expansion.  An application of these methods of matching
continuously to $C^{2}$ for the radiation reaction of the
electromagnetic force is given in the Appendix~\ref{appendix}.

\subsubsection{3.1.2.1\ \ \ Matching Continuously to Second Derivative}
\addcontentsline{toc}{subsection}{\ \ \ \ \ \ \ \ \ 3.1.2.1\ \ \
Matching Continuously to Second Derivative}

\noindent With $b_{i}=0$ and $b_{f}=0$, we then find the following
expressions. The time $\tau_{i}$ at which the first segment joins to
the exponential segment is
\begin{equation}
\tau_{i}=s_{i}\ln\left(\frac{3a_{1i}^{\ \ 4}-3a_{2i}^{\ \
4}+C_{i}}{8a_{2i}^{\ \ 4}}\right).
\end{equation}
The constant $a(t_{1})$ in Eq.~(\ref{eq:infl}) is
\begin{equation}
a(t_{1})=\left(\frac{-3a_{1i}^{\ \ 4}-3a_{2i}^{\ \
4}+C_{i}}{2}\right)^{1/4}.
\end{equation}
Because the maximum value of $H(t)$ in the first segment must equal
$H_{\rm infl}$, we find that
\begin{eqnarray}
H_{\rm infl}=&&\left[\frac{2^{3/4}\left(-a_{1i}^{\ \ 4}+a_{2i}^{\ \
4}\right)}{a_{2i}^{\ \ 4}\left(11a_{1i}^{\ \ 4}-3a_{2i}^{\ \
4}+C_{i}\right)^{2}s_{i}}\right]
\nonumber\\
&&\times\left(-3a_{1i}^{\ \ 4}-3a_{2i}^{\ \ 4}+C_{i}\right)^{1/4}
\nonumber\\
&&\times\left(3a_{1i}^{\ \ 4}-3a_{2i}^{\ \ 4}+C_{i}\right),
\end{eqnarray}
where
\begin{equation}
C_{i}\equiv\sqrt{9a_{1i}^{\ \ 8}+46a_{1i}^{\ \ 4}a_{2i}^{\ \
4}+9a_{2i}^{\ \ 8}}.
\end{equation}

Once we choose values for $a_{1f}$ and $a_{2f}$, the remaining
constants are determined to have the following values:
\begin{eqnarray}
s_{f}=&&\left[\frac{2^{3/4}\left(-a_{1f}^{\ \ 4}+a_{2f}^{\ \
4}\right)}{a_{2f}^{\ \ 4}\left(11a_{1f}^{\ \ 4}-3a_{2f}^{\ \
4}+C_{f}\right)^{2}H_{\rm infl}}\right]
\nonumber\\
&&\times\left(-3a_{1f}^{\ \ 4}-3a_{2f}^{\ \ 4}+C_{f}\right)^{1/4}
\nonumber\\
&&\times\left(3a_{1f}^{\ \ 4}-3a_{2f}^{\ \ 4}+C_{f}\right).
\end{eqnarray}
We denote the parameter $\tau$ of Eq.~(\ref{eq:aHG}) as $\tau'$ in
the final segment.  At the time $\tau'_{f}$ when the exponential
segment joins to the final segment, we find that
\begin{equation}
\tau'_{f}=s_{f}\ln\left(\frac{3a_{1f}^{\ \ 4}-3a_{2f}^{\ \
4}+C_{f}}{8a_{2f}^{\ \ 4}}\right).
\end{equation}
The corresponding proper time $t$ at which the exponential segment
joins to the final segment is
\begin{equation}
t_{2}=\frac{1}{4H_{\rm infl}}\ln\left(\frac{-3a_{1f}^{\ \
4}-3a_{2f}^{\ \ 4}+C_{f}}{-3a_{1i}^{\ \ 4}-3a_{2i}^{\ \
4}+C_{i}}\right)+t_{1},
\end{equation}
where
\begin{equation}
C_{f}\equiv\sqrt{9a_{1f}^{\ \ 8}+46a_{1f}^{\ \ 4}a_{2f}^{\ \
4}+9a_{2f}^{\ \ 8}}.
\end{equation}

See Fig.~\ref{fig:diagram} for a schematic diagram of how we match
our segments of the scale factor together.

Fig.~\ref{fig:patch} shows
\begin{figure}[hbtp]
\includegraphics[scale=2.5]{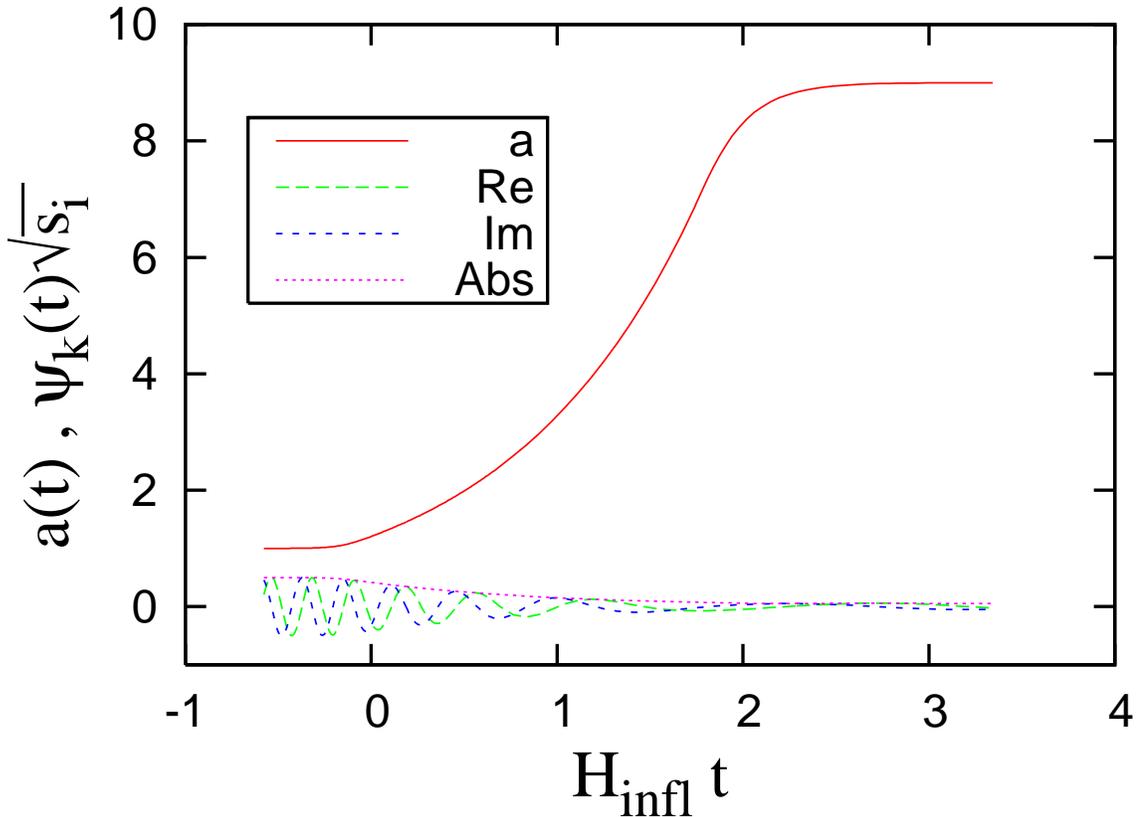}
\caption{\label{fig:patch} Matching Boundary Conditions.}
\end{figure}
an example of our composite scale factor and a particular
dimensionless solution to the evolution equation, where both are
plotted versus dimensionless time. This example shows our composite
scale factor over a moderate expansion of $\sim2$ e-folds.  The
scale factor, $a(t)$, is continuous, as are $\dot{a}(t)$ and
$\ddot{a}(t)$. The parameters for the first asymptotically flat
segment are $a_{1i}=1$, $a_{2i}=2$, and $s_{i}=1$.  The free
parameters of the end asymptotically flat segment are $a_{2f}=9$ and
$a_{1f}=6$. We choose $b_{i}=b_{f}=0$. We plot the $k=2$ Fourier
mode of $\sqrt{s_{i}}\psi_{k}$ alongside the scale factor to show
how this representative evolution solution changes with respect to
the scale factor.  The real part of $\sqrt{s_{i}}\psi_{k}$, ``Re,"
the imaginary part of $\sqrt{s_{i}}\psi_{k}$, ``Im," and the
magnitude of $\sqrt{s_{i}}\psi_{k}$, ``Abs," are all plotted.

\subsubsection{3.1.2.2\ \ \ Avoidance of Divergent Energy Density}
\addcontentsline{toc}{subsection}{\ \ \ \ \ \ \ \ \ 3.1.2.2\ \ \
Avoidance of Divergent Energy Density}

\noindent We have checked our method against known mathematical
theorems. One such theorem is that in an oscillator with a changing
frequency, the quantity $E/\omega$ is conserved if the changes in
frequency are made continuously in all derivatives with respect to
time; however, if any of the derivatives of the frequency with
respect to time are discontinuous, then this introduces changes to
the conserved quantity of order $N$, where the $N$-th derivative is
the first discontinuous derivative \cite{Kulsrud}.  It is also shown
by \cite{Littlewood} that for adiabatic changes, the changes to the
conserved quantity fall off with increasing frequency faster than
any power of the frequency. We find in this conserved quantity a
close analogy with the average number of particles created per mode
for high-energy particles, which are those particles whose
wavelengths have not yet exited the Hubble radius before the end of
inflation. It is found in Ref.~\cite{Parker1}, that when the scale
factor is changed adiabatically, the amount of particle production
falls off with frequency faster than any power of the frequency. The
dependence of high-frequency particle production upon the continuity
of the scale factor is also noted in \cite{Chung}. The scale factor
must maintain continuity in the zeroth, first, and second
derivatives to avoid an ultraviolet divergence in the energy
density.  This is the reason why we choose matching conditions that
are continuous in $a(t)$, $H(t)$, and $\dot{H}(t)$.  We could in
principle maintain continuity in higher derivatives of our composite
scale factor, as well, which would further reduce the amount of
high-energy particle production. This further reduction in the
high-energy particles would not appreciably improve upon any of our
qualitative or quantitative results. The need for $C^{2}$ matching
conditions when trying to calculate a finite energy density was
previously realized by \cite{Ford1}. In the work of
\cite{Allen3,Yajnik} upon the creation of gravitons during
inflation, the scale factor is not $C^{2}$, and both authors adopt a
UV-cutoff frequency. The author of \cite{Allen3} recognizes the
dependence of high-energy particle production upon the transition
from de Sitter space to a radiation dominated universe, and he
attributes the entire amount of high-energy particle production to
the instantaneous change in the Ricci scalar curvature given by
Eq.~(\ref{eq:ricciscalar}) from $12H^{2}$ during inflation to $0$ in
a radiation dominated universe. In \cite{Parker3}, Parker has shown
that massless gravitons satisfying a conformally invariant spin-2
field would not be produced for any $a(t)$. However, an Einstein
graviton that instead satisfied a weak field approximation such as
Eq.~(\ref{eq:lineargrav}), which in vacuum would lead to
$\Box\bar{h}_{ab}=0$, is not conformally invariant. (We use here the
definition $\bar{h}_{ab}\equiv h_{ab}-\frac{1}{2}h\eta_{ab}$, and we
work in the Lorentz gauge where $\bar{h}^{ab}{}_{,\beta}=0$, which
means $\bar{h}_{,\beta}=-h_{,\beta}=0$ \cite{Misner}.) This is
analogous to a massless, minimally-coupled Klein-Gordon field
equation of the form of Eq.~(\ref{eq:kleingordon}), except for the
two polarizations ($h_{+}$ and $h_{\times}$) of gravitational waves
\cite{Lifshitz,Grishchuk1,Grishchuk2,Ford2}. This means that for
quanta of this linear field, we would expect the same results for
average number of quanta created per mode for each polarization;
therefore, $\left|\beta_{k}\right|^{2}_{\rm Einstein\
graviton}=2\left|\beta_{k}\right|^{2}_{\rm scalar}$.

\section{Solutions to the Evolution Equation}
\label{sec:evoasymp}

Consider an inflaton field composed of a spatially homogeneous term
plus a first order perturbation,
\begin{equation}
\phi(\vec{x},t)=\phi^{(0)}(t)+\delta\phi(\vec{x},t).
\end{equation}
We investigate, in units of $\hbar=c=1$, a minimally-coupled scalar
field that obeys Eq.~(\ref{eq:evolutionequation}), which we will
refer to as the evolution equation:
\begin{equation}
\partial_{t}^{\
2}\delta\phi+3H\partial_{t}\delta\phi-a^{-2}(t)\sum_{i=1}^{3}\partial_{i}^{\
2}\delta\phi+m(\phi^{(0)})^{2}\delta\phi=0.
\end{equation}
The mass term is related to the inflationary potential by
\begin{equation}
m(\phi^{(0)})^{2}=\frac{d^{2}V}{d(\phi^{(0)})^{2}}.
\end{equation}
For simplicity, we take $m(\phi^{(0)})^{2}$ as a constant, $m^{2}$.
This is an effective mass, and from now on $m^{2}$ will refer only
to this effective mass, which may or may not be the same as the mass
of the scalar field, which we will call $m_{\rm scalar}$. In
Eq.~(\ref{eq:vprimeprime}), we show how $m^{2}$ could incorporate a
scalar coupling to the background curvature.  In what follows, we
will assume the minimally coupled case of $\xi=0$, even though the
$m^{2}$ term could include a non-zero coupling term if the curvature
were also constant.  (In the asymptotically flat segments of our
composite scale factor the Ricci scalar curvature is not a
constant.) We note that the massless, conformally-coupled case of
$m_{\rm scalar}=0$ and $\xi=1/6$ (in a 4-dimensional spacetime)
would be conformally-invariant. In the conformally-invariant case
the metric tensor and field can be deformed continuously at all
points as
\begin{eqnarray}
g_{ab}(x)&\rightarrow&\tilde{g}_{ab}(x)=\Omega(x)^{2}g_{ab}(x),\\
\phi(x)&\rightarrow&\tilde{\phi}(x)=\Omega(x)^{\rm const}\phi(x),
\end{eqnarray}
where $\Omega(x)^{2}$ is a continuous, finite, real, scalar
function; in the conformally-invariant case, no particle production
occurs \cite{Parker1,Parker2,Parker3,Birrell1,Parker5}.

The quantized field $\delta\phi$ can be written in terms of the
early time creation and annihilation operators,
$A_{\vec{k}}^{\dagger}$ and $A_{\vec{k}}$, as
\begin{equation}
\delta\phi=\sum_{\vec{k}}\left(A_{\vec{k}}f_{\vec{k}}+
A_{\vec{k}}^{\dagger}f_{\vec{k}}^{*}\right), \label{eq:phi}
\end{equation}
where
\begin{equation}
f_{\vec{k}}=V^{-\frac{1}{2}}e^{i\vec{k}\cdot\vec{x}}\psi_{k}(t(\tau)).
\label{eq:wave}
\end{equation}
We are imposing periodic boundary conditions upon a cubic coordinate
volume, $V=L^{3}$. In the continuum limit $L$ would go to infinity.
The function $\psi_{k}(t)$ satisfies
\begin{equation}
\partial_{t}^{2}\psi_{k}(t)+3H\partial_{t}\psi_{k}(t)+
\frac{k^{2}}{a^{2}(t)}\psi_{k}(t)+m^{2}\psi_{k}(t)=0,
\label{eq:evot}
\end{equation}
where $k=2\pi n/L$, with $n$ an integer. Because the creation and
annihilation operators in Eq.~(\ref{eq:phi}) correspond to particles
at early times, we require that $\psi_{k}$ satisfies the early-time
positive frequency condition
\begin{equation}
\lim_{\tau\rightarrow -\infty}\psi_{k}(t(\tau)) \sim
\frac{1}{\sqrt{2a_{1i}{}^{3}\, \omega_{1i}(k)} }
e^{-ia_{1i}{}^{3}\,\omega_{1i}(k)\,\tau}, \label{eq:minkowski}
\end{equation}
where  $\omega_{1i}(k) \equiv \sqrt{(k/a_{1i})^2 + m^2}$.

At late times, this solution will have the asymptotic form
\begin{eqnarray}
\lim_{\tau'\rightarrow \infty}\psi_{k}(t(\tau')) &\sim&
\frac{1}{\sqrt{2a_{2f}{}^{3}\, \omega_{2f}(k)} }
\bigg[\alpha_{k}e^{-ia_{2f}{}^{3}\,\omega_{2f}(k)\,\tau'}
\nonumber\\
&&+ \beta_{k}e^{ia_{2f}{}^{3}\,\omega_{2f}(k)\,\tau'}\bigg],
\label{eq:alphabeta}
\end{eqnarray}
where $\omega_{2f}(k) \equiv \sqrt{(k/a_{2f})^2 + m^2}$.

\subsection{Joining Conditions}

\label{sec:match}

Consider a spacetime composed of three segments of the scale factor,
$a(t)$, in a homogeneous background metric given by
Eq.~(\ref{eq:metric}). For an example, see Figs.~\ref{fig:diagram}
and~\ref{fig:patch}. The first and second segments are joined at the
time $t_{1}$, and the second and third segments are joined at the
time $t_{2}$.

The quantities $\psi_{k}$ and $d\psi_{k}/dt$ are continuous across
the joining regions given a continuity of the scale factor of at
least $C^{1}$. Using Eq.~(\ref{eq:massiveevotau}), it is possible to
show the conservation of the Wronskian. Multiplying
Eq.~(\ref{eq:massiveevotau}) by its conjugate leads to
\begin{equation}
\frac{d^{2}\psi_{k}(t)^{*}}{d\tau^2}\psi_{k}(t)=\frac{d^{2}\psi_{k}(t)}{d\tau^2}\psi_{k}(t)^{*}.
\end{equation}
Integrating by parts shows
\begin{equation}
\left[\psi_{k}(t)\frac{d\psi_{k}(t)^{*}}{d\tau}-\psi_{k}(t)^{*}\frac{d\psi_{k}(t)}{d\tau}\right]_{\rm
boundary}=0.
\end{equation}
Since the boundary conditions are arbitrary, it follows with Eq.
(\ref{eq:tvtau}) that the Wronskian,
\begin{equation}
a(t)^3\left[\psi_{k}(t)\frac{d\psi_{k}(t)^{*}}{dt}-\psi_{k}(t)^{*}\frac{d\psi_{k}(t)}{dt}\right],
\end{equation}
is a constant.  Using Eq.~(\ref{eq:minkowski}), we see that this
constant is just $i$; and using Eq.~(\ref{eq:alphabeta}), we see
that $i\alpha_{k}\alpha_{k}{}^{*}-i\beta_{k}\beta_{k}{}^{*}=i$, or
\cite{Parker1}
\begin{equation}
\left|\alpha_{k}\right|^{2}-\left|\beta_{k}\right|^{2}=1.
\label{eq:check}
\end{equation}

We have two linearly independent solutions to the evolution equation
in both the second segment, with solutions $h_{1}(t)$ and
$h_{2}(t)$; and the third segment, with solutions $g_{1}(t)$ and
$g_{2}(t)$; for a total of four separate functions. These functions
are multiplied by constant coefficients that we must determine.
During the second segment, from $t_{1}$ to $t_{2}$, we have:
\begin{eqnarray}
\psi_{k}(t)=Ah_{1}(t)+Bh_{2}(t), \\
\nonumber \psi_{k}'(t)=Ah_{1}'(t)+Bh_{2}'(t).
\end{eqnarray}
For $t>t_{2}$, we have:
\begin{eqnarray}
\psi_{k}(t)=Cg_{1}(t)+Dg_{2}(t), \\
\nonumber \psi_{k}'(t)=Cg_{1}'(t)+Dg_{2}'(t).
\end{eqnarray}
If we require that $\psi_{k}(t)$ and $\psi_{k}'(t)$ be continuous at
$t_{1}$ and $t_{2}$. This imposes 4 matching conditions:
\begin{eqnarray}
Ah_{1}(t_{1})+Bh_{2}(t_{1})=\psi_{k}(t_{1}), \\
\nonumber Ah_{1}'(t_{1})+Bh_{2}'(t_{1})=\psi_{k}'(t_{1}), \\
\nonumber Cg_{1}(t_{2})+Dg_{2}(t_{2})=Ah_{1}(t_{2})+Bh_{2}(t_{2}), \\
\nonumber
Cg_{1}'(t_{2})+Dg_{2}'(t_{2})=Ah_{1}'(t_{2})+Bh_{2}'(t_{2}).
\end{eqnarray}

Given the values of $\psi_{k1}$ and $\psi_{k1}'$, and the matching
conditions
\begin{eqnarray}
\label{eq:A1}
Ah_{1}(t_{1})+Bh_{2}(t_{1})&=&\psi_{k}(t_{1})=\psi_{k1}, \\
\nonumber Ah_{1}'(t_{1})+Bh_{2}'(t_{1})&=&\psi_{k}'(t_{1})=\psi_{k1}', \\
\nonumber Cg_{1}(t_{2})+Dg_{2}(t_{2})&=&Ah_{1}(t_{2})+Bh_{2}(t_{2}), \\
\nonumber
Cg_{1}'(t_{2})+Dg_{2}'(t_{2})&=&Ah_{1}'(t_{2})+Bh_{2}'(t_{2}),
\end{eqnarray}
we wish to calculate the constant coefficients $C$ and $D$ in terms
of the functions $h_{1}(t)$, $h_{2}(t)$, $g_{1}(t)$, and $g_{2}(t)$;
and the values of $\psi_{k1}$, $\psi_{k1}'$, $t_{1}$, and $t_{2}$.
(Here a prime denotes derivative with respect to $t$.) Rearranging
the first two matching conditions leads to
\begin{eqnarray}
B=\left[\frac{\psi_{k1}-Ah_{1}}{h_{2}}\right]_{t=t_{1}}, \\
\nonumber
A=\left[\frac{\psi_{k1}'-Bh_{2}'}{h_{1}'}\right]_{t=t_{1}}.
\end{eqnarray}
Combining these two equations leads to
\begin{eqnarray}
A=\left[\frac{\psi_{k1}'h_{2}-\psi_{k1}
h_{2}'}{h_{1}'h_{2}-h_{1}h_{2}'}\right]_{t=t_{1}},
\nonumber\\
B=\left[\frac{\psi_{k1}'h_{1}-\psi_{k1}
h_{1}'}{h_{2}'h_{1}-h_{2}h_{1}'}\right]_{t=t_{1}}.
\end{eqnarray}
At the time $t_{2}$ we have:
\begin{eqnarray}
\psi_{k}(t_{2})&=&Ah_{1}(t_{2})+Bh_{2}(t_{2})
\nonumber\\
&=&\bigg\{\left[\frac{\psi_{k1}'h_{2}-\psi_{k1}
h_{2}'}{h_{1}'h_{2}-h_{1}h_{2}'}\right]_{t=t_{1}}h_{1}(t_{2})
\nonumber\\
&&+\left[\frac{\psi_{k1}'h_{1}-\psi_{k1}
h_{1}'}{h_{2}'h_{1}-h_{2}h_{1}'}\right]_{t=t_{1}}h_{2}(t_{2})\bigg\},
\end{eqnarray}
and
\begin{eqnarray}
\psi_{k}'(t_{2})&=&Ah_{1}'(t_{2})+Bh_{2}'(t_{2})
\nonumber\\
&=&\bigg\{\left[\frac{\psi_{k1}'h_{2}-\psi_{k1}
h_{2}'}{h_{1}'h_{2}-h_{1}h_{2}'}\right]_{t=t_{1}}h_{1}'(t_{2})
\nonumber\\
&&+\left[\frac{\psi_{k1}'h_{1}-\psi_{k1}
h_{1}'}{h_{2}'h_{1}-h_{2}h_{1}'}\right]_{t=t_{1}}h_{2}'(t_{2})\bigg\}.
\end{eqnarray}
Let us also define $\psi_{k2}\equiv\psi_{k}(t_{2})$ and
$\psi_{k2}'\equiv\psi_{k}'(t_{2})$.  In terms of $\psi_{k2}$ and
$\psi_{k2}'$ the last two boundary conditions in Eq.~(\ref{eq:A1})
become
\begin{eqnarray}
C&=&\left(\frac{\psi_{k2}'g_{2}-\psi_{k2}
g_{2}'}{g_{1}'g_{2}-g_{1}g_{2}'}\right)_{t=t_{2}},
\nonumber\\
D&=&\left(\frac{\psi_{k2}'g_{1}-\psi_{k2}
g_{1}'}{g_{2}'g_{1}-g_{2}g_{1}'}\right)_{t=t_{2}}.
\end{eqnarray}
Substituting for $\psi_{k2}$ and $\psi_{k2}'$ yields
\begin{eqnarray}
C=\left(\frac{[Ah_{1}'+Bh_{2}']g_{2}-[Ah_{1}+Bh_{2}]
g_{2}'}{g_{1}'g_{2}-g_{1}g_{2}'}\right)_{t=t_{2}},
\nonumber\\
D=\left(\frac{[Ah_{1}'+Bh_{2}']g_{1}-[Ah_{1}+Bh_{2}]
g_{1}'}{g_{2}'g_{1}-g_{2}g_{1}'}\right)_{t=t_{2}}.
\end{eqnarray}
Finally, expressing $A$ and $B$ in terms of the given values of
$\psi_{k1}$ and $\psi_{k1}'$ specified at $t_{1}$ leads to
\begin{eqnarray}
C&=&\frac{1}{\left(g_{1}'g_{2}-g_{1}g_{2}'\right)_{t=t_{2}}}
\nonumber\\
&&\times\bigg\{\left[\frac{\psi_{k1}'h_{2}-\psi_{k1}
h_{2}'}{h_{1}'h_{2}-h_{1}h_{2}'}\right]_{t=t_{1}}\left(h_{1}'g_{2}-h_{1}g_{2}'\right)_{t=t_{2}}
\nonumber\\
&&+\left[\frac{\psi_{k1}'h_{1}-\psi_{k1}
h_{1}'}{h_{2}'h_{1}-h_{2}h_{1}'}\right]_{t=t_{1}}\left(h_{2}'g_{2}-h_{2}g_{2}'\right)_{t=t_{2}}\bigg\},
\end{eqnarray}
and
\begin{eqnarray}
D&=&\frac{1}{\left(g_{2}'g_{1}-g_{2}g_{1}'\right)_{t=t_{2}}}
\nonumber\\
&&\times\bigg\{\left[\frac{\psi_{k1}'h_{2}-\psi_{k1}
h_{2}'}{h_{1}'h_{2}-h_{1}h_{2}'}\right]_{t=t_{1}}\left(h_{1}'g_{1}-h_{1}g_{1}'\right)_{t=t_{2}}
\nonumber\\
&&+\left[\frac{\psi_{k1}'h_{1}-\psi_{k1}
h_{1}'}{h_{2}'h_{1}-h_{2}h_{1}'}\right]_{t=t_{1}}\left(h_{2}'g_{1}-h_{2}g_{1}'\right)_{t=t_{2}}\bigg\},
\end{eqnarray}
which are the combined joining conditions for $\psi_{k}$ and
$\psi_{k}'$.

We find $\psi_{k1}$ and $\psi_{k1}'$ from the solution to the
evolution equation in the initial asymptotically flat segment of the
scale factor.  In the massless case, this solution is given by
Eq.~(\ref{eq:hypergeometric}).  The functions $h_{1}(t)$ and
$h_{2}(t)$ are to be related to the evolution equation solutions in
the inflationary middle segment of the scale factor. Comparing this
with Eqs.~(\ref{eq:infleigen}) and~(\ref{eq:massinfleigen}) shows
$A=E(k)$ and $B=F(k)$. Similarly, the functions $g_{1}(t)$ and
$g_{2}(t)$ are to be related to to the evolution equation solutions
in the final asymptotically flat segment of the scale factor, and we
will later make the identification $C=N_{1}(k)$ and $D=N_{2}(k)$,
where the coefficients $N_{1}(k)$ and $N_{2}(k)$ are defined through
their use in Eq.~(\ref{eq:hypergeometric2}).

\subsection{Exact Massless Solutions}

\label{sec:masslesspc}

We will first consider the case, $m=0$.  Rewriting the evolution
equation, Eq.~(\ref{eq:evot}), in terms of $\tau$ instead of $t$
leads to
\begin{equation}
\frac{d^{2}\psi_{k}}{d\tau^2}=-k^{2}a^{4}\psi_{k}. \label{eq:evotau}
\end{equation}

For the first segment of our composite scale factor, the solution of
(\ref{eq:evotau}) having positive frequency form
(\ref{eq:minkowski}) at early times is the hypergeometric function
\cite{Parker4,Parker5,Parker8,Parker9}
\begin{eqnarray}
\psi_{k}(t(\tau))&=&\frac{1}{\sqrt{2a_{1i}{}^{2}k}}
e^{-ika_{1i}{}^{2}\tau} F(-ika_{1i}{}^{2}s_{i} +ika_{2i}{}^2 s_{i},
\nonumber\\
&&-ika_{1i}{}^{2}s_{i}-ika_{2i}{}^{2}s_{i};1-2ika_{1i}{}^2
s_{i};-e^{\frac{\tau}{s_{i}}}),
\label{eq:hypergeometric}
\end{eqnarray}
where $F(a,b;c;d)$ is the hypergeometric function as defined in
\cite[see 15.1.1]{Abramowitz}:
\begin{equation}
F(a,b;c;z)=\frac{\Gamma(c)}{\Gamma(a)\Gamma(b)}\sum_{n=0}^{\infty}\frac{\Gamma(a+n)\Gamma(b+n)}{\Gamma(c+n)}\,\frac{z^{n}}{n!}.
\end{equation}

For the exponentially expanding segment of the scale factor in the
massless case ($V''=0$ in Eq.~(\ref{eq:bunchdavies}) above)
\begin{equation}
\psi_{k}(t)=a(t)^{-\frac{3}{2}}\left[E(k)H_{\frac{3}{2}}^{(1)}\left(\frac{k}{a(t)H_{\rm
infl}}\right)+F(k)H_{\frac{3}{2}}^{(2)}\left(\frac{k}{a(t)H_{\rm
infl}}\right)\right],
\label{eq:infleigen}
\end{equation}
where $H^{(1)}$ and $H^{(2)}$ are the Hankel functions of the first
and second kind.  The variables $t$ and $\tau$ are related by
Eq.~(\ref{eq:tvtau}). The coefficients $E(k)$ and $F(k)$ are
determined by the matching conditions of the first joining point at
$t=t_{1}$. We note that the finite period of exponential inflation
lacks the full symmetries of a de Sitter universe. In the pure de
Sitter case, as shown in \cite{Allen2}, the $k=0$ mode has to be
chosen in a special way to avoid infrared divergences. For our
$a(t)$, infrared divergences do not arise (see
Sec.~\ref{sec:masslesspc}).

For the final segment of our composite scale factor, the solution of
the evolution equation (\ref{eq:evotau}) is a linear combination of
hypergeometric functions \cite{Parker4,Parker5,Parker8,Parker9}:
\begin{eqnarray}
\psi_{k}(t(\tau'))
&=&N_{1}(k)e^{-ika_{1f}{}^{2}\tau'}F(-ika_{1f}{}^{2}s_{f}
+ika_{2f}{}^{2} s_{f},
\nonumber\\
&&-ika_{1f}{}^{2}s_{f}-ika_{2f}{}^{2}s_{f};1-2ika_{1f}{}^{2}
s_{f};-e^{\frac{\tau'}{s_{f}}})
\nonumber\\
&&+N_{2}(k)e^{ika_{1f}{}^{2}\tau'}F(ika_{1f}{}^{2}s_{f}
+ika_{2f}{}^{2} s_{f},
\nonumber\\
&&ika_{1f}{}^{2}s-ika_{2f}{}^{2}s_{f};1+2ika_{1f}{}^{2}
s_{f};-e^{\frac{\tau'}{s_{f}}}),
\label{eq:hypergeometric2}
\end{eqnarray}
where the coefficients $N_{1}(k)$ and $N_{2}(k)$ are determined by
the matching conditions of the second joining point at $t=t_{2}$. An
example of the evolution for a particular mode is plotted for a
specific choice of parameters using our composite scale factor in
Figs.~\ref{fig:patch} and~\ref{fig:fluc}.

Fig.~\ref{fig:fluc} shows
\begin{figure}[hbtp]
\includegraphics[scale=2.25]{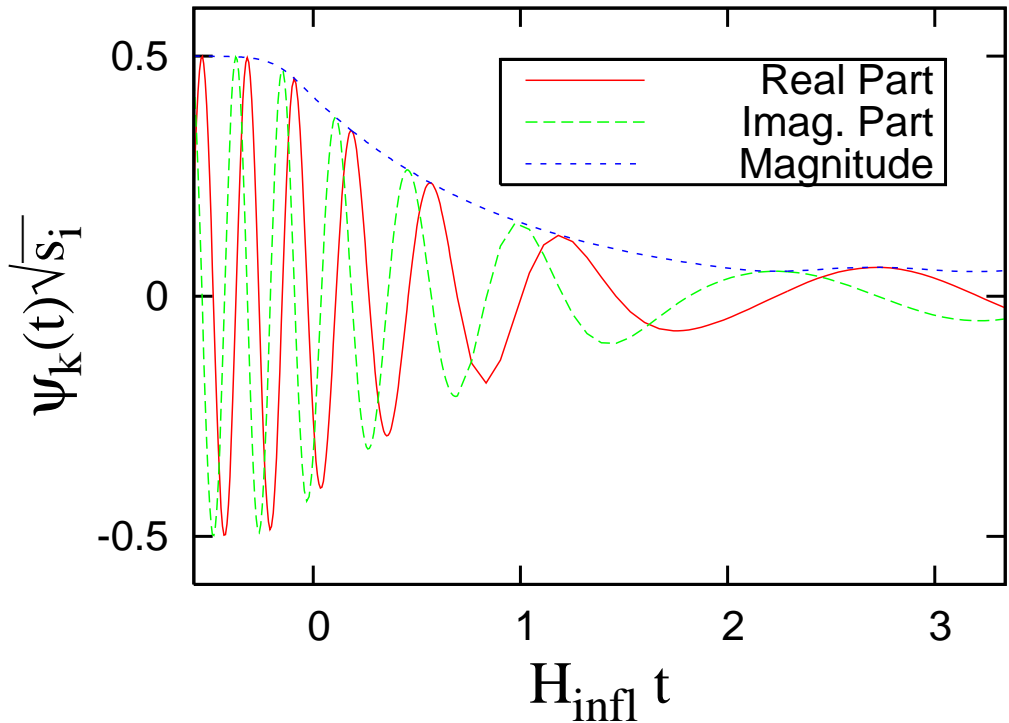}
\caption{\label{fig:fluc} A Dimensionless Solution to the Evolution
Equation.}
\end{figure}
a dimensionless solution to the massless evolution equation, where
the $k=2$ Fourier mode is plotted versus dimensionless time for the
same composite scale factor used in Fig.~\ref{fig:diagram}. The real
part of $\sqrt{s_{i}}\psi_{k}$, the imaginary part of
$\sqrt{s_{i}}\psi_{k}$, and the magnitude of $\sqrt{s_{i}}\psi_{k}$
are all plotted.

With joining conditions for the segments of the scale factor, the
derived solution to the evolution equation can be matched up with
the known solution for the exponential expansion of an inflationary
segment by matching $\delta\phi_{k}(t)$ and its time derivative
across the boundary conditions.  See Figure~\ref{fig:invar}
\begin{figure}[hbtp]
\includegraphics[scale=2.25]{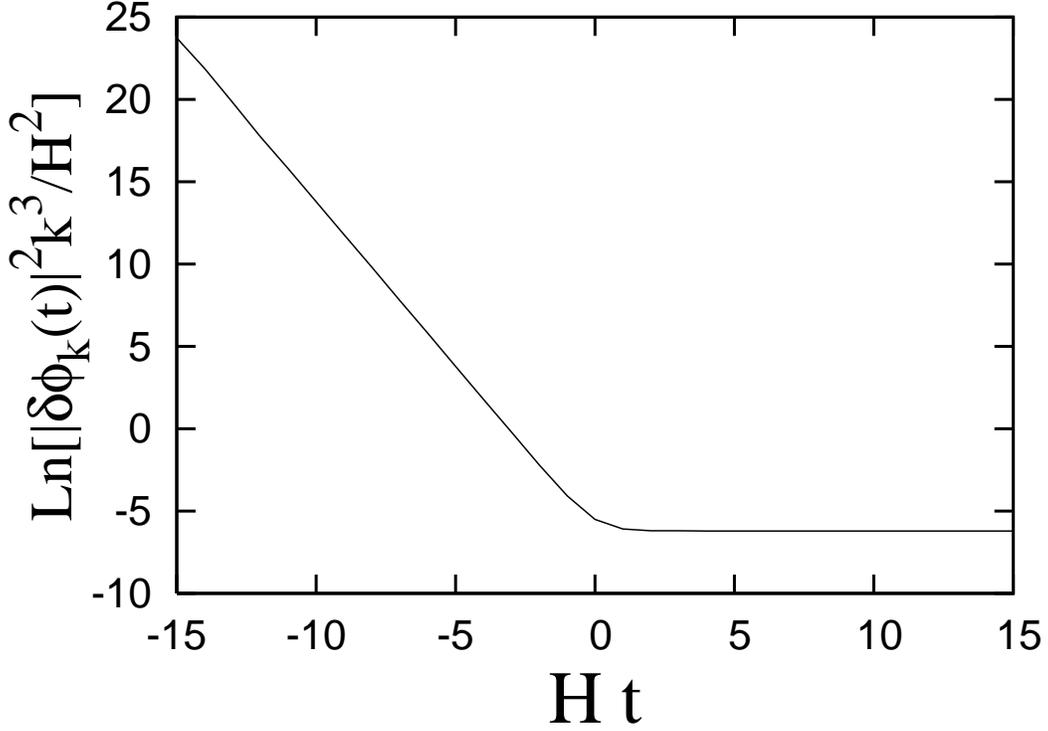}
\caption{\label{fig:invar} Scale-Invariance of Inflaton
Perturbations in Continuum Limit.}
\end{figure}
for the evolution of modes in the middle of a long inflationary
period for the massless case. The time $t$ is taken to be zero when
$k=a(t)H$ (when the plotted mode exits the Hubble radius) and
depends on the mode number $k$.  Multiplied by $k^{3}/H_{\rm
inflation}^{2}$ and plotted against this mode-dependent time, all of
the different fluctuation modes align along the same curve in this
graph.  This shows, in the massless case, the scale-invariance of
the spectrum for those modes that exit the Hubble radius during a
period of constant $H(t)$.

\subsection{Approximations to Massive Solution}

In the case of a massive scalar field, the evolution equation,
Eq.~(\ref{eq:evot}), can be written in terms of $\tau$ as
\begin{equation}
\frac{d^{2}\psi_{k}}{d\tau^2}=-(k^{2}a^{4}+m^{2}a^{6})\psi_{k}.
\label{eq:massiveevotau}
\end{equation}
For the middle, inflationary segment of our scale factor, our
solution given by Eq.~(\ref{eq:bunchdavies}) is
\begin{equation}
\psi_{k}(t)=a(t)^{-\frac{3}{2}}\left[E(k)H_{\sqrt{\frac{9}{4}-m_{H}^{\
2}}}^{(1)}\left(\frac{k}{a(t)H_{\rm
infl}}\right)+F(k)H_{\sqrt{\frac{9}{4}-m_{H}^{\
2}}}^{(2)}\left(\frac{k}{a(t)H_{\rm infl}}\right)\right],
\label{eq:massinfleigen}
\end{equation}
where we define $m_{H}$ in terms of the effective mass by
\begin{equation}
m_{H}\equiv\frac{m}{H_{\rm infl}}.
\end{equation}

We know the solution to the evolution equation for the region of the
scale factor given by Eq.~(\ref{eq:infl}) exactly, but we do not
have an analytic solution for an asymptotically flat segment of our
scale factor except for the trivial case of a constant scale factor.
We instead use one of two different approximations that we find
reduce to the same numerical solutions in their mutual realms of
applicability: the effective-k approach and the dominant-term
approach.

\subsubsection{3.2.2.1\ \ \ Effective-k Approach}
\addcontentsline{toc}{subsection}{\ \ \ \ \ \ \ \ \ 3.2.2.1\ \ \
Effective-k Approach}

\noindent In the first of these approximations, the effective-k
approach, we choose our initial and final asymptotically flat
segments of the scale factor such that $a_{1i}\simeq a_{2i}$ and
$a_{1f}\simeq a_{2f}$.  The middle segment of our scale factor,
under these conditions, is thus where almost all of the change in
the scale factor occurs, and we make use of our exact solution in
this region.  In the beginning and final asymptotically flat
segments we make the transformation $k\rightarrow k_{\rm eff}$,
where $k_{\rm eff}$ is an effective $k$ defined in the initial
region as
\begin{equation}
k_{i\,\rm eff}\equiv\sqrt{k^{2}+m^{2}a_{1i}^{\ \ 2}},
\end{equation}
and in the final region by
\begin{equation}
k_{f\,\rm eff}\equiv\sqrt{k^{2}+m^{2}a_{2f}^{\ \ 2}}.
\end{equation}
In the limit that $a_{2}=a_{1}$ in a given segment, the
approximation becomes exact and reduces to the known Minkowski flat
space solution of
\begin{equation}
\psi_{k}(t(\tau))=\frac{1}{\sqrt{2a^{3}\omega}}\left[\alpha_{k}e^{-ia^{3}\omega\tau}+\beta_{k}e^{ia^{3}\omega\tau}\right],
\label{eq:psi}
\end{equation}
where $\omega$ is given by
\begin{equation}
\omega\equiv\sqrt{\frac{k^{2}}{a^{2}}+m^{2}}.
\end{equation}
The closer the ratio $a_{2}/a_{1}$ comes to unity in an
asymptotically flat segment of the scale factor, the more
trustworthy the effective-k approach becomes. If the two parameters
are precisely equal, however, then the scale factor becomes a
constant in time and derivatives of the scale factor are equal to
zero. In such a case where $a_{2}=a_{1}$, we cannot join to the
inflationary middle segment continuously in any derivatives of the
scale factor. When $a_{1f}\simeq a_{2f}$ in the end segment of our
composite scale factor, we observe ultraviolet particle production
due to the rapid breaking, or deceleration, of the scale factor's
expansion. This is true regardless of effective mass, because this
``extended" region of particle production occurs where the mass is
negligible and $(k/a(t))^{2}\gg m^{2}$.

\subsubsection{3.2.2.2\ \ \ Dominant Term Approach}
\addcontentsline{toc}{subsection}{\ \ \ \ \ \ \ \ \ 3.2.2.2\ \ \
Dominant Term Approach}

\noindent The Effective-$k$ Approach works very well| especially for
the case where the final asymptotically flat scale factor is
parameterized such that $a_{1f}\simeq a_{2f}$.  The Effective-$k$
Approach need not be as accurate when $a_{1f}\ll a_{2f}$, and for
this situation we introduce an alternate massive approximation, that
of the Dominant Term Approach.  In this case we introduce a new
asymptotically flat scale factor that yields an exact solution in
the limit that $k\rightarrow0$.  For a fixed mass, this
approximation becomes exceedingly close to the exact solution
whenever $|m|\gg k/a(t)$. In the Dominant Term Approach, when
$k/a\gg |m|$, we use the asymptotically flat scale factor given
above along with the massless solution; and when $|m|\gg k/a(t)$, we
use a new asymptotically flat scale factor and its associated zeroth
Fourier mode solution.  These two solutions can be matched up for
the case of modes in the intermediary-$q_{2}$ region, where we would
use the massless solution for the initial asymptotically flat scale
factor and the massive solution for the final asymptotically flat
scale factor. The Dominant Term Approach is suspect at the interface
between the small- and intermediary-$q_{2}$ behaviors and at the
interface between the intermediary- and large-$q_{2}$ behaviors,
where the justification for neglecting either the $m$-term or the
$k/a$-term is weakest.  Depending upon which term is neglected,
however, this method provides tight upper and lower limits on the
average particle production per mode even at these interfaces. When
an abrupt transition from the exponential inflation of the middle
scale factor segment to the asymptotically flat final scale factor
segment is taken to make a fair comparison, the Dominant Term
Approach is in excellent agreement with the Effective-$k$ Approach|
even at the interfaces of $q_{2}\simeq1$ and
$q_{2}\simeq\exp(-N_{e})$.  When the final transition between the
second and third scale factor segments is not taken to be abrupt,
the upper- and lower-limits place the results of the Dominant Term
Approach very close to the Effective-$k$ Approach| even at the
interfaces| and they differ only in their descriptions of the
large-$q_{2}$ behavior.  This is because the Effective-$k$ Approach
requires an abrupt end to inflation and is not a contradiction
between the two approaches, but rather is a result of the previously
mentioned fact that an abrupt transition at the end of inflation
produces a high-energy region of residual particle production.

\subsubsection{Inflaton Field of Fixed Mass and Zeroth Fourier Mode}

In units of $\hbar=c=1$, the perturbations to the inflaton field
satisfy the evolution equation for mode-$k$
\begin{equation}
\ddot{\delta\phi_{k}}+3H(t)\dot{\delta\phi_{k}}+\frac{k^{2}}{a(t)^{2}}\delta\phi_{k}+m^{2}\delta\phi_{k}=0;
\end{equation}
where a dot represents a derivative with respect to the proper time;
where $a(t)$ is the scale factor; where $H(t)\equiv\dot{a(t)}/a(t)$
is the Hubble constant, which may vary with time; and where $m$ is
taken to be a constant effective inflaton mass, which is equal to
the square root of the second derivative of the inflationary
potential with respect to the homogeneous, background part of the
inflaton field. With a change of variables from the proper time,
$t$, to a new time variable that satisfies the relationship
$d\tau\equiv a(t)^{-3}dt$; and examining the zeroth Fourier mode,
where $k=0$, which can in fact can be taken to be approximately
correct whenever $k/a(t)\ll m$, the evolution equation becomes
\begin{equation}
\frac{d^{2}\delta\phi_{0}}{d\tau^{2}}=-m^{2}a(\tau)^{6}\delta\phi_{0}.
\label{eq:evoRamopithecus}
\end{equation}
Using an analysis patterned after that which Epstein used to model
the scattering of radio waves off the ionosphere \cite{Epstein} and
that which Eckart used to model potential energy in one-dimensional
scattering in quantum mechanics \cite{Eckart}, we define a scale
factor that is asymptotically flat in both the past- and future-time
infinities as
\begin{equation}
a(\tau)=\bigg\{a_{1}^{\ 6}+e^{\tau/s}[(a_{2}^{\ 6}-a_{1}^{\
6})(e^{\tau/s}+1)+b](e^{\tau/s}+1)^{-2}\bigg\}^{\frac{1}{6}}.
\label{eq:aHGmassive}
\end{equation}
The form of this scale factor is modeled after the scale factor
first introduced by Parker \cite{Parker4,Parker5,Parker8,Parker9}
which has four adjustable parameters $a_{1}$, $a_{2}$, $s$, and $b$
that allow one to approximate a wide range of possible scale factors
$a(\tau)$. The field equation, Eq.~(\ref{eq:evoRamopithecus}), with
this scale factor, $a(\tau)$, has exact solutions in terms of
hypergeometric functions \cite{Epstein,Eckart}. With this scale
factor, Eq.~(\ref{eq:evoRamopithecus}) becomes
\begin{equation}
\frac{d^{2}\delta\phi_{0}}{d\tau^{2}}=-m^{2}\left\{a_{1}^{\
6}+e^{\tau/s}[(a_{2}^{\ 6}-a_{1}^{\
6})(e^{\tau/s}+1)+b](e^{\tau/s}+1)^{-2}\right\}\delta\phi_{0}.
\end{equation}
A change of variables to $u\equiv e^{\tau/s}$ leads to
\begin{equation}
\frac{d^{2}\delta\phi_{0}}{d(s\ln u)^{2}}=-m^{2}\left\{a_{1}^{\
6}+u[(a_{2}^{\ 6}-a_{1}^{\
6})(u+1)+b](u+1)^{-2}\right\}\delta\phi_{0}.
\end{equation}
With the chain rule, we use
\begin{eqnarray}
\frac{d^{2}\delta\phi_{0}}{d(s\ln
u)^{2}}&=&\frac{1}{s^{2}}\left(\frac{d\ln
u}{du}\right)^{-1}\frac{d}{du}\left[\left(\frac{d\ln
u}{du}\right)^{-1}\frac{d}{du}\delta\phi_{0}\right]\nonumber\\
&=&\frac{u}{s^{2}}\frac{d}{du}\left[u\frac{d}{du}\delta\phi_{0}\right]\nonumber\\
&=&\frac{u^{2}}{s^{2}}\frac{d^{2}}{du^{2}}\delta\phi_{0}+\frac{u}{s^{2}}\frac{d}{du}\delta\phi_{0}
\end{eqnarray}
to write, with a prime denoting a derivative with respect to the
variable $u$,
\begin{equation}
\delta\phi_{0}''+\frac{\delta\phi_{0}'}{u}+\frac{s^{2}m^{2}}{u^{2}}\left\{a_{1}^{\
6}+u[(a_{2}^{\ 6}-a_{1}^{\
6})(u+1)+b](u+1)^{-2}\right\}\delta\phi_{0}=0. \label{eq:genevo}
\end{equation}
Without having yet made any assumption as to the reality of
$\tau/s$, the variable $u$ may range from $-\infty$ to $+\infty$ on
the complex plane.  Portions of this evolution equation can be seen
to become infinite at $u=0$ and $u=-1$.  For the case of $u=0$,
where the evolution equation becomes
\begin{equation}
\delta\phi_{0}''+\frac{\delta\phi_{0}'}{u}+\frac{s^{2}m^{2}}{u^{2}}a_{1}^{\
6}\delta\phi_{0}=0,
\end{equation}
we use the chain rule to change variables to $v=\ln u$, where
$\partial_{u}=u^{-1}\partial_{v}$, to get
\begin{equation}
e^{-v}\partial_{v}\left(e^{-v}\partial_{v}\delta\phi_{0}\right)+e^{-2v}\partial_{v}\delta\phi_{0}+e^{-2v}s^{2}m^{2}a_{1}^{\
6}\delta\phi_{0}=0,
\end{equation}
which simplifies to
\begin{equation}
\partial_{v}{}^{2}\delta\phi_{0}=-s^{2}m^{2}a_{1}^{\
6}\delta\phi_{0},
\end{equation}
the solution of which is,
\begin{equation}
\delta\phi_{0}=e^{\pm isma_{1}{}^{3}v}=u^{\pm isma_{1}{}^{3}}.
\label{eq:solutu}
\end{equation}
For the case of $u=-1$, where the evolution equation becomes
\begin{equation}
\delta\phi_{0}''-\delta\phi_{0}'+s^{2}m^{2}\left\{a_{1}{}^{6}-b(u+1)^{-2}\right\}\delta\phi_{0}=0,
\end{equation}
we test the analog of the solution found in Eq.~(\ref{eq:solutu}) to
look for a solution of the form
\begin{equation}
\delta\phi_{0}=(u+1)^{x},
\end{equation}
and insert this into the evolution equation for the case of $u=-1$
to find
\begin{equation}
x(x-1)(u+1)^{x-2}-x(u+1)^{x-1}+s^{2}m^{2}a_{1}{}^{6}(u+1)^{x}-s^{2}m^{2}b(u+1)^{x-2}=0.
\end{equation}
Because $(u+1)=0$, the factors with the lowest exponential power of
$(u+1)^{x-2}$ dominate this equation, and at the point of $u=-1$ the
evolution equation obeys
\begin{equation}
x(x-1)(u+1)^{x-2}=s^{2}m^{2}b(u+1)^{x-2},
\end{equation}
or
\begin{equation}
x(x-1)=s^{2}m^{2}b,
\end{equation}
with solutions
\begin{equation}
x_{\pm}=\frac{1\pm\sqrt{1+4s^{2}m^{2}b}}{2},
\end{equation}
so at $u=-1$
\begin{equation}
\delta\phi_{0}=(u+1)^{x_{\pm}}.
\end{equation}
A second order differential equation has at most two distinct
solutions; therefore, our test has found all the solutions for the
case of $u=-1$. To write the $u=0$ case in an equivalent form, we
define
\begin{equation}
p_{1}\equiv isma_{1}{}^{3},
\end{equation}
such that for the $u=0$ case
\begin{equation}
\delta\phi_{0}=u^{\pm p_{1}},
\end{equation}
and define for later use
\begin{equation}
p_{2}\equiv isma_{2}{}^{3}.
\end{equation}
To find the general solution of $\delta\phi_{0}(u)$, we write
\begin{equation}
\delta\phi_{0}=(1+u)^{x_{-}}u^{-p_{1}}f[u],
\end{equation}
where the function $f[u]$ is defined by this equation.  We insert
this expression for $\delta\phi_{0}$ back into Eq.~(\ref{eq:genevo})
to get
\begin{eqnarray}
0&=&\left((1+u)^{x_{-}}u^{-p_{1}}f[u]\right)''+\frac{\left((1+u)^{x_{-}}u^{-p_{1}}f[u]\right)'}{u}\\
&&+\frac{s^{2}m^{2}}{u^{2}}\left\{a_{1}^{\ 6}+u[(a_{2}^{\
6}-a_{1}^{\
6})(u+1)+b](u+1)^{-2}\right\}(1+u)^{x_{-}}u^{-p_{1}}f[u]\nonumber,
\end{eqnarray}
which, with $s^{2}m^{2}a_{1,2}{}^{6}=-p_{1,2}{}^{2}$ and
$s^{2}m^{2}b=x_{-}x_{+}=x_{-}{}^{2}-x_{-}$, becomes
\begin{eqnarray}
0&=&x_{-}(x_{-}-1)(1+u)^{x_{-}-2}u^{-p_{1}}f[u]-p_{1}x_{-}(1+u)^{x_{-}-1}u^{-p_{1}-1}f[u]+x_{-}(1+u)^{x_{-}-1}u^{-p_{1}}f'[u]\nonumber\\
&&-p_{1}x_{-}(1+u)^{x_{-}-1}u^{-p_{1}-1}f[u]+p_{1}(p_{1}+1)(1+u)^{x_{-}}u^{-p_{1}-2}f[u]-p_{1}(1+u)^{x_{-}}u^{-p_{1}-1}f'[u]\nonumber\\
&&+x_{-}(1+u)^{x_{-}-1}u^{-p_{1}}f'[u]-p_{1}(1+u)^{x_{-}}u^{-p_{1}-1}f'[u]+(1+u)^{x_{-}}u^{-p_{1}}f''[u]\nonumber\\[2mm]
&&+\frac{x_{-}(1+u)^{x_{-}-1}u^{-p_{1}}f[u]-p_{1}(1+u)^{x_{-}}u^{-p_{1}-1}f[u]+(1+u)^{x_{-}}u^{-p_{1}}f'[u]}{u}\nonumber\\
&&+\frac{1}{u^{2}}\left\{-p_{1}{}^{2}+u[(-p_{2}{}^{2}+p_{1}{}^{2})(u+1)+x_{-}x_{+}](u+1)^{-2}\right\}(1+u)^{x_{-}}u^{-p_{1}}f[u],
\end{eqnarray}
multiplying by $(1+u)^{-x_{-}+1}u^{p_{1}+1}$ produces
\begin{eqnarray}
0&=&x_{-}(x_{-}-1)(1+u)^{-1}uf[u]-p_{1}x_{-}f[u]+x_{-}uf'[u]\nonumber\\
&&-p_{1}x_{-}f[u]+p_{1}(p_{1}+1)(1+u)u^{-1}f[u]-p_{1}(1+u)f'[u]\nonumber\\
&&+x_{-}uf'[u]-p_{1}(1+u)f'[u]+(1+u)uf''[u]\nonumber\\[2mm]
&&+\frac{x_{-}uf[u]-p_{1}(1+u)uf[u]+(1+u)uf'[u]}{u}\nonumber\\
&&+\frac{u+1}{u}\left\{-p_{1}{}^{2}+u[(-p_{2}{}^{2}+p_{1}{}^{2})(u+1)+x_{-}x_{+}](u+1)^{-2}\right\}f[u],
\end{eqnarray}
which can be simplified to
\begin{eqnarray}
0&=&u(u+1)f''+[2x_{-}u-2p_{1}(1+u)+(1+u)]f'\\
&&+\bigg[x_{-}(x_{-}-1)(1+u)^{-1}u-2p_{1}x_{-}+x_{-}+[p_{1}(p_{1}+1)-p_{1}](1+u)u^{-1}\nonumber\\
&&+\frac{u+1}{u}\left\{-p_{1}{}^{2}+u[(-p_{2}{}^{2}+p_{1}{}^{2})(u+1)+x_{-}x_{+}](u+1)^{-2}\right\}\bigg]f\nonumber,
\end{eqnarray}
which can be further simplified to
\begin{eqnarray}
0&=&u(u+1)f''+[2x_{-}u-2p_{1}(1+u)+(1+u)]f'\\
&&+\bigg[([x_{-}{}^{2}-x_{-}]u+x_{-}x_{+})(u+1)^{-1}+(p_{1}{}^{2}-p_{1}{}^{2})u^{-1}\nonumber\\
&&+(-2p_{1}x_{-}+p_{1}{}^{2}-p_{2}{}^{2}+p_{1}{}^{2}-p_{1}{}^{2}+x_{-})\bigg]f\nonumber,
\end{eqnarray}
then to
\begin{eqnarray}
0&=&u(u+1)f''+[2x_{-}u-2p_{1}(1+u)+(1+u)]f'\\
&&+\left(-2p_{1}x_{-}+p_{1}{}^{2}-p_{2}{}^{2}+x_{-}{}^{2}\right)f\nonumber,
\end{eqnarray}
and finally to
\begin{eqnarray}
0&=&u(u+1)f''+[(2x_{-}-2p_{1}+1)u+(1-2p_{1})]f'\\
&&+(x_{-}-p_{1}+p_{2})(x_{-}-p_{1}-p_{2})f\nonumber.
\end{eqnarray}
This is a hypergeometric equation and can be solved in terms of the
hypergeometric function
$f=F(x_{-}-p_{1}+p_{2},x_{-}-p_{1}-p_{2};1-2p_{1};-u)$, using the
notation of \cite{Abramowitz}.

\subsubsection{Joining Scale Factors Continuously to Second
Derivative}

To achieve a finite energy density we must maintain the continuity
of the composite scale factor to $C^{2}$ at the matching points of
the individual scale factor segments. Sec.~\ref{sec:dmem} discusses
further the need for $C^{2}$ joining conditions. See
Figure~\ref{fig:itFigures}
\begin{figure}[hbtp]
\includegraphics[scale=2.5]{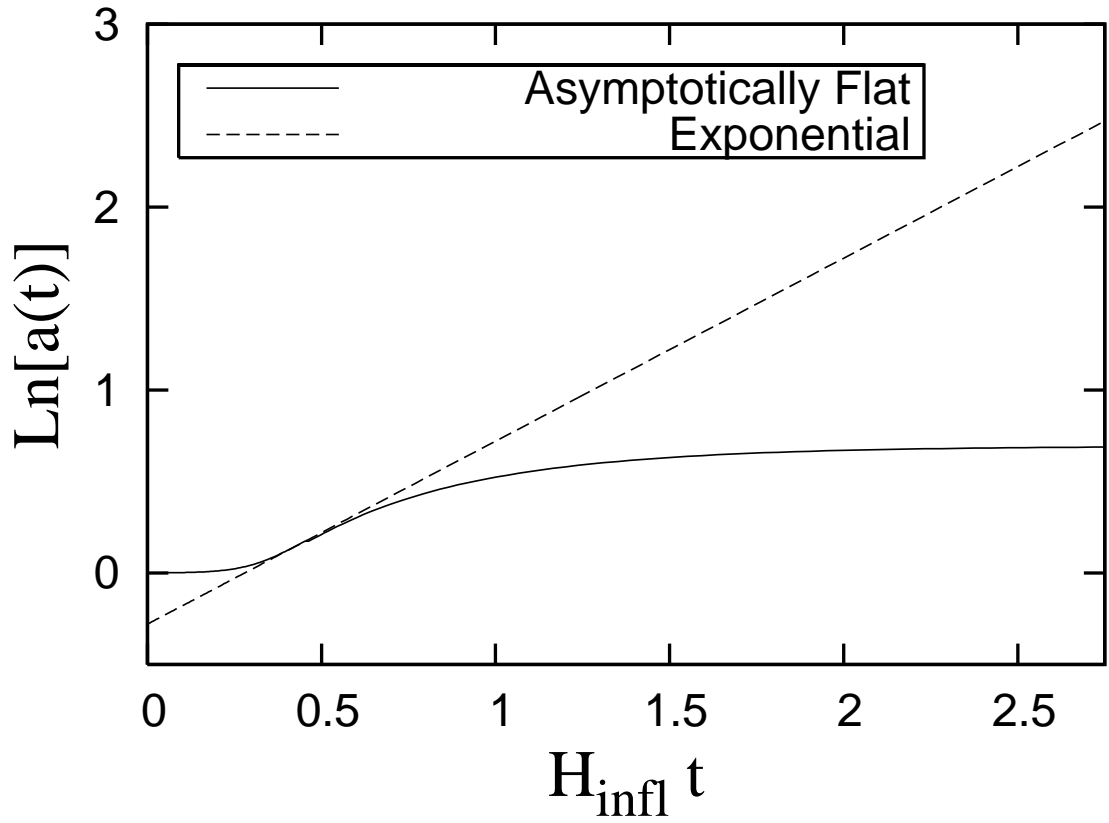}
\caption{\label{fig:itFigures} Joining Segments of Scale Factor
Continuously to $C^{2}$.}
\end{figure}
for an example of the asymptotically flat scale factor described in
the previous section joined to a region of inflation where the scale
factor grows exponentially with respect to proper time.  This graph
shows how an asymptotically flat region could be joined onto the
beginning or end of an exponential region.

To join these different scale factors continuously to the second
derivative, we note that an exponentially growing scale factor, of
the form $a(t)=a_{0}\exp(Ht)$, has a time-independent Hubble
constant.  To find a point in the asymptotically flat scale factor
described above where $\dot{H}=0$, we must find a local extremum of
$H(t)$. When $b=0$, there is a unique maximum value of $H(t)$. In a
simpler scale factor of the form $a(t)\propto t^{n}$, which
describes a radiation- or matter-dominated universe, no such point
would exist.  Using the relationship $d\tau\equiv a(t)^{-3}dt$, the
Hubble constant is $H(t)\equiv
a(t)^{-1}(da/dt)=a(\tau)^{-4}(da/d\tau)$, and its time-derivative is
$\dot{H}(t)=a(\tau)^{-3}\partial_{\tau}[a(\tau)^{-4}\partial_{\tau}a(\tau)]=a(\tau)^{-7}\partial_{\tau}{}^{2}a(\tau)-4a(\tau)^{-8}[\partial_{\tau}a(\tau)]^{2}$.
This is zero when
$(d^{2}a(\tau)/d\tau^{2})=4a(\tau)^{-1}(da/d\tau)^{2}$; in other
words, when
\begin{eqnarray}
&&\left[\frac{\left\{\frac{1}{3}-\frac{1}{6}(1+e^{-\tau/s})e^{\tau/s}\right\}(1+e^{-\tau/s})^{-3}(a_{2}^{\
6}-a_{1}^{\ 6})e^{-2\tau/s}}{s^{2}\left[a_{1}^{\
6}+(1+e^{-\tau/s})^{-1}(a_{2}^{\ 6}-a_{1}^{\
6})\right]^{\frac{5}{6}}}\right]\\[2mm]
&&+\left[\frac{-\frac{5}{36}(1+e^{-\tau/s})^{-4}(a_{2}^{\
6}-a_{1}^{\ 6})^{2}e^{-2\tau/s}}{s^{2}\left[a_{1}^{\
6}+(1+e^{-\tau/s})^{-1}(a_{2}^{\ 6}-a_{1}^{\
6})\right]^{\frac{11}{6}}}\right]\nonumber\\[2mm]
&=&4\left[a_{1}^{\ 6}+(1+e^{-\tau/s})^{-1}(a_{2}^{\ 6}-a_{1}^{\
6})\right]^{-\frac{1}{6}}\left[\frac{\frac{1}{6}(1+e^{-\tau/s})^{-2}(a_{2}^{\
6}-a_{1}^{\ 6})e^{-\tau/s}}{s\left[a_{1}^{\
6}+(1+e^{-\tau/s})^{-1}(a_{2}^{\ 6}-a_{1}^{\
6})\right]^{\frac{5}{6}}}\right]^{2},\nonumber
\end{eqnarray}
where the parameter $b$ in Eq.~(\ref{eq:genevo}) has been taken to
be zero so that there might be a unique maximum value of the Hubble
constant. To simplify this, we multiply both sides of the equation
by
$12s^{2}a(\tau)^{\frac{11}{6}}(a_{2}{}^{6}-a_{1}{}^{6})^{-1}(1+e^{-\tau/s})^{4}e^{3\tau/s}$
to get
\begin{eqnarray}
&&\left(\left\{4e^{\tau/s}-2(1+e^{-\tau/s})e^{2\tau/s}\right\}\left[a_{1}^{\
6}(1+e^{-\tau/s})+(a_{2}^{\ 6}-a_{1}^{\
6})\right]\right)\nonumber\\
&&+\left(-\frac{5}{3}(a_{2}^{\ 6}-a_{1}^{\
6})e^{\tau/s}\right)=\left(\frac{4}{3}(a_{2}^{\ 6}-a_{1}^{\
6})e^{\tau/s}\right),
\end{eqnarray}
which can be expressed as
\begin{equation}
2a_{2}{}^{6}e^{2\tau/s}+(a_{2}{}^{6}-a_{1}{}^{6})e^{\tau/s}-2a_{1}{}^{6}=0.
\end{equation}
This is a quadratic equation with two roots for $e^{\tau/s}$.  The
ratio $\tau/s$ is now taken to be real, which means $e^{\tau/s}$ is
non-negative; this leaves only the positive root solution of
\begin{equation}
e^{\tau/s}=\frac{a_{1}{}^{6}-a_{2}{}^{6}+\sqrt{a_{1}{}^{12}+14a_{1}{}^{6}a_{2}{}^{6}+a_{2}{}^{12}}}{4a_{2}{}^{6}}.
\end{equation}
Once that is found, the $C^{2}$ matching conditions for $\tau$,
$a(\tau$), and $H$ are
\begin{eqnarray}
\tau&=&s\ln\left[\frac{a1_{1}^{\ 6}-a_{2}^{\
6}+\sqrt{a_{1}{}^{12}+14a_{1}{}^{6}a_{2}{}^{6}+a_{2}{}^{12}}}{4a_{2}^{\
6}}\right],\\
a(\tau)&=&\left(\frac{a_{2}^{\ 6}(5a_{1}^{\ 6}-a_{2}^{\
6}+\sqrt{a_{1}{}^{12}+14a_{1}{}^{6}a_{2}{}^{6}+a_{2}{}^{12}})}{a_{1}^{\ 6}+3a_{2}^{\ 6}+\sqrt{a_{1}{}^{12}+14a_{1}{}^{6}a_{2}{}^{6}+a_{2}{}^{12}}}\right)^{\frac{1}{6}},\\
H&=&\left(\frac{\sqrt{2}(-a_{1}^{\ 6}+a_{2}^{\ 6})}{3a_{2}^{\
6}(5a_{1}^{\ 6}-a_{1}^{\
6}+\sqrt{a_{1}{}^{12}+14a_{1}{}^{6}a_{2}{}^{6}+a_{2}{}^{12}})^{2}\
s}\right)
\nonumber\\
&&\times(a_{1}^{\ 6}-a_{2}^{\
6}+\sqrt{a_{1}{}^{12}+14a_{1}{}^{6}a_{2}{}^{6}+a_{2}{}^{12}})\nonumber\\
&&\times\sqrt{-a_{1}^{\ 6}-a_{2}^{\
6}+\sqrt{a_{1}{}^{12}+14a_{1}{}^{6}a_{2}{}^{6}+a_{2}{}^{12}}}.
\end{eqnarray}

\section{Particle Creation}

At late times, our solution to the evolution equation will have the
asymptotic form given by Eq.~(\ref{eq:alphabeta}).  The early- and
late-time vacua are related through a Bugoliubov Transformation
\cite{Parker1} (alternately Romanized in the literature from the
Cyrillic as Bugolubov or Bugolyubov or Bogoliubov), where the
early-time creation and annihilation operators
($A_{\vec{k}}^{\dagger}$ and $A_{\vec{k}}$) are related to the
late-time creation and annihilation operators
($a_{\vec{k}}^{\dagger}$ and $a_{\vec{k}}$) through
\begin{equation}
a_{\vec{k}}=\alpha_{k}A_{\vec{k}}+\beta_{k}^{*}A_{\vec{k}}^{\dagger},
\end{equation}
where $\alpha_{k}$ and $\beta_{k}$ are the Bugoliubov coefficients
given by Eq.~(\ref{eq:alphabeta}) and satisfying
Eq.~(\ref{eq:check}). Because our scale factor is asymptotically
Minkowskian, the meaning of particles at early and late times has no
ambiguity. At late times, the number operator is
\begin{equation}
\langle
N_{\vec{k}}\rangle_{t\rightarrow\infty}=\langle0|a_{\vec{k}}^{\dagger}a_{\vec{k}}|0\rangle=\left|\beta_{k}\right|^{2},
\end{equation}
where $\left|0\right>$ is the state annihilated by the early-time
annihilation operators $A_{\vec{k}}$. For the rest of this chapter,
the notation $\left|\delta\phi_{k}\right|^{2}$ is defined as
$\langle0\left|\delta\phi_{k}\delta\phi_{k}\right|0\rangle =
|f_{\vec{k}}|^2$. In the continuum limit, this reduces to
$(2\pi)^{-3}|\psi_k|^2$. Thus, $\left|\beta_{k}\right|^{2}$ is the
average number of particles in mode-${\vec k}$ created by the
expansion of the scale factor from a state that initially has no
particles \cite{Parker1,Parker3}.

\subsection{Dependence on Mode, Expansion, and Mass}
\label{sec:dmem}

In the absence of units, the magnitudes of $k$, $a$, $H$, and $m$
have no inherent significance.  The ratio of the Hubble radius,
$H^{-1}$, to wavelength, $a/k$, however, does have significance.
This combination of $k/aH$ is what we call $q_{2}$ when we take the
particular values of $a=a_{2f}$ and $H=H_{\rm infl}$. The other
relevant dimensionless ratios are $m_{H}\equiv m/H_{\rm infl}$ and
$N_{e}$. Transformations that simultaneously leave the values of
$k/(a(t)H(t))$ and $m_{H}$ intact do not change the arguments of any
of the evolution solutions used in our composite scale factor. See
Eq.~(\ref{eq:massinfleigen}) for the inflationary middle segment of
our composite scale factor. For an asymptotically flat scale factor
of either the form described by Eq.~(\ref{eq:aHG}) or the form
described by Eq.~(\ref{eq:aHGmassive}), no matter how we scale
$a=a(\tau/s,a_{1},a_{2})$, the ratio $a_{2}/a_{1}$ remains a
constant; furthermore, when keeping the particular value of $\tau/s$
fixed, $H\propto1/(sa_{1}^{\ 3})\propto1/(sa_{2}^{\ 3})$.  For
example, if we multiply $k$ by a constant and multiply $a(t)$ by
that same constant, we don't change the wavelength of our mode.  If
we don't alter $H$, this rescaling won't change
$\left|\beta_{q_{2}}\right|^{2}$.  When $b=0$, we see that this
transformation is
\begin{eqnarray}
k\rightarrow&& k*x
\nonumber\\
a_{1}\rightarrow&& a_{1}*x
\nonumber\\
a_{2}\rightarrow&& a_{2}*x
\nonumber\\
s\rightarrow&& s*x^{-3}.
\end{eqnarray}
For a second example, rescaling $k$, $H_{\rm infl}$, and $m$ by the
same factor is equivalent to
\begin{eqnarray}
k\rightarrow&& k*y
\nonumber\\
s\rightarrow&& s*y^{-1}
\nonumber\\
m\rightarrow&& m*y.
\end{eqnarray}
This second example won't change the average number of particles
created per mode, either.  We note that in the massless case the
coefficient $1/\sqrt{2a_{1i}{}^{2}k}$ from
Eq.~(\ref{eq:hypergeometric}) may change in invariant
transformations, but $\left|\beta_{q_{2}}\right|^{2}$ does not
change because Eqs.~(\ref{eq:alpha}) and~(\ref{eq:beta}) contain
factors that compensate for the change in $N_{1}$.  The same is true
in the massive case under the transformation
$k/a(t)\rightarrow\sqrt{(k/a(t))^{2}+m^{2}}$.

\subsubsection{In the massless case we find the following:}

For our choice of the final asymptotically flat segment given by
Eq.~(\ref{eq:aHG}), where we use Eq.~(\ref{eq:hypergeometric2}) to
define our functions $g_{1}(t)$ and $g_{2}(t)$ in terms of the
relationship $\psi_{k}(t)=N_{1}g_{1}(t(\tau))+N_{2}g_{2}(t(\tau))$,
we find the coefficients $\alpha_{k}$ and $\beta_{k}$ of
Eq.~(\ref{eq:alphabeta}) from the large argument asymptotic forms
\cite{Parker4,Parker5,Parker8,Parker9,Abramowitz}. With $b_{f}=0$,
$c_{1}\equiv iks_{f}a_{1f}^{\ \ 2}$, and $c_{2}\equiv
iks_{f}a_{2f}^{\ \ 2}$, we have
\begin{equation}
\alpha_{k}=\sqrt{2ka_{2f}^{\ \ 2}}\left[\frac{C\
\Gamma(1-2c_{1})\Gamma(-2c_{2})}{\Gamma(1-c_{1}-c_{2})\Gamma(-c_{1}-c_{2})}+\frac{D\
\Gamma(1+2c_{1})\Gamma(-2c_{2})}{\Gamma(1+c_{1}-c_{2})\Gamma(c_{1}-c_{2})}\right],
\label{eq:alpha}
\end{equation}
and
\begin{equation}
\beta_{k}=\sqrt{2ka_{2f}^{\ \ 2}}\left[\frac{C\
\Gamma(1-2c_{1})\Gamma(2c_{2})}{\Gamma(1-c_{1}+c_{2})\Gamma(-c_{1}+c_{2})}+\frac{D\
\Gamma(1+2c_{1})\Gamma(2c_{2})}{\Gamma(1+c_{1}+c_{2})\Gamma(c_{1}+c_{2})}\right].
\label{eq:beta}
\end{equation}
Recall that $C$ and $D$ and the functions $g_1(t)$ and $g_2(t)$ were
defined in Sec.~\ref{sec:match}. A useful check of our method is the
test of whether Eq.~(\ref{eq:check}) is validated, which we find to
be true in all our numerical calculations.

The variable $\left|\beta_{k}\right|^2$ is the average number of
particles created in the mode k, as measured at late times, from the
expansion of the scale factor through $N_{e}$ number of e-folds,
starting from a universe that is initially in a vacuum state that is
asymptotically Minkowskian. We use the dimensionless variable
\begin{equation}
q_{2}\equiv\frac{k}{a_{2f}H_{\rm infl}}, \label{q2}
\end{equation}
where k is the wave number, $a_{2f}$ is the asymptotically flat
late-time scale factor, and $H_{\rm infl}$ is the constant value of
$\left({\dot{a}(t)}/{a(t)}\right)$| where the dot represents a
derivative with respect to proper time| during the exponential
expansion of the middle segment.  We express our results using
$q_{2}$ instead of the wave number, $k$, because we find that
$|\beta_{q_{2}}|^{2}$ is an invariant quantity (see
Fig.~\ref{fig:b2vq2}), whereas $|\beta_{k}|^{2}$ depends on the
arbitrary value of the scale factor.  By $|\beta_{q_{2}}|^{2}$, we
refer to the average number of particles created in the mode given
by $k=q_{2}H_{\rm infl}a_{2f}$. See the end of Sec.~\ref{sec:mass}
for a discussion of invariant transformations.

We define three regions of $q_{2}$.  Values of
$q_{2}\lesssim\exp(-N_{e})$ are in the small-$q_{2}$ region.  Values
of $\exp(-N_{e})\lesssim q_{2}\lesssim1$ are in the
intermediary-$q_{2}$ region.  Values of $1\lesssim q_{2}$ are in the
large-$q_{2}$ region.

Fig.~\ref{fig:b2vq2} shows the
\begin{figure}[hbtp]
\includegraphics[scale=2.5]{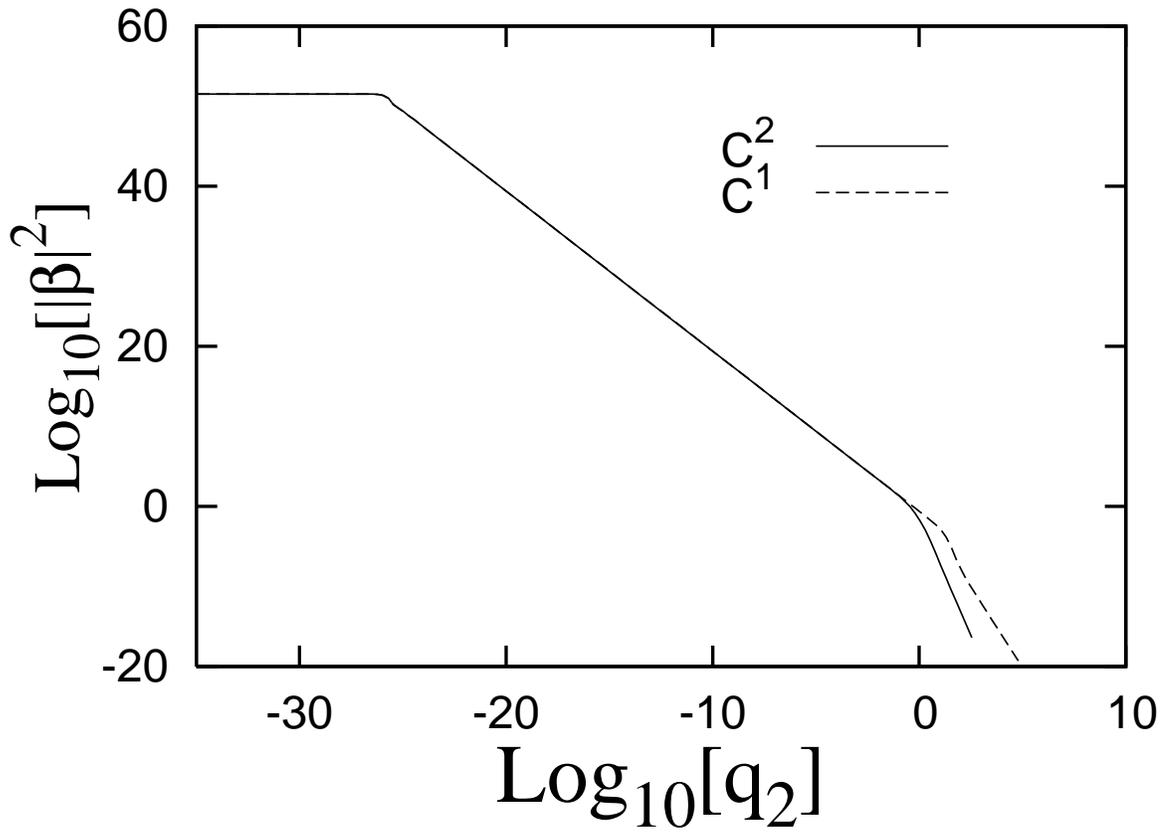}
\caption{\label{fig:b2vq2} Particle Production in the Massless
Case.}
\end{figure}
average late time particle number per mode ($|\beta_{q_{2}}|^{2}$)
versus $q_{2} = k/(a_{2f}H_{\rm infl})$ for 60 e-folds of inflation.
Two cases are plotted for the massless case based on the behavior at
the matching conditions: the scale factor continuous in 0th, 1st,
and 2nd derivatives ($C^{2}$); and the scale factor continuous in
0th and 1st derivatives ($C^{1}$). Note that in the $C^{1}$ case,
$\left|\beta_{q_{2}}\right|^{2}$ transitions from a $q_{2}^{\ -2}$
dependence at the end of the intermediary-$q_{2}$ region all the way
to a $q_{2}^{\ -6}$ dependence, temporarily parallel to the $C^{2}$
large-$q_{2}$ regime, before settling down into its ultraviolet
$q_{2}^{\ -4}$ behavior. For the wiggles near the transition from
the small-$q_{2}$ region to the intermediary-$q_{2}$ region at
$q_{2}=e^{-N_{e}}$, compare with the graph of the dispersion
spectrum in Fig.~\ref{fig:2pt}.

When $a(t)$ is $C^{1}$ or $C^{2}$, i.e. when $H_{\rm infl}$ is
continuous, we find numerically that the particle production per
mode in the small-$q_{2}$ region, ($q_{2}\lesssim e^{-N_{e}}$), is
\begin{equation}
\beta_{q_{2}}=\sinh[N_{e}]. \label{eq:smallqbeta}
\end{equation}
We also find this to be the case, analytically, by taking the limit
$k\rightarrow0$.  This analytical limit can be seen as follows.
Eq.~(\ref{eq:evotau}), in the $k\rightarrow0$ limit tells us that
$d\psi_{k}(\tau)/d\tau$ is constant.  From Eq.~(\ref{eq:minkowski}),
we see that at early times $\psi_{k}(\tau)=1/\sqrt{2ka_{1i}{}^{2}}$
and $d\psi_{k}(\tau)/d\tau=-i\sqrt{ka_{1i}{}^{2}/2}\rightarrow0$ in
the $k\rightarrow0$ limit.  Because $d\psi_{k}(\tau)/d\tau$ is both
constant and zero, so must $\psi_{k}(\tau)$ be constant. Matching
$\psi_{k}(\tau)$ and $d\psi_{k}(\tau)/d\tau$ with the late-time
conditions| which do not make any assumptions about the changing
scale factor before the late-time asymptotically flat region of
spacetime is reached| leads to two boundary conditions:
\begin{equation}
1/\sqrt{2ka_{1i}{}^{2}}=(\alpha_{k}+\beta_{k})/\sqrt{2ka_{2f}{}^{2}},
\end{equation}
\begin{equation}
-i\sqrt{ka_{1i}{}^{2}/2}=(-i\alpha_{k}+i\beta_{k})\left(\sqrt{ka_{2f}{}^{2}/2}\right).
\end{equation}
This leads to
\begin{equation}
\alpha_{k}+\beta_{k}=e^{N_{e}},
\end{equation}
\begin{equation}
\alpha_{k}-\beta_{k}=e^{-N_{e}}.
\end{equation}
The solution to this is
\begin{equation}
\alpha_{k}=\cosh N_{e},
\end{equation}
\begin{equation}
\beta_{k}=\sinh N_{e}.
\end{equation}
In the limit of $k\rightarrow0$, both coefficients happen to be
real, and we can see that Eq.~(\ref{eq:check}) is naturally
satisfied. Although this result was derived in the $k\rightarrow0$
limit, it is valid in the massless case whenever $k/(a_{1i}H_{\rm
infl})\ll1$. This small-$q_{2}$ limit holds for arbitrary
expansions, besides those described by our parameterized composite
scale factor, provided they initiate from a Minkowski vacuum state.
We find that the requirement for an alternative to the Bunch-Davies
state for the $k=0$ mode in de Sitter space would be a consequence
of taking $N_{e}\rightarrow\infty$ in this analytical limit.

For at least a moderate number of e-folds, this simplifies to
\begin{equation}
|\beta_{q_{2}}|^{2}\simeq\frac{1}{4}e^{2N_{e}}.
\end{equation}
The dependence in the intermediary-$q_{2}$ region
($e^{-N_{e}}\lesssim q_{2}\lesssim 1$) for the $C^{2}$ or $C^{1}$
massless case is
\begin{equation}
|\beta_{q_{2}}|^{2}\simeq\frac{1}{4}q_{2}^{\ -2}.
\end{equation}

When $N_{e}$ is finite, with our composite scale factor there are no
infrared divergences.  For infinite inflation, where
$N_{e}\rightarrow\infty$, we find the infrared divergences of a de
Sitter universe.  This problem is resolved for a true de Sitter
universe in \cite{Allen2}.  Our composite scale factor is different
from a purely de Sitter universe in that our initial conditions are
specified by our initial asymptotically flat region of the scale
factor.

Discontinuities in the derivatives of the scale factor at the
matching points introduce additional particle production for modes
in the large-$q_{2}$ (or $q_{2}\gtrsim1$) region. For the $C^{1}$
case, where the scale factor and $H=\dot{a}(t)/a(t)$ are both
continuous, the large-$q_{2}$ region goes like
\begin{equation}
|\beta_{q_{2}}|^2=n_{4}q_{2}^{-4}.
\end{equation}
For the $C^{2}$ case, where the scale factor and $H=\dot{a}(t)/a(t)$
and $\dot{H}(t)$ are all continuous, the large-$q_{2}$ region goes
like
\begin{equation}
|\beta_{q_{2}}|^2=n_{6}q_{2}^{-6}.
\end{equation}
Here $n_{4}$ and $n_{6}$ are constant coefficients, with
$n_{4}\simeq n_{6}\simeq\mathcal{O}(1/4)$ for a gradual end to
inflation.  For a sufficiently abrupt end to inflation, $n_{4}$ and
$n_{6}$ can be made to be arbitrarily large. See
Sec.~\ref{sec:reheat}.

In the $C^{0}$ case, $H(t)$ is not continuous, and we find quite a
different behavior. The evolution equation, Eq.~(\ref{eq:evot}), may
be written \cite{Parker3}
\begin{equation}
\frac{d^{2}\psi_{k}(t)}{dt^{2}}+\left[\frac{k^{2}}{a(t)^{2}}+m^{2}-\frac{3}{4}\left(\frac{\dot{a}(t)}{a(t)}\right)^{2}-\frac{3}{2}\frac{\ddot{a}(t)}{a(t)}\right]\psi_{k}(t)=0.
\end{equation}
At the discontinuity in $\dot{a}(t)$, if we express the jump as a
step function, then the form of $\ddot{a}(t)$ picks up a
delta-function contribution. Thus, there is a finite jump in
$d\psi_{k}(t)/dt$ across the discontinuity. The Wronskian is still
conserved. In the $C^{0}$ case, $\left|\beta_{q_{2}}\right|^{2}$ is
proportional to $q_{2}^{\ -2}$ in the small- and large-$q_{2}$
regions, and it is proportional to $q_{2}^{\ -4}$ in the
intermediary-$q_{2}$ region. A $C^{0}$ scenario would suffer from
both infrared and ultraviolet divergences, hence we will not
consider it further.

For a non-composite scale factor composed of one asymptotically flat
scale factor defined by Eq.~(\ref{eq:aHG}), at large values of
$q_{2}$ the value of $|\beta_{q_{2}}|^{2}$ falls off faster than any
power of $q_{2}$, and in terms of $k$ we have:
\cite{Parker4,Parker5,Parker8,Parker9}
\begin{equation}
|\beta_{k}|^2=\frac{\sin^{2}\left(\frac{1}{2}[1-\sqrt{1+4k^{2}s^{2}b}]\right)+\sinh^{2}[\pi
ks(a_{1}^{\ 2}-a_{2}^{\ 2})]}{\sinh^{2}[\pi ks(a_{1}^{\ 2}+a_{2}^{\
2})]-\sinh^{2}[\pi ks(a_{1}^{\ 2}-a_{2}^{\ 2})]}.
\end{equation}
In the limit that $k\rightarrow0$ for the case of the scale factor
of Eq.~(\ref{eq:aHG}), which is asymptotically flat at early and
late times and has no exponential segment, we find that
$\lim_{k\rightarrow0}\left|\beta_{k}\right|^{2}=\sinh^2[N_{e}]$,
where in this case $N_{e}$ is $\ln\left(a_{2}/a_{1}\right)$. This is
the same small-$q_{2}$ limit for the average number of particles
created per mode as we found above in Eq.~(\ref{eq:smallqbeta}). The
analog of the intermediary-$q_{2}$ region extends over a range of
$\ln q_{2}$ equal to $2N_{e}$, as opposed to $N_{e}$ for the
particle production associated with our composite scale factor.
Thus, a graph of the average number of particles created per mode
for a single asymptotically flat scale factor would look similar to
Fig~\ref{fig:b2vq2}, except the region analogous to the
intermediary-$q_{2}$ region would be twice as long and would have
half the slope relative to a scale factor dominated by an
exponential expansion.

\subsubsection{In the massive case we find the following:}

In Fig.~\ref{fig:massive},
\begin{figure}[hbtp]
\includegraphics[scale=2.5]{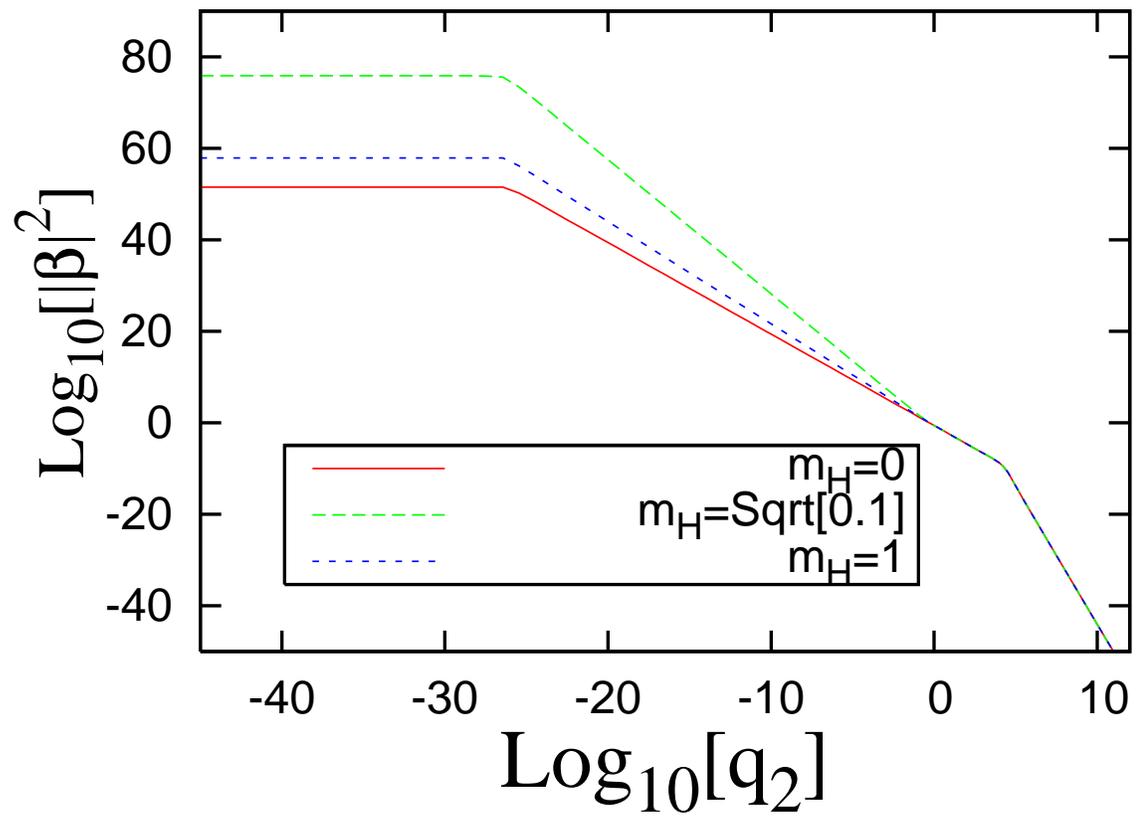}
\caption{\label{fig:massive} Particle Production in the
Effective-$k$ Approximation.}
\end{figure}
the dependence of particle production
($\left|\beta_{q_{2}}\right|^{2}$) on mass is shown for an expansion
of 60 e-folds.  The beginning and end segments are defined by
$a_{2i}=a_{1i}(1+10^{-26})$, $a_{2f}=a_{1i}e^{60}$, and
$a_{1f}=0.9999a_{2f}$.  The massless case can be compared with the
plot in Fig.~\ref{fig:b2vq2} which is continuous up to the second
derivative of the scale factor to see that the two graphs are the
same for $q_{2}\lesssim1$.  In this graph, however, there is an
``extended" region of $\left|\beta_{q_{2}}\right|^{2}\propto
q_{2}^{-2}$ shortly after $q_{2}\simeq1$ that lasts until
$q_{2}\simeq10^{4}$ before the ultraviolet behavior of
$\left|\beta_{q_{2}}\right|^{2}\propto q_{2}^{-6}$ is seen.  The
term ``extended" is defined in Sec.~\ref{sec:reheat}. This is due to
particle creation caused by the rapid transition from the
inflationary region to the asymptotically flat scale factor. The two
approximations, the effective-k approach and the dominant-term
approach, give the same results with this particular
parameterization of inflation. Both of the massive cases shown here
produce more red-shifted particles of low momentum than the massless
case. The case of $m_{H}^{2}=1/10$ produces many more low momentum
particles than the case of $m_{H}^{2}=1$. See also
Figs.~\ref{fig:domterm} and~\ref{fig:pvmh}.

In Fig.~\ref{fig:domterm},
\begin{figure}[hbtp]
\includegraphics[scale=2.5]{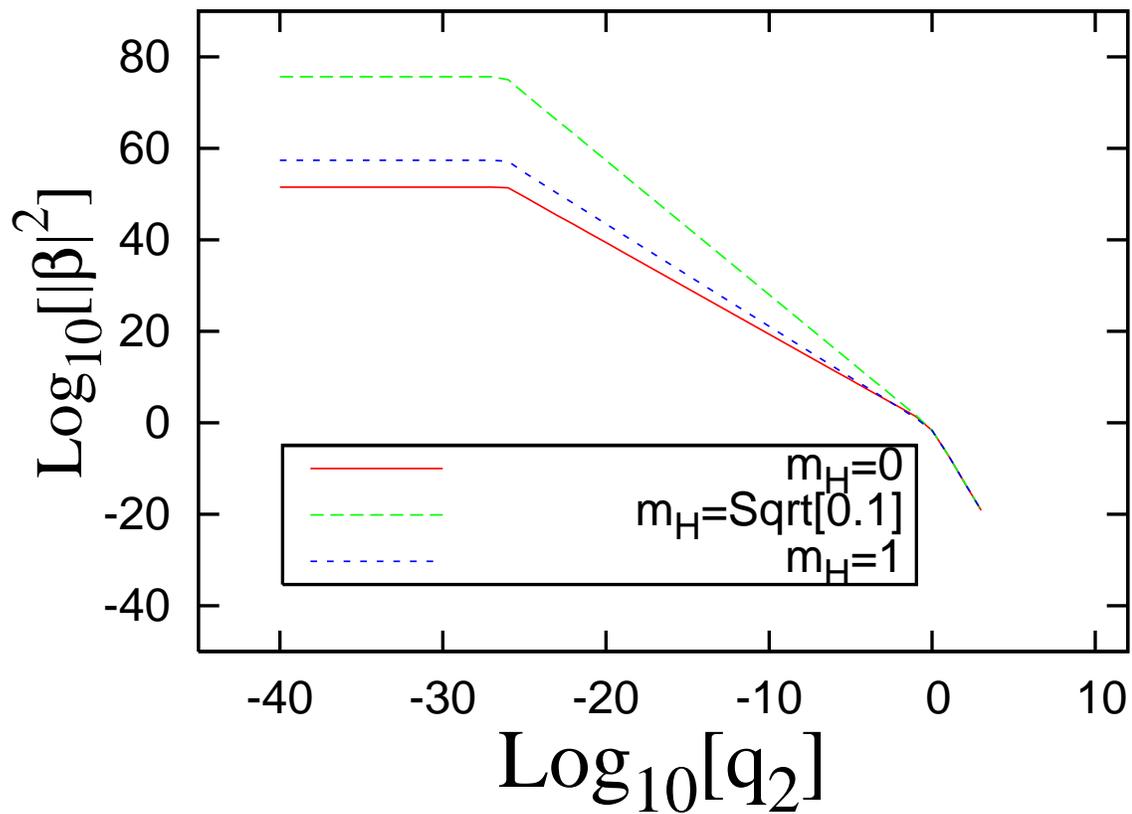}
\caption{\label{fig:domterm} Particle Production in the Dominant
Term Approximation.}
\end{figure}
the dependence of particle production
($\left|\beta_{q_{2}}\right|^{2}$) on mass is shown for an expansion
of 60 e-folds.  This graph is different from Fig.~\ref{fig:massive}
in that the transition from exponential expansion to the final
asymptotic segment of the scale factor is more gradual, happening
over about an e-fold.  Thus, we use the dominant-term approximation.
The effective-k approach, in spite of the gradual transition to an
asymptotically flat scale factor, overlaps with the dominant-term
approach in this graph except very close to $q_{2}=1$.  For values
of $q_{2}\lesssim1$, this graph is identical to that of
Fig.~\ref{fig:massive}.

\subsection{Limit of Negligible Mass with Respect to H}

In Fig.~\ref{fig:tinymass},
\begin{figure}[hbtp]
\includegraphics[scale=2.5]{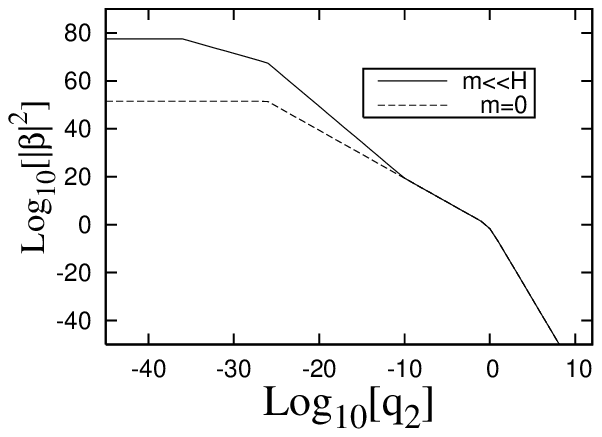}
\caption{\label{fig:tinymass} Non-Zero Mass, Negligible with Respect
to H.}
\end{figure}
particle production as a function of $q_{2}$ is plotted for 60
e-folds for both the massless case and the case of $m=10^{-10}H_{\rm
infl}$, labeled as $m<<H$.  This graph was made using the
dominant-term approximation.  The effective-k approach would overlap
on this graph except very near to $q_{2}=m_{H}=10^{-10}$. It is
always the case that $(k/a(t))^{2}\gg m^{2}$ for $q_{2}>m_{H}$ and
in this region the plot of $m_{H}=10^{-10}$ overlaps with the
massless case.  For $q_{2}<m_{H}\exp(-N_{e})$, relative to the mass
we can take $k=0$, and in this region of $q_{2}$ in the tiny mass
case of $m_{H}\ll1$, the value of $\left|\beta_{q_{2}}\right|^{2}$
approaches the constant $(1/4)q_{2}^{\ 3N_{e}}$.  In the region of
$m_{H}\exp(-N_{e})<q_{2}<m_{H}$, we have $(k/a(t))^{2}\gg m^{2}$ in
the initial asymptotically flat region and $(k/a(t))^{2}\ll m^{2}$
in the final asymptotically flat region. Between $q_{2}\simeq
m_{H}\exp(-N_{e})$ and $q_{2}\simeq \exp(-N_{e})$, we see
$\left|\beta_{q_{2}}\right|^{2}\propto q_{2}^{\ -1}$; and between
$q_{2}\simeq \exp(-N_{e})$ and $q_{2}\simeq m_{H}$, we see
$\left|\beta_{q_{2}}\right|^{2}\propto q_{2}^{\ -3}$.  In light of
these characteristics, a comparison of Eqs.~(\ref{eq:disp})
and~(\ref{eq:massdisp}) can be made with consideration to where
$(k/a(t))^{2}\gg m^{2}$ and to where $(k/a(t))^{2}\ll m^{2}$. Such
an analysis shows that in the tiny mass limit of $m_{H}\ll1$, the
dispersion spectrum reduces to the massless dispersion spectrum. The
tiny mass limit bridges the transition from the massless case to the
case of small, non-negligible $m_{H}$ such as $m_{H}=0.01$, and the
dispersion spectra as a function of $q_{2}$ for all cases changes
continuously when going from massless to tiny mass to small mass.
This is a successful check on our method.

\section{Dispersion Spectrum}
\label{sec:MasslessDispersionSpectrum}

The dispersion spectrum is \cite{Parker5,Parker10}
\begin{equation}
\left<\ \right|\delta\phi^{2}\left|\
\right>=\frac{1}{2(a_{2f}L)^{3}}\sum_{k}\left[\frac{1+2|\beta_{k}|^{2}}{\sqrt{(k/a_{2f})^{2}+m^{2}}}\right].
\end{equation}

We will first consider the massless case where $m=0$.  See below in
Sec.~\ref{sec:mass} for the massive case.  We subtract off the
late-time Minkowski vacuum contribution, which is that part of the
unrenormalized dispersion which would be present in a Minkowski
vacuum without any particles ($\left|\beta_{k}\right|^{2}=0$ for all
$k$), to get the dispersion
\begin{eqnarray}
\left<\ \right|\delta\phi^{2}\left|\
\right>&&=\frac{1}{2(a_{2f}L)^{3}}\sum_{k}\frac{2|\beta_{k}|^{2}}{\sqrt{(k/a_{2f})^{2}}}
\nonumber\\
&&=\frac{1}{a_{2f}^{\ \ 2}L^{3}}\sum_{k}\frac{|\beta_{k}|^{2}}{k},
\end{eqnarray}
which in the continuum limit becomes
\begin{equation}
\left<\ \right|\delta\phi^{2}\left|\ \right>=\frac{1}{a_{2f}^{\ \
2}(2\pi)^{3}}\int_{0}^{\infty}\frac{|\beta_{k}|^{2}}{k}d^{3}k.
\end{equation}
Spherical symmetry, where $d^{3}k=4\pi k^{2} dk$, gives us
\begin{equation}
\left<\ \right|\delta\phi^{2}\left|\
\right>=\frac{1}{2\pi^{2}a_{2f}^{\ \
2}}\int_{0}^{\infty}k|\beta_{k}|^{2}dk.
\end{equation}
With $k=q_{2}a_{2f}H_{\rm infl}$ and $dk=dq_{2}a_{2f}H_{\rm infl}$,
we have
\begin{eqnarray}
\left<\ \right|\delta\phi^{2}\left|\ \right>&&=\frac{a_{2f}^{\ \
2}H_{\rm infl}^{2}}{2\pi^{2}a_{2f}^{\ \
2}}\int_{0}^{\infty}q_{2}|\beta_{q_{2}}|^{2}dq_{2}
\nonumber\\
&&=\frac{H_{\rm
infl}^{2}}{2\pi^{2}}\int_{0}^{\infty}q_{2}|\beta_{q_{2}}|^{2}dq_{2}.
\end{eqnarray}
In the massless case, the dispersion spectrum amplitude is thus
\begin{equation}
Z\equiv\frac{q_{2}|\beta_{q_{2}}|^{2}H_{\rm infl}^{2}}{2\pi^{2}}.
\label{eq:disp}
\end{equation}

We plot $Z/H_{\rm infl}^{2}$ in Fig.~\ref{fig:2pt}. We see that in
both the case where  $a(t)$, $\dot{a}(t)$, and $\ddot{a}(t)$ are all
continuous; and the case where $a(t)$ and $\dot{a}(t)$ are
continuous; $\left<\ \right|\delta\phi^{2}\left|\ \right>$ is finite
without the need for any renormalization beyond subtracting off the
Minkowski vacuum terms. When none of the derivatives of the scale
factor is continuous, then the dispersion spectrum does not
converge.\begin{figure}[hbtp]
\includegraphics[scale=2.5]{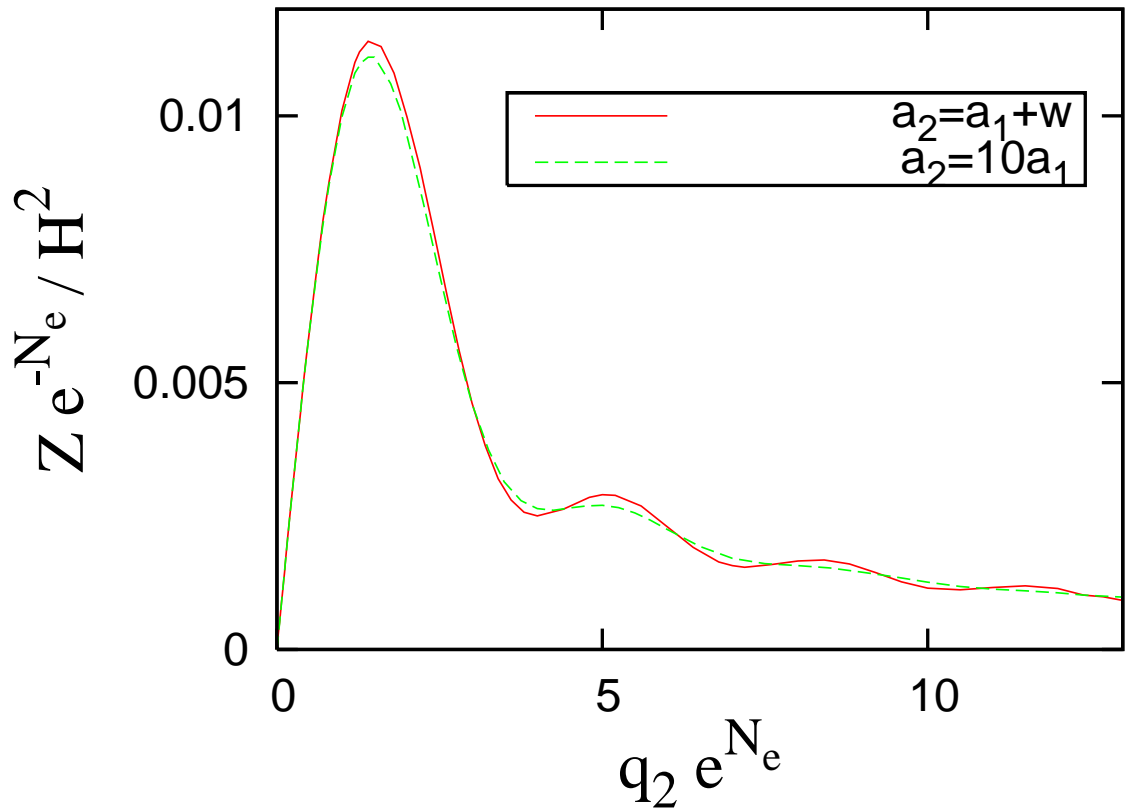}
\caption{\label{fig:2pt} Massless Dispersion Spectrum.}
\end{figure}

Fig.~\ref{fig:2pt} shows the dispersion spectrum $Z/H_{\rm
infl}^{2}$ given by Eq.~(\ref{eq:disp}) for our composite scale
factor continuous in $a(t)$, $\dot{a}(t)$, and $\ddot{a}(t)$ over an
expansion of 60 e-folds.  The y-axis, $Z/H_{\rm infl}^{2}$, is shown
multiplied by a factor of $e^{-N_{e}}$; and the x-axis, $q_{2}$, is
shown multiplied by a factor of $e^{N_{e}}$. When using this
scaling, the region plotted in this graph would look identical for
an expansion of 10 e-folds, and it would look identical for an
expansion of 80 e-folds. In the case of $a_{2i}=a_{1i}+w$, where
$w\equiv10^{-26}a_{1i}$, we see marked peaks in the dispersion
spectrum. When we change the parameters in the initial
asymptotically flat region to $a_{2i}=10a_{1i}$, these peaks are
damped as shown.  The ending conditions of the final asymptotically
flat segment do not affect these peaks.

A calculation of the dispersion spectrum in the massive case leads
to an equation analogous to Eq.~(\ref{eq:disp}):
\begin{equation}
Z\equiv\frac{q_{2}|\beta_{q_{2}}|^{2}H_{\rm
infl}^{2}}{2\pi^{2}\sqrt{1+\frac{m_{H}^{\ 2}}{q_{2}^{\ 2}}}}.
\label{eq:massdisp}
\end{equation}

\subsection{Sensitivity to Initial Conditions}
\label{sec:KBSENS}

We take for our initial conditions a quantum state to be asymptotic
at early times to that of a Minkowski vacuum spacetime for all
modes. This is a consequence of our asymptotically flat scale factor
and our assumption that no particles are initially present.  It is
more common in the literature to take instead the Bunch-Davies state
for quantum fluctuations, that is to assume a de Sitter spacetime.
As pointed out by \cite{Allen3}, this leads to an infrared
divergence of the two-point function, where the two-point function
is another name for our dispersion spectrum, and the cause of this
divergence is correctly diagnosed as being due to the choice of
initial conditions in \cite{Yajnik}. Both of the authors of
\cite{Allen3,Yajnik} handle these infrared divergences with a cutoff
frequency that omits modes that are currently outside the Hubble
radius of our observable universe.

The use of de Sitter initial conditions is equivalent to supposing
an inflationary period that extends over an infinite number of
e-folds, or $N_{e}\rightarrow\infty$.  If we assume a finite
$N_{e}$, and if we assume that in the future our universe will be
approximately matter-dominated for all times, which means neglecting
any dark energy or cosmological constant, then eventually every mode
that exited the Hubble radius during inflation would eventually
re-enter the Hubble radius of our universe after inflation if it has
not already done so.

Both Figs.~\ref{fig:2pt} and~\ref{fig:massive2pt} show additional
peaks after the primary peak, where the primary peak roughly
indicates the interface between small-$q_{2}$ and
intermediary-$q_{2}$ behavior.  These minor peaks are caused by
phase differences between modes with similar wavelengths as they
exit the Hubble radius near the beginning of inflation.  The modes
that exit the Hubble radius with a large amplitude| either a
positive real amplitude, a negative real amplitude, a positive
imaginary amplitude, or a negative imaginary amplitude| quickly have
this large amplitude translated into a near constant value outside
of the Hubble radius.  Those modes that exit the Hubble radius with
relatively small amplitudes are frozen into evolutions of relatively
small magnitudes outside of the Hubble radius; these relatively
low-amplitude modes have a relatively high change in amplitude with
respect to time, but this initial excess in the derivative of the
amplitude with respect to time is rapidly redshifted away during
inflation. With an abrupt transition from an asymptotically
Minkowski vacuum to an exponential inflation of the scale factor, by
which we mean that $a(t_{1})\simeq a_{1i}$, where $a(t_{1})$ is the
scale factor at the transition from the initial asymptotically flat
segment to the exponentially growing segment of inflation, and where
$a_{1i}$ is the scale factor at early times, we see that the minor
peaks are more pronounced. With a more gradual transition from the
initial asymptotically flat segment of the scale factor to inflation
(when $a(t_{1})\simeq1.2 a_{1i}$), these minor peaks are damped out.
If these modes were observable in our universe, that is if they have
already re-entered our Hubble radius, their measurement might tell
us something about initial conditions before the beginning of
inflation: whether there had been a phase transition from the very
early universe to inflation, how rapidly the very early universe had
been expanding (or contracting) relative to the expansion of
inflation, and what the dominant contribution to the evolution of
our universe might have been before the start of inflation. Because
measuring the contribution of these minor wiggles to the scale
dependence of large-scale structure would be experimentally
challenging (if not impossible), this is in some sense speculation,
but that does not change the fact that the two dispersion spectra
shown in Fig.~\ref{fig:2pt} are different, and this difference| if
observed| would tell us about our pre-inflationary universe.

\subsection{Sensitivity to Sub-Planck Length Physics}

Consider a quantum fluctuation of the particular mode that, at the
beginning of inflation, has a wavelength equal to the Planck length.
By the time this wavelength has been stretched to the point that the
mode is exiting the Hubble radius, it will have a wavelength the
size of the Hubble radius.  For this to happen, the scale factor
must increase by a factor of $H_{\rm infl}^{-1}/\ell_{\rm Planck}$.

With $\hbar=c=1$, the Planck length is $\ell_{\rm
Planck}=\sqrt{G}=8\times10^{-20}\ {\rm (GeV)}^{-1}$. Using the value
of $H_{\rm infl}=7\times10^{13}$ GeV given in
Eq.(\ref{eq:quadpoth}), we find $H_{\rm infl}^{-1}/\ell_{\rm
Planck}\simeq2\times10^{5}$, which corresponds to a mode exiting the
Hubble radius $\ln(10^{7})\simeq12$ e-folds after the start of
inflation. All higher frequency modes, that is for $q_{2}\gtrsim
e^{-N_{e}+12}$, will have originated from trans-Planckian modes
during inflation. With $N_{e}=60$ e-folds of inflation, if we use
the estimate of the number of e-folds before the end of inflation in
which the observable modes of the CMB are exiting the Hubble radius
given by Eq.~(\ref{eq:quadpote}) (50 e-folds) or by
Eq.~(\ref{eq:galacmode}) (53 e-folds), then it might be possible to
observe the difference in amplitudes between those modes that were
initially super-Planckian quantum fluctuations and those that were
initially sub-Planckian quantum fluctuations before the start of
inflation. With either a smaller value of the Hubble constant during
inflation or with a larger number of total e-folds of inflation, the
re-entry of the first trans-Planckian modes back into our Hubble
radius after inflation could be postponed to epochs of our universe
much later than recombination.

\subsection{Model in Terms of Expansion and Mass}

\label{sec:mass}

Fig.~\ref{fig:massive2pt} shows
\begin{figure}[hbtp]
\includegraphics[scale=2.5]{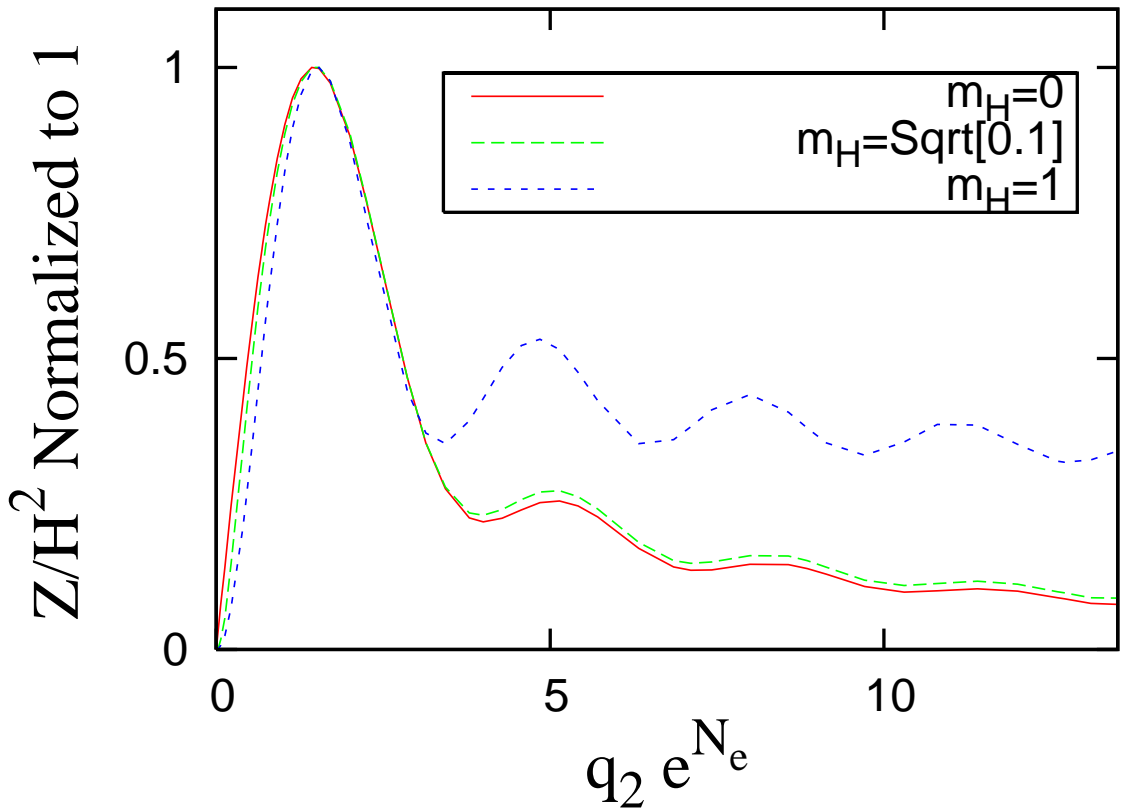}
\caption{\label{fig:massive2pt} Massive Dispersion Spectrum.}
\end{figure}
a comparison of the dispersion spectrum ($Z/H_{\rm infl}^{2}$) given
by Eq.~(\ref{eq:massdisp}) and normalized to 1 for our composite
scale factor continuous in $a(t)$, $\dot{a}(t)$, and $\ddot{a}(t)$
over an expansion of 60 e-folds for various masses. The values of
$Z/H_{\rm infl}^{2}$ were divided by the maximum value of the
primary peak for each located at $q_{2}\simeq\exp(-N_{e})$. To
normalize these peaks, $Z/H_{\rm infl}^{2}$ was divided by the
following factors: $1.3\times10^{24}$ for the massless case,
$2.3\times10^{22}$ for $m_{H}^{\ 2}=0.1$, and $2.2\times10^{4}$ for
$m_{H}^{\ 2}=1$.

The dispersion spectrum is plotted for three different cases of
$m_{H}$ in Fig.~\ref{fig:massive2pt}.  The effective-k approach is
useful for this approximation.  This approach demands that in the
initial asymptotically flat segment of the scale factor, $a(t)$ must
always be approximately equal to $a_{1i}$. Because $a(t_{1})\simeq
a_{1}$ if either $a_{2}\simeq a_{1}$ or $a_{2}\gg a_{1}$, however,
this approach can be used with a wide range of initial conditions.
Specifically, when $a_{2}\gg a_{1}$ in Eq.~(\ref{eq:aHG}), we have
$a(t_{1})\simeq(7/3)^{(1/4)}$.  Although we are not at the moment
considering the case of $a_{2}\gg a_{1}$ in
Eq.~(\ref{eq:aHGmassive}), for comparison we note that it would lead
to $a(t_{1})\simeq3^{(1/6)}$. In both cases
$a(t_{1})\simeq1.2a_{1}$. We have found that, even in the massive
case, the observed humps are dependent only upon the initial
conditions.  In the region shown in this figure, the graph would not
be significantly altered by using $C^{1}$ joining conditions instead
of our $C^{2}$ matching conditions. The effective-k approximation
plotted on this graph would overlap with the exact solution, if an
exact solution were available.  The shapes of the curves are fixed
above a moderate number of e-folds. We define the variable $J$ such
that the maximum value of $Z/H_{\rm infl}^{2}$ for the major peak,
which is the peak located nearest to $q_{2}=e^{-N_{e}}$, is $J\
e^{(P-1)N_{e}}$ in the massless case and is $J\ e^{(P-2)N_{e}}$ in
the massive case. Then, the normalization factor scales like
$e^{(P-1)N_{e}}$ in the massless case, as can be seen from
Eq.~(\ref{eq:disp}); and the normalization factor scales like
$e^{(P-2)N_{e}}$ in the massive case, as can be seen from
Eq.~(\ref{eq:massdisp}), where we define the exponent $P$ in the
following way:
\begin{equation}
\left|\beta_{q_{2}}\right|^{2}\simeq\frac{1}{4}q_{2}^{\ -P}
\label{eq:P}
\end{equation}
in the region of intermediary-$q_{2}$ ($e^{-N_{e}}\lesssim
q_{2}\lesssim1$), and
\begin{equation}
\left|\beta_{q_{2}}\right|^{2}\simeq\frac{1}{4}e^{PN_{e}}
\end{equation}
in the small-$q_{2}$ region ($q_{2}\lesssim e^{-N_{e}}$). The
exponent $P$ is well described by a $q_{2}$-independent value in the
case of $m=0$ and in the case of $0.01\lesssim m_{H}^{\
2}\lesssim9/4$.

In the massless case, $P=2$, so the height of the major peak in the
graph of the massless case in Fig.~\ref{fig:massive2pt} grows with
an increasing number of e-folds as $e^{N_{e}}$, while the widths of
the peaks narrow with an increasing number of e-folds as
$e^{-N_{e}}$.  The area under an individual peak in the massless
graph therefore does not change appreciably when changing the number
of e-folds of expansion, provided there are at least a few e-folds
of inflation.  For the massive cases, we see that $P=2.93358$ when
$m_{H}^{\ 2}=0.1$, and that $P=2.23607$ when $m_{H}=1$.

Fig.~\ref{fig:pvmh} shows
\begin{figure}[hbtp]
\includegraphics[scale=2.5]{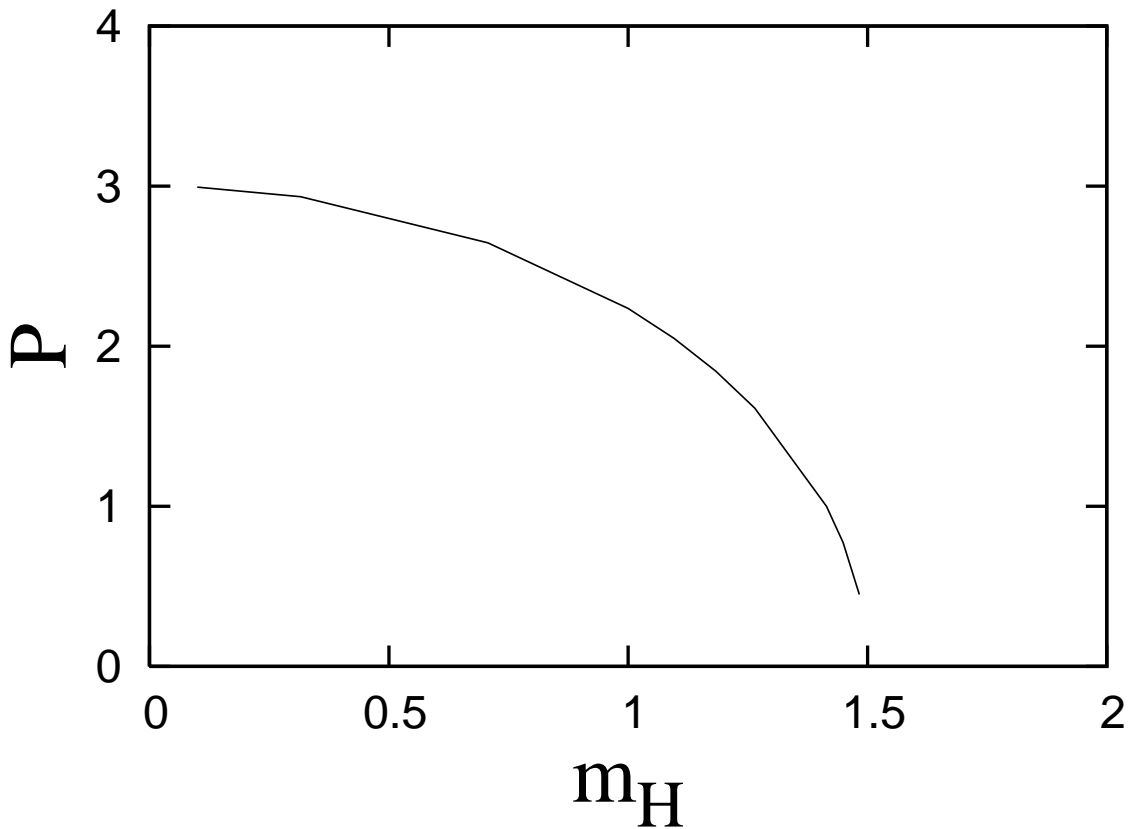}
\caption{\label{fig:pvmh} Inflaton Spectrum Characterized in Terms
of Inflaton Mass.}
\end{figure}
the dependence of the variable $P$, as defined in Eq.~(\ref{eq:P}),
upon $m_{H}=m/H_{\rm infl}$. The calculated data points shown lie on
the curve $P=\sqrt{9-4m_{H}^{\ 2}}$.  Outside of the region plotted,
however, $P$ does not have a constant, $q_{2}$-independent value.
For $m_{H}>1.5$, the argument, $\sqrt{(9/4)-m_{H}^{\ 2}}$, of the
Hankel functions becomes imaginary, and
$\left|\beta_{q_{2}}\right|^{2}$ oscillates with changing $q_{2}$.
For an example of a non-zero mass much smaller than $H_{\rm infl}$,
see Fig.~\ref{fig:tinymass}.

We wish now to approximate the dependence of the configuration space
dispersion $\left<\ \right|\delta\phi^{2}\left|\ \right>/H_{\rm
infl}^{2}$ with regard to the number of e-folds. In our
approximation we neglect the minor peaks; we assume that the major
peak is located exactly at $q_{2}=e^{-N_{e}}$, that $Z/H^{2}$
increases linearly with $q_{2}$ up to the major peak, and that
$Z/H^{2}$ decreases as $(q_{2}e^{N_{e}})^{2-P}$ in the massive case|
or as $(q_{2}e^{N_{e}})^{1-P}$ in the massless case| until the onset
of large-$q_{2}$ behavior at $q_{2}=1$, which effectively serves as
a cut-off point. The maximum of the major peak is given by
$\textrm{height}=J\ e^{(P-2)N_{e}}$, in the massive case; and
$\textrm{height}=J\ e^{(P-1)N_{e}}$, in the massless case. We find
$J\simeq0.01$ for all three cases. In this simple approximation, the
configuration space dispersion is given by TABLES~\ref{tab:model}
and~\ref{tab:area}.
\begin{table}
\caption{Approximation} \label{tab:model}
\begin{tabular}
{ c | c |} & $\left<\ \right|\delta\phi^{2}\left|\ \right>/H_{\rm
infl}^{2}$\\  \hline
\hspace{1.2cm}$0.01\lesssim m_H\lesssim1.49$\hspace{1.2cm} & \hspace{1.5cm}$\frac{1}{2}J\ e^{(P-3)N_{e}}+\int_{e^{-N_{e}}}^{1}dq_{2}\ J\ q_{2}^{2-P}$\hspace{1.5cm}\ \\
\hline $m_H=0$ & $\frac{1}{2}J+\int_{e^{-N_{e}}}^{1}dq_{2}\ J\ q_{2}^{1-P}$\\
\hline
\end{tabular}
\end{table}
\begin{table}
\caption{Configuration Space Dispersion} \label{tab:area}
\begin{tabular}
{ c | c |} & $\left<\ \right|\delta\phi^{2}\left|\ \right>/H_{\rm
infl}^{2}$\\  \hline
\hspace{1.2cm}$0.01\lesssim m_H\lesssim1.49$\hspace{1.2cm} & \hspace{2.1cm}$\left(\frac{1}{3-P}+\frac{1-P}{6-2P}e^{(P-3)N_{e}}\right)J$\hspace{2.1cm}\ \\
\hline $m_H=0$ & $\left(\frac{1}{2}+N_{e}\right)J$\\
\hline
\end{tabular}
\end{table}
The small mass limit of $P\rightarrow3$ and the massless case of
$P=2$, both reduce to the same limit of
\begin{equation}
\left|\delta\phi\right|\simeq\frac{H_{\rm
infl}}{10}\sqrt{N_{e}+\frac{1}{2}}.
\end{equation}
For further discussion of the small mass limit reducing to the
massless dispersion spectrum, see Fig.~\ref{fig:tinymass}.

\section{Scalar Spectral Index and Scale Invariance}

\label{sec:spec}

We define a given mode of $\delta\phi_{k}$ to be crossing the Hubble
radius when $k/(a(t)H(t))=1$.  We define a mode of $k$ to be inside
the Hubble radius when $k>a(t)H(t)$, and we define a mode of $k$ to
be outside the Hubble radius when $k<a(t)H(t)$. Modes in the
intermediary-$q_{2}$ range exit during inflation to eventually
re-enter the Hubble radius at some time after inflation has ended,
provided any cosmological constant or dark energy can be taken to be
negligible. Using our composite scale factor, we note that after a
few e-folds of inflation, the quantum perturbations that are exiting
the Hubble radius are found numerically to satisfy
\begin{eqnarray}
\left|\psi_{k}\right|^{2}=\frac{H_{\rm infl}^{2}}{k^{3}D(m_{H})},
\label{eq:scaleinv}
\end{eqnarray}
where $\left|\psi_{k}\right|^{2}$ is the time-dependent part of
$\left|\delta\phi_{k}\right|^{2}$, as given by Eqs.~(\ref{eq:phi})
and~(\ref{eq:wave}). The variable
$D(m_{H})\simeq(1+\frac{1}{5}m_{H}^{\ 2})^{2}$ is a constant of
order 1 that we have evaluated numerically to be
\begin{eqnarray}
D(m_{H}=0)&=&1.00,
\nonumber\\
D(m_{H}=\sqrt{0.1})&=&1.04,
\nonumber\\
D(m_{H}=1)&=&1.45.
\end{eqnarray}
Thus, our spectrum of $\left|\delta\phi_{k}\right|^{2}$, if
evaluated at the time of exiting the Hubble radius, is
scale-invariant, regardless of effective mass. By
Eqs.~(\ref{eq:specdef}) and~(\ref{eq:inflationspectra}) we have
\begin{equation}
\mathcal{P}_{\mathcal{R}}\propto
k^{3}\left|\delta\phi_{k}\right|^{2}. \label{eq:spectrum}
\end{equation}
The scalar spectral index given by Eq.(\ref{eq:scalarspectralindex})
is
\begin{equation}
n_{s}=1+\frac{d\ln\mathcal{P}_{\mathcal{R}}}{d\ln k}.
\label{eq:spectral}
\end{equation}
We see that when taken at the time of crossing the Hubble radius,
the spectrum, which is proportional to
$k^{3}\left|\delta\phi_{k}\right|^{2}$, has no k-dependence because
we have shown the spectrum is proportional to $k^{0}\,H_{\rm
infl}^{2}/D(m_{H})$. Evaluating the scalar spectral index at the
time of exiting the Hubble radius thus leads to $n_{s}=1$, which can
be used as the definition of a scale-invariant spectrum.

The modes that exit the Hubble radius at the very beginning of
inflation, however, along with those that exit the Hubble radius
before reaching the middle segment of our composite scale factor
where $a(t)$ begins to grow exponentially with respect to $t$, are
not described by Eq.~(\ref{eq:scaleinv}).  These modes in the
small-$q_{2}$ region are not scale-invariant; therefore, well after
the end of inflation, long-wavelength modes that are not
scale-invariant would eventually re-enter the Hubble radius of a
matter-dominated universe. If the total number of e-folds of
inflation is sufficiently small, it would be possible to observe a
transition from the scale-invariance to a scale-dependence of
large-scale structure.  See Figs.~\ref{fig:b2vq2},
\ref{fig:massive}, and~\ref{fig:domterm}. Because the small-$q_{2}$
modes of large enough wavelength exit the Hubble radius before
evolving away from the early-time conditions specified by
Eq.~(\ref{eq:minkowski}), we would expect a massless inflaton to
generate a spectral index of $n_{s}=3$ in the small-$q_{2}$ region,
and we would expect a massive inflaton to generate a spectral index
of $n_{s}=4$ in the small-$q_{2}$ region.  If scale-invariance
continued indefinitely for large wavelength modes, the dispersion
would be infrared divergent, so this eventual end to
scale-invariance is not an artifact of our initial conditions. The
modes responsible for the galaxy-size structure of today left the
Hubble radius approximately 45 e-folds before the end of inflation
\cite[p. 285]{Kolb}, so if $N_{e}$ were not too much larger than
this, we would expect it to be possible to measure the end of
scale-invariance in our observable universe.

In Fig.~\ref{fig:exit} we plot six scenarios depicting the behavior
of $\left|\psi_{k}\right|^{2}$ after exiting the Hubble radius.  The
first two cases, A and B, are for $m_{H}=0$.  In both of these
cases, an expansion of 20 total e-folds is plotted. In case A, there
is a gradual end to inflation spanning one e-fold; and in case B,
there is an abrupt end to inflation. Because the mode has exited the
Hubble radius, neither of these end conditions changes
$\left|\delta\phi_{k}\right|^{2}$, and the two lines overlap. Here,
and in general for the massless case,
$\left|\delta\phi_{k}\right|^{2}$ reaches a constant value a few
e-folds after exiting the Hubble radius, and this constant value is
close to the value at the time of exit.  In the massless case, we
find a scale-invariant spectrum even when the spectrum is defined in
terms of the value of $\left|\delta\phi_{k}\right|^{2}$ at the end
of inflation. This can be found by noting that at the time a
particular mode is crossing the Hubble radius, the value of
$\left|\psi_{k}\right|^{2}$ given by Eq.~(\ref{eq:infleigen}) is
approaching a constant value as the argument $k/[a(t)H_{\rm infl}]$
becomes much less than 1.

For cases labeled C, D, E, and F; we use $m_{H}=\sqrt{0.1}$. In the
massive cases, $\left|\psi_{k}\right|^{2}$ never reaches a constant
value, although it changes much more slowly after exiting the Hubble
radius.  Cases C and D are the massive analogs of cases A and B,
respectively.  In case E, we end inflation gradually over the length
of one e-fold, starting just as our specific mode crosses the Hubble
radius.  In case F, we end inflation abruptly just as our specific
mode crosses the Hubble radius.

From \cite{Abramowitz}, \{Eq. 9.1.9\}, we see that in the small
argument limit of the Hankel functions
\begin{equation}
\left|H_{v}^{(1)}(z)\right|^{2}\simeq\left|H_{v}^{(2)}(z)\right|^{2}\simeq\left(\frac{\Gamma(v)}{\pi}\right)^{2}\left(\frac{1}{2}z\right)^{-2v},
\end{equation}
when the real part of the parameter $v$ is positive and non-zero. In
Eq.~(\ref{eq:massinfleigen}), we find numerically that
$E(k)\sim-(i/2)\sqrt{\pi/H_{\rm infl}}$ and $F(k)\sim0$ for modes of
intermediary-$q_{2}$, which are the modes that exit during the
exponential expansion of our composite scale factor. Long after
these modes have exited the Hubble radius, when $k/[a(t)H_{\rm
infl}]\ll1$, we expect Eq.~(\ref{eq:massinfleigen}) to approach
$\left|\psi_{k}\right|^{2}\simeq
a^{-3}\left|H_{v}^{(1)}(z)\right|^{2}\propto a^{-3}z^{-2v}$, where
$z=k/[a(t)H_{\rm infl}]$ and $v=\sqrt{(9/4)-m_{H}^{\ 2}}$. Using
this small argument approximation with Eqs.~(\ref{eq:spectrum})
and~(\ref{eq:spectral}) leads to
\begin{equation}
\frac{d\ln\mathcal{P}_{\mathcal{R}}}{d\ln k}=3-\sqrt{9-4m_{H}^{\
2}},
\end{equation}
under the assumption that we evaluate
$\left|\delta\phi_{k}\right|^{2}$ at late times, in which case
\begin{equation}
n_{s}=4-\sqrt{9-4m_{H}^{\ 2}}. \label{eq:blue}
\end{equation}

In determining the spectrum, $\left|\delta\phi_{k}\right|^{2}$ is
often evaluated at the time when a mode exits the Hubble radius
\cite{Liddle1, Dodelson}.  In this case, we would get $n_{s}=1$,
exactly. The WMAP results \cite{Dunkley} find $n_{s}\simeq0.96$,
which would suggest a  value of $m_{H}\simeq i/4$ to go along with
our assumption of a constant value of $H_{\rm infl}$, provided the
end of inflation is the appropriate time to evaluate
$\mathcal{P}_{\mathcal{R}}$.  This value of $m_{H}\simeq i/4$ is
what we found in Eq.~(\ref{eq:tachmass}) for the Coleman-Weinberg
potential. An imaginary $m_{\rm scalar}$ would lead to tachyonic
behavior \cite{Anderson1}, but here $m$ is an effective mass, so
this need not be a problem.

A different method of calculating Eq.~(\ref{eq:blue}), involves
combining Eq.~(\ref{eq:P}) and $q_{2}\equiv k/a_{2f}H_{\rm infl}$
with Eq.~(\ref{eq:timeavgphi}). To renormalize
Eq.~(\ref{eq:timeavgphi}) would involve dropping the Minkowski
vacuum term such that
$(1+2\left|\beta_{k}\right|^{2})\rightarrow2\left|\beta_{k}\right|^{2}$;
although, in the case of the intermediary-$q_{2}$ modes, the term to
be subtracted off is already negligible compared with the particle
number per mode.  To simplify the massive case, we treat $|m|\gg
k/a(t)$ in the intermediary-$q_{2}$ region of modes that exit the
Hubble radius during inflation. Then Eqs.~(\ref{eq:spectrum})
and~(\ref{eq:spectral}), together with the relationship given in
Fig.~\ref{fig:pvmh} of $P=\sqrt{9-4m_{H}^{\ 2}}$, or $P=2$ in the
massless case, give us the same result as in Eq.~(\ref{eq:blue}).

If $H_{\rm infl}$ were not constant, but were slowly decreasing
during inflation, then we would find a red-tilted spectrum.  We
could incorporate this effect into our exact calculation by taking
the adiabatic approach and using the value of $H_{\rm infl}(t_{1})$
for our first matching conditions and the value of $H_{\rm
infl}(t_{2})$ for our second joining. Combining
Eqs.~(\ref{eq:inflationspectra}), (\ref{eq:phi}), (\ref{eq:wave}),
(\ref{eq:scaleinv}), and~(\ref{eq:spectral}), we find
\begin{equation}
n_{s}=1+\frac{d}{d\ln k}\ln\left(\frac{H_{\rm
infl}(t)^{4}}{\dot{\phi}^{2}}\right),
\end{equation}
which, with Eq.~(\ref{eq:slowroll2}), becomes
\begin{equation}
n_{s}=1+\frac{d}{d\ln k}\left(6\ln[H_{\rm infl}(t)]-2\ln[V']\right),
\end{equation}
where a dot denotes a derivative with respect to time, and a prime
denotes a derivative with respect to $\phi$.  Then, using $d/d\ln
k=H^{-1}d/dt$, we have
\begin{equation}
n_{s}=1+6\frac{\dot{H}_{\rm infl}(t)}{H^{2}}-2\frac{\dot{V}'}{V'H},
\end{equation}
which, through the chain rule and with
Eq.~(\ref{eq:slowroll2}), $\dot{V}' = \dot{\phi}V''$\\
$=-V'V''/(3H_{\rm infl}(t))$, so that, finally, with
Eq.~(\ref{eq:slowroll1}), we have
\begin{equation}
n_{s}=1-6\frac{-\dot{H}_{\rm infl}(t)}{H_{\rm
infl}(t)^{2}}+2\frac{1}{8\pi G}\left(\frac{V''}{V}\right),
\end{equation}
which we write in terms of Eqs.~(\ref{eq:slowrolleps})
and~(\ref{eq:slowrolleta}) to get
\begin{equation}
n_{s}=1-6\epsilon+2\eta,
\end{equation}
which is equivalent to Eq.~(\ref{eq:slowrollspectralindex}) first
shown by \cite{Liddle2}.

\section{Density Perturbations}

Fig.~\ref{fig:exit} shows that the maximum difference between the
late-time values of $\left|\delta\phi_{k}\right|$ in all six of the
cases plotted is about 40\%.  We conclude that when $m_{H}\ll1$, the
value of $\left|\delta\phi_{k}\right|^{2}$ at late times is a
reasonably good indicator of the value of
$\left|\delta\phi_{k}\right|^{2}$ at Hubble radius exit.  For the
rest of this section we will adopt the assumption that the late time
value of $\left|\delta\phi_{k}\right|^{2}$ is indicative of the
value of $\left|\delta\phi_{k}\right|^{2}$ at the time of exiting
the Hubble radius.  This assumption allows us to extrapolate our
method of late-time renormalization in Minkowski space to a time of
curved spacetime in lieu of applying a more rigorous analysis that
would require a more complex method of curved spacetime
renormalization such as in \cite{Parker6}.

The final conditions do not affect the value of
$\left|\delta\phi_{k}\right|^{2}$ much once a given mode has crossed
the Hubble radius.  Thus, we could end inflation just after a mode
has exited the Hubble radius to find that the value of
$\left|\delta\phi_{k}\right|^{2}$ will be very close to its
late-time value.  At late times, Eqs.~(\ref{eq:wave})
and~(\ref{eq:alphabeta}) show that the time averaged expectation
value|
\begin{eqnarray}
\langle\overline{\left|\delta\phi_{k}\right|^{2}}\rangle&=&\frac{1}{2L^{3}a_{2f}^{\
\
3}\omega_{2f}}\left(\left|\alpha_{k}\right|^{2}+\left|\beta_{k}\right|^{2}\right)
\nonumber\\
&=&\frac{1}{2L^{3}a_{2f}^{\ \
3}\omega_{2f}}\left(1+2\left|\beta_{k}\right|^{2}\right).
\label{eq:timeavgphi}
\end{eqnarray}

This value of $\langle\left|\delta\phi_{k}\right|^{2}\rangle$
obtained from Eqs.~(\ref{eq:phi}), (\ref{eq:wave}),
and~(\ref{eq:scaleinv}), however, is un-renormalized.  To use the
renormalized values, we take
$(1+2\left|\beta_{q_{2}}\right|^{2})\rightarrow
2\left|\beta_{q_{2}}\right|^{2}$.

Although \cite[p. 285]{Kolb} identifies the scale factor,
$a_{gal}=a_{2f}\ e^{-45}$, as the one in which the k-modes
responsible (by seeding the density perturbations) for the formation
of galaxies are exiting the Hubble radius; we note that when
$m_{H}\ll1$ there is a relative constancy of
$\left|\delta\phi_{k}\right|^{2}$ after a mode crosses the Hubble
radius, and thus our subsequent method is widely applicable to the
range of intermediary-$q_{2}$ modes. In our assumption described
above, a mode defined by $q_{2}=1$ at late times is an excellent
indication of the state of any mode just after crossing the Hubble
radius when $m_{H}\ll1$. We can assume for the moment that inflation
ends abruptly just as the mode $k=a_{2f}H_{\rm infl}$ exits the
Hubble radius.  This abrupt ending does not change
$\left|\beta_{q_{2}}\right|^{2}$ for the $q_{2}=1$ mode, because we
have found that the ending conditions do not affect modes of
$q_{2}\lesssim1$. In this case,
$\left|\delta\phi_{q_{2}}\right|^{2}$ isn't changing from its value
at Hubble radius crossing (or is roughly equal to the late-time
value it would have reached a few e-folds after crossing the Hubble
radius), the late-time value of $\left|\beta_{q_{2}}\right|^{2}$ for
the $q_{2}=1$ mode isn't changing (because there is no more
inflation and the mode $q_{2}=1$ is insensitive to other factors),
and the scale factor isn't changing; therefore the renormalized
value of $\left|\delta\phi_{q_{2}}\right|^{2}$ is not changing. This
argument wouldn't hold for modes of large-$q_{2}$, because they are
sensitive to the time-derivatives of the scale factor, but we find
that the late-time dispersion spectra for the mode $q_{2}=1$ is a
good approximation to the renormalized value of
$\left|\delta\phi_{q_{2}}\right|^{2}$ at the time any mode exits the
Hubble radius.  For an analysis of the instantaneous renormalized
value of $\delta\phi_{k}$ that does not rely on a late-time
argument, see \cite{Parker6}.

We next consider the curvature perturbation given by
Eq.~(\ref{eq:curvpert}) and defined at the time of Hubble radius
crossing as
\begin{equation}
\mathcal{R}_{k}=-\frac{H}{\dot{\phi}}\delta\phi_{k}.
\end{equation}
The variable $\dot{\phi}$ is the rate of change of the homogeneous
background scalar field.  The quantum perturbations we have
considered so far, $\delta\phi_{k}$, are assumed to be much smaller
in magnitude than the zeroth-order field.

\subsection{Hybrid Combination with Slow Roll Approximation}

So far our method in this chapter has not been linked to any
particular potential or model of inflation.  In what comes next, we
choose a simple potential, which is found to be in good agreement
with the 3-Year WMAP data \cite{Spergel2}, and we use a hybrid
combination of our method and the slow roll approximation. The
remainder of this section is intended to be of a more speculative
nature than the rest of this dissertation. For our example, we use
the Linde quadratic chaotic-inflation potential \cite{Linde,Habib}
\begin{equation}
V=\frac{1}{2}m^{2}\phi^{2}. \label{eq:pot}
\end{equation}
From Eqs.~(\ref{eq:slowroll1}) and~(\ref{eq:slowroll2}), the two
slow roll conditions are
\begin{equation}
H^{2}\simeq\frac{8\pi G}{3}V,
\end{equation}
and
\begin{equation}
\dot{\phi}\simeq-\frac{dV/d\phi}{3H}.
\end{equation}
We combine these two slow roll equations with the potential
specified in Eq.~(\ref{eq:pot}) to find
\begin{equation}
\dot{\phi}\simeq-m\sqrt{\frac{2}{3}}\ \frac{1}{\sqrt{8\pi G}}.
\end{equation}
We rewrite this as
\begin{equation}
\dot{\phi}\simeq-H_{\rm infl}^{2}\ m_{H}\sqrt{\frac{2}{3}}\
\frac{1/\sqrt{8\pi G}}{H_{\rm infl}}.
\end{equation}
In our notation, with $\delta\phi_{k}$ taken from
Eq.~(\ref{eq:massdisp}),
\begin{equation}
\mathcal{R}_{k}=-\left[\frac{H_{\rm infl}}{-H_{\rm
infl}^{2}m_{H}\sqrt{\frac{2}{3}}\frac{1/\sqrt{8\pi G}}{H_{\rm
infl}}}\right]\sqrt{\frac{(q_{2}\rightarrow
1)(|\beta_{q_{2}}|^{2}\rightarrow\frac{1}{4})H_{\rm
infl}^{2}}{2\pi^{2}\sqrt{1+\frac{m_{H}^{\ 2}}{\left(q_{2}^{\
2}\rightarrow1\right)}}}};
\end{equation}
therefore, with $1/\sqrt{8\pi G}\simeq2.436\times10^{18}$ GeV,
\begin{equation}
\mathcal{R}_{k}=\frac{1}{4\pi}\sqrt{\frac{3}{m_{H}^{\
2}\sqrt{1+m_{H}^{\ 2}}}}\left(\frac{H_{\rm
infl}}{2.436\times10^{18}{\rm\ GeV}}\right).
\end{equation}
The magnitude of the curvature perturbation has been shown to be a
conserved quantity outside of the Hubble radius
\cite{Bardeen2,Sasaki}, and the curvature perturbation can be
related to the amplitude of density perturbations at the time of
re-entry, when once again $k/[a(t)H(t)]=1$.  In a matter-dominated
universe this relationship is \cite{Liddle1}
\begin{equation}
\frac{\delta\rho_{k}}{\rho}\equiv\delta_{k}=\frac{2}{5}\mathcal{R}_{k}.
\end{equation}

\begin{table}
\caption{Comparison of $\delta_{H}$ for
$V=\frac{1}{2}m^{2}\phi^{2}$} \label{tab:fluc}
\begin{tabular}{ l | c | c | c | c |} & $H=10^{12}\textrm{\ GeV}$ & $H=10^{14}\textrm{\ GeV}$ & $H=10^{16}\textrm{\ GeV}$\\  \hline
$m_H=0.0001$\hspace{1.0cm} & \hspace{0.5cm}$2.263\times
10^{-4}$\hspace{0.5cm} & \hspace{0.5cm}$2.263\times
10^{-2}$\hspace{0.5cm} & \hspace{0.5cm}$2.263\times 10^{0}$\hspace{0.5cm}\ \\
\hline $m_H=0.01$ & $2.263\times 10^{-6}$ & $2.263\times 10^{-4}$
& $2.263\times 10^{-2}$\\
\hline $m_H=0.1$ & $2.258\times 10^{-7}$ & $2.258\times 10^{-5}$ &
$2.258\times 10^{-3}$\\
\hline $m_H=0.25$ & $8.917\times 10^{-8}$ & $8.917\times 10^{-6}$ &
$8.917\times 10^{-4}$\\
\hline $m_H=1$ & $1.903\times 10^{-8}$ & $1.903\times 10^{-6}$ &
$1.903\times 10^{-4}$\\  \hline
\end{tabular}
\end{table}

See TABLE~\ref{tab:fluc} for sample values of $\delta_{H}$, the
density contrast defined in \cite{Liddle1}, at the time of re-entry
into the Hubble radius and for the potential given by
Eq.~(\ref{eq:pot}).

\subsection{Relative Constancy of Modes Outside Hubble Radius}

In Fig.~\ref{fig:exit},
\begin{figure}[hbtp]
\includegraphics[scale=2.5]{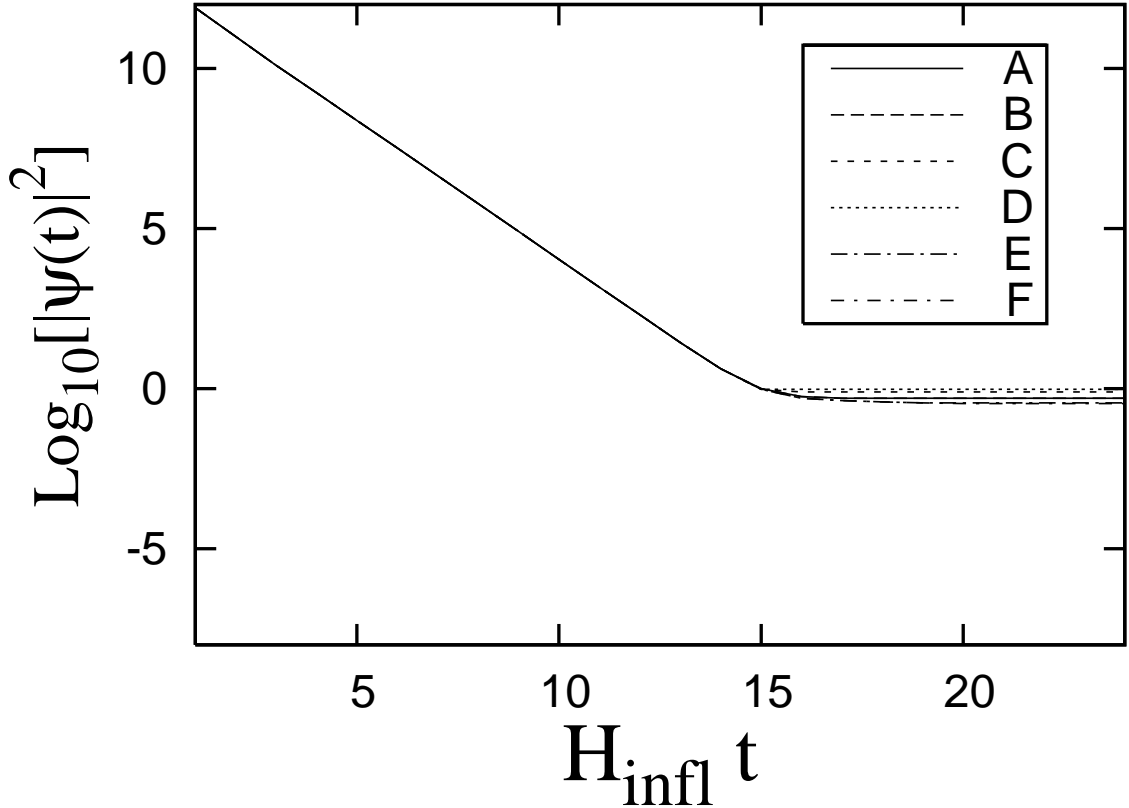}
\caption{\label{fig:exit} Modes Exiting the Hubble Radius.}
\end{figure}
the value of $\left|\psi_{k}\right|^{2}$, in units of $H_{\rm
infl}^{2}/k^{3}$ and for the mode $k=(a(t_{1})e^{15})H_{\rm infl}$,
is plotted versus dimensionless time $H_{\rm infl}\,t$. The graph
shows the relative constancy of $|\delta\phi_{k}|^{2}$ for modes
that have exited the Hubble radius during inflation.  For additional
discussion of this graph and the difference between cases A-F, see
Sec.~\ref{sec:spec}.

The inflaton perturbations only approach a true constant well
outside the Hubble radius for the massless case.  During an
exponential expansion in the massive case, $m_{H}\gg k/[a(t)H_{\rm
infl}]$, and we may rewrite Eq.~(\ref{eq:evot}) as
\begin{equation}
\partial_{t}^{2}\psi_{k}(t)+3H\partial_{t}\psi_{k}(t)+
m^{2}\psi_{k}(t)=0.
\end{equation}
The two linearly independent solutions to this are
\begin{equation}
\psi_{\pm}\propto\exp\left[-\frac{1}{2}\left(3\pm P\right)H_{\rm
infl}t\right],
\end{equation}
where $P$ is defined as in Fig~\ref{fig:pvmh}.  In the small mass
limit of $P\rightarrow3$, well outside the Hubble radius one of
these linearly independent solutions approaches a constant value
with respect to $t$, while the other solution decays exponentially.
With $m_{H}$ of order 1, both linearly independent solutions decay
exponentially outside the Hubble radius. Although the magnitude of
these massive perturbations are constant when the scale factor is
constant in time, the rates of their decay well outside the Hubble
radius depends on how the scale factor is changing. For our
composite scale factor with an abrupt end to inflation, where almost
all of the expansion occurs in the exponentially growing segment of
our scale factor, there is more small-$q_{2}$ particle production
than in our composite scale factor with a relatively gradual end to
inflation, where more of the total expansion of the scale factor
takes place in the final asymptotically flat segment of our scale
factor.  We turn now to tracing a particular mode as it exits the
Hubble radius until it re-enters our observable universe.

Consider, as an example, the $k$-modes responsible for large-scale
structure formation.  As an approximation, take the following three
epochs to be simultaneous: recombination (the time light was emitted
from the surface of last scattering), the transition from a
radiation-dominated universe to a matter-dominated universe, and the
re-entry of the modes that would provide the density perturbations
to seed galaxies.  Furthermore, also as an approximation, assume a
transition to a radiation-dominated universe, where $a(t)=Ct^{1/2}$,
immediately after the end of inflation such that $H(t)$ is
continuous.  Call the time of the end of inflation $t_{f}$, and call
the time of re-entry and recombination $t_{r}$.  Turner and Kolb
give the temperature of inflation and the temperature at
recombination as $10^{14}$ GeV and $1$ eV, respectively \cite{Kolb}.
In a radiation-dominated universe, the energy density| neglecting
particle production| is related to the scale factor as $\rho_{\rm
rad}\propto a(t)^{-4}$, and the temperature is related to the energy
density as $T\propto\rho_{\rm rad}^{1/4}$, so the temperature is
related to the scale factor as $T\propto a(t)^{-1}$ after radiation
and matter have decoupled and are no longer in thermal equilibrium.
Thus we know that $a(t_{r})=10^{23}a(t_{f})$. The
radiation-dominated scale factor then gives us
$(t_{r}/t_{f})^{1/2}=10^{23}$, or $t_{r}=10^{46}t_{f}$.  The hubble
constant in the radiation-dominated universe is
$H(t)=a(t)^{-1}da(t)/dt=(2t)^{-1}$. Because $H(t_{f})=H_{\rm infl}$,
we have $H(t_{r})=10^{-46}H_{\rm infl}$. When a mode exits the
Hubble radius during inflation, we have
$k/(a(t)H(t))=k/(a(t_{f})e^{-K_{e}}H_{\rm infl})$, where the
variable $K_{e}$ is the number of e-folds before the end of
inflation at which a mode exits the Hubble radius.  When our example
mode re-enters the Hubble radius after inflation, we have
$k/(a(t)H(t))=k/(a(t_{r})H(t_{r}))=1$.  By equating the relations
for exit and re-entry, we have $k/(a(t_{f})e^{-K_{e}}H_{\rm
infl})=k/(10^{23}a(t_{f})10^{-46}H_{\rm infl})$, or
$e^{-K_{e}}=10^{-23}$.  This means in our approximation
\begin{equation}
K_{e}\simeq53\ {\rm e\frac{\ }{\ }folds}. \label{eq:galacmode}
\end{equation}
Turner and Kolb find, with a more detailed calculation, a value of
$45$ for this number \cite[p. 285]{Kolb}.  The simplification of
treating recombination, matter-radiation equality, and galaxy
seeding as concurrent is a relatively useful approximation.  The
radiation-dominated universe transitions to a matter-dominated
universe at a temperature roughly one order of magnitude higher than
the temperature of recombination, which means the mode that will
later re-enter the Hubble radius at the time of radiation-matter
equality exits the Hubble radius during inflation roughly 2 e-folds
later than the mode that will later re-enter the Hubble radius at
recombination.  The exact relationship between these two events with
the Hubble crossing for the modes responsible for seeding galaxy
formation depends on the nature of dark matter: the current size of
galaxies does not lead to a simple estimate of their size in the
past, because their size does not scale with the size of the
universe once they have become gravitationally bound. Baryonic
matter will clump to structure initiated by cold dark matter, but
not until after recombination, when radiation pressure overcomes
gravitational attraction; dark matter will start clumping earlier
than this, at the epoch when it decouples from the dominant
radiation background \cite{Kolb,Dodelson,Liddle1}.  The
approximation of an immediate transition from inflation to a
radiation-dominated universe is less certain, as the validity of
this approach could vary based on the specific inflationary
potential being considered.

\section{Reheating}
\label{sec:reheat}

Our analysis of this particle creation reveals a mechanism for
Reheating, which is a return to the temperatures and densities that
are responsible for the successes of the Big Bang model.  We find
that the energy density present after inflation depends on how
abrupt the transition is from the inflationary middle segment of
exponential growth to the final asymptotically flat region of the
scale factor.

\subsection{Energy Density from Abrupt End to Inflation}

Our scale factor can be made to be continuous to the scale factor
and two of its derivatives, but no more, so we see additional
particle production caused by discontinuities of higher derivatives.
When we maintain continuity of $a(t)$, $\dot{a}(t)$, and
$\ddot{a}(t)$, the particle number is proportional to $q_{2}^{\ -6}$
for large-$q_{2}$. The energy of a particle of mode-$k$ at late
times is $\omega_{2f}=\sqrt{\left(k/a_{2f}\right)^{2}+m^{2}}$. The
energy per mode in the large-$q_{2}$ regime is then proportional to
$(k/a_{2f})q_{2}^{\ -6}=q_{2}^{\ -5}H_{\rm infl}$.

With a gradual transition between segments of $a(t)$, the
large-$q_{2}$ behavior in which $\left|\beta_{q_{2}}\right|^{2}$
falls off as $q_{2}{}^{-6}$ starts around $q_{2}\simeq1$.  With an
arbitrarily abrupt transition from the end of inflation to our final
asymptotically flat scale factor, however, this transition can be
prolonged to an arbitrarily high value of $q_{2}$, which we denote
by $q_{2 \rm cut-off}$. We find empirically that $q_{2 \rm cut-off}
\simeq a_{2f}/(a_{2f}-a_{1f})$. We define the region between
$1\lesssim q_{2}\lesssim q_{2 \rm cut-off}$ as the ``extended"
region. In the ``extended" region the fall off of
$\left|\beta_{q_{2}}\right|^{2}\propto q_{2}{}^{-2}$ is extended
from $q_{2}\simeq1$ to larger values of $q_{2}$, such as the value
of $q_{2}\simeq10^{4}$ shown in Fig.~\ref{fig:massive}, in which
$a_{2f}-a_{1f}\ll a_{2f}$ as $a(t)$ makes a rapid transition to
flatness. This extension is caused by the production of particles of
higher momenta by the rapid change in $H(t)$ after inflation. When
the transition of $a(t)$ is gradual, one finds, as in
Fig.~\ref{fig:domterm}, that beyond $q_{2}\simeq1$, the quantity
$\left|\beta_{q_{2}}\right|^{2}$ falls off more rapidly, eventually
going as $q_{2}{}^{-6}$ if the function $a(t)$ is $C^2$.  With
sufficient extension, the particle number per mode in the
``extended" region is proportional to $q_{2}^{\ -2}$, regardless of
the value of $P$ in the intermediary-$q_{2}$ region, so for both the
massless and massive cases the contribution to the total energy
density is dominated by these ``extended" modes, and we neglect both
the red-shifted modes and the ultraviolet modes. When $a(t)$ is
$C^2$ and there exists a significant ``extended" region, the
contribution to the energy density from values of $q_{2}> q_{2 \rm
cut-off}$ is negligible. The energy density associated with the
``extended" region, which dominates the total energy density when
$a_{2f}-a_{1f}\ll a_{2f}$, is
\begin{eqnarray}
\left<E\right>&&\simeq\frac{1}{(2\pi a_{2f})^{3}}\int_{a_{2f}H_{\rm
infl}}^{a_{2f}H_{\rm infl}\ q_{2 \rm
cut-off}}\frac{k}{a_{2f}}\left|\beta_{q_{2}}\right|^{2}d^{3}k
\nonumber\\
&&=\int_{1}^{q_{2 \rm cut-off}}\frac{q_{2}^{\
3}\left|\beta_{q_{2}}\right|^{2}H_{\rm infl}^{4}}{2\pi^{2}}dq_{2}
\nonumber\\
&&=\int_{1}^{q_{2 \rm cut-off}}\frac{q_{2}H_{\rm
infl}^{4}}{8\pi^{2}}dq_{2}.
\end{eqnarray}
When $q_{2 \rm cut-off}\gg1$, we find for the energy density
\begin{equation}
\left<E\right>\simeq\frac{H_{\rm
infl}^{4}\left(\frac{a_{2f}}{a_{2f}-a_{1f}}\right)^{2}}{16\pi^{2}}.
\label{eq:abruptengdens}
\end{equation}
Because $q_{2 \rm
cut-off}\simeq\left(a_{2f}/[a_{2f}-a_{1f}]\right)$, we can see that
an abrupt end to inflation can lead to energy densities large enough
to produce reheating.  For particle production as the cause of
reheating, see also \cite{Ford1}.

\subsection{Associated Temperature}

In units of $\hbar=c=k_{B}=1$, the temperature is $T=(\langle
E\rangle/\sigma)^{1/4},$ where $\sigma$ is the Stefan-Boltzmann
constant.  Then the energy density attributable to an abrupt end to
inflation given by Eq.~(\ref{eq:abruptengdens}) leads to an
effective temperature of
\begin{equation}
T\simeq\sqrt{\frac{a_{2f}/\sqrt{\sigma}}{a_{2f}-a_{1f}}}\
\frac{H_{\rm infl}}{2\sqrt{\pi}}.
\end{equation}
This approximation holds for any relatively abrupt transition and
does not depend on any discontinuities of the scale factor.

In an expansion governed by the asymptotically flat scale factor of
Eq.~(\ref{eq:aHG}) with no exponential middle segment, the large-$k$
behavior| in both the massless case and the effective-$k$ approach|
follows a thermal spectrum given by
\cite{Parker4,Parker5,Parker8,Parker9}
\begin{equation}
T=\frac{1}{4\pi sa_{<}^{\ 2}a_{2}}.
\end{equation}
When $a_{1}\simeq a_{2}$, we use
\begin{equation}
H_{\rm max}\simeq\frac{1-\frac{a_{1}^{\ 4}}{a_{2}^{\ 4}}}{16a_{2}^{\
3}s},
\end{equation}
to show that in our notation this is equivalent to
\begin{equation}
T\simeq\frac{4H_{\rm infl}}{\pi (1-\frac{a_{1f}^{\ \ 4}}{a_{2f}^{\ \
4}})}
\end{equation}
for a single asymptotically flat scale factor with $a_{1}\simeq
a_{2}$. In the large-$q_{2}$ regime of our composite scale factor
with $a_{1f}\simeq a_{2f}$, we would expect to find the temperature
approaching this same value, regardless of mass, of $P$, and of the
number of e-folds; but only if we were able to maintain continuity
with the previous segments of the scale factor across an infinite
number of derivatives.

With a gradual transition between segments of $a(t)$, the
large-$q_{2}$ behavior in which $\left|\beta_{q_{2}}\right|^{2}$
falls off as $q_{2}{}^{-6}$ starts around $q_{2}\simeq1$.  For such
a gradual transition, we find a late-time temperature| which is
red-shifted after the end of inflation by the expansion of the final
asymptotically flat segment of the scale factor| that is comparable
to the Gibbons-Hawking temperature of $H/(2\pi)$ \cite{Gibbons}.

It is tempting to imagine the temperature varying continuously from
the Gibbons-Hawking temperature describing a de Sitter state| or
from an approximate Gibbons-Hawking temperature associated with the
approximate de Sitter state in our case| to the near Gibbons-Hawking
temperature equivalent at late times in our asymptotically flat
space, but this is perhaps unwarranted. At late times, the average
number of particles created per mode from an early-time vacuum is
well defined.  This is not necessarily the case during inflation,
when a choice must be made whether to make a measurement rapidly or
slowly. If the measurement were made quickly, then by the
time-energy uncertainty relationship, particles would be created
through the act of measurement; if the measurement were made slowly,
then the size of the scale factor would change appreciably during
the measurement process, which could change the outcome
\cite{Parker1}. Just as an observer accelerating through a Minkowski
vacuum measures particles
\cite{Parker5,Rindler,Fulling2,Fulling3,Unruh,Davies}, so would a
temperature-measuring device be excited in de Sitter space; however,
unlike a thermal bath in flat spacetime, a moving observer in de
Sitter space would register no red-shifting in any direction. In
fact, the authors of \cite{Anderson3} find that for a massless,
minimally-coupled scalar field in de Sitter space, no particles
would be produced, and the associated effective temperature from
these particles would be zero.  With our composite scale factor, and
using our late-time evaluation method alone, it is difficult to say
whether particles are present during the exponential expansion, or
whether they are created by the changing Hubble constant at the end
of inflation.  It is likely that during the expansion, the
long-wavelength modes that have exited the Hubble radius correspond
to real, low-energy particles, while the high-frequency modes that
have not left the Hubble radius correspond to virtual particles
whose promotion to real particles depends upon the future evolution
of the universe| such as our matching conditions| but to say
conclusively whether particles exist during inflation would require
a quantum field renormalization in curved spacetime, such as the
adiabatic method given by \cite{Parker6}.

By showing that the particle production of certain predicted
particle species would cause conditions incompatible with
observations in our universe, high-energy particle physics may be
able to constrain the amount of reheating.  Because we have shown
how reheating| subject to ending conditions| is general to
large-$H_{\rm infl}$ inflationary models, this can similarly be used
to place model-dependent constraints on predictions for new
particles, such as theorized supersymmetric partners of observed
particles, under particular values of $H_{\rm infl}$. In one such
analysis \cite{Pradler}, if the gravitino $\tilde{G}$ is the
lightest supersymmetric particle (LSP), then this constrains the
maximum reheating temperature to be less than $10^{7}$ GeV.  If the
$\tilde{G}$ is not the LSP, and if its mass might be expected to be
$\sim100$ GeV, then the maximum reheating temperature may still be
less than or about $10^{7}$ GeV \cite{Cyburt}. Another example of a
constraint on reheating is for the particle creation of scalar
moduli, which may be present in supersymmetry and string theories:
if the magnitude of the effective mass of the moduli field is less
than $H_{\rm infl}$, then the upper limit on the reheating
temperature could be as low as $100$ GeV \cite{Giudice}.  This
constraining works both ways. If evidence were found for the
existence of such a reheating-constraining particle, this could
eliminate those models of inflation that predict a large, nearly
constant value of $H_{\rm infl}$ along with a rapid end to
inflation. Those models that would be in agreement with such a low
reheating temperature would be those with either a relatively small
value of $H_{\rm infl}$, or those with a final period of inflation
at which the inflationary potential has reached a near-minimum
value, but at which it remains the dominant influence on the
evolution of the scale factor, so that the initial high-energy
particle production is greatly red-shifted and so that any unwanted
relic particles are sufficiently attenuated such that they do not
interfere with later early-universe processes, such as Big Bang
Nucleosynthesis.

\newpage

\

\

\

\

\

\

\

\

\

\

\

\noindent\textbf{\Huge Part II:}

\

\noindent\textbf{\huge Binary System of}
\addcontentsline{toc}{chapter}{Part II - Binary System of Compact
Masses}

\

\noindent\textbf{\huge Compact Masses}

\newpage
\thispagestyle{fancy}
\chapter{Unequal Mass Binary Solution in a Post-Minkowski Approximation}
\thispagestyle{fancy}
\pagestyle{fancy}

\label{sec:binar}

In \cite{Friedman}, Friedman and Ury\=u investigate a particular
system of binary point masses that acquires a helical symmetry by
taking the half-advanced plus half-retarded fields from the
linearized Einstein equation.  This time-invariant system in the
co-rotating reference frame provides for an action at a distance
theory, as has been previously discussed by
\cite{Fokker,Wheeler1,Wheeler2}.  It allows for a single action
integral that depends on the dynamical variables and trajectories of
each particle, without requiring a description of the force field
acting on the particles.  Such an action is called a Fokker action
\cite{Friedman}.  The Fokker action is not a true action, as the
variation of the Fokker action integral depends on the boundary
conditions and it involves integrals over each point mass's
parameter time. When, however, a limit is taken after the variation
of the Fokker action, in which its endpoints are taken at times of
$-\infty$ and $+\infty$, the variation yields the correct equations
of motion. The conserved energy and angular momentum associated with
the Fokker action remain finite, even though energy and angular
momentum of the field are infinite due to radiation from the system
occurring over an infinite amount of time.

In the post-Minkowski (PM) approximation, the metric is assumed to
be flat with small perturbations of the form
$g_{ab}=\eta_{ab}+h_{ab}$, where to linear order $h_{ab}$ is the
half-advanced plus half-retarded field of each particle. Unlike the
post-Newtonian (PN) approximation, however, $v/c\ll1$ need not be
the case \cite{Ledvinka}. For the rest of this chapter we will use
units of $c=G=1$.  Friedman and Ury\=u note that in zeroth order PM
approximation $T^{ab}=\rho u^{a}u^{b}=0$, and particles travel on
flat space geodesics. A naive first order perturbation would then
lead to $\delta T^{ab}=\delta\rho u^{a}u^{a}+\rho\delta
u^{a}u^{b}+\rho u^{a}\delta u^{b}=\delta\rho u^{a}u^{a}$, which,
because $u^{a}$ is the unperturbed straight-line motion, does not
allow for bound orbits. In \cite{Friedman}, this is avoided by
considering a parameterized family of solutions to
$T^{ab}(s)=\rho(s)u^{a}(s)u^{b}(s)+p(s)[g^{ab}(s)+u^{a}(s)u^{b}(s)]$
that corresponds to flat space for $s=0$.  In a radiation gauge,
\begin{equation}
-2G_{ab}^{(1)}\equiv\Box(h_{ab}-\frac{1}{2}\eta_{ab})h=-16\pi
T_{ab}^{(1)}, \label{eq:lineargrav}
\end{equation}
where $h$ is the trace $h^{a}{}_{a}$, the first-order stress-tensor
is constructed from the first-order $u^{a}$, from the first-order
$\rho$, and from the flat-space metric.  In the binary solution, to
first order the motion of each mass is given by the linear field of
the other, and the self-force serves only to renormalize the mass as
a self-energy. Furthermore, Friedman and Ury\=u note of their
post-Minkowski solution that it is correct to Newtonian order (0PN),
the radiation field of the linearized metric is correct to 2.5PN,
and a correction term to the equations of motion is necessary to
have the orbits agree with the 1PN solutions. For the case of the
electromagnetic force, a specific example, in which the self force
and radiation reaction are calculated, is given in greater detail in
the Appendix~\ref{appendix}. For the case of gravity, instead of
photons the radiation takes the form of gravitational waves. The
measurement of the energy loss due to this radiation in a particular
binary system which contained a pulsar earned Hulse and Taylor a
Nobel Prize in 1993.

Although in linearized gravity non-linear terms are dropped that are
of the same PN-order as linear terms that are kept, which means the
next highest PM-order will have terms of equal magnitude to those
used at linear PM-order, the post-Minkowski approximation may be
helpful in evaluating solutions involving the full Einstein
equations in General Relativity that use helically symmetric initial
data sets. Such initial conditions neglect the radial velocities
associated with the radiation-reaction force, but a second-order
post-Minkowski framework might lead to a better understanding of
requirements for initial data in full-GR simulations.

Fig.~\ref{fig:circle}
\begin{figure}[hbtp]
\includegraphics[scale=4]{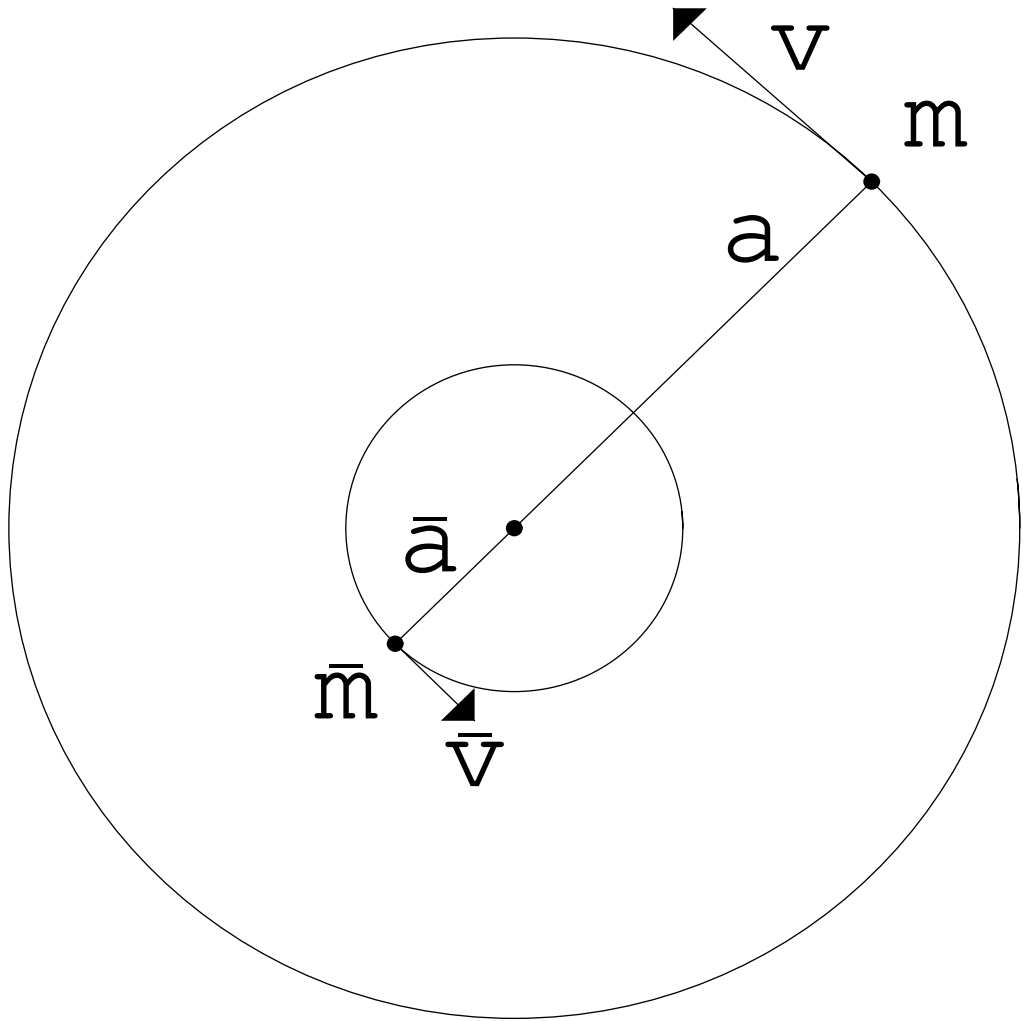}
\caption{\label{fig:circle} Binary in Circular Motion.}
\end{figure}
shows the two point masses, $m$ and $\bar{m}$, with respective
velocities $v$ and $\bar{v}$.  The radial parameters can be
expressed as $a\equiv v/\Omega$ and $\bar{a}\equiv\bar{v}/\Omega$,
where $\Omega$ is the angular velocity shared by both point masses.
Accounting for relativistic velocities, the radial parameter is not
equal to the  $1/(2\pi)$ times the circumference observed in the
particle's co-moving frame.  The position vectors are
$x^{a}=tt^{a}+a\varpi^{a}$ and
$\bar{x}^{a}=\bar{t}t^{a}+\bar{a}\varpi^{a}$. The trajectory of $m$
is tangent to the helical Killing vector $k^{a}=t^{a}+\Omega
a\hat{\phi}^{a}$, and the trajectory of $\bar{m}$ is tangent to the
helical Killing vector $\bar{k}^{a}=\bar{t}^{a}+\Omega
\bar{a}\hat{\phi}^{a}$, where $\gamma\equiv dt/d\tau$.

In Fig.~\ref{fig:angle}
\begin{figure}[hbtp]
\includegraphics[scale=4]{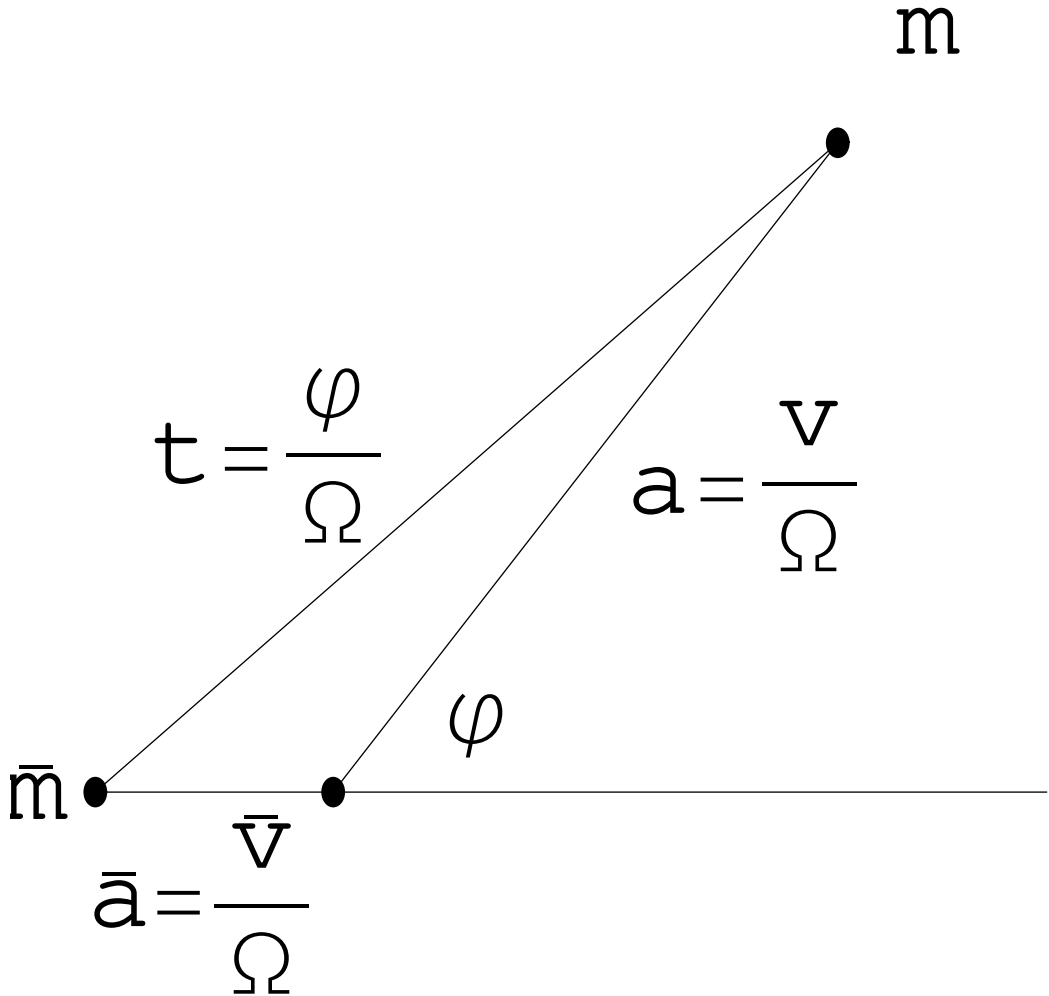}
\caption{\label{fig:angle} Retarded Angle $\varphi$.}
\end{figure}
the Law of Cosines relates
$t^{2}=a^{2}+\bar{a}^{2}-2a\bar{a}\cos(\pi-\varphi)$, or
$(\varphi/\Omega)^{2}=(v/\Omega)^{2}+(\bar{v}/\Omega)^{2}+2(v/\Omega)(\bar{v}/\Omega)\cos(\varphi)$,
so that the retarded angle, which is equal in magnitude to the angle
associated with the advanced position, is given by the positive root
of the transcendental equation
$\varphi^{2}=v^{2}+\bar{v}^{2}+2v\bar{v}\cos\varphi$.

From two types of Fokker action, a parametrization invariant action
with a post-Newtonian correction and an affinely parametrized
action, the equations of motion and expressions for conserved energy
and angular momentum are derived following the variational
calculation of Ref. \cite{Dettman}.  In the Affine case, we
parameterize the trajectories using the perturbed flat-space metric
as $(\eta_{ab}+h_{ab})\dot{x}^{a}\dot{x}^{b}=-1$ and
$(\eta_{ab}+\bar{h}_{ab})\dot{\bar{x}}{}^{a}\dot{\bar{x}}{}^{b}=-1$,
where the dots represent derivatives with respect to the parameter
times of $x(\tau)$ and $\bar{x}(\bar{\tau})$. This leads to
$\gamma=(1-v^{2}-h_{ab}k^{a}k^{b})^{-1/2}$ and
$\bar{\gamma}=(1-\bar{v}^{2}-\bar{h}_{ab}\bar{k}{}^{a}\bar{k}{}^{b})^{-1/2}$.
In the parameter-invariant case, we parameterize the trajectories
using the flat-space metric as $\eta_{ab}\dot{x}^{a}\dot{x}^{b}=-1$
and $\eta_{ab}\dot{\bar{x}}{}^{a}\dot{\bar{x}}{}^{b}=-1$. This leads
to $\gamma=(1-v^{2})^{-1/2}$ and
$\bar{\gamma}=(1-\bar{v}^{2})^{-1/2}$.  The affine parameterization
is characterized by the following: the parameter times of geodesics
are the proper times of the perturbed metric; the PM-form of the
geodesic equation
$(\eta_{ab}+h_{ab})\ddot{x}^{b}+C_{abc}\dot{x}^{b}\dot{x}^{c}$
applies, where
$C_{abc}\equiv(1/2)(\nabla_{b}h_{ac}+\nabla_{c}h_{ba}-\nabla_{a}h_{bc})$;
and, finally, the 4-velocity is orthogonal to the 4-acceleration, or
$U^{a}\nabla_{a}U^{b}=0$, that is the particles travel along
geodesics. The linear post-Minkowski approximation is not at this
point accurate to 1PN order, but Friedman and Ury\=u give two
different adjustments to the parametrization-invariant case: the
simplest correction consistent with 1PN (called PN where confusion
will not arise) and a correction that is both
parameterization-invariant and special-relativistically covariant
(SPN), where results are given in \cite{Friedman} for the deDonder
gauge.  They show also that for both of the Fokker actions the form
of the first law of thermodynamics $dE=\Omega dL$ holds, and this
law can be used to check for the presence of an Innermost Stable
Circular Orbit (ISCO).

We find a solution describing a helically symmetric circular orbit
in the post-Minkowski approximation (with post-Newtonian
corrections) that is analogous to the circular solution of two
charges obtained by Schild for the electromagnetic interaction
\cite{Schild}. In \cite{Glenz2} we report results supplementing
those of \cite{Friedman}: numerically computed solution sequences
for unequal mass particles, and analytic formulas in the extreme
mass ratio limit. The latter results agree with the first
post-Newtonian (1PN) formulas; hence a consistency of our model is
confirmed in this limit.

We present a set of formulas governing the helically symmetric
circular orbits of two point particles, $\{m,v\}$ and
$\{\bar{m},\bar{v}\}$, and derive analytic expressions in the
extreme mass ratio limit $q\equiv m/\bar{m} \rightarrow 0$. The set
of algebraic equations is solved numerically for a fixed binary
separation to specify each circular orbit. The result for the
unequal mass binary orbit is presented in Sec. \ref{sec:numerical}.

We compute the solution to the equation of motion numerically for
three mass ratios: q=1.0, q=0.1, and q=0.001.  We solve the equation
of motion for each mass ratio in the PM+PN model, the PM+SPN model,
and the affine model.  We also calculate the solution for the
\begin{math}q\rightarrow 0\end{math} limit analytically in each of
the three models, for which see Sec.~\ref{sec:qzero}. Whenever the
analytical solution is plotted along with the q=0.001 numerical
solution, the two lines overlap in the graphs given here.

\section{Numerical Solutions}
\label{sec:numerical}

We discuss solutions to the post-Minkowski approximation in the case
of parametrization-invariant plus 1PN correction terms, and then we
discuss solutions in the affine case.  For the analytical solution
in the $q\rightarrow0$ limit, see Sec.~\ref{sec:qzero}.

\subsubsection{Parameter Invariant Circular Solution}
\label{sec:pinveqs}

We first list the result from \cite{Friedman} for the
parametrization invariant model with 1PN correction terms.  After
integration, the equations of motion for particles $m$ and $\bar{m}$
are written in terms of the velocities, $v$ and $\bar{v}$, of
particles $m$ and $\bar{m}$, which are related to the orbital radius
by $a\equiv v/\Omega$ and $\bar{a}\equiv \bar{v}/\Omega$, through
the equations:
\begin{eqnarray} -m\gamma^2 v\Omega &=& - m\bar{m}\gamma^2\bar{\gamma} \Omega^2
\big[\,F(\varphi,v,\bar{v})+(m+\bar{m})\Omega\, {F_{\,\rm I}}
(\varphi,v,\bar{v},\gamma,\bar{\gamma})\, \big], \label{eq:eompn}
\\[2mm]
-\bar{m}\bar{\gamma}^2 \bar{v}\Omega &=& -
m\bar{m}\gamma\bar{\gamma}^2 \Omega^2
\big[\,\bar{F}(\varphi,v,\bar{v})+(m+\bar{m})\Omega\, {\bar F_{\,\rm
I}} (\varphi,v,\bar{v},\gamma,\bar{\gamma})\, \big].
\label{eq:eompnbar}
\end{eqnarray}
As shown below, $\{\varphi,v,\bar{v},\gamma,\bar{\gamma}\}$ are not
independent. The functions
$F(\varphi,\bar{v},v)=\bar{F}(\varphi,v,\bar{v})$ are the
post-Minkowski terms, while ${F_{\,\rm
I}}(\varphi,\bar{v},v,\bar{\gamma},\gamma) = {\bar F_{\,\rm I}}
(\varphi,v,\bar{v},\gamma,\bar{\gamma})$ is either of two
alternative 1PN correction terms that agree at 1PN order: ${F_{\,\rm
I}} = {F_{\rm PN}}(\varphi,v,\bar{v},\gamma,\bar{\gamma})$ derived
from a non-relativistic correction, or ${F_{\,\rm I}} = {F_{\rm
SPN}}(\varphi,v,\bar{v},\gamma,\bar{\gamma})$ derived from a special
relativistically invariant correction. These are
\begin{eqnarray} F(\varphi,v,\bar{v}) \,&\equiv&\, - 4
\frac1{(\varphi+v\bar{v}\sin\varphi)^2}
\bigg\{(1+v\bar{v}\cos\varphi)\bar{v}
\nonumber\\
&& \times (\varphi\cos\varphi-v^2\sin\varphi) +\frac12
v(1-\bar{v}^2)(\varphi+v\bar{v}\sin\varphi)
\nonumber\\
&&- \frac12\big[\bar{v}\sin\varphi(\varphi+v\bar{v}\sin\varphi) +
(1+v\bar{v}\cos\varphi)(v+\bar{v}\cos\varphi)
\nonumber\\
&&- \frac{v}{1-v^2}(\varphi+v\bar{v}\sin\varphi)^2
\big]\Phi(\varphi,v,\bar{v}) \bigg\},
\\[2mm]
{F_{\rm PN}}(\varphi,v,\bar{v},\gamma,\bar{\gamma}) \,&\equiv&\, -
\frac{1}{\gamma^2 \bar{\gamma}^2 (v+\bar{v})^{3}} \left[
1+\frac12\gamma ^2 v(v+\bar{v})\right],
\\[2mm]
{F_{\rm SPN}}(\varphi,v,\bar{v},\gamma,\bar{\gamma}) \,&\equiv&\,
-\frac1{(\gamma \bar{\gamma})^{5/2}} \frac{1}{\left( \varphi
+v\bar{v}\sin \varphi \right)^2}\bigg\{ \frac34\gamma ^{2}v
+\frac{\bar{v}\sin \varphi } {\varphi +v\bar{v}\sin \varphi
}\nonumber\\
&&+\frac{\left( 1+v\bar{v}\cos \varphi \right) \left( v+\bar{v}\cos
\varphi \right) }{\left( \varphi +v\bar{v}\sin \varphi
\right)^2}\bigg\}.
\end{eqnarray} The function $\Phi(\varphi,v,\bar{v})$ is defined by \begin{equation}
\Phi(\varphi,v,\bar{v}) \,\equiv\,\frac{(1+v\bar{v}\cos\varphi)^2 -
\frac12(1-v^2)(1-\bar{v}^2)} {\varphi+v\bar{v}\sin\varphi}.
\label{eq:fncPhi} \end{equation} For the parametrization invariant
models, $\gamma$ and $\bar{\gamma}$ are derived from a flat-space
normalization of the four-velocity, \begin{equation} \gamma =
(1-v^2)^{-\frac12},\ \ \ \ \ \ \bar{\gamma} =
(1-\bar{v}^2)^{-\frac12}. \end{equation} The retarded angle
$\varphi$ is the positive root of $\varphi^2 = v^2 + \bar{v}^2 +2v
\bar{v} \cos\varphi$.

In Fig.~\ref{fig:PNOV}
\begin{figure}[hbtp]
\includegraphics[scale=2.5]{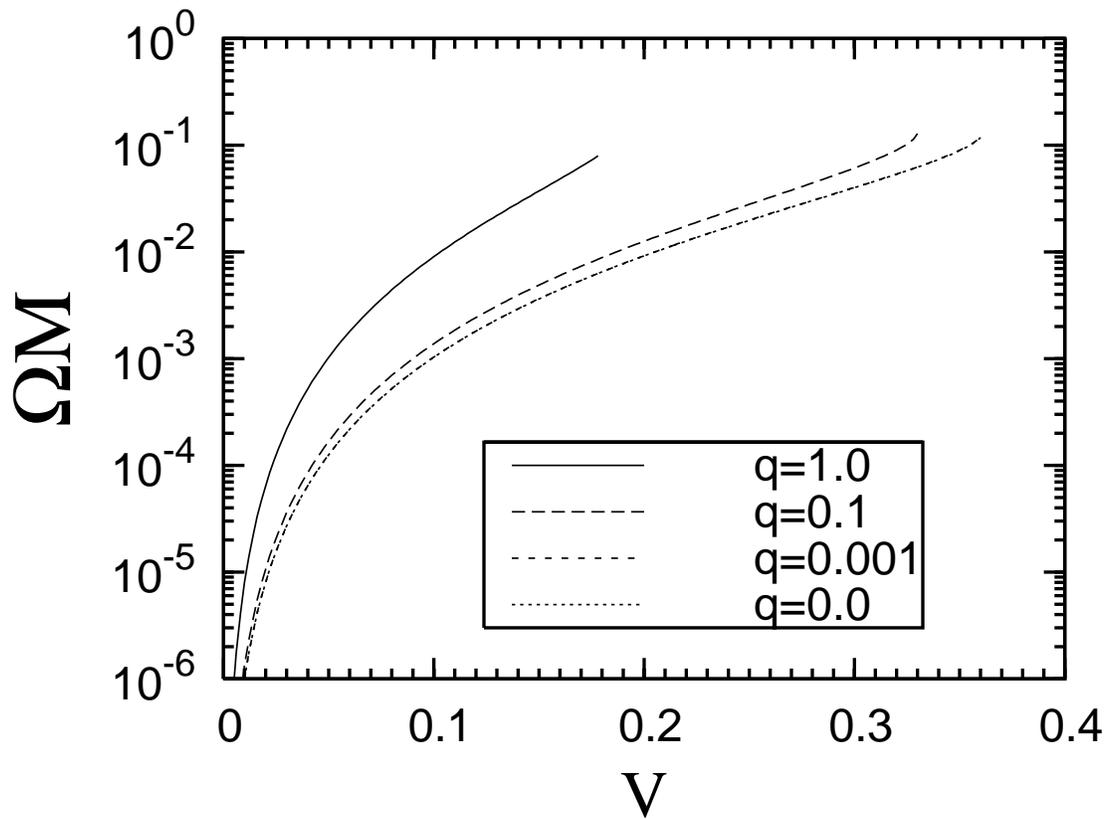}
\caption{\label{fig:PNOV} Parametrization-Invariant (PN) Omega
versus Velocity.}
\end{figure}
the angular velocity, in dimensionless form $\Omega M$, is plotted
against the velocity of the lighter particle for 3 mass ratios and
the $q\rightarrow 0$ limit for the parametrization invariant model
with PN correction. Curves of the analytic solution for $q
\rightarrow 0$ and that of $q=0.001$ overlap each other in the plot.
The inflection displayed in the logarithmic plot changes near the
cutoff velocity for the small mass ratio cases. In
Fig.~\ref{fig:SPNOV}
\begin{figure}[hbtp]
\includegraphics[scale=2.5]{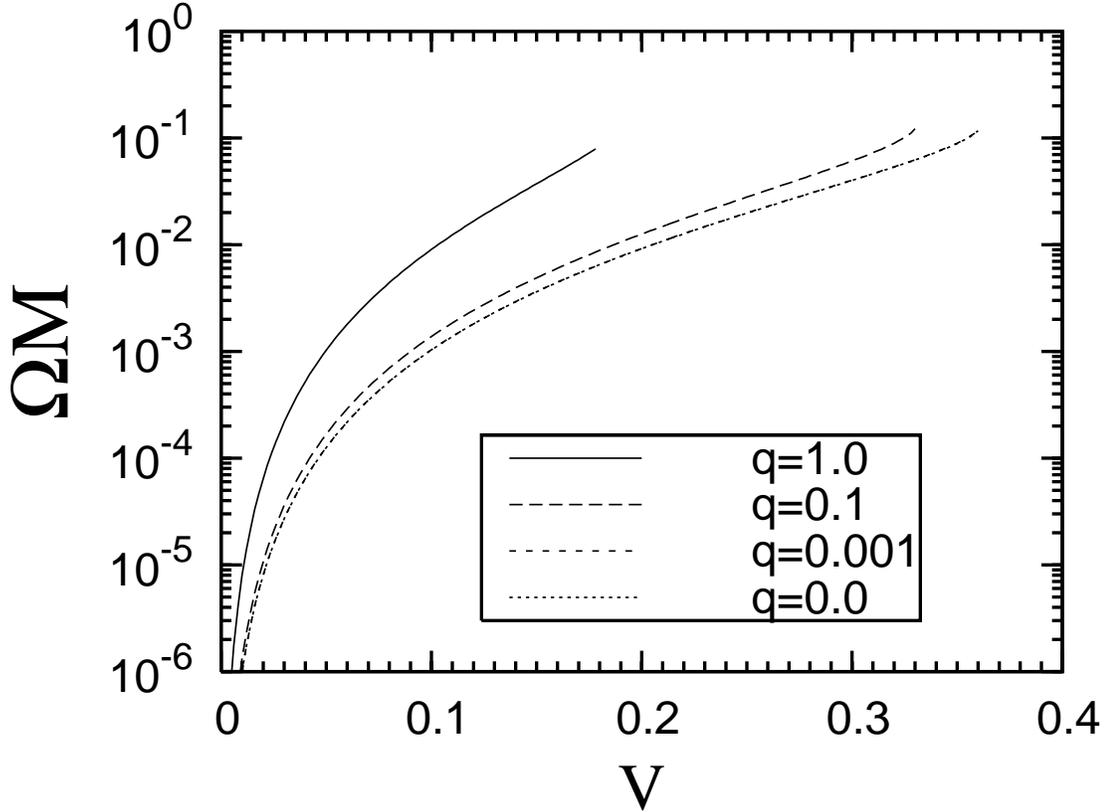}
\caption{\label{fig:SPNOV} Parametrization-Invariant (SPN) Omega
versus Velocity.}
\end{figure}
the angular velocity, in dimensionless form $\Omega M$, is plotted
against the velocity of the lighter particle for 3 mass ratios and
the $q\rightarrow 0$ limit for the parametrization invariant model
with SPN correction. Curves of the analytic solution for $q
\rightarrow 0$ and that of $q=0.001$ overlap each other in the plot.
The inflection displayed in the logarithmic plot changes near the
cutoff velocity for the small mass ratio cases.

In this notation Kepler's Law, $(T_{\rm
period})^{2}=4\pi^{2}a^{3}/\bar{m}$, may be written as
$(\bar{m}\Omega)^{2}=(\bar{m}/a)^{3}$ \cite{Glenz2}. In this form it
may be compared with Eq.~(\ref{eq:eompn}), when written as,
\begin{equation}
(\Omega\bar{m})^2 = \left(\frac{\bar{m}}{a}\right)^3 \,
\left\{v^2\bar{\gamma}\left[F+(m+\bar{m})\Omega\,{F_{\,\rm
I}}\right]\right\}. \label{eq:kepler}
\end{equation}
To see how this post-Minkowski approximation is related to Newtonian
gravity in the non-relativistic ($v<<1$) limit, see
Sec.~\ref{sec:qzero}.

\subsubsection{Parameter Invariant Energy and Angular Momentum}

In Figs.~\ref{fig:PNEO} (PN) and~\ref{fig:SPNEO} (SPN),
\begin{figure}[hbtp]
\includegraphics[scale=2.5]{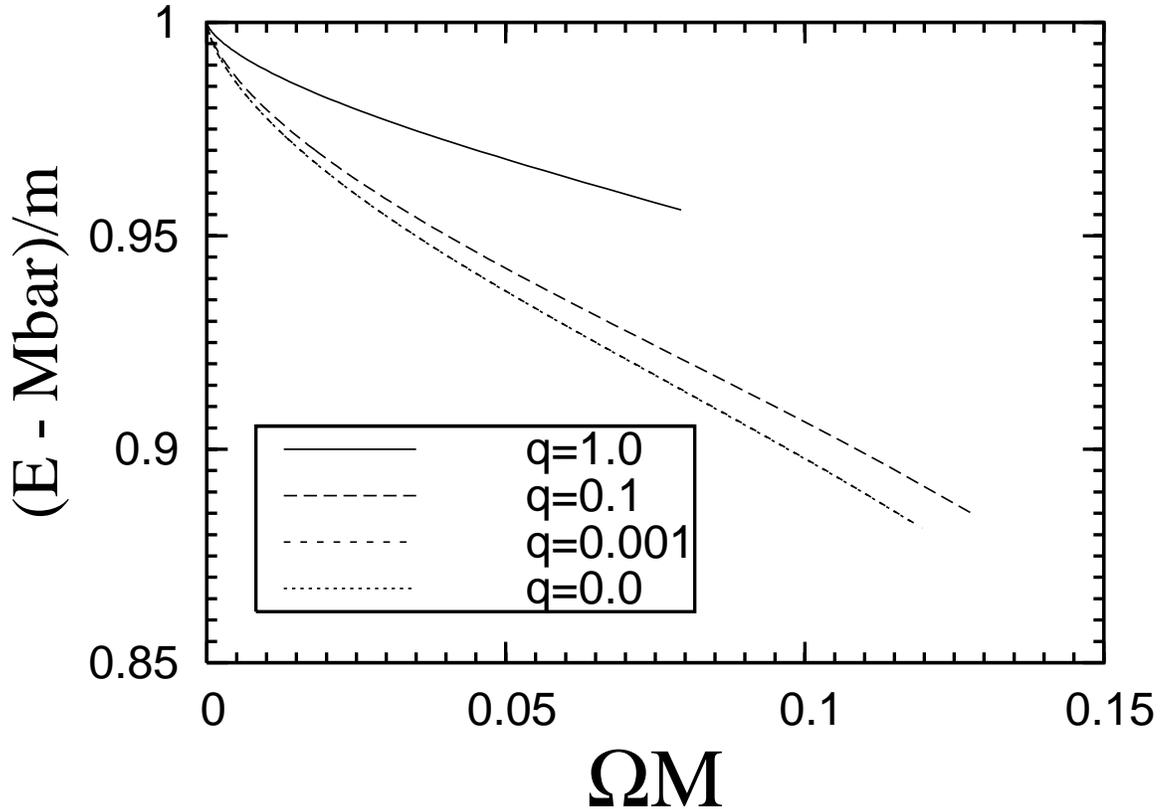}
\caption{\label{fig:PNEO} Parametrization-Invariant (PN) Energy
versus Omega.}
\end{figure}
\begin{figure}[hbtp]
\includegraphics[scale=2.5]{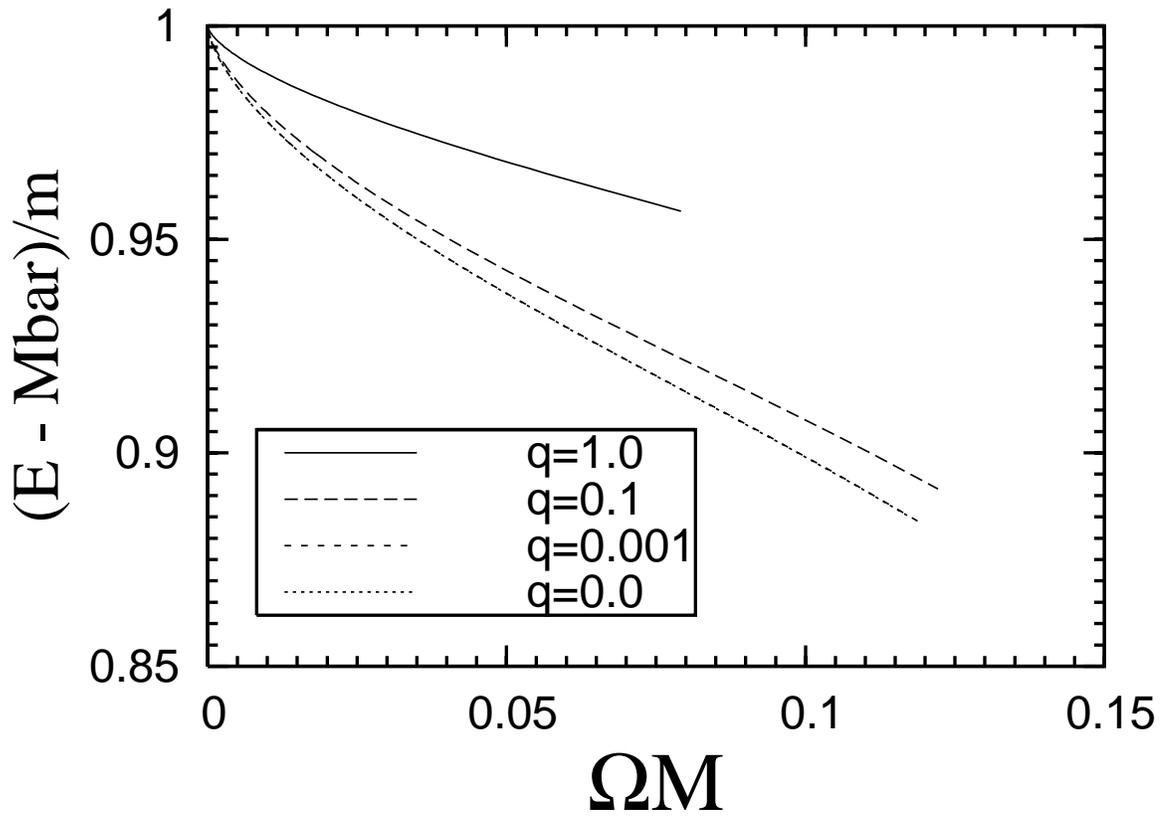}
\caption{\label{fig:SPNEO} Parametrization-Invariant (SPN) Energy
versus Omega.}
\end{figure}
the unit energy of the lighter particle, in dimensionless form
$\hat{E}/m$, where $\hat{E}=E-\bar{m}$, is plotted against $\Omega
M$.  In the limit that the particle approaches becoming unbound,
$v\rightarrow0$, the unit energy of the lighter mass approaches 1,
its rest mass energy.  As it becomes more tightly bound, its energy
decreases below the rest mass energy it would have in flat space.

The conserved energy and angular momentum for the parametrization
invariant model are written \begin{equation} E \,=\, {E_{\rm PM}} +
{e_{\rm I}}, \quad {\rm and }\quad L \,=\, {L_{\rm PM}} + {\ell_{\rm
I}},
\end{equation} where ${E_{\rm PM}}$ and ${L_{\rm PM}}$ are the
post-Minkowski terms
\begin{eqnarray} {E_{\rm PM}} &=& \frac{m}{\gamma} +
\frac{\bar{m}}{\bar{\gamma}} \label{eq:SchildEnergy}
\\
{L_{\rm PM}} &=&
2m\bar{m}\gamma\bar{\gamma}\,\Phi(\varphi,v,\bar{v}),
\end{eqnarray}
and ${e_{\rm I}}$ and ${\ell_{\rm I}}$ are the parametrization
invariant 1PN corrections ${e_{\rm I}} = {e_{\rm PN}}$ and
${\ell_{\rm I}}={\ell_{\rm PN}}$, or those of the special
relativistically invariant model ${e_{\rm I}} = {e_{\rm SPN}}$ and
${\ell_{\rm I}}={\ell_{\rm SPN}}$ given by
\begin{eqnarray}
e_{\rm PN} &=& \frac12\Omega \ell_{\rm PN},
\label{epn}\\
e_{\rm SPN} &=& \frac12 \Omega\ell_{\rm SPN},
\label{espn}\\
\ell_{\rm PN} &=& -\frac{m\bar m(m+\bar
m)\Omega}{\gamma\bar\gamma(v+\bar v)^2},
\label{lpn}\\
\ell_{\rm SPN} &=& -\frac{m\bar m(m+\bar
m)\Omega}{(\gamma\bar\gamma)^{3/2}} \frac1{(\varphi+v\bar v
\sin\varphi)^2}. \label{lspn}\end{eqnarray}

In Figs.~\ref{fig:PNJO} (PN) and~\ref{fig:SPNJO} (SPN);
\begin{figure}[hbtp]
\includegraphics[scale=2.5]{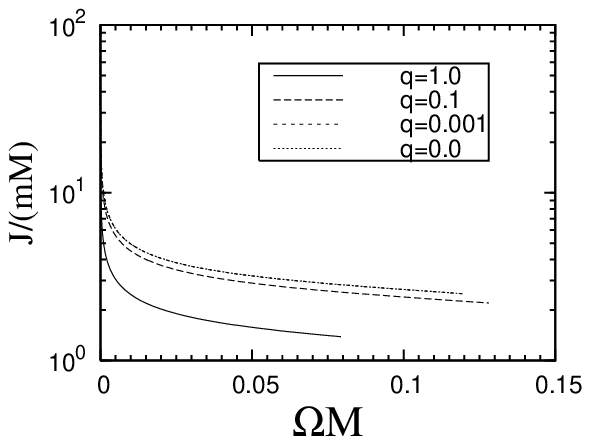}
\caption{\label{fig:PNJO} Parametrization-Invariant (PN) Angular
Momentum versus Omega.}
\end{figure}
\begin{figure}[hbtp]
\includegraphics[scale=2.5]{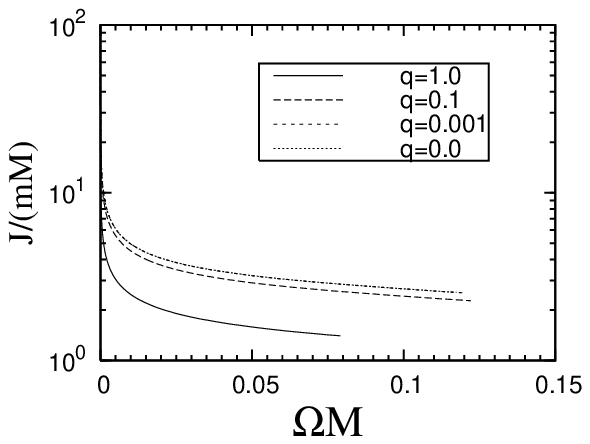}
\caption{\label{fig:SPNJO} Parametrization-Invariant (SPN) Angular
Momentum versus Omega.}
\end{figure}
angular momentum, in dimensionless form $J/(mM)$, where $M$ is the
total mass of both particles and $m$ is the mass of the lighter
particle having velocity, v; is plotted against $\Omega M$ for 3
mass ratios and the $q\rightarrow 0$ limit. There is no ISCO, but at
the maximum value of $v$ for each mass ratio, beyond which there are
no further solutions, there is an Innermost Circular Orbit (ICO).
One possible explanation for the termination of solutions can be
found by looking at Eq.~(\ref{eq:kepler}), which may be written as a
quadratic equation in terms of $(\bar{m}\Omega)$.  Beyond the
maximum value of $v$, the solutions for $(\bar{m}\Omega)$ become
imaginary.

\newpage
\subsubsection{Affine Circular Solution}
\label{sec:affineqs}

In Fig.~\ref{fig:AFOV}
\begin{figure}[hbtp]
\includegraphics[scale=2.5]{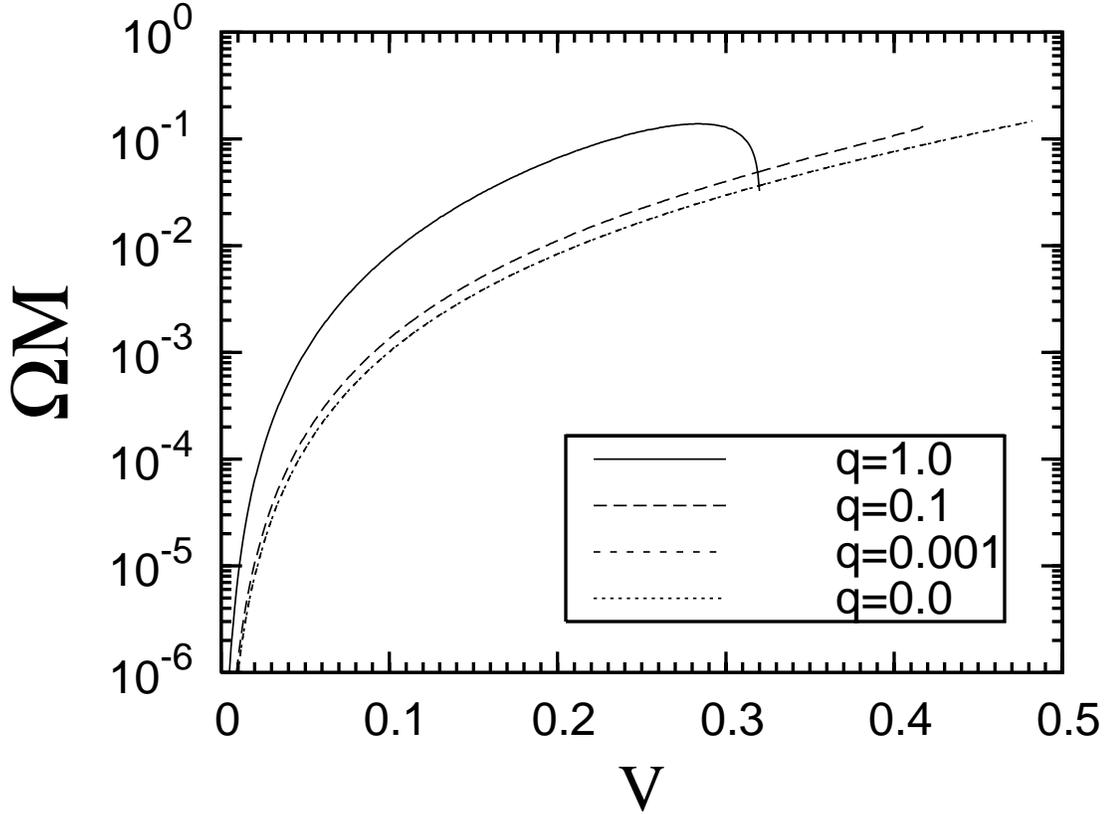}
\caption{\label{fig:AFOV} Full Affine Case of Omega versus
Velocity.}
\end{figure}
The angular velocity, in dimensionless form $\Omega M$, where
$M\equiv(m+\bar{m})$, is plotted against the velocity of the lighter
particle for 4 mass ratios in the affine model. The behavior for
solutions existing beyond $v_{\rm isco}$, the velocity at which the
minimum energy and angular momentum occur, is most prominently
displayed for the q=1.0 case. In the $q=1$ case the ISCO occurs at
$v\sim$0.184. Fig.~\ref{fig:AOV}
\begin{figure}[hbtp]
\includegraphics[scale=2.5]{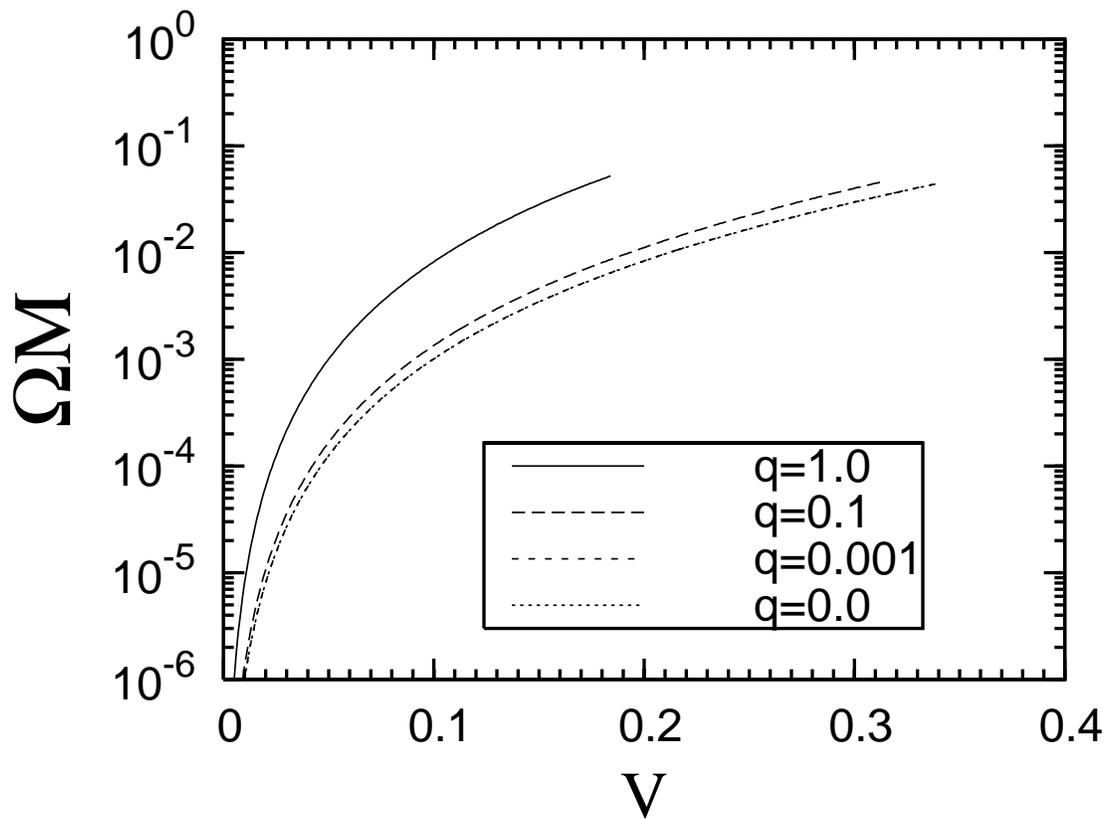}
\caption{\label{fig:AOV} Truncated Affine Case of Omega versus
Velocity.}
\end{figure}
shows the same data as Fig.~\ref{fig:AFOV}, but it only shows
solutions up to the ISCO.

For the affinely parametrized post-Minkowski model, analogous forms
of Eqs.~(\ref{eq:eompn}) and (\ref{eq:eompnbar}) are written
\begin{eqnarray} -m\gamma^2 v\Omega &=& - m\bar{m}\gamma^2\bar{\gamma}
\Omega^2 F^A(\varphi,v,\bar{v}), \label{eq:eomaf}
\\
-\bar{m}\bar{\gamma}^2 \bar{v}\Omega &=& -
m\bar{m}\gamma\bar{\gamma}^2 \Omega^2 \bar{F}^A(\varphi,v,\bar{v}),
\label{eq:eombaraf} \end{eqnarray} where the function
$F^A(\varphi,\bar{v},v) = \bar{F}^A(\varphi,v,\bar{v})$ is defined
as
\begin{eqnarray} F^A(\varphi,v,\bar{v}) \,&\equiv&\, - 4
\frac1{(\varphi+v\bar{v}\sin\varphi)^2}\bigg\{(1+v\bar{v}\cos\varphi)\bar{v}(\varphi\cos\varphi-v^2\sin\varphi)
\nonumber\\
&& +\frac12 v(1-\bar{v}^2)(\varphi+v\bar{v}\sin\varphi) -
\frac12\big[\bar{v}\sin\varphi(\varphi+v\bar{v}\sin\varphi)
\nonumber\\
&& + (1+v\bar{v}\cos\varphi)(v+\bar{v}\cos\varphi)
\big]\Phi(\varphi,v,\bar{v}) \bigg\}. \end{eqnarray} For the
affinely parametrized world line, $\gamma$ and $\bar{\gamma}$
satisfy
\begin{eqnarray} -\gamma^2(1-v^2) +
4\bar{m}\gamma^2\bar{\gamma}\Omega\,\Phi(\varphi,v,\bar{v}) = -1,
\label{eq:norm1}
\\
-\bar{\gamma}^2(1-\bar{v}^2) +
4m\gamma\bar{\gamma}^2\Omega\,\Phi(\varphi,v,\bar{v}) = -1.
\label{eq:norm2} \end{eqnarray}

\subsubsection{Affine Energy and Angular Momentum}

In Fig.~\ref{fig:AFEO}
\begin{figure}[hbtp]
\includegraphics[scale=2.5]{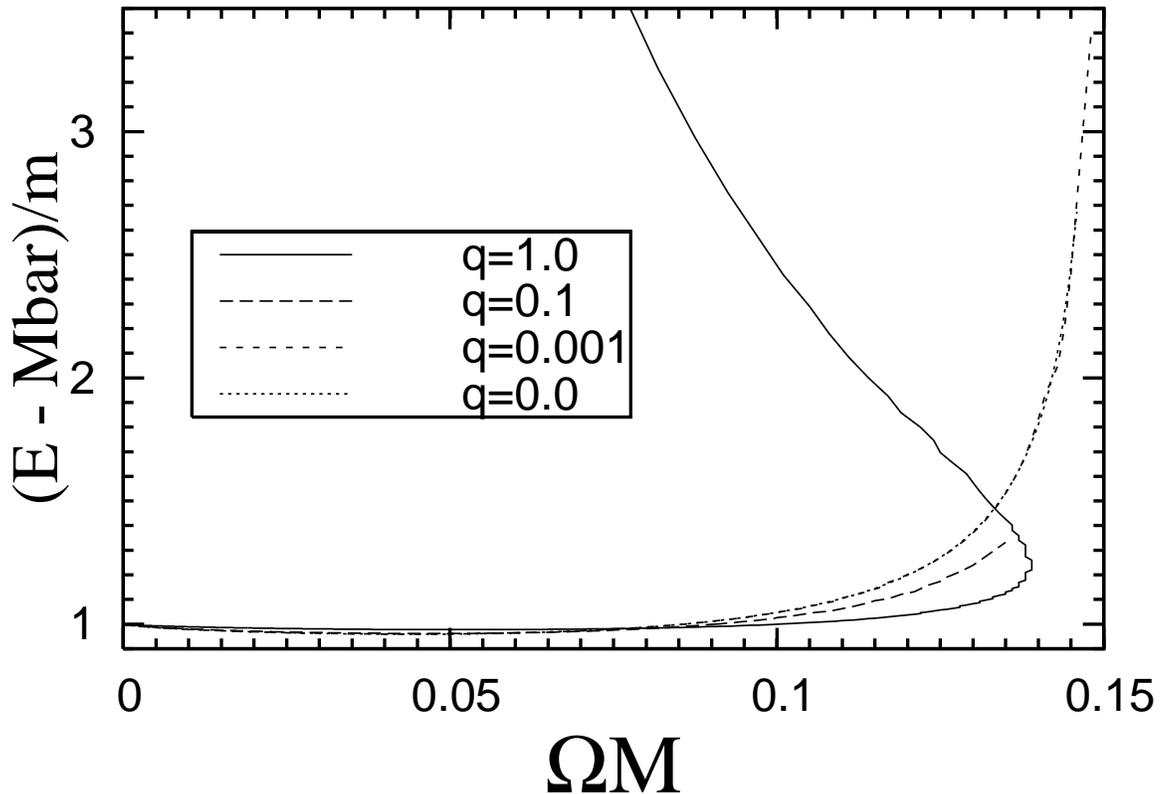}
\caption{\label{fig:AFEO} Full Affine Case of Energy versus Omega.}
\end{figure}
the lighter particle's unit energy per mass, in dimensionless form
$\hat{E}/m$, where $\hat{E}=E-\bar{m}$, is plotted against $\Omega
M$ for the affinely-parameterized case. In Fig.~\ref{fig:AFJO}
\begin{figure}[hbtp]
\includegraphics[scale=2.5]{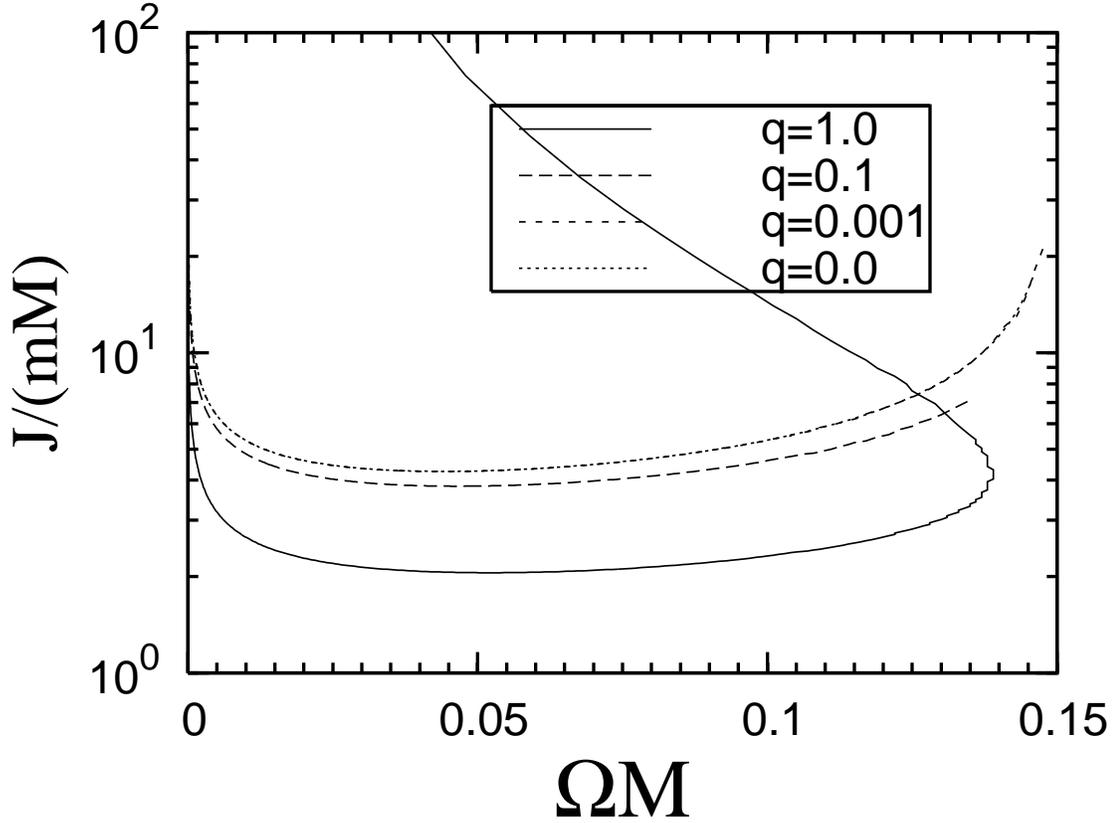}
\caption{\label{fig:AFJO} Full Affine Case of Angular Momentum
versus Omega.}
\end{figure}
angular momentum, in dimensionless form $J/(mM)$, where $M$ is the
total mass of both particles and $m$ is the mass of the lighter
particle having velocity, $v$; is plotted against $\Omega M$ for 3
mass ratios and the $q\rightarrow 0$ limit in the affine model.
Minima of each curve corresponds to the ISCO.

The conserved energy and angular momentum for the affinely
parametrized model are written \begin{eqnarray} E &=&
\frac{m}{\gamma}+\frac{\bar{m}}{\bar{\gamma}} +
4m\bar{m}\gamma\bar{\gamma}\Omega\,\Phi(\varphi,v,\bar{v}),
\nonumber\\
&=& m\gamma(1-v^2)+\bar{m}\bar{\gamma}(1-\bar{v}^2)
- 4m\bar{m}\gamma\bar{\gamma}\Omega\,\Phi(\varphi,v,\bar{v}),
\\
L &=& 2m\bar{m}\gamma\bar{\gamma}\,\Phi(\varphi,v,\bar{v}),
\nonumber\\
&=&
2m\bar{m}\gamma\bar{\gamma}\,\frac{(1+v\bar{v}\cos\varphi)^2-\frac12(1-v^2)(1-\bar{v}^2)}
{\varphi+v\bar v\sin\varphi}, \label{eq:angsolaffine} \end{eqnarray}
where the form of $\Phi(\varphi,v,\bar{v})$ is the same as that of
the parametrization invariant model (\ref{eq:fncPhi}). Using
Eq.(\ref{eq:norm1}) and (\ref{eq:norm2}), the energy can be
rewritten \begin{equation} E = \frac12\frac{m}{\gamma} + \frac12
m\gamma(1-v^2) +\frac12\frac{\bar{m}}{\bar{\gamma}} + \frac12
\bar{m}\bar{\gamma}(1-\bar{v}^2). \end{equation}  This can be
compared with Eq.~(\ref{eq:SchildEnergy}), noting the different
definitions of $\gamma$ in the parametrization-invariant and affine
models.

Figs.~\ref{fig:AEO} and~\ref{fig:AJO}
\begin{figure}[hbtp]
\includegraphics[scale=2.5]{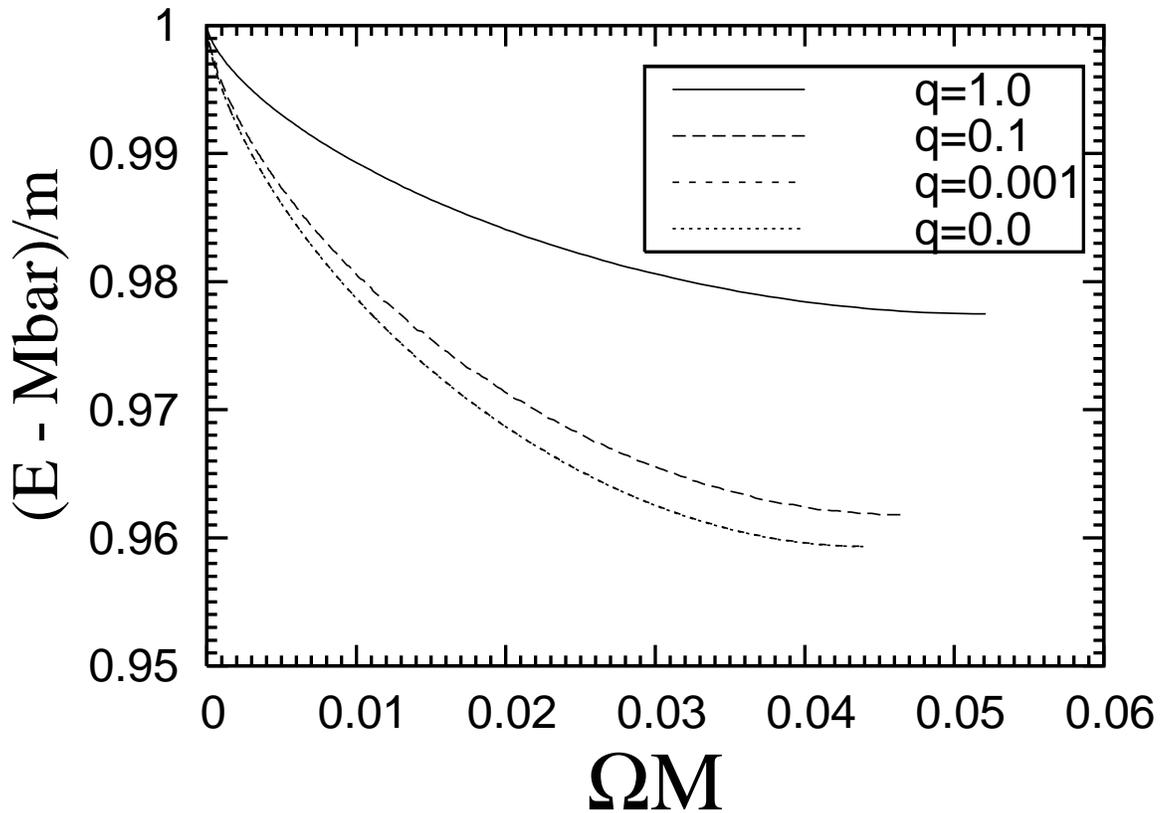}
\caption{\label{fig:AEO} Truncated Affine Case of Energy versus
Omega.}
\end{figure}
\begin{figure}[hbtp]
\includegraphics[scale=2.5]{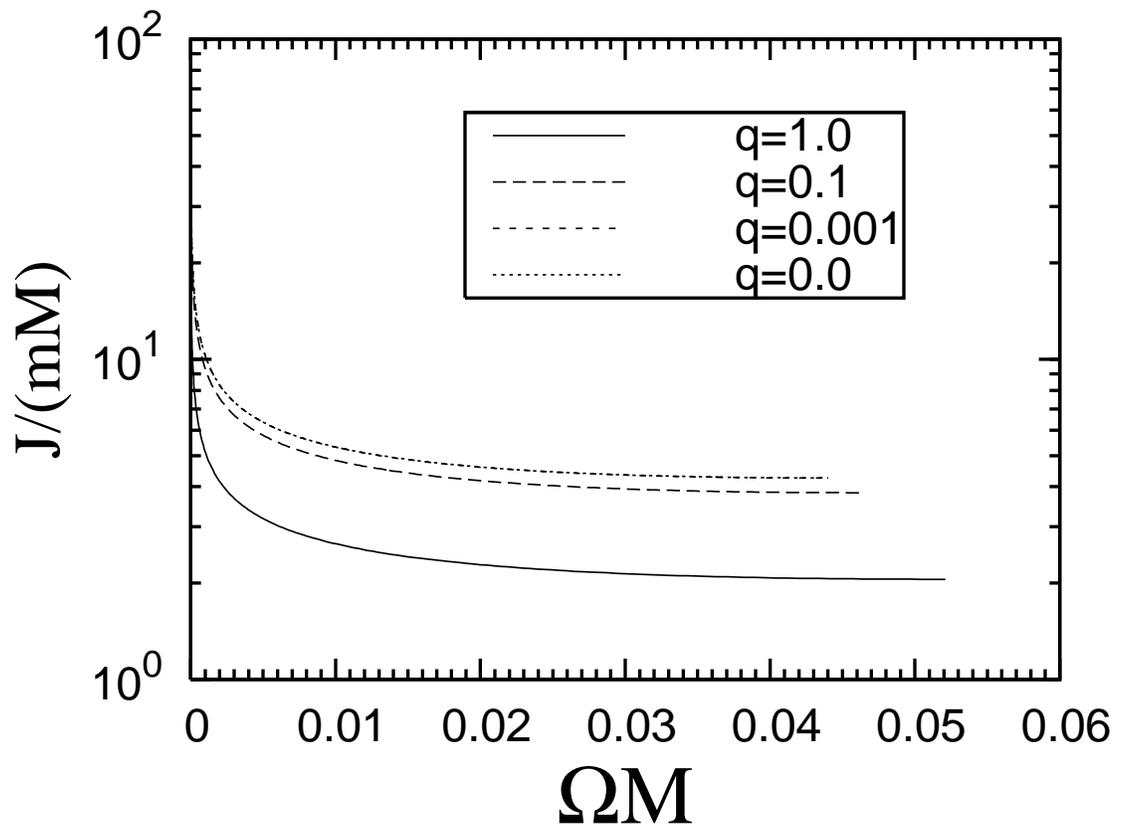}
\caption{\label{fig:AJO} Truncated Affine Case of Angular Momentum
versus Omega.}
\end{figure}
show the same data as Figs.~\ref{fig:AFEO} and~\ref{fig:AFJO},
respectively, except that only the solutions where $0\leq v\leq
v_{\rm isco}$ are plotted.

In the affine case, for any mass ratio $q\in[0,1]$ we find a
simultaneous minima in the energy and angular momentum which
corresponds to the ISCO.  The values of the normalized angular
velocity, angular momentum, and energy that occur at the ISCO in the
affine model vary monotonically from q=1 to 1=0.  With q ranging
from 1 to 0, $\Omega M$ decreases from $\approx 0.0521$ to $\approx
0.0440$, $L/(mM)$ increases from $\approx 2.0558$ to $\approx
4.2617$, and $\hat{E}/m$ decreases from $\approx 0.9775$ to $\approx
0.9593$.

\subsubsection{Numerical Solutions for Unequal Mass Circular
Orbit}

A circular solution is calculated from algebraic equations given in
Eqs.~(\ref{eq:eompn}) and~(\ref{eq:eompnbar}) for the
parametrization invariant model or Eqs.~(\ref{eq:eomaf})
and~(\ref{eq:eombaraf}) for the afinely parametrized model.  One
method of solving for a fixed ratio $q=m/\bar{m}$ is| (1) assume a
ratio of velocities $v/\bar{v}$ and determine the corresponding mass
ratio from the equations of motion, then (2) change the velocity
ratio to adjust the value of the mass ratio to a fixed value (using
the bisection method, for example). The mass ratio, $q$, can be
determined by multiplying both sides of either Eqs.~(\ref{eq:eompn})
and~(\ref{eq:eompnbar}) or Eqs.~(\ref{eq:eomaf})
and~(\ref{eq:eombaraf}) by $\bar{v}\bar{\gamma}^{2}/(v\gamma^{2})$
and then dividing both sides of the first equation listed in either
of these pairs of equations with the second equation to yield an
expresion for the mass ratio $q$.

Another method involves solving the relationship
$M\,\Omega[q,v,\bar{v}]$=$M\,\Omega[(1/q),\bar{v},v]$ by varying the
parameters $v$ and $\bar{v}$.  For a description of an efficient
means of sampling a parameter space and fine-tuning the optimal
result, see Sec.~\ref{sec:randMeth}.

\section{Analytical Formulas for Extreme Mass Ratio}
\label{sec:qzero}

For the Extreme Mass Ratio, the mass of the lighter particle is
negligible relative to that of the more massive particle.  This
would be appropriate for a test mass orbiting in the spherically
symmetric gravitation field of a much more massive object.  Whereas
the signal from a merger of identical black holes, each with a mass
on the order of one solar mass, would fall within the sensitivity of
LIGO's (Laser Interferometer Gravity-wave Observatory) frequency
band; the inspiral of a solar-mass black hole into a
billion-solar-mass black hole, such as those predicted to be at the
centers of many galaxies, would fall into the most sensitive
frequency band of LISA (Laser Interferometer Space Antenna)
\cite{LISA}.  The Extreme Mass Ratio is an excellent approximation
to this latter scenario of a mass ratio on the order of $10^{-9}$.

\subsubsection{Extreme Mass Ratio Limit}

The extreme mass ratio limit $q\equiv m/\bar{m} \rightarrow 0$ is
identical to the limit $\bar{v}\rightarrow 0$ with $\Omega$ fixed.
In the limit $\bar{v}\rightarrow 0$, we may assume that $v$ and
$\bar{m}$ remain finite. Consequently, we have $\bar{\gamma}
\rightarrow 1$, $\varphi \rightarrow v$, and $\bar{m} \rightarrow
M$, where $M\equiv m+\bar{m}$ is the total mass. With $v$ and
$\Omega$ regarded as independent variables, Eq.~(\ref{eq:eompn}) is
a quadratic equation for $\Omega M$, whose $q=0$ form is
\begin{equation} {F_{\,\rm I}}\,(\Omega M)^2+ F\,(\Omega M)-v=0,
\end{equation} with physical solution
\begin{equation} \Omega M = \frac1{2{F_{\,\rm I}}}\left(-F+\sqrt{F^2+4{F_{\,\rm
I}} v}\right). \label{eq:omesol} \end{equation} The functions $F$
(the post-Minkowski term), ${F_{\,\rm I}}={F_{\rm PN}}$ (the
simplest 1PN correction), and ${F_{\,\rm I}}={F_{\rm SPN}}$ (the
special-relativistically covariant 1PN correction) for $q=0$ become
\begin{eqnarray} \!\!\!\!\!\!\! F(\varphi,v,\bar{v}) \,&=&\,
\frac{1-3v^2}{v^2(1-v^2)},
\\[2mm]
\!\!\!\!\!\!\! {F_{\rm PN}}(\varphi,v,\bar{v},\gamma,\bar{\gamma})
\,&=&\, -\, \frac1{v^{3}}\left(1-\frac12 v^2\right),
\\[2mm]
\!\!\!\!\!\!\! {F_{\rm
SPN}}(\varphi,v,\bar{v},\gamma,\bar{\gamma})\,&=&\,
-\,\frac{(1-v^2)^{1/4}}{v^3} \left(1-\frac14 v^2 \right),
\end{eqnarray}
where $\Phi$ has the form
\begin{equation}
\Phi(\varphi,v,\bar{v}) \,=\, \frac{1+v^2}{2\,v}.
\end{equation}
Without the 1PN correction, the parametrization-invariant
post-Minkowski model is given by setting ${F_{\,\rm I}}=0$, and
therefore $\Omega M = v/F$. In the $q \rightarrow 0$ limit, this is
\begin{equation}
\Omega M \,=\, \frac{v^3(1-v^2)}{1-3v^2}. \label{eq:omqzero}
\end{equation}

\subsubsection{Parameter Invariant Solution Sequence in $q \rightarrow 0$ Limit}

In the $q\rightarrow 0$ limit, the conserved energy and angular
momentum normalized by the mass remain finite. Subtracting the mass
of the heavier particle from the post-Minkowski energy,
$\widehat{E}_{\rm PM} \equiv {E_{\rm PM}} - \bar{m}$, and taking the
limit $\bar{v}\rightarrow 0$ with $\bar{m}\rightarrow M$, we have
\begin{eqnarray} \frac{\widehat{E}_{\rm PM}}{m} \,&=&\,
(1-v^2)^{1/2},
\\[2mm]
\frac{{L_{\rm PM}}}{mM} \,&=&\,  \frac{1+v^2}{v(1-v^2)^{1/2}},
\\[2mm]
\frac{e_{\rm PN}}{m} &=& \frac12 \frac{\ell_{\rm PN}}{mM}\Omega M,
\\[2mm]
\frac{e_{\rm SPN}}{m} &=& \frac12 \frac{\ell_{\rm SPN}}{mM}\Omega M,
\\[2mm]
\frac{\ell_{\rm PN}}{mM} \,&=&\, -\, \frac{(1-v^2)^{1/2}}{v^2}\Omega
M, \label{lpn}
\\[2mm]
\frac{\ell_{\rm SPN}}{mM} \,&=&\, -\,
\frac{(1-v^2)^{3/4}}{v^2}\Omega M. \label{lspn}
\end{eqnarray}

In \cite{Friedman}, it is shown that the first law of thermodynamics
that relates the changes in the conserved energy and the angular
momentum, $dE = \Omega dL$, is satisfied by binary solutions derived
from the parametrization invariant Fokker action. This relation is
used to cross check both the analytic formula in the $q\rightarrow
0$ limit as well as the numerical solutions obtained in
Sec.~\ref{sec:numerical} by calculating
${d\widehat{E}}/{dv}=\Omega{dL}/{dv}$, where
$\widehat{E}\equiv\widehat{E}_{\rm PM} + {e_{\rm I}}$.

In the parametrization invariant post-Minkowski model without a 1PN
correction, the normalized angular velocity, $\Omega M$, is defined
in an interval $0\le v < 1/\sqrt{3}$ for $q\rightarrow0$, and
$\Omega M$ becomes infinite at $v=1/\sqrt{3}$. With the 1PN
correction ${F_{\,\rm I}}={F_{\rm PN}}$, the range of finite $\Omega
M$ is approximately $0\le v \lesssim 0.361598$, and with the special
relativistic invariant 1PN correction ${F_{\,\rm I}}={F_{\rm SPN}}$,
it is $0\le v \lesssim 0.36166$. Newtonian point particles have no
innermost stable circular orbit (ISCO), but adding a 1PN correction
to the Newtonian orbit recovers the ISCO that is present in the
exact theory of general relativity.  In the post-Minkowski
framework, we find that the existence of an ISCO depends on our
choice among actions that are equivalent to first post-Minkowski
order. In particular, we find that the parametrization-invariant
action leads to sequences with no ISCO even when 1PN terms are
included. This is plausibly due to the fact that the sequences
associated with the parametrization-invariant action terminate
before reaching the angular velocity of an ISCO.  For the 1PN
formalism given in \cite{Blanchet}, an Extreme Mass Ratio ISCO
occurs at the unrealistically high value of $\Omega M = 0.544$.  The
2PN and 3PN values for the $q=0$ ISCO are $\Omega M = 0.124$ and
$0.0867$, respectively \cite{Blanchet}. Below we show that sequences
associated with the affinely parametrized action do have an ISCO;
however, the $q\rightarrow0$ ISCO of the affine case occurs at an
unrealistically small value of $\Omega M$.

In Eq.~(\ref{eq:omesol}), an expansion of $\Omega M$ in the small
$v$ limit becomes $\Omega M = v^3 + 3 v^5 + {O}(v^7)$ for both PN
and SPN models, and this is inverted to write $v$ in terms of small
$\Omega M$ as
\begin{equation}
v = (\Omega M)^{1/3} -  \Omega M +{O}\big((\Omega M)^{5/3}\big).
\end{equation}
Substituting this into the energy and angular momentum formulas, the
leading two terms agree with the post-Newtonian formulas (see e.g.
\cite{Blanchet}) up to the 1PN order for the extreme mass ratio
$q\rightarrow 0$,
\begin{eqnarray}
\label{eq:ehatm} \frac{\widehat{E}}{m} &=& 1-\frac12(\Omega M)^{2/3}
+ \frac38(\Omega M)^{4/3} +{O}\big((\Omega M)^{2}\big),
\\
\frac{L}{mM} \!\!&=& \!\! \frac1{(\Omega
M)^{1/3}}\left[1+\frac32(\Omega M)^{2/3} +{O}\big((\Omega
M)^{4/3}\big) \right]. \label{eq:lmm}
\end{eqnarray}

\subsubsection{Affine Solution Sequence in $q \rightarrow 0$ Limit}

Eq.(\ref{eq:eomaf}) implies
\begin{equation} \Omega\bar{m} = v\bar{\gamma}^{-1}\, F^A(\varphi,v,\bar{v})^{-1},
\end{equation} where $\bar{\gamma}$ is evaluated from Eqs.~(\ref{eq:norm1}) and
(\ref{eq:norm2}).

In the limit of $q\rightarrow 0$ (or more directly
$\bar{v}\rightarrow 0$),
\begin{equation} F^A(\varphi,v,\bar{v})
\,=\,\frac{1-v^2}{v^2}. \end{equation}

From Eq.~(\ref{eq:eomaf}) and (\ref{eq:norm1}), we have
\begin{equation} \gamma
\,=\, \left(\frac{1-v^2}{1-4v^2-v^4}\right)^{1/2},
\end{equation}
while in Eq.~(\ref{eq:norm2}), taking $\bar{v}\rightarrow 0$ and $m
\rightarrow 0$ yields $\bar{\gamma} \rightarrow 1$. As a result we
have in the extreme mass ratio,
\begin{equation}
\Omega M \,=\, \frac{v^3}{1-v^2}. \label{eq:afomqzero}
\end{equation}

In the $q\rightarrow 0$ limit, the energy without the rest mass of
the heavier particle, $\widehat{E}\,\equiv\, E-\bar{m}$, and the
angular momentum become \begin{eqnarray} && \frac{\widehat{E}}{m}
\,=\, \frac{(1-3v^2)}{[(1-v^2)(1-4v^2-v^4)]^{1/2}},
\\
&& \frac{L}{mM} \,=\,
\frac{1+v^2}{v}\left(\frac{1-v^2}{1-4v^2-v^4}\right)^{1/2}.
\end{eqnarray}

The first law $\delta E=\Omega\delta L$ is also satisfied for the
affinely parametrized model, and hence one can cross check formulas
in the $q\rightarrow 0$ limit using the relation
$d\widehat{E}/dv=\Omega dL/dv$. Although the lighter particle's
normalized angular velocity, $\Omega M$, is finite in an interval
$v\in[0,1)$, the redshift factor $\gamma$ as well as conserved
quantities $E$ and $L$ become infinite at
$v=\sqrt{\sqrt{5}-2}\approx0.485868$, which corresponds to $ \Omega
M = {\left(\sqrt{5}-2\right)^{3/2}}/({3-\sqrt{5}})\approx 0.150142.
$

In this interval, $v\in\big[0,\sqrt{\sqrt{5}-2}\big)$, the energy
and angular momentum have a simultaneous minima at $
v=\sqrt{({1+2^{4/3}-2^{5/3}})/{3}}\approx 0.339136, $ which
corresponds to
\begin{equation}
\Omega M= \frac{\left(1+2^{4/3}-2^{5/3}\right)^{3/2}}
{2\sqrt{3}\left(1-2^{1/3}+2^{2/3}\right)}\approx 0.0440743.
\end{equation}
The Schwarzschild ISCO occurs at $\Omega
M=6^{(-3/2)}\sqrt{2}\simeq0.096$. In terms of this exact solution,
the ISCO of the affine parametrization has an error of 54\%, whereas
the ISCO of the 1PN approximation given by \cite{Blanchet} has an
error of 465\%.

\subsubsection{Radial Parameter in $q\rightarrow0$ Limit}

With the definition $a=v/\Omega$, we can write $a/M=v/(M\Omega)$,
where $M=m+\bar{m}$ in the $q\rightarrow0$ limit is just
$M=\bar{m}$.  Then we insert into Eqs.~(\ref{eq:omqzero}) (0PN
parametrization-invariant without 1PN correction term),
(\ref{eq:omesol}) (PN and SPN cases), or~(\ref{eq:afomqzero}) (0PN
affine case) the maximum velocity (parametrization-invariant) or the
ISCO velocity (affine).  These cutoff velocities are| $1/\sqrt{3}$
(PM), $0.361598$ (PN), $0.36166$ (SPN), or $0.485868$ (Affine). This
leads to a minimum radial parameter for a circular orbit. In units
of $M^{-1}$, these minimum radial parameters are as follows: $0$
(PM), $2.67$ (PN and SPN), and $3.24$ (Affine).  Note that the ISCO
of the affine case occurs at $v=0.339136$, which corresponds to
$a/M\simeq7.69$.

The parametrization-invariant model without a 1PN correction has no
minimum radius, which is the same as Newtonian gravity.  The affine
case has an ISCO on the order of the $6M$ that is predicted by the
full theory of general relativity.  In the case of the PM+1PN
correction term model, and in the affine case without a correction,
the minimum radial parameter occurs on the order of the $2M$ event
horizon for a Schwarzschild black hole.

\subsubsection{1PN Energy and Angular Momentum}

For comparison with our post-Minkowski analysis, we list the 1PN
equations of \cite{Blanchet}, where Blanchet's $\nu$ is our
$q(1+q)^{-2}$.  In our notation,
\begin{equation}
\frac{E}{M}=-\frac{1}{2}\frac{q}{(q+1)^{2}}(M\Omega)^{2/3} \left[
1-\left(\frac{3}{4}+\frac{1}{12}\frac{q}{(1+q)^{2}}\right)(M\Omega)^{2/3}
\right],
\end{equation}
or,
\begin{equation}
\frac{\hat{E}}{m}=1-\frac{1}{2}\frac{1}{q+1}(M\Omega)^{2/3} \left[
1-\left(\frac{3}{4}+\frac{1}{12}\frac{q}{(1+q)^{2}}\right)(M\Omega)^{2/3}
\right]; \label{eq:E1PN}
\end{equation}
and
\begin{equation}
\frac{L}{M^{2}}=\frac{q}{(1+q)^{2}}\frac{1}{(M\Omega)^{1/3}}\left[1+\left(\frac{3}{2}+\frac{1}{6}\frac{q}{(1+q)^{2}}\right)(M\Omega)^{2/3}\right],
\end{equation}
or
\begin{equation}
\frac{L}{mM}=\frac{1}{1+q}\,\frac{1}{(M\Omega)^{1/3}}\left[1+\left(\frac{3}{2}+\frac{1}{6}\frac{q}{(1+q)^{2}}\right)(M\Omega)^{2/3}\right].
\label{eq:J1PN}
\end{equation}
Thus we show explicitly in the limit $q\rightarrow0$ that in the
parametrization-invariant case with a first-order post-Newtonian
correction term, the energy and angular momentum
Eqs.~(\ref{eq:ehatm}) and~(\ref{eq:lmm}) agree with
Eqs.~(\ref{eq:E1PN}) and~(\ref{eq:J1PN}).

Agreement between the energy and angular momentum formulas of the
1PN circular solution, and those of the parametrization invariant
post-Minkowski model with post-Newtonian correction, is exhibited
explicitly for the extreme mass ratio limit. For an arbitrary mass
ratio one needs to expand the retarded angle $\varphi$ to the next
order in the velocities, $v$ and $\bar{v}$, as $\varphi \approx
(v+\bar{v})(1-v\bar{v}/2)$, and the rest of the calculation closely
parallels that of the $q = 0$ case.

\newpage

\

\

\

\

\

\

\

\

\

\

\noindent\textbf{\Huge Part III:}

\

\noindent\textbf{\huge Production and Decay}
\addcontentsline{toc}{chapter}{Part III - Production and Decay of
Small Black Holes at the TeV Scale}

\

\noindent\textbf{\huge of Small Black Holes}

\

\noindent\textbf{\huge at the TeV Scale}

\newpage
\thispagestyle{fancy}
\chapter{TeV-Scale Black Hole Production at the South Pole}
\thispagestyle{fancy}
\pagestyle{fancy}

\

We discuss the possibility of observing TeV-scale black holes
produced at the IceCube Neutrino Telescope \cite{Anchordoqui1}.
After giving a brief summary of the IceCube experiment, we explain
what TeV-scale black holes are.  We then examine a gravitational
interaction between a neutrino and a nucleon.  Because a nucleon is
not a point particle, we rely on the parton model, which describes
the nucleon as a collection of quarks and gluons. Following this, we
describe our method for modeling Parton Distribution Functions
(PDFs), we evaluate the cross section for the interaction of
neutrino+nucleon$\rightarrow$black hole, and then we calculate
IceCube's detection sensitivity for observing TeV-scale black holes.

\section{IceCube Neutrino Telescope}

In the Standard Model a neutrino can interact with a nucleon through
both charge current (CC) interactions and neutral current (NC)
interactions \cite{Anchordoqui2,Mandl,Eisberg}. In a CC interaction,
a neutrino (anti-neutrino) interacts with a quark to become a lepton
(anti-lepton), conserving electron-, muon-, and tau-lepton number.
In this interaction a $W^{+}$ ($W^{-}$) particle is exchanged with a
down-quark (up-quark), which becomes an up-quark (down-quark).  In a
NC interaction, a neutrino exchanges a $Z^{0}$ with a quark and
neither the neutrino nor the quark changes flavor.

Because neutrinos experience only gravity and the weak force, they
may travel astronomical distances without interactions. Thus, they
preserve information about the environment in which they were
produced.  The corollary to this is that a sufficiently large
detector must be used to observe these cosmic neutrinos here on
Earth.

The IceCube Neutrino Telescope is composed of approximately one
cubic kilometer of Antarctic ice ranging from 1400 meters in depth
to 2400 meters in depth below the surface near the Amundsen-Scott
Station located at the geographic South pole \cite{Karle}. IceCube
is already taking data, and it is scheduled to be fully operational
by 2009-2010.  At that time, it will consist of 80 strings, each a
kilometer long, of 60 evenly spaced PhotoMultiplier Tubes (PMT)
each, for a total of 4800 PMT.  The strings are 125 meters apart,
and each interior string will be surrounded by six equidistant
neighbors \cite{Anchordoqui3}.

When high energy charged particles move faster than the local speed
of light through the ultraclear Antarctic ice, in which the
absorption length of the relevant wavelengths is greater than 100
meters \cite{Halzen}, they emit Cherenkov radiation. This radiation,
within a range that includes visible light and some UV light, can be
detected by the PMT used in IceCube, and the time at which this
happens| including the time for the signal to register| can be
recorded within an accuracy of a few nanoseconds \cite{IceCube}. The
paths of these charged particles may be dominated by jets from a
high energy muon or tau.  They may also be diffused throughout a
shower. With sufficient data, these paths can be used to reconstruct
the particle interactions that have taken place.  This requires the
energy of the incident neutrino to be greater than 100 GeV. In the
case of a series of interactions caused by a single incident
particle, the total energy| provided that it is contained within the
volume of IceCube and is less than $10^{10}$ GeV so that it does not
saturate the detector| can be measured \cite{Halzen}.

When measuring neutrino interactions, one must contend with a
background event rate of charged particles, such as muons produced
by cosmic rays hitting the atmosphere \cite{Lipari}.  Examining
upward going tracks, or particles that have passed through a
significant fraction of the Earth, effectively restricts the
progenitor particle of an interaction to a neutrino, which, because
it is only weakly interacting, is able to easily penetrate the
Earth, whereas charged particles are not.  A horizontally traveling
neutrino passing through the center of IceCube travels through 150
kilometers of the Earth \cite{Alvarez-Muniz}. There is also a
background trigger rate for IceCube's PMT of less than one kilohertz
\cite{Karle,Halzen}. That is, in the absence of a signal a PMT will
discharge on average no more than once every millisecond. This is
not a problem, because the transit time across IceCube for those
particles that produce Cherenkov radiation is on the order of a few
microseconds and the PMT recording time is accurate to within a few
nanoseconds. Although the volume of IceCube is one cubic kilometer,
the effective volume for detecting neutrinos is larger, because
muons may be produced outside the IceCube volume and still travel
inside to be measured \cite{Kowalski}.

Another useful veto is the IceTop surface array of 160 Cherenkov
detectors of 2.7 meter diameter tanks of ice spread out over one
square kilometer of area \cite{Anchordoqui4}. IceTop helps reject
background events and is also useful for calibration.

Amanda, the prototype of IceCube that proved the viability of
detecting neutrinos in polar ice caps, is still running.  Because
its volume overlaps with the volume of IceCube, it can either
contribute to IceCube's sensitivity, or it can serve as a check on
IceCube detections, depending upon whether data from the two
experiments is examined collectively or independently
\cite{IceCube}.

\section{Black Hole Production in Higher Dimensions}

In the standard model (SM), gravity is by far the weakest of the
four fundamental forces.  It has been theorized that this weakness
is due to the presence of extra dimensions beyond the 4 familiar
dimensions of our spacetime \cite{Arkani-Hamed}.  If gravitons
propagated into the extra dimensions while SM fields were confined
to our brane of 3+1 dimensions, then gravity thus diluted would
appear much weaker than the other forces.  In this case, gravity
might become much stronger at small distances than a 4-dimensional
theory would predict.

We will investigate the possibility that the distance at which
gravity and the electromagnetic force have the same strength is at
$\sim10^{-19}$ m, the distance at which the electromagnetic and weak
forces unify as the electro-weak force. This would mean that for the
small distances at which gravity matched the electro-weak force in
strength, there would be a fundamental $D$-dimensional Planck mass
of about 1 TeV, in which case our 4-dimensional Planck mass would
just be an effective Planck mass over macroscopic dimensions.

The strength of TeV-scale gravity at small distances could
potentially make it easier for interacting particles to form black
holes.  This can be qualitatively understood via Gauss's Law
\cite{Myers,Argyres}.  The surface area of a sphere in $D$
dimensions, where there is 1 time dimension and $D-1$ spatial
dimensions, is proportional to $r^{D-2}$. The magnitude of a
$D$-dimensional Newtonian gravitational force acting between two
masses would be proportional to $M_{1}m_{2}G_{D}$, where $G_{D}$ is
the $D$-dimensional gravitational constant.  Spread evenly over the
surface area of a sphere, this force would be proportional to
$M_{1}m_{2}G_{D}r^{2-D}$.  The value of the potential energy at a
separation $r$ between the two masses would be proportional to
$M_{1}m_{2}G_{D}r^{3-D}$.  Taking $M_{1}$ as the primary mass and
$m_{2}$ as a test mass, then using a non-relativistic argument to
relate the maximum kinetic energy of a test mass moving near the
speed of light ($mc^{2}/2$) to the potential energy, places an event
horizon at $r\propto(M_{1}G_{D}/c^{2})^{(1/[D-3])}$.  The
dimensionality of $G_{D}$ is ${\rm length}^{D-1}\ {\rm mass}^{-1}\
{\rm time}^{-2}$.  The $D$-dimensional Planck mass, $M_{D}$, is then
proportional to $(\hbar^{D-3}c^{5-D}G_{D}{}^{-1})^{(1/[D-2])}$.  In
units of $\hbar=c=1$, then $G_{D}\propto M_{D}^{2-D}$, and the event
horizon would be
$r\propto(M_{1}G_{D})^{(1/[D-3])}\propto(M_{1}M_{D}^{2-D})^{(1/[D-3])}\propto(1/M_{D})(M_{1}/M_{D})^{(1/[D-3])}$.

It has been suggested that the Large Hadron Collider (LHC) could
easily produce such black holes in this scenario
\cite{Dimopolous,Giddings,Ringwald}.  If the LHC would be powerful
enough to detect this sort of black hole interaction, then cosmic
rays would also be energetic enough to produce this interaction.  In
particular, we will discuss the possibility that neutrinos produce
black holes in the ice of the south pole and can be detected by the
IceCube Neutrino Telescope.

To model the gravitational interaction between a neutrino, which is
a point particle, and a nucleon, which is an object of finite extent
and which has an internal structure attributable to constituent
point particles, we turn to Parton Distribution Functions.

\subsection{Modeling Parton Distribution Functions}

In high energy interactions between a neutrino and a nucleon, the
neutrino interacts primarily with a single parton, a quark or a
gluon.  For these collisions, the proton and neutron are not just an
up-up-down and an up-down-down, but are composed of these and other,
virtual particles that are continually created and annihilated
through the time-energy uncertainty relationship.

Similarly, for low energy interactions, a nucleon acts as a single
particle of rest mass energy $m_{N}$ in its rest frame.  For high
energy interactions between a neutrino and a parton, in the
nucleon's rest frame the parton will have have some fraction of the
total energy rest-mass of the nucleon.  This fraction is denoted as
$x$, where $x$ ranges from 0 to 1, or from none of the nucleon's
total energy to all of it \cite{Halzen2}.

A Parton Distribution Function (PDF) describes the probability of
finding a given parton| up ($u$), anti-up ($\bar{u}$), down ($d$),
anti-down ($\bar{d})$, strange ($s$), anti-strange ($\bar{s}$),
charm ($c$), anti-charm ($\bar{c}$), bottom ($b$), anti-bottom
($\bar{b}$), or gluon ($g$)| with a fraction $x$ of the total rest
energy of $m_{N}$. The contributions from the PDFs for the
super-massive top and anti-top within the nucleons at rest are
negligible, and we neglect them. Thus, the probability that an $i$th
species of parton exists with a fractional energy between $x_{1}$
and $x_{2}$ is
\begin{equation}
\mathcal{P}=\int_{x_{1}}^{x_{2}}f_{i}(x,Q)dx.
\end{equation}
The variable $Q$ is the momentum transfer, where we choose $Q\simeq
r_{s}^{-1}$ \cite{Emparan}, and the PDFs are somewhat insensitive to
changing $Q$ \cite{Anchordoqui5}.  We thus use
\begin{equation}
Q={\rm min}\{r_{s}^{-1},10\ {\rm TeV}\}.
\end{equation}

The PDFs cannot be calculated analytically from first principles in
the Standard Model.  They must be fitted to experimental data.  We
use the CTEQ6D PDFs \cite{Pumplin}.  The largest uncertainty in the
PDFs exists for large-$x$ gluons, where $f(x,Q)_{\rm gluon}$ may be
off by more than a factor of 2 \cite{Stump}.  At small $x$, where
the PDFs are much more certain, the gluon quickly comes to dominate
the neutrino-parton interactions through its high probability of
being available for a collision.

\begin{figure}[hbtp]
\includegraphics[scale=0.8]{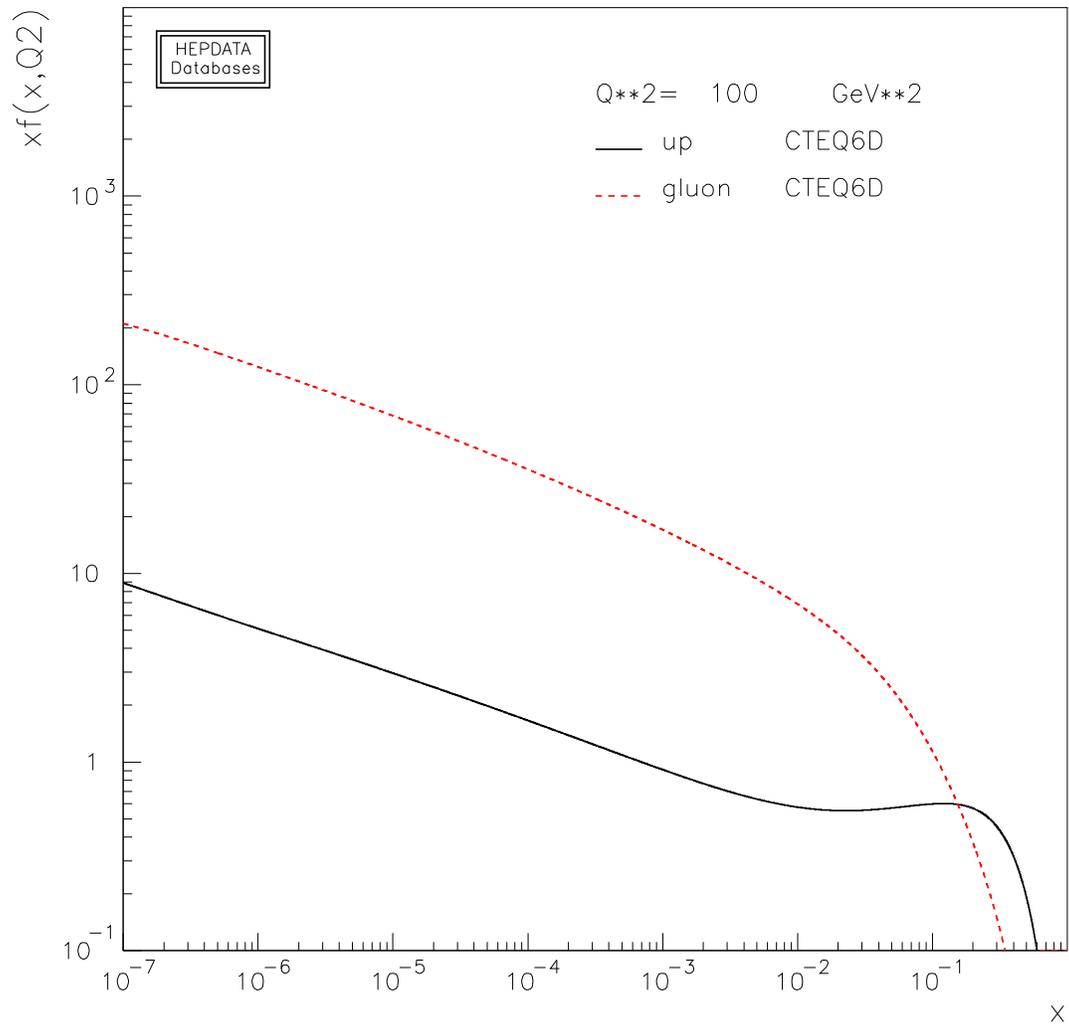}
\caption{\label{fig:pdf1} Parton Distribution Functions: Lower
Momentum Transfer.}
\end{figure}
\begin{figure}[hbtp]
\includegraphics[scale=0.8]{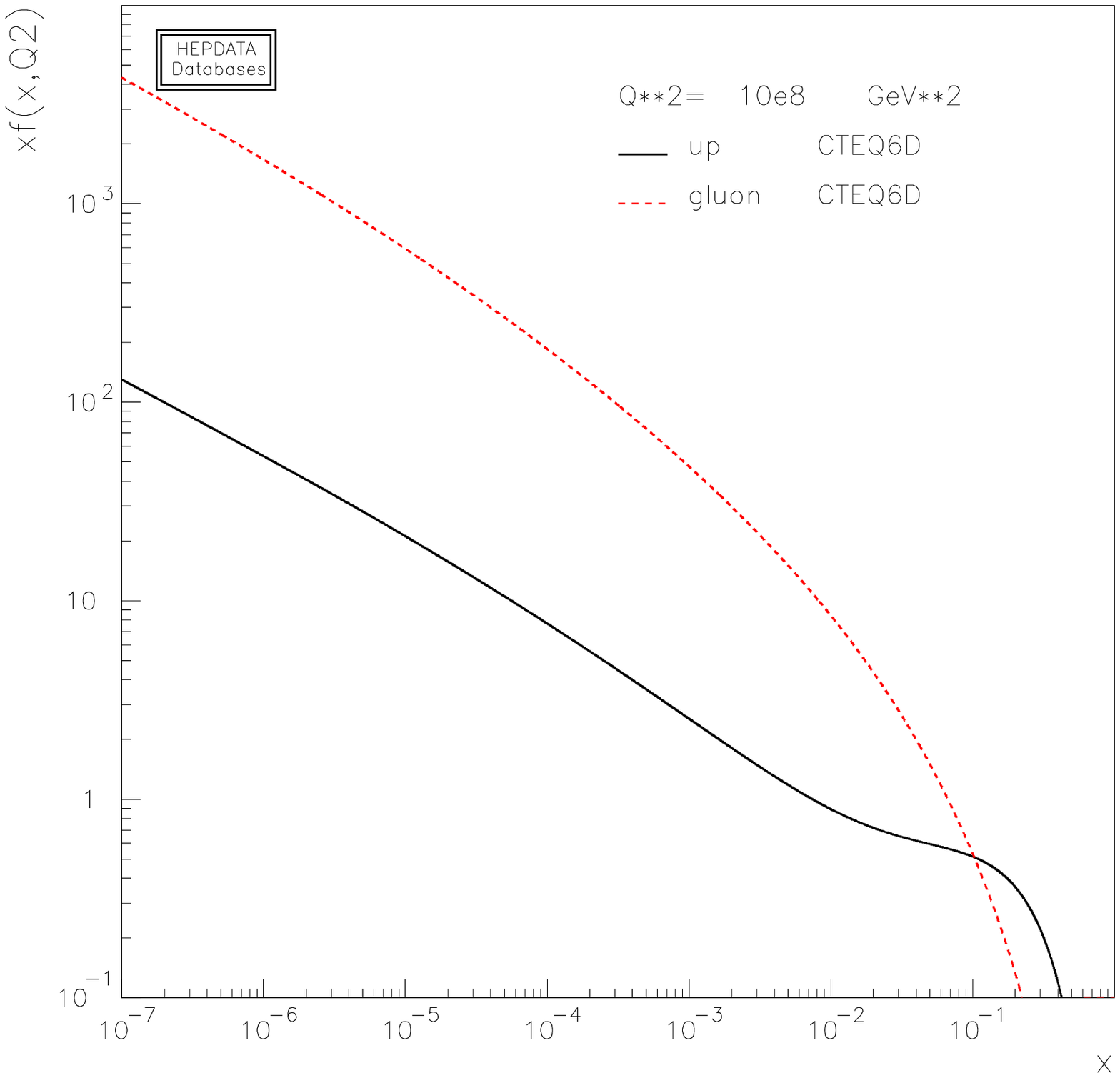}
\caption{\label{fig:pdf2} Parton Distribution Function: Higher
Momentum Transfer.}
\end{figure}

Fig.~\ref{fig:pdf1} plots $\log(x)$ versus $\log(xf_{i}(x,Q))$ for a
representative quark, the up, and for a gluon, both of which for the
relatively low $Q$ of 10 GeV, or $Q^{2}=$ 100 (GeV)${}^{2}$
\cite{Durham}. Fig.~\ref{fig:pdf2} plots $\log(x)$ versus
$\log(xf_{i}(x,Q))$ for a representative quark, the up, and for a
gluon, both of which for the relatively high $Q$ of 10 TeV, or
$Q^{2}=$ 100,000,000 (GeV)${}^{2}$ \cite{Durham}.  The variable
$Q^{2}$ changes by six orders of magnitude between these two cases,
but the PDFs shown only change by about an order of magnitude.  For
$x$ less than about $10^{-3}$, the graphs of the PDFs are nearly
linear in these log-log plots. For this reason we use different
models of these PDFs for small $x$ and large $x$.  We also use a
different modeling of PDFs for the ranges $Q>$ 10 TeV, 10 TeV $>Q>$
1 TeV, 1 TeV $>Q>$ 100 GeV, 100 GeV $>Q>$ 10 GeV, 10 GeV $>Q>$ 1
GeV, and 1 GeV $>Q$.  These different regimes of PDFs lead to the
almost imperceptible bulge between $E_{\nu}=10^{10}$ GeV and
$E_{\nu}=10^{11}$ GeV in Fig.~\ref{fig:cross}.

For these different regions of $x$ and $Q$, we make use of simple
approximations to the PDFs by fitting the CTEQ6D data to a form of
\begin{equation}
f_{i}(x,Q)=Ax^{n},
\label{eq:pdf}
\end{equation}
where, for example, in the small-$x$ and large-$Q$ regime,
$n\sim-1.4$ for all the partons and $A\sim0.2$ for quarks and
$A\sim3.5$ for gluons. Because we use different PDFs for the
different regions, this form of $Ax^{n}$ is a good approximation
that is simple to use when we integrate the cross section for a
black hole interaction.

What follows is a brief description of our numerical method.  To
accurately fit both the variables $A$ and $n$, we refine our best
guess and also sample the two dimensional parameter space. At a
fixed value of $Q$ and given a two dimensional array relating
$f_{i}(x)$ to $x$, we start with a reasonable guess for the
variables $A$ and $n$ and a reasonable value for our step variable.
At each iteration, we compare our previous lowest result for the sum
of the squares of the difference between the given data points and
$Ax^{n}$ for all the points in the array with new values of the
variables $A$ and $n$. We try altering our current best values of
$A$ and $n$ by increasing or decreasing one or the other or both in
tandem or opposition for a total of 8 different combinations.  If
one of these combinations results in a better fit, then we store
these new values of $A$ and $n$ as our new current best values, and
we retain the new sum of the squares of the difference between the
given data points and the new $Ax^{n}$ as the new best target, and
then we repeat the eight combinations.  If we do not find a better
fit, then we decrease the size of our step variable and repeat the
above algorithm.  If we reach a sufficiently small step variable, we
do not yet give up: there are local quasi-minima in the parameter
space that are not good fits, such as $A=0$ and $n\ll-1$.  We
instead pick a new value of our step variable that is large enough
to jump to unexplored, and potentially rewarding, areas of the
parameter space.  To prevent getting stuck with the same poor choice
iteration after iteration, we choose a random number between 0 and a
reasonable maximum for our new step variable. At this point it is
better to choose a step variable that is too big, or big enough to
jump to a new region of parameter space, than too small to be
effective.  The subsequent shrinking of the step variable will take
care of any initial excess.  After a set number of loops of this
entire process where the best fit remains unchanged, we take the
resulting best values of $A$ and $x$ as our best fit.

\label{sec:randMeth} A simpler method than the one just described
involves simultaneously changing both variables (or more, if a
problem requires sampling a higher dimensional parameter space) with
different random steps weighted towards zero.  This method was not
used for the project described here, but I have used it elsewhere
with success. Each parameter is changed by a different step value,
and each step value involves two layers of randomness. The first
layer determines the order of magnitude of the step, with a sizable
probability it will be insignificantly small or zero, and the second
layer determines the coefficient and sign associated with the order
of magnitude.  This simpler method effectively encapsulates the
whole of the important parts of the above method in very few lines
of code and is faster at searching for and homing in on the best
solution.

To check our results, we plot our best $Ax^{n}$ against the CTEQ6D
PDF data to ensure our data is a good fit.  For an appropriately
nearby value of $Q$, the benefit to using a simple function over an
array of data is that we can calculate $f_{i}(x)$ for any given x,
and we avoid both having to maintain in our program memory a
complicated series of PDF arrays and having to interpolate between
data points in these arrays.  Our fits are an excellent
approximation to the PDFs being used and are certainly well within
the uncertainty of the PDFs, themselves.
\begin{figure}[hbtp]
\includegraphics[scale=2.5]{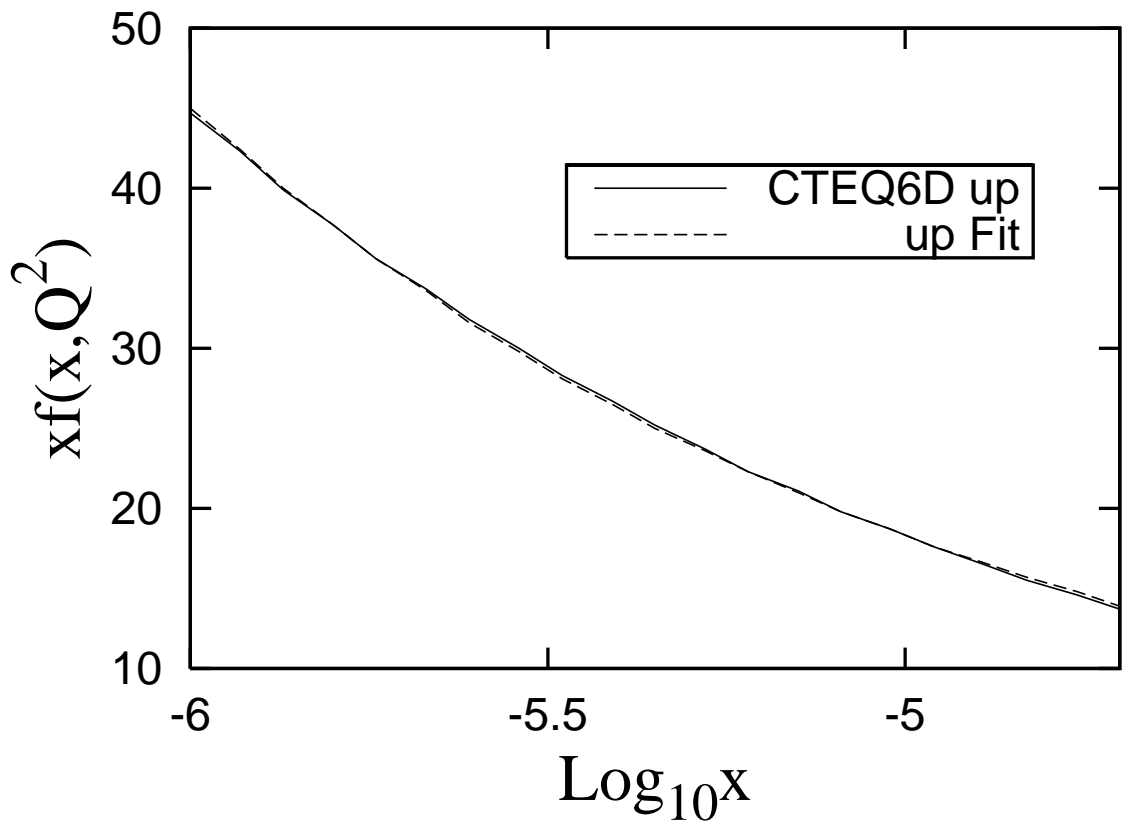}
\caption{\label{fig:goodfit} An Accurate Fit for a Small Range of
Data.}
\end{figure}
\begin{figure}[hbtp]
\includegraphics[scale=2.5]{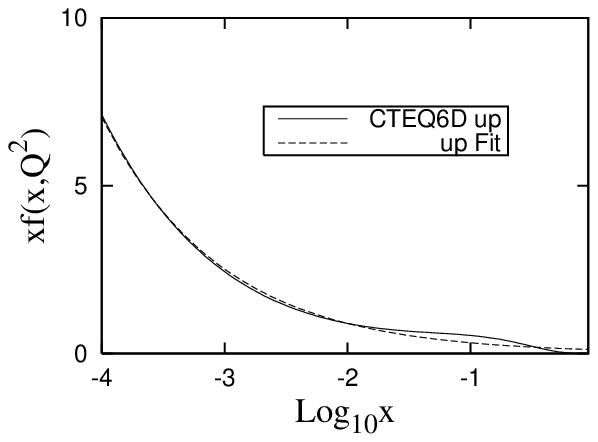}
\caption{\label{fig:reasonablefit} A Reasonable Fit for a Large
Range Data.}
\end{figure}

For examples of these PDF fits for the $u$-quark, see
Figs.~\ref{fig:goodfit} and~\ref{fig:reasonablefit}.  The first of
these shows why the PDF fits are broken up across the range of the
parton momentum fraction, $x$, and the resulting excellent fit.  The
second of these figures shows how the PDF fit can drift from the
exact PDF when overextending the fit over too large a range of $x$.
This may also result from overextending a particular fit over too
wide a range of $Q^{2}$.  In both of these figures a value of
$Q^{2}=10^{8}{\rm\ (GeV)}^{2}$ is used for the CTEQ6D PDFs.  Our fit
in Fig.~\ref{fig:goodfit} is given by Eq.~(\ref{eq:pdf}) with
$A=0.201144$ and $n=-1.391611$.  Our fit in
Fig.~\ref{fig:reasonablefit} is given by $A=-0.391611$ and
$n=-1.447867$.

\subsection{Cross Section of Black Hole Interaction}

In its simplest form, calculating the cross section for the
interaction of ${\rm neutrino} + {\rm nucleon} \rightarrow {\rm
Black\ Hole}$ involves using the Thorne Hoop Conjecture
\cite{Thorne} and checking to see if the neutrino and parton come
close enough together to be within the radius of the Schwarzschild
black hole that would be formed from their combined center-of-mass
energy. At this stage of our simple approximation of checking to see
if the impact parameter, $b$, is smaller than the Schwarzschild
radius, $r_{s}$, for our cross section, we would simply have the
area $\pi r_{s}^{2}$ of a disk.

To find the center-of-mass energy, $E_{CM}$, we use the conservation
of relativistic 4-momentum, $P_{\rm
total}{}^{a}=P_{\nu}{}^{a}+P_{\rm parton}{}^{a}$.  We define our
metric as $\eta_{ab}={\rm diag}(-1,1,1,1)$. In the lab frame of
IceCube, $P_{\nu}{}^{a}=(E_{\nu},p_{\nu,x},p_{\nu,y},p_{\nu,z})$ and
$P_{\rm parton}{}^{a}=(xm_{N},0,0,0)$, where the variable $x$ is the
fraction of the total rest-mass energy of the nucleon present in the
parton at the time of the interaction.  Squaring the 4-momentum,
\cite{Schutz}
\begin{eqnarray}
P_{\rm total}{}^{a}P_{\rm total}{}_{\ a}&=&(P_{\rm
parton}{}^{a}+P_{\nu}{}^{a})(P_{\rm parton}{}_{\ a}+P_{\nu a})\nonumber \\
&=&P_{\rm parton}{}^{a}P_{\rm parton}{}_{\ a}+2P_{\rm
parton}{}^{a}P_{\nu a}+P_{\nu a}P_{\nu}{}^{a},
\end{eqnarray}
and using
\begin{equation}
P_{a}P^{a}=-E_{CM}^{2},
\end{equation}
we have
\begin{equation}
-E_{CM}^{2}=-x^{2}m_{N}^{2}+2(-xm_{N}E_{\nu}+0\cdot\vec{p}_{\nu})-m_{\nu}^{2}.
\end{equation}
Because we are interested in energies where $E_{\nu}\gg$ $m_{N}$ and
$m_{\nu}$, we find
\begin{equation}
E_{CM}^{2}=2xm_{N}E_{\nu}.
\end{equation}
We denote this quantity by
\begin{equation}
\hat{s}\equiv2xm_{N}E_{\nu}.
\label{eq:shat}
\end{equation}

In terms of the variables $\sqrt{\hat{s}}$, the neutrino-parton
center-of-mass energy; $D$, the total number of dimensions of
spacetime; and $M_{D}$, the $D$-dimensional Planck scale; we express
the Schwarzschild radius as \cite{Myers,Argyres}
\begin{equation}
r_{s}(\sqrt{\hat{s}},D,M_{D})=\frac{1}{M_{D}}\left[\frac{\sqrt{\hat{s}}}{M_{D}}\right]^{\frac{1}{D-3}}\left[\frac{2^{D-4}\pi^{(D-7)/2}\Gamma(\frac{D-1}{2})}{D-2}\right]^{\frac{1}{D-3}}.
\end{equation}
From here on we will work within the assumption of string theory
that $D=10$, and we will have $M_{D}=M_{10}$, which we will
eventually take to be near 1 TeV \cite{Antoniadis}.  In 10
dimensions, we then have
\begin{equation}
r_s(\sqrt{\hat{s}}, M_{\rm 10}) = \frac{1}{M_{10}} \left[
\frac{\sqrt{\hat s}}{M_{10}} \, 8 \, \pi^{3/2} \,\, \Gamma(9/2)
\right]^{1/7}
\end{equation}
for the Schwarzschild radius.

The actual radius of the black hole will differ from the
Schwarzschild radius $r_{s}$, due to factors such as angular
momentum and the geometry of spacetime, and we will call this
corrected cross sectional area $F\pi r_{s}^{2}$, where the variable
$F$ is a prefactor used to correct for differences from an exact
Schwarzschild metric.  We define the inelasticity as
\cite{Anchordoqui6}
\begin{equation}
y\equiv\frac{M_{BH}}{\sqrt{\hat{s}}},
\label{eq:inelasticity}
\end{equation}
which is a measure of how much of the center-of-mass energy is
available to the black hole for Hawking radiation
\cite{Hawking2,Hawking3,Hawking4}.  The energy difference, the
deficit between the final mass of the black hole after its ring-down
phase \cite{Frolov1,Frolov2,Frolov3,Frolov4} and the center-of-mass
energy initially present in the collision, is carried off via
incoming shock wave multipole moments radiating gravitational waves
\cite{Aichelburg,Penrose,Eath1,Eath2,Eath3}. The inelasticity $y$
depends on the impact parameter $b$, and we define
\begin{equation}
z\equiv\frac{b}{b_{\rm max}},
\end{equation}
where $b_{\rm max}=\sqrt{F}r_{s}$.  The values of $F$ and $y(z)$
calculated depend upon the slicing of spacetime used to determine
whether or not an apparent horizon is present.  In the work of
\cite{Yoshino1,Yoshino2,Eardley}, it is found for $D=10$ that
$F=1.819$ and we approximate their findings for the inelasticity as
$y(z)=0.59-0.57z^{2}$.  In the later work of \cite{Yoshino3}, in
which a slicing on the future light cone is used, it is found for
$D=10$ that $F=3.09$ and we approximate their findings for the
inelasticity as $y(z)=0.59-0.59z^{2}+0.234z^{3}$.  We will refer to
these two different slicings as the ``old slice'' and the ``new
slice,'' respectively.

The prefactor $F$ and the inelasticity $y(z)$ were derived using
classical general relativity.  Since we don't yet have a quantum
theory of gravity, we need to make sure we stay within a
semi-classical regime.  We expect a thermal distribution of Hawking
radiation \cite{Parker11,Wald2,Hawking5} for
\begin{equation}
M_{BH}\geq x_{\rm min}M_{10},
\label{eq:xmin}
\end{equation}
where $x_{\rm min}=3$ ensures a well-defined resonance not dominated
by the 3-brane tension \cite{Anchordoqui6,Preskill}, and thus
$M_{BH}\geq3$ TeV. The thermal distribution of Hawking radiation is
a Planckian spectrum, where the emission rate per degree of particle
freedom $i$ of particles of spin $s$ with initial total energy
between $\omega$ and $\omega + d \omega$ is~\cite{Han}
\begin{equation}
\frac{\dot{N}_i}{d \omega} = \frac{\sigma_s (\omega) \Omega_{D-3}
\omega^{D-2}}{(D-2) (2\pi)^{D-1}} \left[ e^{\omega/T} - (-1)^{2s}
\right]^{-1}, \label{rate}
\end{equation}
where
\begin{equation}
T = \frac{D-3}{4\,\pi\,r_{s}}
\end{equation}
is the instantaneous Hawking temperature,
\begin{equation}
\Omega_{D-3} = \frac{2\,\pi^{(D-2)/2}}{\Gamma[(D-2)/2]}
\end{equation}
is the volume of a unit $(D-3)$-sphere, and $\sigma_s (\omega)$ is
the greybody factor that accounts for the backscattering of part of
the outgoing radiation into the black hole~\cite{Page}.  Note that a
rough estimate of the instantaneous Hawking temperature can be found
from the first law of black hole thermodynamics (which is analogous
to the combined first and second law of thermodynamics):
$T=dE/dS\simeq(dA/dM)^{-1}$ \cite{Argyres}. Combining
Eqs.~(\ref{eq:shat}), (\ref{eq:inelasticity}), and~(\ref{eq:xmin})
shows that
\begin{equation}
\chi\equiv\frac{(x_{\rm min}M_{10})^{2}}{2m_{N}E_{\nu}y^{2}(z)}\leq
x,
\end{equation}
where to find the cross section we integrate the PDFs over the
parton momentum fraction $x$ and use $\chi$ as our lower limit of
integration.

In addition to integrating the PDFs over the parton momentum
fraction, we also integrate over $z$ for an impact
parameter-weighted average over parton cross sections.  The area of
a thin ring of inner radius $z$ and thickness of $dz$ is
proportional to $zdz$.  We multiply this by a factor of 2, so that
when we integrate $\int_{0}^{1}zdz$ alone, we get a factor of 1;
therefore, if $y(z)$ did not depend on $z$, this weighted average
could be neglected.  Because the value of $y(z)$ does in fact depend
on $z$, the weighted average ensures we use the correct lower limit
of integration, $\chi$, when integrating over the parton momentum
fraction, $x$.

The final expression for the $\nu N \to {\rm BH}$ cross section is
\cite{Anchordoqui7}
\begin{equation}
\sigma =  \int_0^1 2 z \,dz \int_{{\cal X}}^1 dx \, F \, \pi
r_s^2(\sqrt{\hat{s}}, M_{\rm 10}) \,\sum_i f_i(x,Q),
\end{equation}
where $i$ labels parton species, and the $f_i(x,Q)$ are PDFs.

\begin{figure}[hbtp]
\includegraphics[scale=0.8]{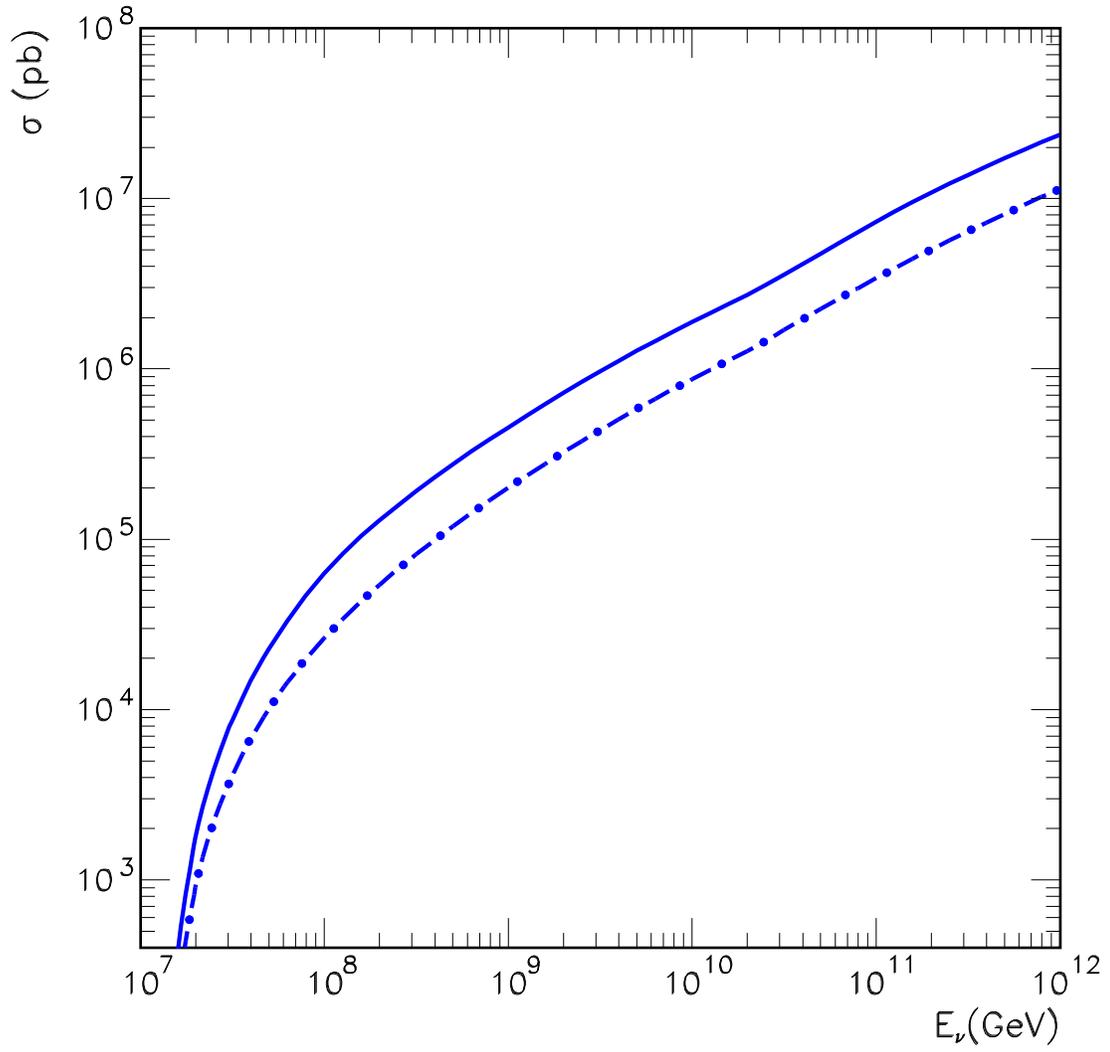}
\caption{\label{fig:cross} Cross Section: New and Old Slicing.}
\end{figure}
\begin{figure}[hbtp]
\includegraphics[scale=2.5]{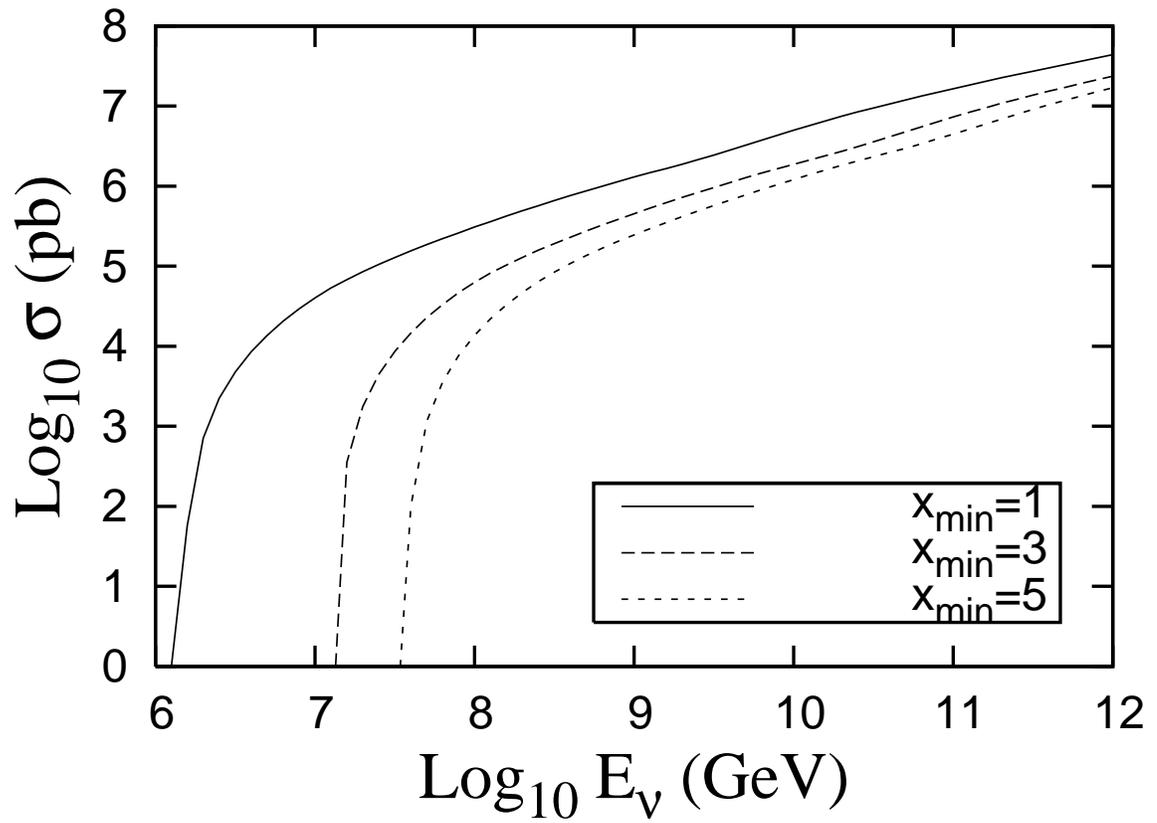}
\caption{\label{fig:xmin} Cross Section: Varying Semi-Classical
Regime.}
\end{figure}

Fig.~\ref{fig:cross} shows $\log(\sigma)$ plotted versus
$\log(E_{\nu})$ for both the case of apparent horizons on the ``old
slice'' (dot-dash line) and the ``new slice'' (solid line).  The
cross section is given in units of picobarns (pb) and the energy of
the incoming neutrino is given in units of GeV.  We use $x_{\rm
min}=3$ and $M_{10}=1$ TeV.

Fig.~\ref{fig:xmin} shows $\log(\sigma)$ plotted versus
$\log(E_{\nu})$ on a log-log scale for different values of $x_{\rm
min}$ using the new slicing. The cross section is given in units of
picobarns (pb) and the energy of the incoming neutrino is given in
units of GeV, where $M_{10}=1$ TeV and $Q={\rm min}\{r_{s}^{-1},10\
{\rm TeV}\}$.

The cross sections were integrated with a variable step size with
respect to the parton momentum function $x$.  The dominant
contribution from the PDFs comes from the small-$x$ region, which is
only probed when the lower limit of integration $\chi$ is
sufficiently small.  This happens with large enough values of the
incoming neutrino energy $E_{\nu}$.  We keep the step variable of
integration smaller than $\chi/100$ for $x<10^{-3}$ and equal to
$1/100$ for $x>10^{-3}$.  This gives us excellent accuracy and a
fast numerical calculation of the cross section.

\section{Detection Sensitivity}

One of the major outstanding questions that IceCube is hoped to be
able to answer is| what is the flux rate of cosmic neutrinos?  A
good estimate involves a consideration of the number of neutrinos
expected to be created in association with the observed flux of
charged cosmic ray particles: this is the Waxman-Bahcall (WB) flux
\cite{Waxman} of
\begin{equation}
\phi_{\nu}\simeq6.0\times10^{-8}(E_{\nu}/{\rm GeV})^{-2}\ {\rm
GeV}^{-1}{\rm cm}^{-2}{\rm s}^{-1}{\rm sr}^{-1},
\end{equation}
including all species of neutrinos.  Another estimated flux assumes
that extragalactic cosmic rays dominate the spectrum at energies
above $\sim10^{8.6}$ GeV and that additional neutrinos are to be
expected from sources opaque to ultra-high energy cosmic rays; this
is the AARGHW flux \cite{Ahlers} of
\begin{equation}
\phi_{\nu}\simeq3.5\times10^{-3}(E_{\nu}/{\rm GeV})^{-2.54}\ {\rm
GeV}^{-1}{\rm cm}^{-2}{\rm s}^{-1}{\rm sr}^{-1},
\end{equation}
including all species of neutrinos.

To confirm the existence of black hole interactions amidst the
background noise of standard model (SM) interactions, we pick out a
signal that has a high likelihood for the relatively democratic
Hawking radiation and a low likelihood for charge current (CC)
interactions: we search for soft muons, or muons with less than 20\%
of the incident neutrino energy.  In SM CC interactions, a produced
muon will generally carry away at least 80\% of the incident energy.
We only consider interactions with at least 4 secondary particles,
where at least one of them is a muon \cite{Dimopolous}.  The cross
section for the SM CC interaction producing a soft muon is
\cite{Anchordoqui1}
\begin{equation}
\sigma_{\rm CC}^{y>0.8} \simeq 1.2 \ \ (E_\nu/{\rm
GeV})^{0.358}~{\rm pb} \,\, .
\end{equation}
For incident neutrino energies larger than $10^{7}$ GeV, the
background number of SM CC interactions meeting these criteria for
the AARGHW flux, which produces more events than the WB flux, is 10
events over the 15 year lifetime of IceCube.  For $E_{\nu}>10^{8}$
GeV, the expected event rate for the SM CC interaction over
IceCube's lifetime is less than 1 event.

\begin{table}
\caption{Probability of Signal.} \label{table:SignalProbability}
\begin{tabular}{|c|c|}
\hline \hline $\hspace{3.2cm} M_{\rm BH} \hspace{3.2cm}$ &
$\hspace{3.2cm} \mathcal{P}_{\rm sig} \hspace{3.2cm} $   \\
\hline
$3\ M_{10}$ & 0.078203  \\
$4\ M_{10}$ & 0.122514  \\
$5\ M_{10}$ & 0.161455  \\
$6\ M_{10}$ & 0.196967  \\
$7\ M_{10}$ & 0.230733  \\
\hline \hline
\end{tabular}
\end{table}

The probability that a black hole interaction produces the criteria
we propose to search for depends on the mass of the black hole
formed \cite{Anchordoqui1}.  See Table~\ref{table:SignalProbability}
for some values of the signal probability versus the size of the
black hole created from a neutrino-parton interaction.  In this
probability we neglect the gravitons radiated into the bulk of the
compactified dimensions, but these are thought to carry away less
than $15\%$ of the radiated energy when $D=10$
\cite{Cardoso1,Cardoso2,Cardoso3}.

With the probability of signal given as a function of black hole
mass, we need a way to determine the expected number of TeV-scale
black holes formed within a given mass range.  We do this by
dividing the expected number of black holes produced into bins at
$0.1\,M_{10}$ mass intervals.  We vary our value of $x_{\rm min}$,
and repeat our calculation for the expected number of black holes
created at IceCube.  For example, should a rate of 235 TeV-scale
black holes be created at IceCube for $x_{\rm min}=3.1$, and 246
created for $x_{\rm min}=3.0$, then we could assign 11 black holes
to the $M_{BH}=3.0$ TeV bucket.  Each of these eleven black holes
has a probability of about 0.078 to produce our signal, so this
means our expected detection rate for this bin is approximately 0.86
TeV-scale black hole signals over the 15-year lifetime of the
IceCube experiment.  When we calculate the rates and associated
signal probabilities for all of our buckets in bins of $x_{\rm
min}\geq3$, we get our cumulative totals.

We will integrate, with respect to energy, the neutrino flux over
the 15 year lifetime of the IceCube experiment, or
$T\simeq4.7\times10^{8}$ seconds. At the energies of interest the
Earth is opaque to neutrinos.  Hence, we will only consider
neutrinos passing down through the Antarctic ice, and we will only
accept measurements from this half of the available directions,
which makes for $2\pi$ steradians of solid angle for observation.
The background rate of non-neutrino events at such high energies is
entirely negligible. IceCube's effective volume is 1 ${\rm km}^{3}$
\cite{Anchordoqui4}, which at a density of 900 kg/${\rm m}^{3}$
means the number of nucleons available for neutrino interaction
targets is $n_{T}\simeq5.4\times10^{38}$.  Our upper limit of
integration is an energy of $10^{10}$ GeV, because beyond this the
IceCube detector will be saturated and unable to resolve all the
details of the interaction \cite{Halzen}.  The total number of black
hole signal events over the life of the IceCube experiment is
\begin{equation}
{\cal N}_{\rm sig} = 2 \pi\, n_{\rm T}\, T\, \int dE_\nu\,\, \sigma
(E_\nu)\,\, \phi_\nu (E_\nu)\ \mathcal{P}_{\rm sig}\,  .
\end{equation}

\begin{table}
\caption{Number of Signal Events.} \label{table:SignalEvents}
\begin{tabular}{|c|c|c|}
\hline \hline \hspace{1.0cm} $x_{\rm min}$ \hspace{1.0cm} &
\hspace{1.2cm} ${\cal N}_{\rm BH}$ [WB] \hspace{1.2cm} &
 \hspace{1.2cm} ${\cal N}_{\rm BH}$ [AARGHW] \hspace{1.2cm}   \\
\hline
 3 & 43 \ \ \ \ \ (19) & 69 \ \ \ \ \ (30) \\
 4 & 34 \ \ \ \ \ (15) & 43 \ \ \ \ \ (19) \\
 5 & 27 \ \ \ \ \ (12) & 28 \ \ \ \ \ (12) \\
 6 & 22 \ \ \ \ \ \ (9) & 20 \ \ \ \ \ \ (9) \\
\hline \hline
\end{tabular}
\end{table}

In Table~\ref{table:SignalEvents} we calculate the expected number
of black hole signals over the lifetime of IceCube.  With a lower
limit of integration of $10^{7}$ GeV, we fix $M_{10}=1$ TeV, but we
allow $x_{\rm min}$ to vary. We compare the number of events for the
WB flux to the AARGHW flux. For each flux, we have calculated the
number of events using both the ``new slice,'' which is given
without parentheses; and the ``old slice,'' which is given inside
parentheses.

\begin{table}
\caption{10-Dimensional Planck Mass Sensitivity.}
\label{table:PlanckSensitivity}
\begin{tabular}{|c|c|c|}
\hline \hline \hspace{0.9cm} $x_{\rm min}$ \hspace{0.9cm} &
\hspace{0.8cm} $M_{10}/$TeV [WB] \hspace{0.8cm} &
 \hspace{0.8cm} $M_{10}/$TeV [AARGHW] \hspace{0.8cm}   \\
\hline
 3 & 1.5 \ \ \ \ \ (1.2) & 1.5 \ \ \ \ \  (1.2) \\
 5 & 1.3 \ \ \ \ \ (1.1) & 1.3 \ \ \ \ \  (1.1) \\
 7 & 1.2 \ \ \ \ \ (1.0) & 1.2 \ \ \ \ \  (1.0) \\
 9 & 1.1 \ \ \ \ \ (1.0) & 1.1 \ \ \ \ \  (0.9) \\
\hline \hline
\end{tabular}
\end{table}

In Table~\ref{table:PlanckSensitivity} we calculate the maximum
10-dimensional Planck mass for which we would expect be able to
observe the interaction at the $3\sigma$ level.  With a lower limit
of integration of $10^{8}$ GeV, and for differing values of $x_{\rm
min}$, we find the corresponding value of $M_{10}$.  We do this for
both the WB flux and the AARGHW flux.  For each flux, we have
calculated the number of events using both the ``new slice,'' which
is given without parentheses; and the ``old slice,'' which is given
inside parentheses.

\begin{figure}[hbtp]
\includegraphics[scale=0.8]{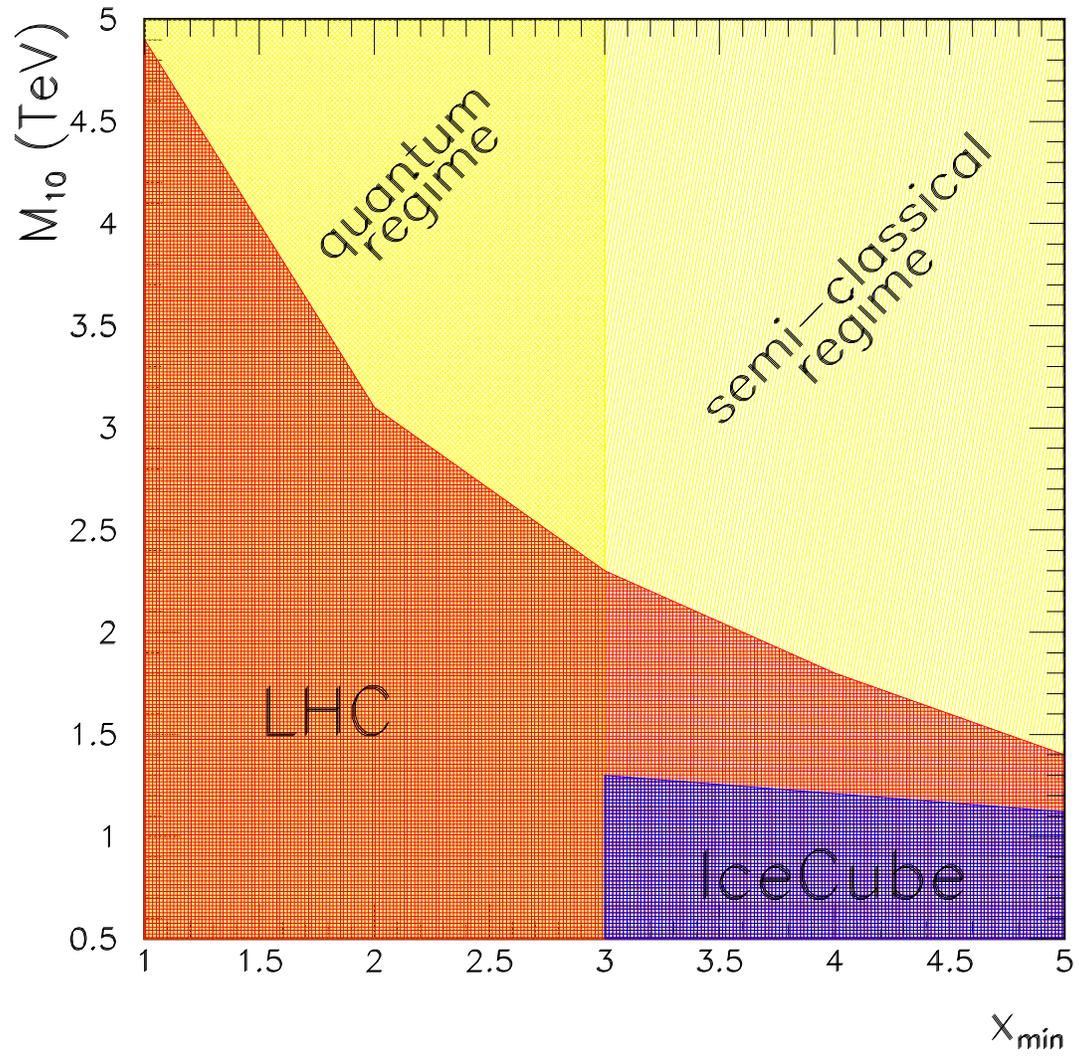}
\caption{\label{fig:LHC} IceCube and LHC Discovery Reach.}
\end{figure}

In Fig.~\ref{fig:LHC} we plot the TeV-scale discovery reach for both
IceCube and the Large Hadron Collider (LHC) \cite{Anchordoqui6},
assuming a cumulative integrated luminosity of 1 ${\rm ab}^{-1}$
over the life of the collider. We calculate the maximum value of
$M_{10}$ that could be observed at the $5\sigma$ level versus
$x_{\rm min}$, and we use a lower limit on the energy integral of
$10^{7}$ GeV. We plot the IceCube discovery reach only in the
semi-classical regime of $x_{\rm min}\geq3$; however, the LHC could
potentially be focused on superstring resonances
\cite{Anchordoqui8,Anchordoqui9,Anchordoqui10}, and could thus be
able to probe the quantum regime \cite{Anchordoqui1}.

\newpage
\thispagestyle{fancy}
\chapter{Conclusion}
\thispagestyle{fancy}
\pagestyle{fancy}

Using the de Sitter Bunch-Davies state for modes of
intermediary-$q_{2}$ and large-$q_{2}$ is valid in the exponentially
growing region of the scale factor, but imposing the Bunch-Davies
state on modes of small-$q_{2}$ leads to infrared divergences in the
dispersion spectrum.  Maintaining continuity of the scale factor to
$C^{2}$ is necessary to prevent ultraviolet divergences of the
energy density of particles created during inflation.  The
asymptotically Minkowskian regions of our composite scale factor do
not affect the near scale-invariance of the intermediate-$q_{2}$
region of particle production, but it does allow for an unambiguous
interpretation of the number of particles produced versus $q_{2}$,
and it allows for flat-space renormalization.  An asymptotically
flat scale factor segment may be joined continuously to $C^{2}$ with
an exponentially growing segment of scale factor, whereas a simple
power law such as $a(t)\propto t^{n}$ may not.  Both of our massive
approximations are trustworthy in their respective regimes: little
growth of the composite scale factor outside of the exponentially
growing region for the effective-$k$ approach, and with modes not at
the interface between the small- and intermediate-$q_{2}$ behavior
and not at the interface between the intermediate- and large-$q_{2}$
behavior for the dominant-term approach.  In our model, the average
number of particles created per mode can be characterized in terms
of three parameters: the number of e-folds, $N_{e}$; the ratio of
the mass to the Hubble constant during inflation, $m_{H}$; and the
dimensionless mode number, $q_{2}$.  We find a scale-invariant
spectrum when $H_{\rm infl}$ and $m_{H}$ are both constant, provided
modes are converted individually into curvature perturbations soon
after exiting the Hubble radius. The spectral index can be shifted
towards a blue spectrum if all the curvature perturbations are
created around the same time or at a time after the end of
inflation.  The spectral index can be shifted towards a red spectrum
by taking into consideration a changing value of $H_{\rm infl}$ or
$\dot{\phi}$. We find that an abrupt end to inflation leads to a
boosted production of high-energy particles and an associated high
temperature. If monopoles, or certain other exotic particles, were
found to be created copiously at low temperatures| at the LHC, for
instance| it could place rigorous constraints on the characteristics
of inflation.

The predicted energy and angular momentum in the post-Minkowski
approximation for our binary point mass system with helical symmetry
agrees to first post-Newtonian order in the case of
parametrization-invariant action plus either of the 1PN correction
terms. With $q\rightarrow0$, we can make a comparison with the
Schwarzschild solution of General Relativity.  Here, both the affine
case and the parametrization-invariant with a 1PN correction term
have an Innermost Circular Orbit at about $3M$, which is outside the
event horizon of GR located at $2M$.  Only the affine case has an
Innermost Stable Circular Orbit, and it occurs at $\sim7.69\,M$,
which is outside of the ISCO predicted by GR located at $6M$.  These
discrepancies may be due to the linear order of the post-Minkowski
approximation, or they may be due to the radiation being pumped into
the binary system by the half-advanced plus half-retarded helical
symmetry. A form of the first law of thermodynamics $dE/dv=\Omega\,
dL/dv$ is satisfied, and this serves as a useful check on the
analytical and numerical results.

With a flux of cosmic neutrinos at the Waxman-Bahcall rate, over its
15 year lifespan the IceCube Neutrino Telescope could detect
TeV-scale black holes at the $5\,\sigma$ level up to a maximum
10-dimensional Planck mass of 1.3 TeV.  Our analysis shows that PDFs
can be approximated well by fits to $x\,f(x)=A\,x^{n}$, provided the
range of the parton momentum fraction, $x$, for each fit is
restricted to a few decades of variation on a $\log_{10}$ scale. The
fitting of the parameters $A$ and $n$ can best be accomplished by
simultaneously varying each, and by sampling a large enough area of
parameter space to ensure a false minimum deviation is avoided.  The
integration involved in calculating the cross section of the
gravitational interaction between a parton and a cosmic neutrino is
most efficiently carried out with a variable step size of
integration.  Values of the parton momentum fraction closest to the
lower limit of integration dominate the cross section, so care must
be taken to use a small enough step size in this range so that these
values are not over-weighted in the cross section.  A convenient way
of associating events with a given value of $M_{BH}$ is to
recalculate the number of lifetime events for different values of
$x_{\rm min}$, and then subtract the difference between the events
from incremental values of $x_{\rm min}$ into bins.

In the three parts of this dissertation, we have focused on the
topics in inflationary cosmology and astrophysics described in three
papers: \cite{Glenz1,Glenz2,Anchordoqui1}.

\newpage
\thispagestyle{fancy}
\begin{appendix}
\chapter{Composite History of an Exact Reaction-Force
Solution}
\label{appendix}
\thispagestyle{fancy}
\pagestyle{fancy}

This Appendix is motivated by and based on the work of
\cite{Wiseman}.  What follows is an application of the more general
techniques presented in Sec.~\ref{sec:joinc2} for matching
continuously to second derivative in what could be taken as either a
scale factor on the one hand or as a particle's velocity on the
other.  It is hoped that this example serves to illustrate some
aspects of the self force and radiation reaction mentioned in
Chapter~\ref{sec:binar}.  We begin with a charged particle that in
its rest frame emits radiation when accelerated as given by the
Larmor formula (in Gaussian units) of
\begin{equation}
P=\frac{2}{3}\frac{e^{2}}{c^{3}}\dot{v}^{2},
\end{equation}
which leads to, in addition to the external force, a
radiation-reaction force of the form \cite{Lorentz,Poisson,Jackson}
\begin{equation}
\vec{F}_{\rm
applied}=m\dot{\vec{v}}-\frac{2}{3}\frac{e^{2}}{c^{3}}\ddot{\vec{v}},
\end{equation}
as perceived by the particle in its momentarily-comoving rest frame.
In this example, we will consider only rectilinear motion, so we
rewrite this as
\begin{equation}
F_{\rm applied}=m\dot{v}-m\tau\ddot{v},
\end{equation}
where
\begin{equation}
\tau\equiv\frac{2}{3}\frac{e^{2}}{mc^{3}}.
\end{equation}
We thus define the reaction force, or self force, as
\begin{equation}
F_{\rm self}\equiv-m\tau\ddot{v}.
\end{equation}

For constant acceleration, we have
\begin{eqnarray}
v_{c}(t)&=&a_{c}(t-t_{0}),\\
\dot{v}_{c}(t)&=&a_{c},\\
\ddot{v}_{c}(t)&=&0,\\
F_{c\rm\ self}(t)&=&0,\\
F_{c\rm\ applied}(t)&=&ma_{c},\\
P_{c\rm\ radiated}(t)&=&m\tau a_{c}{}^{2}.
\end{eqnarray}
We then introduce a two similar velocity histories given by a
hyperbolic tangent in analog with Sec.~\ref{sec:joinc2},
\begin{eqnarray}
v_{i}(t)&=&v_{0i}+\Delta_{i}\tanh\frac{t-t_{i}}{s_{i}},\\
v_{f}(t)&=&v_{0f}+\Delta_{f}\tanh\frac{t-t_{f}}{s_{f}},
\end{eqnarray}
where $\Delta_{i}$ is twice the difference between early- and
late-time velocities for the first velocity history, $t_{i}$ is the
time at which $v_{i}(t)=v_{0i}$, and $s_{i}$ is a throttling
parameter that decreases the change in velocity with respect to time
as it increases in magnitude; and where $\Delta_{f}$ is twice the
difference between early- and late-time velocities for the final
velocity history, $t_{f}$ is the time at which $v_{f}(t)=v_{0f}$,
and $s_{f}$ is a throttling parameter that decreases the change in
velocity with respect to time as it increases in magnitude. We will
take both $s_{i}$ and $s_{f}$ to be $\geq0$. Then we have
\begin{eqnarray}
v_{i}(t)&=&v_{0i}+\Delta_{i}\tanh\frac{t-t_{i}}{s_{i}},\\
\dot{v}_{i}(t)&=&\frac{\Delta_{i}}{s_{i}}\left(1-\tanh^{2}\frac{t-t_{i}}{s_{i}}\right),\\
\ddot{v}_{i}(t)&=&2\frac{\Delta_{i}}{s_{i}{}^{2}}\left(\tanh\frac{t-t_{i}}{s_{i}}\right)\left[\left(\tanh^{2}\frac{t-t_{i}}{s_{i}}\right)-1\right],\\
F_{i\rm\
self}(t)&=&-2m\tau\frac{\Delta_{i}}{s_{i}{}^{2}}\left(\tanh\frac{t-t_{i}}{s_{i}}\right)\left[\left(\tanh^{2}\frac{t-t_{i}}{s_{i}}\right)-1\right],\\
F_{i\rm\
applied}(t)&=&m\frac{\Delta_{i}}{s_{i}}\left(1-\tanh^{2}\frac{t-t_{i}}{s_{i}}\right)\nonumber\\
&&-2m\tau\frac{\Delta_{i}}{s_{i}{}^{2}}\left(\tanh\frac{t-t_{i}}{s_{i}}\right)\left[\left(\tanh^{2}\frac{t-t_{i}}{s_{i}}\right)-1\right],\\
P_{i\rm\
radiated}(t)&=&m\tau\left[\frac{\Delta_{i}}{s_{i}}\left(1-\tanh^{2}\frac{t-t_{i}}{s_{i}}\right)\right]^{2},
\end{eqnarray}
and
\begin{eqnarray}
v_{f}(t)&=&v_{0f}+\Delta_{f}\tanh\frac{t-t_{f}}{s_{f}},\\
\dot{v}_{f}(t)&=&\frac{\Delta_{f}}{s_{f}}\left(1-\tanh^{2}\frac{t-t_{f}}{s_{f}}\right),\\
\ddot{v}_{f}(t)&=&2\frac{\Delta_{f}}{s_{f}{}^{2}}\left(\tanh\frac{t-t_{f}}{s_{f}}\right)\left[\left(\tanh^{2}\frac{t-t_{f}}{s_{f}}\right)-1\right],\\
F_{f\rm\
self}(t)&=&-2m\tau\frac{\Delta_{f}}{s_{f}{}^{2}}\left(\tanh\frac{t-t_{f}}{s_{f}}\right)\left[\left(\tanh^{2}\frac{t-t_{f}}{s_{f}}\right)-1\right],\\
F_{f\rm\
applied}(t)&=&m\frac{\Delta_{f}}{s_{f}}\left(1-\tanh^{2}\frac{t-t_{f}}{s_{f}}\right)\nonumber\\
&&-2m\tau\frac{\Delta_{f}}{s_{f}{}^{2}}\left(\tanh\frac{t-t_{f}}{s_{f}}\right)\left[\left(\tanh^{2}\frac{t-t_{f}}{s_{f}}\right)-1\right],\\
P_{f\rm\
radiated}(t)&=&m\tau\left[\frac{\Delta_{f}}{s_{f}}\left(1-\tanh^{2}\frac{t-t_{f}}{s_{f}}\right)\right]^{2}.
\end{eqnarray}
At times $t_{i}$ and $t_{f}$, respectively, we have
\begin{eqnarray}
v_{i}(t_{i})&=&v_{0i},\\
\dot{v}_{i}(t_{i})&=&\frac{\Delta_{i}}{s_{i}},\\
\ddot{v}_{i}(t_{i})&=&0,\\
F_{i\rm\ self}(t_{i})&=&0,\\
F_{i\rm\ applied}(t_{i})&=&m\frac{\Delta_{i}}{s_{i}},\\
P_{i\rm\
radiated}(t)&=&m\tau\left(\frac{\Delta_{i}}{s_{i}}\right)^{2},
\end{eqnarray}
and
\begin{eqnarray}
v_{f}(t_{f})&=&v_{0f},\\
\dot{v}_{f}(t_{f})&=&\frac{\Delta_{f}}{s_{f}},\\
\ddot{v}_{f}(t_{f})&=&0,\\
F_{f\rm\ self}(t_{f})&=&0,\\
F_{f\rm\ applied}(t_{f})&=&m\frac{\Delta_{f}}{s_{f}},\\
P_{f\rm\
radiated}(t)&=&m\tau\left(\frac{\Delta_{f}}{s_{f}}\right)^{2}.
\end{eqnarray}

We then specify a composite velocity history by matching the
velocity histories of $v_{i}(t)$ to $v_{c}(t)$ to $v_{f}(t)$. We can
maintain $C^{2}$ joining conditions| meaning the velocity,
acceleration, and radiation-reaction force are all kept continuous|
by joining the initial segment to the start of a region of constant
acceleration at $t=t_{i}$, and by joining the final segment to the
end of a region of constant acceleration at $t=t_{f}$.

See Fig.~\ref{fig:velvet},
\begin{figure}[hbtp]
\includegraphics[scale=2.5]{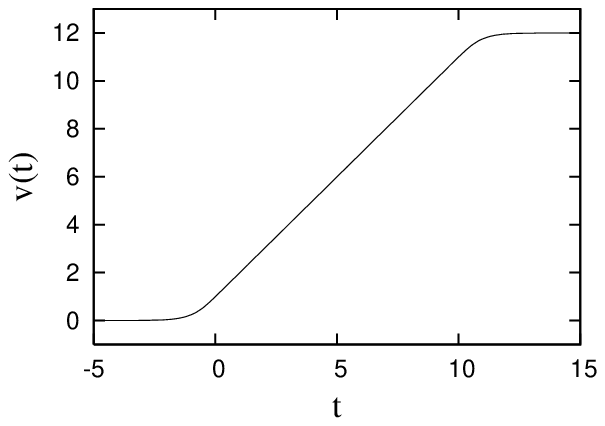}
\caption{\label{fig:velvet} Velocity versus Time.}
\end{figure}
where we plot a dimensionless example of a composite velocity where
$\Delta_{i}=\Delta_{f}=s_{i}=s_{f}=a_{c}=1$.  In this example we
take $t_{i}=0$, $t_{f}=10$, $v_{0i}=1$, and $v_{0f}=11$.

Maintaining continuity of the velocity history up to its second
derivative imposes, in addition to the two conditions of matching
times, the following boundary conditions
\begin{equation}
\frac{\Delta_{i}}{s_{i}}=a_{c}=\frac{\Delta_{f}}{s_{f}},
\end{equation}
\begin{equation}
v_{0i}=\lim_{t\rightarrow-\infty}[v(t)]+\Delta_{i},
\end{equation}
\begin{equation}
v_{0f}=\lim_{t\rightarrow+\infty}[v(t)]-\Delta_{f}.
\end{equation}
We find that $t_{0}=t_{i}-v_{0i}/a_{c}$, and the duration of
constant acceleration is $t_{a}\equiv t_{f}-t_{i}$.

See Fig.~\ref{fig:forvet},
\begin{figure}[hbtp]
\includegraphics[scale=2.5]{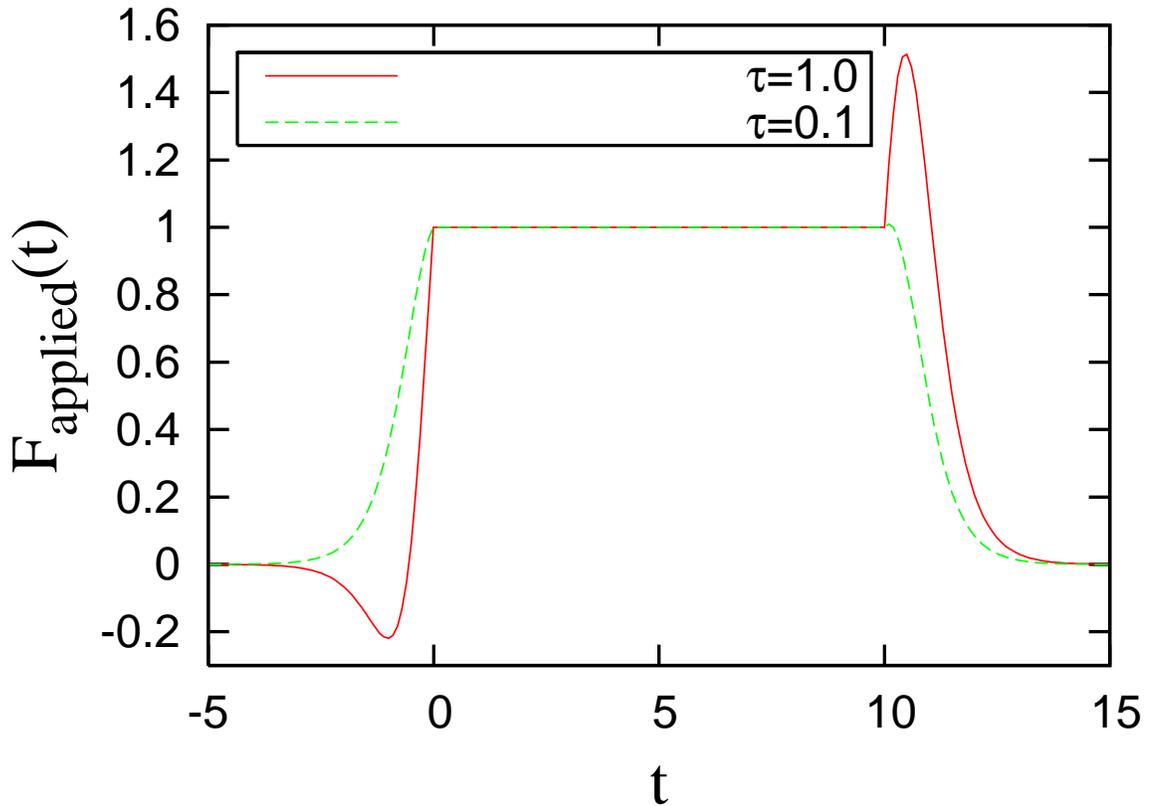}
\caption{\label{fig:forvet} External Force versus Time.}
\end{figure}
where we plot a dimensionless example of the applied force necessary
to maintain the motion of the particle shown in
Fig.~\ref{fig:velvet} for two different dimensionless values of
$\tau$.  In the small-$\tau$ limit, we get a Newtonian 2nd Law of
$F=ma$.  In the large-$\tau$ limit, we note some peculiarities of
the self-force.  To initiate the acceleration, a force must
initially be applied opposite to the direction of motion| this is to
be compared with the pre-acceleration found for radiation-reaction
forces that eliminates runaway-acceleration solutions.  To end the
period of constant acceleration, the force must be increased in the
direction of motion.  As will be shown below, this additional work
is needed to compensate for the energy dissipated by the radiation
emitted.

The total change in kinetic energy of the particle is
\begin{eqnarray}
\Delta
KE&=&\frac{1}{2}m\left\{\left(\lim_{t\rightarrow+\infty}[v(t)]\right)^{2}-\left(\lim_{t\rightarrow-\infty}[v(t)]\right)^{2}\right\}\nonumber\\
&=&\frac{1}{2}m\left\{\left(v_{0i}+a_{c}t_{a}+\Delta_{f}\right)^{2}-\left(v_{0i}-\Delta_{i}\right)^{2}\right\}\nonumber\\
&=&m\,a_{c}\,v_{0i}\left(t_{a}+s_{i}+s_{f}\right)+\frac{1}{2}ma_{c}{}^{2}\left(t_{a}{}^{2}+2t_{a}s_{f}+s_{f}{}^{2}-s_{i}{}^{2}\right).
\end{eqnarray}
The total power radiated is
\begin{eqnarray}
P_{\rm radiated\
total}&=&\left(\int_{-\infty}^{t_{i}}P_{i}\,dt\right)+\left(\int_{t_{i}}^{t_{f}}P_{c}\,dt\right)+\left(\int_{t_{f}}^{+\infty}P_{f}\,dt\right)\nonumber\\
&=&\frac{2}{3}m\tau\frac{\Delta_{i}{}^{2}}{s_{i}}+m\tau
a_{c}{}^{2}t_{a}+\frac{2}{3}m\tau\frac{\Delta_{i}{}^{2}}{s_{i}}\nonumber\\
&=&m\tau
a_{c}{}^{2}\left(\frac{2}{3}s_{i}+t_{a}+\frac{2}{3}s_{f}\right).
\end{eqnarray}
The total work done on the particle by the external force is
\begin{eqnarray}
W_{\rm total}&=&\left(\int_{-\infty}^{t_{i}}F_{i\rm\ applied}\,v_{i}(t)\,dt\right)+\left(\int_{t_{i}}^{t_{f}}F_{c\rm\ applied}\,v_{c}(t)\,dt\right)\nonumber\\
&&+\left(\int_{t_{f}}^{+\infty}F_{f\rm\ applied}\,v_{f}(t)\,dt\right)\nonumber\\
&=&\left(ma_{c}\left[v_{0i}(s_{i}-\tau)+s_{i}a_{c}\left\{\frac{2}{3}\tau-\frac{1}{2}s_{i}\right\}\right]\right)+\left(m\,a_{c}\,v_{0i}\,t_{a}+\frac{1}{2}m\,a_{c}{}^{2}\,t_{a}{}^{2}\right)\nonumber\\
&&+\left(ma_{c}\left[v_{0i}(s_{f}+\tau)+a_{c}\left(\frac{1}{2}s_{f}{}^{2}+s_{f}t_{a}+\frac{2}{3}s_{f}\tau+t_{a}\tau\right)\right]\right)\nonumber\\
&=&ma_{c}v_{0i}\left(s_{i}+t_{a}+s_{f}\right)+\frac{1}{2}ma_{c}{}^{2}\left(t_{a}{}^{2}+2t_{a}s_{f}+s_{f}{}^{2}-s_{i}{}^{2}\right)\nonumber\\
&&+m\tau
a_{c}{}^{2}\left(\frac{2}{3}s_{i}+t_{a}+\frac{2}{3}s_{f}\right).
\end{eqnarray}
We find that
\begin{equation}
W_{\rm total}-P_{\rm radiated}-\Delta KE=0,
\end{equation}
and thus energy is conserved at early and late times. See
Fig.~\ref{fig:enevet},
\begin{figure}[hbtp]
\includegraphics[scale=2.5]{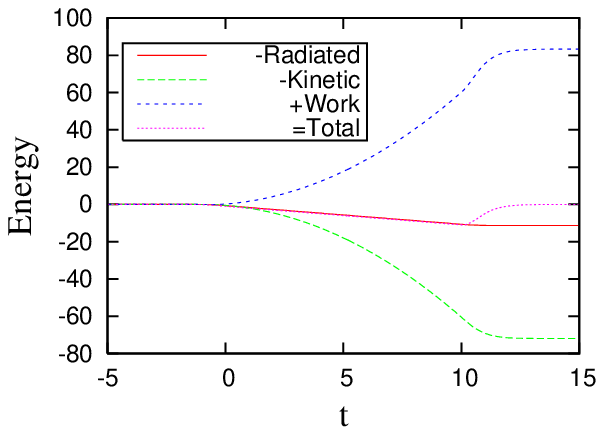}
\caption{\label{fig:enevet} Energy versus Time.}
\end{figure}
for the case of energy conservation between early and late times.
The velocity history is given in Fig.~\ref{fig:velvet}, and we
choose $\tau=1$.  The energy deficit that develops is primarily due
to the energy dissipated through the emitted radiation during the
phase of constant acceleration.  This negative energy must be
balanced by an additional amount of work applied to the particle to
end the acceleration.  If additional energy is not provided to the
system, Wiseman has proven that the kinetic energy of the particle
decreases to compensate \cite{Wiseman}.  In the limit of
$s_{i}\rightarrow0$ and $s_{i}\rightarrow0$, we see that the work
associated with overcoming the reaction force at the initial and
final joining points is $-m\,a_{c}\,v_{0i}\,\tau$ and
$m\,a_{c}\,v_{0f}\,\tau$, respectively. Because in this velocity
history $\ddot{v}=0$ if $t\neq t_{i}$ and $t\neq t_{f}$, and because
$W_{\rm self}=\int F_{\rm self}(t)\,v(t)\,dt$, we see that in the
instantaneous limit, $F_{\rm
self}(t)=ma_{c}\tau\left[\delta(t-t_{f})-\delta(t-t_{i})\right]$,
where $\delta(t)$ is the Dirac delta-function.

\end{appendix}

\thispagestyle{fancy}
\newpage
\thispagestyle{fancy}
\pagestyle{fancy}

\newpage
\thispagestyle{fancy}

\begin{center}
\onehalfspacing CURRICULUM VITAE \\
\textbf{Matthew Glenz}
\end{center}
\textbf{EDUCATION}
\begin{tabbing} Ph.D.,\ \ \= Physics\ \ \ \ \ \ \=
University of Wisconsin|Milwaukee\ \ \
\ \ \ \ \ \ \ \ \ \ \ \ \ \ \ \ \, \= Dec. 2008\\
B.S.,\>Physics\>Iowa State University, Honors\>May 6, 2000\\
\>\>Studied Abroad at Lancaster University, England\>1998-1999
\end{tabbing}
\textbf{EMPLOYMENT}
\begin{tabbing}
Research Assistant\ \ \ \ \ \ \ \= University of
Wisconsin|Milwaukee\ \ \ \ \ \ \ \ \ \ \ \ \ \ \ \= 2006-2008\\
Teaching Assistant\>University of Wisconsin|Milwaukee\>2004-2006\\
Technical Services\>Epic Systems Corporation, Madison,
WI\>2001-2004\\
Support Technician\>Gundersen-Lutheran Hospital, LaCrosse,
WI\>2000-2001\\
Research Aide\>U.S. Dept. of Energy, Iowa State
University\>1997-1998\\
Head Cook/Supervisor\>Boy Scout Camp Decorah, Holmen, WI\>1997\\
Nature Counselor\>Boy Scout Camp Decorah, Holmen, WI\>1996\\
Scout Craft Director\>Boy Scout Camp Decorah, Holmen, WI\>1995
\end{tabbing}
\textbf{AWARDS}
\begin{tabbing}
American Physical Society Travel Grant\ \ \ \ \ \ \ \ \ \ \ \ \ \ \
\ \ \ \ \ \ \ \ \ \ \ \ \ \ \ \ \ \ \ \ \ \ \ \ \ \ \ \= 2008\\
Papastamatiou Scholarship\>2008\\
NASA / Wisconsin Space Grant Consortium Fellowship\>2007-2008\\
Bradley Fellowship, Lynde and Harry Bradley Foundation\>2006-2008\\
UWM  Chancellor's Fellowship\>2004-2008\\
ISU  Foreign Language Student of the Year\>1998\\
ISU  Dedicated Service Award\>1997\\
National Merit Scholarship\>1996
\end{tabbing}
\begin{tabbing}
\textbf{PUBLICATIONS}\\
L.A. Anchordoqui, M.M. Glenz, and L. Parker,
``Black Holes at the
IceCube\\
neutrino telescope," Phys. Rev. \textbf{D 75}, 024011 (2007).\\[2mm]
M.M. Glenz and K. Uryu, ``Circular solution of two unequal mass particles in\\
Post-Minkowski approximation,"  Phys. Rev. \textbf{D
76}, 027501 (2007).\\[2mm]
M.M. Glenz and L. Parker, ``Study of the Spectrum of Inflaton
Perturbations,"\\
to be submitted.
\end{tabbing}
\begin{tabbing}
\textbf{PRESENTATIONS}\\
``Probing TeV Scale Black Hole Production at the South Pole," at
16th Midwest\\
Relativity Meeting, Washington University, November 17, 2006.\\[2mm]
``Study of the Spectrum of Inflaton Fluctuations," at 2008 April APS
Meeting,\\
St. Louis, Missouri, April 14, 2008.\\[2mm]
``Regularization-Independent Inflaton Spectrum," at Pheno 2008
Symposium,\\
University of Wisconsin-Madison, April 29, 2008.\\[2mm]
``Particles Created from Quantum Fields in Cosmological Inflation,"
at 18th\\
Wisconsin Space Conference, UW-Fox Valley, August 14, 2008.\\[2mm]
``Dispersion Spectrum of Inflaton Perturbations Calculated
Numerically with\\
Reheating," at Cosmo 2008, Madison, Wisconsin, August 28,
2008.\\[2mm]
``Post-Minkowski Approximation to Binary Point Mass System with
Helical\\
Symmetry," at University of Wisconsin-Milwaukee, September 12,
2008.\\[2mm]
``Cosmological Inflation with Particle Production," at University of Wisconsin-\\
Milwaukee, October 10, 2008.\\[2mm]
``Early Universe Evolution Characterizes Three Regimes of Spectral Perturbations,"\\
at 18th Midwest Relativity Meeting, University of Notre Dame,
October 24, 2008.
\end{tabbing}
\ \\
\begin{center}
\hrule \vskip 0.2em Major Professor \hfill Date\ \ \ \ \ \ \ \ \ \\
\end{center}

\end{document}